\ifdefined\kanjiskip
  \documentclass[reqno, 12pt, dvipdfmx]{amsart}
\else
  \documentclass[reqno, 12pt]{amsart}
\fi

\IfFileExists{latexml.sty}{\usepackage{latexml}}{\newif\iflatexml}
\iflatexml
  \lxRequireResource[type=text/css]{figures/diagrams.css}
\fi

\usepackage[colorlinks, linkcolor=blue, citecolor=red, urlcolor=cyan,pagebackref]{hyperref}

\allowdisplaybreaks

\usepackage{amsthm,amssymb,amscd,amsmath}
\usepackage{mathrsfs}
\usepackage{bbm}
\usepackage{graphicx}
\iflatexml\else
\usepackage{epic,eepic}
\fi
\iflatexml\else
\usepackage{tikz-cd}%for diagrams
\fi
\usepackage{stmaryrd}%for sslash
\usepackage{mathtools}
\usepackage{stmaryrd}%for llbracket
\usepackage{diagbox}
\usepackage{float}

\iflatexml\else
\usepackage{ascmac}%for screen environment
\fi
\usepackage[normalem]{ulem}

\usepackage{accents}

\usepackage{here}
\usepackage{bm}
\iflatexml\else
\usepackage[all]{xy}
\fi

\usepackage{xcolor} 
\definecolor{mygreen}{rgb}{0,0.7,0.3}
\definecolor{myblue}{rgb}{0,0.50,1.20}%cyan={0,1.74,2.39}
\definecolor{myorange}{rgb}{1,0.5,0.1}

\definecolor{fillred}{rgb}{1,0.9,0.9}
\definecolor{fillgreen}{rgb}{0.9,1,0.9}

 \definecolor{refkey}{rgb}{0,0.7,0.3}
 \definecolor{labelkey}{rgb}{1,0,0}

\iflatexml\else
\usepackage{tikz}
\usetikzlibrary{calc,decorations.markings}
\usetikzlibrary{arrows.meta, nfold}
\usetikzlibrary{positioning}
\usetikzlibrary{cd, intersections, calc, decorations.pathmorphing, arrows, matrix, decorations.pathreplacing}
\usetikzlibrary{braids}
\fi

\numberwithin{equation}{section}

\iflatexml\else
\usepackage{version}
\fi

\usepackage{thmtools} 
\usepackage{cleveref}

\declaretheorem[name=Theorem,numberwithin=section]{thm}
\iflatexml\else
\fi

\declaretheorem[sharenumber=thm,name=Lemma]{lem}

\declaretheorem[sharenumber=thm,name=Proposition]{prop}

\theoremstyle{definition}
\declaretheorem[style=definition,sharenumber=thm,name=Definition]{dfn}

\declaretheorem[style=definition,sharenumber=thm,name=Example]{ex}

\declaretheorem[style=definition,sharenumber=thm,name=Remark]{rmk}

\crefname{thm}{Theorem}{Theorems}
\crefname{cor}{Corollary}{Corollaries}
\crefname{lem}{Lemma}{Lemmas}
\crefname{lemdef}{Lemma-Definition}{Lemma-Definitions}
\crefname{prop}{Proposition}{Propositions}
\crefname{dfn}{Definition}{Definitions}
\crefname{defi}{Definition}{Definitions}
\crefname{ex}{Example}{Examples}
\crefname{claim}{Claim}{Claims}
\crefname{conj}{Conjecture}{Conjectures}
\crefname{conv}{Notation}{Notations}
\crefname{rem}{Remark}{Remarks}
\crefname{rmk}{Remark}{Remarks}
\crefname{prob}{Problem}{Problems}
\crefname{figure}{Figure}{Figures}
\crefname{table}{Table}{Tables}
\crefname{introthm}{Theorem}{Theorems}
\crefname{introconj}{Conjecture}{Conjectures}
\crefname{section}{Section}{Sections}
\crefname{subsection}{Section}{Sections}

\newcommand{\cC}{\mathcal{C}}
\newcommand{\cD}{\mathcal{D}}

\newcommand{\cH}{\mathcal{H}}

\newcommand{\cM}{\mathcal{M}}

\newcommand{\cY}{\mathcal{Y}}

\newcommand{\id}{\text{id}}
\newcommand{\Gray}{\textbf{Gray}}
\newcommand{\Hom}{\text{Hom}}
\newcommand{\D}{\Delta}
\newcommand{\al}{\alpha}
\newcommand{\bt}{\beta}
\newcommand{\gm}{\gamma}
\newcommand{\brg}{\bar{\gamma}}

\iflatexml\else
\newsavebox{\rectboxA}
\newsavebox{\rectboxB}
\newsavebox{\rectboxC}
\newsavebox{\rectboxD}
\newsavebox{\rectboxE}
\newsavebox{\rectboxF}

\pgfkeys{
  /rectlabels/.is family,
  /rectlabels/.cd,
  AB/.store in=\rectlabelAB,
  BC/.store in=\rectlabelBC,
  AD/.store in=\rectlabelAD,
  DE/.store in=\rectlabelDE,
  EF/.store in=\rectlabelEF,
  CF/.store in=\rectlabelCF
}

\newcommand{\DiagramRectangle}[7][]{%
  \begingroup
  \pgfkeys{/rectlabels/.cd,AB=f,BC=g,AD=h,DE=l,EF=j,CF=k}%
  \pgfkeys{/rectlabels/.cd,#1}%
  \sbox{\rectboxA}{#2}%
  \sbox{\rectboxB}{#3}%
  \sbox{\rectboxC}{#4}%
  \sbox{\rectboxD}{#5}%
  \sbox{\rectboxE}{#6}%
  \sbox{\rectboxF}{#7}%
  \begin{tikzpicture}[>=stealth]
    \matrix (M) [
      matrix of nodes,
      ampersand replacement=\&,
      row sep=12mm,
      column sep=20mm,
      inner sep=0pt, 
      nodes={anchor=center,inner sep=2pt,outer sep=0pt}
    ] {
      \usebox{\rectboxA} \& \usebox{\rectboxD} \\
      \usebox{\rectboxB} \& \usebox{\rectboxE} \\
      \usebox{\rectboxC} \& \usebox{\rectboxF} \\
    };

    \draw[|->] (M-1-1.south)
      -- node[midway,right] {$\displaystyle\rectlabelAB$} (M-2-1.north);
    \draw[|->] (M-2-1.south)
      -- node[midway,right] {$\displaystyle\rectlabelBC$} (M-3-1.north);
    \draw[|->] (M-1-1.east)
      -- node[midway,above] {$\displaystyle\rectlabelAD$} (M-1-2.west);
    \draw[|->] (M-1-2.south)
      -- node[midway,right] {$\displaystyle\rectlabelDE$} (M-2-2.north);
    \draw[|->] (M-2-2.south)
      -- node[midway,right] {$\displaystyle\rectlabelEF$} (M-3-2.north);
    \draw[|->] (M-3-1.east)
      -- node[midway,above] {$\displaystyle\rectlabelCF$} (M-3-2.west);
  \end{tikzpicture}%
  \endgroup
}

\newcommand{\shiftfig}[3]{%
  \raisebox{#2}[\height][\depth]{%
    \hspace*{#1}%
    #3%
    \hspace*{-\dimexpr#1\relax}%
  }%
}

\tikzset{condensation/.style={-,
          decoration={markings, 
                    mark=at position -1ex with {\arrow{left hook}},
                    mark=at position 1 with {\arrow[scale=1.5]{>}},
                   },
        preaction={decorate}}}

\tikzset{condensationlarge/.style={-,
          decoration={markings, 
                    mark=at position -1.5ex with {\arrow[scale=1.25]{left hook}},
                    mark=at position 1 with {\arrow[scale=1.75]{>}},
                   },
        preaction={decorate}}}
\tikzset{doublecondensation/.style={double,
        double distance=1.5pt,
        shorten <=6pt, shorten >=6pt, 
        decoration={markings, 
                    mark=at position -11pt with {\arrow[scale=1.25]{left hook}},
                    mark=at position -4.25pt with {\arrow[scale=1.5]{>}},
                   },
        preaction={decorate}}}
\tikzset{doublecondensationlarge/.style={double,
        double distance=1.5pt,
        shorten <=6pt, shorten >=6pt, 
        decoration={markings, 
                    mark=at position -11pt with {\arrow[scale=1.5]{left hook}},
                    mark=at position -4.25pt with {\arrow[scale=2]{>}},
                   },
        preaction={decorate}}}
\fi

\title{Frobenius $2$-algebras and $R$-matrices}

\author[Hibiki Shibuya]{Hibiki Shibuya}
\address{Hibiki Shibuya, Mathematical Institute, Tohoku University, 
6-3 Aoba, Aramaki, Aoba-ku, Sendai, Miyagi 980-8578, Japan.}
\email{shibuya.hibiki.q1@dc.tohoku.ac.jp}

\author[Yuji Terashima]{Yuji Terashima}
\address{Yuji Terashima, Mathematical Institute, Tohoku University, 
6-3 Aoba, Aramaki, Aoba-ku, Sendai, Miyagi 980-8578, Japan.}
\email{yujiterashima@tohoku.ac.jp}

\date{\today}

\begin{document}

\begin{abstract}
  In this paper, introducing Frobenius $2$-algebras, which are
categorical generalizations of Frobenius algebras, we show that every
Frobenius $2$-algebra determines an $R$-matrix satisfying the
Yang–Baxter equation in the monoidal category of endomorphisms of its
underlying object. Moreover, we illustrate this construction through
three examples: a pair of objects in a braided monoidal category, a
Hopf algebra with invertible antipode, and representations of the
Heisenberg double of a finite-dimensional Hopf algebra. In particular,
in the third example, we recover Kashaev's construction of the
universal $R$-matrix. Thus, our construction provides a categorical
generalization of Kashaev's construction.
\end{abstract}

\maketitle

\setcounter{tocdepth}{2} %depth of table of contents
\tableofcontents

\section{Introduction}\label{sec:intro}

The Yang–Baxter equation is fundamental in mathematical physics, with
important applications to integrable systems, quantum groups, and
quantum topology. Its solutions, called $R$-matrices, encode the
exchange relations of interacting particles and provide algebraic
structures underlying various constructions in low-dimensional
topology. A central problem is to understand the algebraic structures
that give rise to solutions of the Yang–Baxter equation and to clarify
the relationships among different constructions of such solutions.

In this paper, we give a construction of $R$-matrices from Frobenius
$2$-algebras which are higher categorical versions of  Frobenius
algebras. More precisely, we show that every Frobenius $2$-algebra in a semistrict monoidal $2$-category determines an $R$-matrix in the
monoidal category of endomorphisms of its underlying object. Our
definition of a Frobenius $2$-algebra is based on the theory of
Frobenius pseudomonoids developed by Street and Lauda \cite{S, L1}, as well as the
coherence axioms formulated by Kerler and Lyubashenko \cite{KL}. In particular,
we start from Lauda's presentation of Frobenius pseudomonoids \cite{L2} and
retain the eight coherence axioms of Kerler–Lyubashenko that do not
involve the unit or counit. This leads to a nonunital and noncounital
version of a Frobenius pseudomonoid, which we call a Frobenius
$2$-algebra. The omission of the unit and counit is essential for
accommodating examples arising from higher-dimensional representations of Heisenberg doubles. Our construction of $R$-matrices is based on the
multiplication and comultiplication of the Frobenius $2$-algebra,
together with four coherent $2$-isomorphisms satisfying eight coherence
axioms. Our main theorem establishes that these eight axioms provide
precisely the categorical relations needed to derive the Yang–Baxter
equation.

The motivation for considering Frobenius $2$-algebras comes from the
growing role of higher Frobenius algebraic structures in mathematical
physics. In particular, Gaiotto and Johnson-Freyd have introduced
condensation monads as higher analogues of special Frobenius algebras,
which are closely related to topological defects and fully extended
topological field theories \cite{GJ}. Such developments suggest that higher
versions of Frobenius algebras will play an important role in the
study of higher-dimensional topological quantum field theories and
related structures. Our construction provides a new connection between
these higher algebraic structures and solutions of the Yang–Baxter
equation.

We demonstrate the applicability of our construction through several
examples. First, a pair of objects in a braided monoidal category
gives rise to a Frobenius $2$-algebra whose associated $R$-matrix is the
braiding on their tensor product. Second, a Hopf algebra with
invertible antipode determines a Frobenius $2$-algebra in the Morita
$2$-category of algebras, bimodules, and bimodule homomorphisms. In this case, the inverse of the resulting $R$-matrix corresponds to the standard braiding on the associated Yetter-Drinfeld module through Hopf bimodule structures.
Finally, we apply our construction to Heisenberg doubles of
finite-dimensional Hopf algebras. In this setting, the four canonical
elements associated with representations of the Heisenberg double and
its opposite algebra determine a Frobenius $2$-algebra. We prove that
the resulting $R$-matrix coincides with the universal $R$-matrix
constructed by Kashaev \cite{K}. See also \cite{Sz} for a geometric interpretation of the construction. Thus, our framework not only recovers
Kashaev's construction but also provides its categorical
generalization, allowing it to be formulated in the broader setting of
Frobenius $2$-algebras and semistrict monoidal $2$-categories.

The interpretation of the $R$-matrix in \cite{K} in terms of the Heisenberg
double of a Hopf algebra became a key tool in \cite{MST2, MST} for
constructing refined quantum invariants of $3$-manifolds from Hopf
algebras. Extending this construction of quantum invariants to
Frobenius $2$-algebras and understanding its relationship with \cite{CM},
where (defect) topological field theories are constructed using similar
algebraic structures, would be an interesting direction for future
research.

This paper is organized as follows. Section~\ref{sec:frob} recalls semistrict monoidal $2$-categories and Frobenius pseudomonoids and introduces Frobenius $2$-algebras. Section~\ref{sec:rmat} gives the $R$-matrix construction and the main theorem. Section~\ref{sec:hopf bimod} describes the Morita $2$-category example and its relation to Hopf bimodules and Yetter–Drinfeld modules. Section~\ref{sec:hsnbrg} shows that our construction coincides with Kashaev's construction in the Heisenberg double example.
\medskip

\noindent\emph{Acknowledgments.}
The authors would like to express our gratitude to Serban Matei Mihalache and Sakie Suzuki for helpful discussions, and Keima Akasaka and Toshinori Takama for helpful discussions on condensation monads and higher categorical structures. The authors are supported by JSPS KAKENHI Grant Numbers 22H01117, 25K06969, and 26K22256. Hibiki Shibuya is also supported by the Tohoku University Advanced Graduate School LEAP Program.

During the preparation of this manuscript, generative AI tools were used for language editing and assistance with checking bibliographic information. All mathematical content was independently developed and proved by the authors, who take full responsibility for the content of this paper.

\section{Frobenius $2$-algebras}\label{sec:frob}
Throughout this section, let $\cC$ be a semistrict monoidal $2$-category \cite{KV, BN}, also called a $\Gray$-monoid \cite{DS}.
By shifting dimensions, we may equivalently regard $\cC$ as a one-object $\Gray$-category \cite{GPS, Kel}, denoted by $\Sigma\cC$.

\subsection{Semistrict monoidal $2$-categories}\label{sec:semi}
We recall the explicit description of a semistrict monoidal $2$-category \cite{KV, BN}.
It consists of a strict $2$-category $\cC$ equipped with the following data, subject to coherence axioms:
\begin{itemize}
  \item An object $\mathbf{1}\in \cC$, called the unit object.
  \item An object $A\otimes B\in \cC$ for any two objects $A,B\in \cC$ (we simply write $A^n$ for the $n$-fold tensor product of $A$).
  \item For any object $A\in \cC$ and $1$-morphism $f:B\rightarrow C$, $1$-morphisms $$A.f=A\otimes f:A\otimes B\rightarrow A\otimes C,
  \qquad f.A=f\otimes A:B\otimes A\rightarrow C\otimes A.$$
  \item For any object $A\in \cC$ and $2$-morphism $\varphi:f\Rightarrow g$, $2$-morphisms $$A.\varphi=A\otimes \varphi:A\otimes f\Rightarrow A\otimes g,
  \qquad \varphi.A=\varphi\otimes A:f\otimes A\Rightarrow g\otimes A.$$
  \item For any two $1$-morphisms $f:A\rightarrow A^{\prime}$ and $g:B\rightarrow B^{\prime}$, a $2$-isomorphism $\theta=\theta_{f,g}$:
  \input{semimonoidal.tex}
  such that the following axioms hold.
  \begin{enumerate}
    \item For any object $A\in \cC$, we have strict $2$-functors 
    $A\otimes -:\cC\rightarrow \cC$ and
    $-\otimes A:\cC\rightarrow \cC$
  with the identities  
  \begin{alignat*}{2}
    \mathbf{1}\otimes - &= -\otimes \mathbf{1}=\id_{\cC}, &\qquad (A\otimes B)\otimes -&=A\otimes (B\otimes -), \\
    -\otimes(A\otimes B)&=(-\otimes A)\otimes B, &\qquad (A\otimes -)\otimes B&=A\otimes (-\otimes B),
  \end{alignat*}
   where $B\in\cC$.
  \item If either $f$ or $g$ is an identity 1-morphism, then $\theta_{f,g}$ is an identity 2-morphism.
  \item For all $1$-morphisms $f$ and $g$, and any object $A\in\cC$,
  $$A\otimes \theta_{f,g}=\theta_{A\otimes f,g}, \quad
  \theta_{f,A\otimes g}=\theta_{f\otimes A,g}, \quad
  \theta_{f,g}\otimes A=\theta_{f,g\otimes A}.$$
  \item For all $1$-morphisms $f,h:A\rightarrow A^{\prime}$ and $1$-morphisms $g,k:B\rightarrow B^{\prime}$, and $2$-morphisms $\varphi:f\Rightarrow h$, $\psi:g\Rightarrow k$,
  \input{axiom1.tex}
  \item For all $1$-morphisms $f:A\rightarrow A^{\prime}$, $g:B\rightarrow B^{\prime}$, $f^{\prime}:A^{\prime}\rightarrow A^{\prime\prime}$, $g^{\prime}:B^{\prime}\rightarrow B^{\prime\prime}$,
  \input{axiom2.tex}
  \end{enumerate}
\end{itemize}

We use the following two examples throughout the paper.
    \begin{enumerate}
        \item Let \((C,\otimes,I,b)\) be a strict braided monoidal category. Its delooping \(\cC=\Sigma C\) is the \(2\)-category with a single object \(*\) and
\[
\operatorname{Hom}_{\Sigma C}(*,*)=C.
\]
%Thus, objects of \(C\) are \(1\)-morphisms of \(\Sigma C\), and morphisms of \(C\) are \(2\)-morphisms of \(\Sigma C\).

We use the convention
\[
Y\circ X:=X\otimes Y.
\]
The identity \(1\)-morphism on \(*\) is \(I\). Vertical composition of \(2\)-morphisms is composition in \(C\), while horizontal composition is given by
\[
g\bullet f:=f\otimes g
\]
for \(f:X\to X'\) and \(g:Y\to Y'\).

The braiding equips \(\Sigma C\) with a semistrict monoidal structure. Its unit object is \(*\), and thus its monoidal product is given by
\[
*\otimes *=*,
\qquad
X\otimes *=*\otimes X=X,
\qquad
f\otimes *=*\otimes f=f.
\]
%The compatibility of \(\boxtimes\) with composition is specified by the invertible \(2\)-morphism
Then the coherent $2$-isomorphism $\theta=\theta_{X,Y}$ is given by the braiding
$$
b_{X,Y}:X\otimes Y\rightarrow Y\otimes X.
$$
% \[
% \Phi_{X',Y';X,Y}:
% (X'\boxtimes Y')\circ(X\boxtimes Y)
% \Longrightarrow
% (X'\circ X)\boxtimes(Y'\circ Y),
% \]
% given in \(C\) by
% \[
% \operatorname{id}_X\otimes b_{Y,X'}\otimes\operatorname{id}_{Y'}:
% X\otimes Y\otimes X'\otimes Y'
% \longrightarrow
% X\otimes X'\otimes Y\otimes Y'.
% \]
% The associativity and unit constraints are identities. Naturality and the hexagon identities for \(b\) imply the required coherence axioms, so that \(\Sigma C\) is a semistrict monoidal $2$-category.
        \item Let \(\mathbb K\) be a field. We implicitly pass to a $\Gray$-monoid strictification \cite{GPS}. The Morita $2$-category $\cC=\mathbf{Mor}\left(\mathbf{Alg}_{\mathbb{K}}\right)$ is defined as follows:

\begin{itemize}
    \item its objects are unital associative \(\mathbb K\)-algebras,
    \item a \(1\)-morphism \(M:A\to B\) is a unital \((A,B)\)-bimodule over \(\mathbb K\), and
    \item a \(2\)-morphism \(f:M\Rightarrow M'\) is an \((A,B)\)-bimodule homomorphism.
\end{itemize}

For \(1\)-morphisms \(M:A\to B\) and \(N:B\to C\), their composite is defined by
\[
N\circ M:=M\otimes_B N.
\]
The identity \(1\)-morphism on \(A\) is the regular bimodule \({}_A A_A\). Vertical composition of \(2\)-morphisms is ordinary composition, and horizontal composition is induced by the relative tensor product. %The associativity and unit constraints are the canonical bimodule isomorphisms.

The tensor product over \(\mathbb K\) equips this $2$-category with a semistrict monoidal structure. Its unit object is \(\mathbb K\), and its tensor product is given by
\[
A\otimes B=A\otimes_{\mathbb K}B,
\qquad
M\otimes A=M\otimes_{\mathbb K}A,
\qquad
f\otimes A=f\otimes_{\mathbb K}\id_A,
\]
with componentwise bimodule actions (and similarly for $A\otimes M$ and $A\otimes f$).

For two bimodules $M:A\rightarrow A^{\prime}$ and $N:B\rightarrow B^{\prime}$, the coherent $2$-isomorphism $\theta=\theta_{M,N}$ is given by the bimodule isomorphism
 \[
 (M\otimes B)\otimes_{A^{\prime}\otimes B}(A^{\prime}\otimes N)
 \xrightarrow{\cong}
 (A\otimes N)\otimes_{A\otimes B^{\prime}}(M\otimes B^{\prime}),
 \]
 \[
(x\otimes b)\otimes(a\otimes y)
\longmapsto
(1\otimes by)\otimes(xa\otimes 1).
\]
% For composable pairs \(M:A\to B\), \(M':B\to C\) and \(N:D\to E\), \(N':E\to F\), the interchange isomorphism is the canonical map
% \[
% (M\otimes N)\otimes_{B\otimes E}(M'\otimes N')
% \xrightarrow{\cong}
% (M\otimes_B M')\otimes(N\otimes_E N'),
% \]
% \[
% (u\otimes v)\otimes(u'\otimes v')
% \longmapsto
% (u\otimes u')\otimes(v\otimes v'),
% \]

% By the coherence theorem for monoidal bicategories, \(\mathsf{Bimod}_{\mathbb K}\) is monoidally biequivalent to a Gray-monoid. We work in a chosen such model, which we call the Morita \(2\)-category. In computations, we retain the notation of algebras and bimodules and suppress the canonical associativity and unit isomorphisms.
    \end{enumerate}

\subsection{Frobenius pseudomonoids}
%In this section, let $\cK$ denote a $\Gray$-category \cite{GPS, Kel}. A $\Gray$-category with a single object is called a semistrict monoidal $2$-category (or called a $\Gray$-monoid \cite{DS}); we denote such a $2$-category by $\cC$. In a $\Gray$-category $\cK$, we use the following coherent $3$-isomorphisms associated with the interchange law:
We begin by recalling the definition of a \emph{pseudomonoid} in a semistrict monoidal $2$-category $\cC$ \cite{DS}. It is also called a \emph{(unital) $E_1$-algebra} in \cite{KY, CM}. Equivalently, a pseudomonoid in $\cC$ is a \emph{pseudomonad} on the unique object of $\Sigma\cC$ \cite{M1, M2, L}. 

\begin{dfn}
A pseudomonoid $A=(A,m,i,\lambda,\rho,\alpha)$ in a semistrict monoidal $2$-category $(\cC,\otimes,\mathbf{1})$ consists of an object $A\in \cC$ together with:
\begin{itemize}
    \item multiplication $m:A^2\rightarrow A$ and unit $i:\mathbf{1}\rightarrow A$, and
    \item coherent $2$-isomorphisms
    \input{p-monad.tex}
\end{itemize}
\end{dfn}

Now, we recall the definition of a Frobenius pseudomonoid as follows \cite{L1, S}.
\begin{dfn}
A pseudomonoid $A=(A,m,i,\lambda,\rho,\alpha)$ in a semistrict monoidal $2$-category $(\cC,\otimes,\mathbf{1})$ is Frobenius if it is equipped with
\begin{itemize}
  \item $1$-morphisms $\varepsilon: A\rightarrow \mathbf{1}$ and $b:\mathbf{1}\rightarrow A^2$, and
  \item coherent $2$-isomorphisms
  \input{p-ad.tex}
\end{itemize}
\end{dfn}

\begin{rmk}
Recall that a Frobenius algebra in a strict monoidal category $C$ can be characterized as a monoid $A\in C$ equipped with a morphism $\varepsilon\colon A\to\mathbf{1}$ such that the Frobenius pairing $\varepsilon m: A^2\rightarrow \mathbf{1}$ is the counit of an adjunction $A\dashv A$ in the delooping $2$-category $\Sigma C$ \cite{S}. Analogously, the definition of a Frobenius pseudomonoid means that the $1$-morphisms $b:\mathbf{1}\rightarrow A^2$ and $e=\varepsilon m:A^2\rightarrow \mathbf{1}$ constitute the unit and counit of a \emph{pseudoadjunction} $A\dashv_p A$ in the $\Gray$-category $\Sigma \cC$ \cite{L1}.

A pseudoadjunction is also referred to as a \emph{locally-adjoint biadjoint pair} \cite{V}. For further details, see the discussion of the \emph{walking pseudoadjunction} (also called the \emph{free-living pseudoadjunction}) defined in \cite{L}.
\end{rmk}

\subsection{Definition and coherence axioms}

We consider a semistrict monoidal $2$-category $(\cC,\otimes, \mathbf{1})$. According to Lauda \cite{L2} and Kerler–Lyubashenko \cite{KL}, if $A$ is a Frobenius pseudomonoid in $\cC$, its Frobenius structure equips the underlying pseudomonoid of $A$ with the following additional data:

\begin{itemize}
    \item comultiplication
    \iflatexml
\begin{center}
  \includegraphics{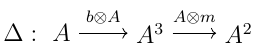}
\end{center}
\else
$$\D:
    \begin{tikzpicture}[baseline=(X.base)]
      \path (0,0) node (X) {$A$}
       (1.5,0) node (Y) {$A^3$}
       (3,0) node (Z) {$A^2$};
      \draw[->] (X) -- node[auto] {$\scriptstyle b\otimes A$} (Y);
      \draw[->] (Y) -- node[auto] {$\scriptstyle A\otimes m$} (Z);
    \end{tikzpicture}
    $$
\fi
    and counit $\varepsilon:A\rightarrow \mathbf{1}$, and
    \item coherent $2$-isomorphisms (listed in \cite[Lemma 32]{L2})
    \input{K-L.tex}
 which satisfy the twenty axioms of diagrams defined in \cite[Definition C.1.4]{KL}. 
\end{itemize}
Of these twenty axioms, eight involve neither the unit nor the counit. These eight axioms suggest a generalization of Frobenius pseudomonoids to the nonunital and noncounital setting. We simply call such a structure a \emph{Frobenius $2$-algebra}.
\begin{dfn}
A Frobenius $2$-algebra $A=(A,m,\D,\al,\bt,\gm,\brg)$ in a semistrict monoidal $2$-category $\cC$ consists of an object $A\in \cC$ together with:
\begin{itemize}
    \item multiplication $m:A^2\rightarrow A$ and comultiplication $\D:A\rightarrow A^2$, and
    \item coherent $2$-isomorphisms
    \begin{center}
        \input{frobenius}
    \end{center}
    such that the following eight coherence axioms hold.
    \input{eight.tex}
\end{itemize}
\end{dfn}

The omission of the unit and counit is motivated by the examples below. In the delooping of the category of finite-dimensional \(\mathbb K\)-vector spaces, a unit for a multiplication represented by a vector space \(U\) would be represented by a vector space \(J\) satisfying \(U\otimes J\cong\mathbb K\). Thus the unit constraints would force \(\dim_{\mathbb K}U=1\). Dually, the counit constraints would force the vector space representing the comultiplication to be one-dimensional. The nonunital and noncounital setting therefore allows us to include the examples in Section~\ref{sec:hsnbrg} arising from higher-dimensional representations of Heisenberg doubles, such as the Fock representation.

\begin{rmk}
  We mention a geometric intuition in terms of \emph{o-graphs} and \emph{MP-moves} \cite{BP, MST2, MSK}. Related graphical interpretations also appear in \cite{CM}.

Normal o-graphs represent branched spines of (combed) $3$-manifolds,
which are dual to branched ideal triangulations.
Under this duality, MP-moves correspond to
$(2,3)$-Pachner moves on ideal triangulations.

\begin{figure}[H]
  \centering
\iflatexml
\begin{center}
  \includegraphics{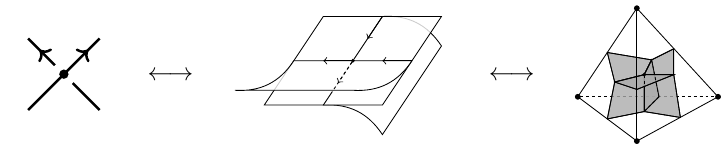}
\end{center}
\else
$$
\vcenter{\hbox{
\begin{tikzpicture}[scale=1.5]
\node (a) at (0,0) {};
\node (b) at (1,0) {};
\node (c) at (0,-1) {};
\node (d) at (1,-1) {};
\draw[very thick, 
postaction={
    decorate,
    decoration={
      markings,
      mark=at position 0.25 with {\arrow{<}}
    }
  }] (a)--(d);
\draw[-, line width=12pt, draw=white] (b)--(c);
\draw[very thick, 
postaction={
    decorate,
    decoration={
      markings,
      mark=at position 0.25 with {\arrow{<}}
    }
  }] (b)--(c);
\fill (0.5,-0.5)
    circle (1.5pt);
\end{tikzpicture}}}
\quad\longleftrightarrow\quad
\vcenter{\hbox{
\includegraphics{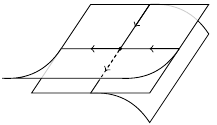}
}}
\quad\longleftrightarrow\quad
\vcenter{\hbox{
\includegraphics{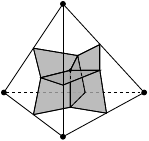}}}
$$
\fi
  \iflatexml\else
    % float has already placed the destination for this caption.
    \makeatletter
    \let\hyper@@anchor\@secondoftwo
    \makeatother
  \fi
  \caption{Correspondence among o-graphs, branched spines, and tetrahedra.}
\end{figure}

The coherence $2$-isomorphisms of a Frobenius $2$-algebra
can be depicted as follows:
\input{o-graph.tex}
From this perspective, the axioms
\eqref{diag:1}--\eqref{diag:8} can be interpreted
as \emph{primary MP-moves} defined in \cite{MSK}.
For example, the axiom \eqref{diag:4} can be visualized by
\input{MPmove.tex}
Moreover, our Yang-Baxter operator \eqref{eq:R}
can be depicted by doubling a crossing:
\input{doubling.tex}
\end{rmk}

\begin{rmk}
  A similar structure appears in a “condensation monad” \cite{GJ}. The coherence $2$-isomorphisms of a Frobenius
  $2$-algebra exhibit a pattern reminiscent of the commuting condensation square:
  \input{cond-sq.tex}
  From this perspective, the cubical pattern of the eight axioms \eqref{diag:1}--\eqref{diag:8} suggests a similar comparison with $3$-dimensional condensation cubes:
  \input{cond-cube.tex}
  For a physical picture, consider a condensation from a gapped topological phase $S$ to a phase $T$ in $(2+1)$ dimensions \cite[Example~2.4.7]{GJ}. On a spatial slice, the interfaces $f\colon S\to T$ and $g\colon T\to S$ are string defects. Their composite $fg$ condenses onto the trivial interface:
$$\varphi:fg\Rightarrow \id_T,\qquad \psi:\id_T\Rightarrow fg.$$
The given condensation induces a condensation monad $A=gf$ on $S$ with coherence data
including condensation squares and cubes.
The condensation of $A^2$ onto $A$ has junctions
$$m=g.\varphi.f:A^2\Rightarrow A, \qquad \Delta=g.\psi.f:A\Rightarrow A^2.$$
In the diagram for $X=m\Delta$, an $A$-strand splits into two
and then rejoins, forming a bubble. In $(1+1)$ dimensions, such a
bubble can be strictly eliminated as $m\Delta=\id_A$.
In $(2+1)$ dimensions, however, this equality is
replaced by a further condensation
$$
  \xi:m\Delta\Rrightarrow\id_A,\qquad
  \eta:\id_A\Rrightarrow m\Delta,\qquad
  \xi\eta=\id_{\id_A}.
$$
Thus, the bubble need not be strictly trivial in $(d+1)$ dimensions if $d\geq 2$.
In this interpretation, our construction \eqref{eq:R}
would provide a Yang-Baxter operator on these bubbles,
which suggests a possible connection to integrable models for bubbles.
\end{rmk}

\section{The $R$-matrix construction}\label{sec:rmat}
Let $(\cC,\otimes,\mathbf{1})$ be a semistrict monoidal $2$-category. For each object $B\in\cC$, the category $\Hom_{\cC}(B,B)$ is monoidal under composition of endomorphisms. In this section, we also denote this monoidal product by $\otimes$ whenever no confusion is possible from the context.

Let $A=(A,m,\D,\al,\bt,\gm,\brg)$ be a Frobenius $2$-algebra in $\cC$,
and set $X:=m\D:A\rightarrow A$. Define the $2$-isomorphism
$R:X^2\Rightarrow X^2$, called the $R$-matrix of $A$, by the following composite:
\begin{equation}\label{eq:R}
\iflatexml
\includegraphics[width=1.6in]{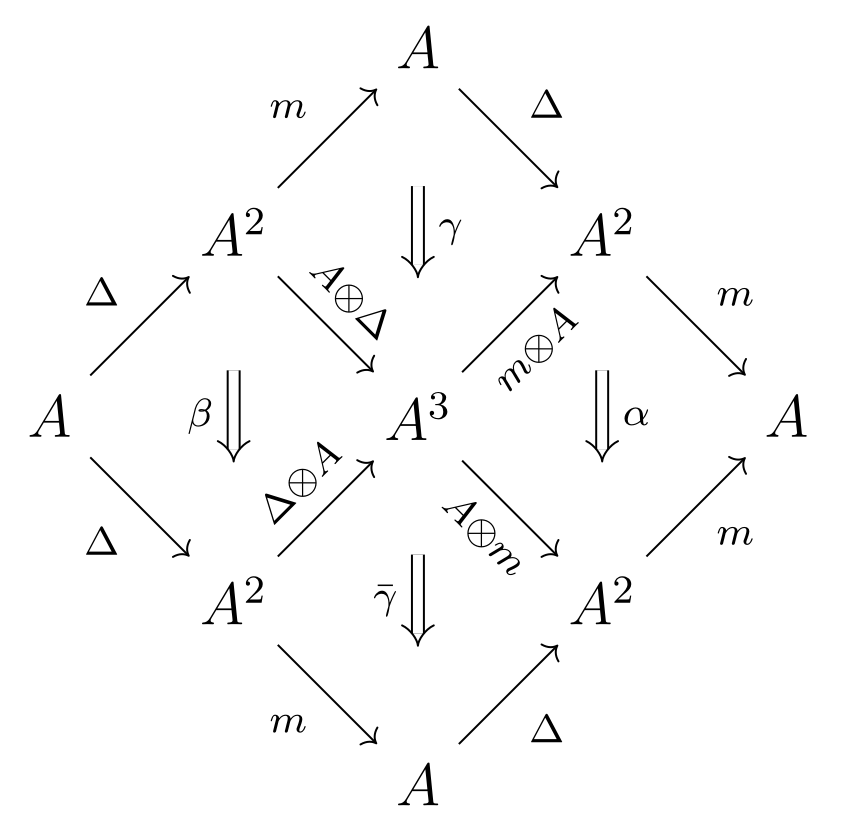}
\else
\vcenter{\hbox{\begin{tikzpicture}[scale=1.3, baseline=(X.base)]
  \path (0,0) node (X) {$A^3$}
     (-1,-1) node (Y1) {$A^2$}
     (1,-1) node (Y2) {$A^2$}
     (0,-2) node (Z) {$A$};
     \draw[double, double distance=2pt, -{Implies}]
  (0,-0.75) -- node[auto,swap] {$\scriptstyle\bar{\gamma}$} (0,-1.25);
  \draw[<-] (X) -- node[auto, sloped] {$\scriptstyle \Delta\otimes A$} (Y1);
  \draw[->] (X) -- node[auto, sloped, swap] {$\scriptstyle A\otimes m$} (Y2);
  \draw[->] (Y1) -- node[auto,swap] {$\scriptstyle m$} (Z);
  \draw[<-] (Y2) -- node[auto] {$\scriptstyle \Delta$} (Z);

    \path (0,2) node (V) {$A$}
     (-1,1) node (W1) {$A^2$}
     (1,1) node (W2) {$A^2$};
     \draw[double, double distance=2pt, -{Implies}]
  (0,1.25) -- node[auto] {$\scriptstyle\gamma$} (0,0.75);
  \draw[<-] (V) -- node[auto,swap] {$\scriptstyle m$} (W1);
  \draw[->] (V) -- node[auto] {$\scriptstyle \Delta$} (W2);
  \draw[->] (W1) -- node[auto, sloped] {$\scriptstyle A\otimes\Delta$} (X);
  \draw[<-] (W2) -- node[auto, sloped, swap] {$\scriptstyle m\otimes A$} (X);

      \path (-2,0) node (X1) {$A$}
     (2,0) node (X2) {$A$};
     \draw[->] (X1) -- node[auto] {$\scriptstyle \Delta$} (W1);
     \draw[->] (X1) -- node[swap, auto] {$\scriptstyle \Delta$} (Y1);
     \draw[->] (W2) -- node[auto] {$\scriptstyle m$} (X2);
     \draw[swap, ->] (Y2) -- node[auto] {$\scriptstyle m$} (X2);

          \draw[double, double distance=2pt, -{Implies}]
  (1,0.25) -- node[auto] {$\scriptstyle\alpha$} (1,-0.25);
       \draw[double, double distance=2pt, -{Implies}]
  (-1,0.25) -- node[auto, swap] {$\scriptstyle\beta$} (-1,-0.25);
\end{tikzpicture}}}
\fi
\end{equation}
\begin{thm}\label{main}
    In the monoidal category $\Hom_{\cC}(A,A)$, the $R$-matrix $R:X\otimes X\rightarrow X\otimes X$ satisfies the Yang-Baxter equation:
    \begin{align}
        (R\otimes \id_X)(\id_X\otimes R)(R\otimes \id_X)=(\id_X\otimes R)(R\otimes \id_X)(\id_X\otimes R).
    \end{align}
\end{thm}

\begin{proof}
 This result follows from the following diagrammatic calculation using the eight axioms \eqref{diag:1}--\eqref{diag:8}.
 \iflatexml
\begin{center}
  \includegraphics{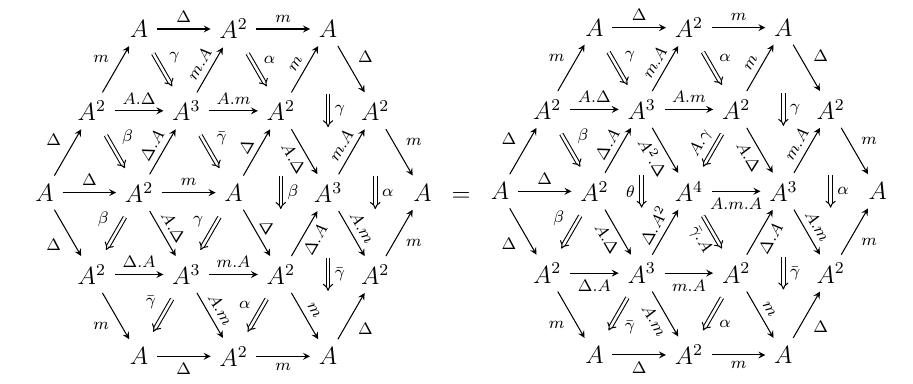}\par
  \includegraphics{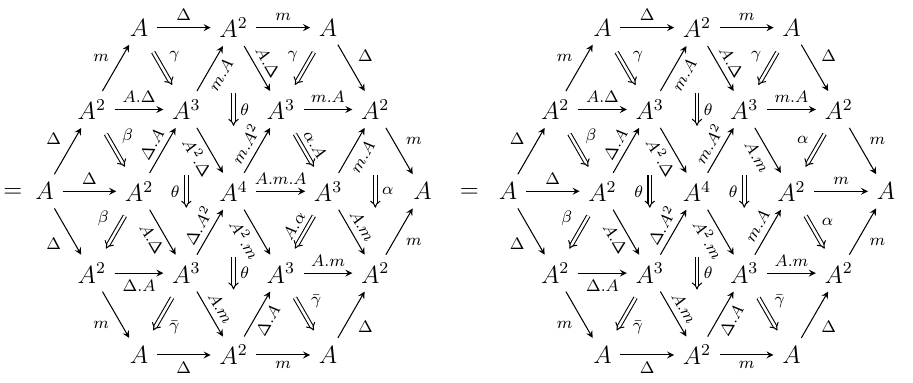}\par
  \includegraphics{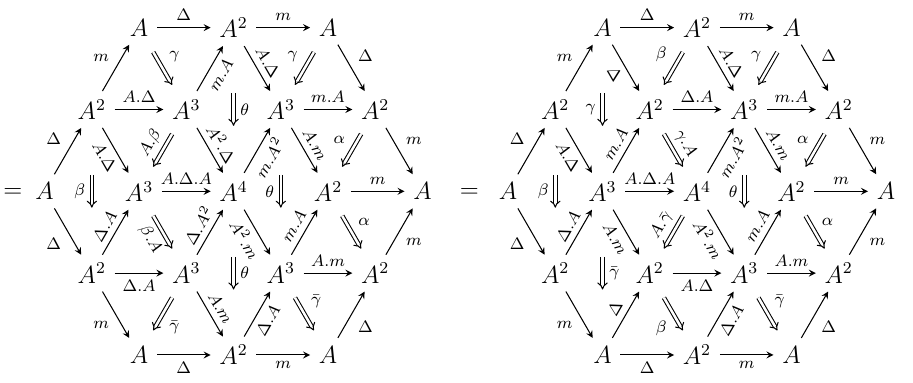}\par
  \includegraphics{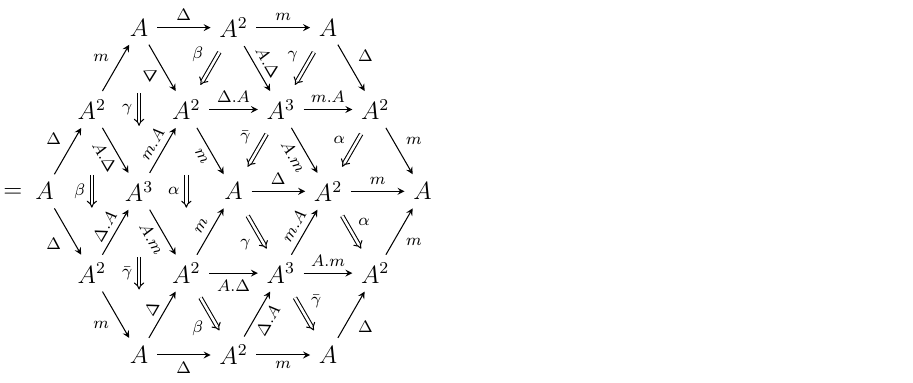}
\end{center}
\else
\begin{align*}
   &\begin{tikzpicture}[
  x=0.8cm, y=0.8cm,
  >=stealth,
  line width=0.5pt,
  every node/.style={inner sep=3pt},
  shorten <=1pt, shorten >=1pt,
  baseline=(v10.base)
]
  \pgfmathsetmacro{\hexh}{sqrt(3)}

  % Top row
  \node (v3)  at (-2,{2*\hexh}) {$A$};
  \node (v8)  at ( 0,{2*\hexh}) {$A^2$};
  \node (v13) at ( 2,{2*\hexh}) {$A$};

  % Upper row
  \node (v1)  at (-3,\hexh) {$A^2$};
  \node (v4)  at (-1,\hexh) {$A^3$};
  \node (v9)  at ( 1,\hexh) {$A^2$};
  \node (v16) at ( 3,\hexh) {$A^2$};

  % Middle row: vertex 10 is the origin.
  \node (v0)  at (-4,0) {$A$};
  \node (v5)  at (-2,0) {$A^2$};
  \node (v10) at ( 0,0) {$A$};
  \node (v14) at ( 2,0) {$A^3$};
  \node (v18) at ( 4,0) {$A$};

  % Lower row
  \node (v2)  at (-3,{-\hexh}) {$A^2$};
  \node (v6)  at (-1,{-\hexh}) {$A^3$};
  \node (v11) at ( 1,{-\hexh}) {$A^2$};
  \node (v17) at ( 3,{-\hexh}) {$A^2$};

  % Bottom row
  \node (v7)  at (-2,{-2*\hexh}) {$A$};
  \node (v12) at ( 0,{-2*\hexh}) {$A^2$};
  \node (v15) at ( 2,{-2*\hexh}) {$A$};

  \draw[->] (v0)  -- node[midway,auto] {$\scriptstyle \D$}  (v1);
  \draw[->] (v0)  -- node[midway,swap,auto] {$\scriptstyle \D$}  (v2);
  \draw[->] (v0)  -- node[midway,auto,sloped] {$\scriptstyle \D$}  (v5);
  \draw[->] (v1)  -- node[midway,auto] {$\scriptstyle m$}  (v3);
  \draw[->] (v1)  -- node[midway,auto,sloped] {$\scriptstyle A.\D$}  (v4);
  \draw[->] (v2)  -- node[midway,auto,sloped] {$\scriptstyle \D.A$}  (v6);
  \draw[->] (v2)  -- node[midway,swap,auto] {$\scriptstyle m$}  (v7);
  \draw[->] (v3)  -- node[midway,auto,sloped] {$\scriptstyle \D$}  (v8);
  \draw[->] (v4)  -- node[midway,auto,sloped] {$\scriptstyle m.A$}  (v8);
  \draw[->] (v4)  -- node[midway,auto,sloped] {$\scriptstyle A.m$}  (v9);
  \draw[->] (v5)  -- node[midway,auto,sloped] {$\scriptstyle \D.A$}  (v4);
  \draw[->] (v5)  -- node[midway,auto,sloped] {$\scriptstyle A.\D$}  (v6);
  \draw[->] (v5)  -- node[midway,auto,sloped] {$\scriptstyle m$}  (v10);
  \draw[->] (v6)  -- node[midway,auto,sloped] {$\scriptstyle m.A$}  (v11);
  \draw[->] (v6)  -- node[midway,auto,sloped] {$\scriptstyle A.m$}  (v12);
  \draw[->] (v7)  -- node[midway,auto,swap] {$\scriptstyle \D$}  (v12);
  \draw[->] (v8)  -- node[midway,auto,sloped] {$\scriptstyle m$}  (v13);
  \draw[->] (v9)  -- node[midway,auto,sloped] {$\scriptstyle m$}  (v13);
  \draw[->] (v9)  -- node[midway,auto,sloped,swap] {$\scriptstyle A.\D$}  (v14);
  \draw[->] (v10) -- node[midway,auto,sloped] {$\scriptstyle \D$}  (v9);
  \draw[->] (v10) -- node[midway,auto,sloped] {$\scriptstyle \D$}  (v11);
  \draw[->] (v11) -- node[midway,auto,sloped,swap] {$\scriptstyle \D.A$}  (v14);
  \draw[->] (v11) -- node[midway,auto,sloped] {$\scriptstyle m$}  (v15);
  \draw[->] (v12) -- node[midway,auto,swap] {$\scriptstyle m$}  (v15);
  \draw[->] (v13) -- node[midway,auto] {$\scriptstyle \D$}  (v16);
  \draw[->] (v14) -- node[midway,auto,sloped] {$\scriptstyle m.A$}  (v16);
  \draw[->] (v14) -- node[midway,auto,sloped] {$\scriptstyle A.m$} (v17);
  \draw[->] (v15) -- node[midway,auto,swap] {$\scriptstyle \D$} (v17);
  \draw[->] (v16) -- node[midway,auto] {$\scriptstyle m$} (v18);
  \draw[->] (v17) -- node[midway,auto,swap] {$\scriptstyle m$} (v18);

  % Double arrows 
  \tikzset{twocell/.style={
    double, double distance=1.2pt,
    -implies,
    shorten <=0pt, shorten >=0pt
  }}

  % Upper-left faces: (1,3,8,4), (4,8,13,9),
  %                   (0,1,4,5), (5,4,9,10).
  \draw[twocell] ($(v3)!0.3!(v4)$)
    -- node[midway,auto] {$\scriptstyle \gm$} ($(v3)!0.7!(v4)$);
  \draw[twocell] ($(v8)!0.3!(v9)$)
    -- node[midway,auto] {$\scriptstyle \al$} ($(v8)!0.7!(v9)$);
  \draw[twocell] ($(v1)!0.3!(v5)$)
    -- node[midway,auto] {$\scriptstyle \bt$} ($(v1)!0.7!(v5)$);
  \draw[twocell] ($(v4)!0.3!(v10)$)
    -- node[midway,auto] {$\scriptstyle \brg$} ($(v4)!0.7!(v10)$);

  % Lower-left faces: (0,5,6,2), (5,10,11,6),
  %                   (2,6,12,7), (6,11,15,12).
  \draw[twocell] ($(v5)!0.3!(v2)$)
    -- node[midway,auto,swap] {$\scriptstyle \bt$} ($(v5)!0.7!(v2)$);
  \draw[twocell] ($(v10)!0.3!(v6)$)
    -- node[midway,auto,swap] {$\scriptstyle \gm$} ($(v10)!0.7!(v6)$);
  \draw[twocell] ($(v6)!0.3!(v7)$)
    -- node[midway,auto,swap] {$\scriptstyle \brg$} ($(v6)!0.7!(v7)$);
  \draw[twocell] ($(v11)!0.3!(v12)$)
    -- node[midway,auto,swap] {$\scriptstyle \al$} ($(v11)!0.7!(v12)$);

  % Right faces: (9,13,16,14), (10,9,14,11),
  %              (14,16,18,17), (11,14,17,15).
  \draw[twocell] ($(v13)!0.4!(v14)$)
    -- node[midway,auto] {$\scriptstyle \gm$} ($(v13)!0.6!(v14)$);
  \draw[twocell] ($(v9)!0.4!(v11)$)
    -- node[midway,auto] {$\scriptstyle \bt$} ($(v9)!0.6!(v11)$);
  \draw[twocell] ($(v16)!0.4!(v17)$)
    -- node[midway,auto] {$\scriptstyle \al$} ($(v16)!0.6!(v17)$);
  \draw[twocell] ($(v14)!0.4!(v15)$)
    -- node[midway,auto] {$\scriptstyle \brg$} ($(v14)!0.6!(v15)$);
\end{tikzpicture} 
   \;=\;\begin{tikzpicture}[
  x=0.8cm, y=0.8cm,
  >=stealth,
  line width=0.5pt,
  every node/.style={inner sep=3pt},
  shorten <=1pt, shorten >=1pt,
  baseline=(v10.base)
]
  \pgfmathsetmacro{\hexh}{sqrt(3)}

  % Top row
  \node (v3)  at (-2,{2*\hexh}) {$A$};
  \node (v8)  at ( 0,{2*\hexh}) {$A^2$};
  \node (v13) at ( 2,{2*\hexh}) {$A$};

  % Upper row
  \node (v1)  at (-3,\hexh) {$A^2$};
  \node (v4)  at (-1,\hexh) {$A^3$};
  \node (v9)  at ( 1,\hexh) {$A^2$};
  \node (v16) at ( 3,\hexh) {$A^2$};

  % Middle row: vertex 10 is the origin.
  \node (v0)  at (-4,0) {$A$};
  \node (v5)  at (-2,0) {$A^2$};
  \node (v10) at ( 0,0) {$A^4$};
  \node (v14) at ( 2,0) {$A^3$};
  \node (v18) at ( 4,0) {$A$};

  % Lower row
  \node (v2)  at (-3,{-\hexh}) {$A^2$};
  \node (v6)  at (-1,{-\hexh}) {$A^3$};
  \node (v11) at ( 1,{-\hexh}) {$A^2$};
  \node (v17) at ( 3,{-\hexh}) {$A^2$};

  % Bottom row
  \node (v7)  at (-2,{-2*\hexh}) {$A$};
  \node (v12) at ( 0,{-2*\hexh}) {$A^2$};
  \node (v15) at ( 2,{-2*\hexh}) {$A$};

  \draw[->] (v0)  -- node[midway,auto] {$\scriptstyle \D$}  (v1);
  \draw[->] (v0)  -- node[midway,swap,auto] {$\scriptstyle \D$}  (v2);
  \draw[->] (v0)  -- node[midway,auto,sloped] {$\scriptstyle \D$}  (v5);
  \draw[->] (v1)  -- node[midway,auto] {$\scriptstyle m$}  (v3);
  \draw[->] (v1)  -- node[midway,auto,sloped] {$\scriptstyle A.\D$}  (v4);
  \draw[->] (v2)  -- node[midway,auto,sloped,swap] {$\scriptstyle \D.A$}  (v6);
  \draw[->] (v2)  -- node[midway,swap,auto] {$\scriptstyle m$}  (v7);
  \draw[->] (v3)  -- node[midway,auto,sloped] {$\scriptstyle \D$}  (v8);
  \draw[->] (v4)  -- node[midway,auto,sloped] {$\scriptstyle m.A$}  (v8);
  \draw[->] (v4)  -- node[midway,auto,sloped] {$\scriptstyle A.m$}  (v9);
  \draw[->] (v5)  -- node[midway,auto,sloped] {$\scriptstyle \D.A$}  (v4);
  \draw[->] (v5)  -- node[midway,auto,sloped,swap] {$\scriptstyle A.\D$}  (v6);
  \draw[->] (v4)  -- node[midway,auto,sloped,swap] {$\scriptstyle A^2.\D$}  (v10);
  \draw[->] (v6)  -- node[midway,auto,sloped,swap] {$\scriptstyle m.A$}  (v11);
  \draw[->] (v6)  -- node[midway,auto,sloped,swap] {$\scriptstyle A.m$}  (v12);
  \draw[->] (v7)  -- node[midway,auto,swap] {$\scriptstyle \D$}  (v12);
  \draw[->] (v8)  -- node[midway,auto,sloped] {$\scriptstyle m$}  (v13);
  \draw[->] (v9)  -- node[midway,auto,sloped] {$\scriptstyle m$}  (v13);
  \draw[->] (v9)  -- node[midway,auto,sloped,swap] {$\scriptstyle A.\D$}  (v14);
  \draw[->] (v6) -- node[midway,auto,sloped] {$\scriptstyle \D.A^2$}  (v10);
  \draw[->] (v10) -- node[midway,auto,swap] {$\scriptstyle A.m.A$}  (v14);
  \draw[->] (v11) -- node[midway,auto,sloped,swap] {$\scriptstyle \D.A$}  (v14);
  \draw[->] (v11) -- node[midway,auto,sloped] {$\scriptstyle m$}  (v15);
  \draw[->] (v12) -- node[midway,auto,swap] {$\scriptstyle m$}  (v15);
  \draw[->] (v13) -- node[midway,auto] {$\scriptstyle \D$}  (v16);
  \draw[->] (v14) -- node[midway,auto,sloped] {$\scriptstyle m.A$}  (v16);
  \draw[->] (v14) -- node[midway,auto,sloped] {$\scriptstyle A.m$} (v17);
  \draw[->] (v15) -- node[midway,auto,swap] {$\scriptstyle \D$} (v17);
  \draw[->] (v16) -- node[midway,auto] {$\scriptstyle m$} (v18);
  \draw[->] (v17) -- node[midway,auto,swap] {$\scriptstyle m$} (v18);

  % Double arrows
  \tikzset{twocell/.style={
    double, double distance=1.2pt,
    -implies,
    shorten <=0pt, shorten >=0pt
  }}

  % Upper-left faces: (1,3,8,4), (4,8,13,9),
  %                   (0,1,4,5), (5,4,9,10).
  \draw[twocell] ($(v3)!0.3!(v4)$)
    -- node[midway,auto] {$\scriptstyle \gm$} ($(v3)!0.7!(v4)$);
  \draw[twocell] ($(v8)!0.3!(v9)$)
    -- node[midway,auto] {$\scriptstyle \al$} ($(v8)!0.7!(v9)$);
  \draw[twocell] ($(v1)!0.3!(v5)$)
    -- node[midway,auto] {$\scriptstyle \bt$} ($(v1)!0.7!(v5)$);
  \draw[twocell] ($(v10)!0.3!(v11)$)
    -- node[midway,auto, sloped,swap] {$\scriptstyle \brg.A$} ($(v10)!0.7!(v11)$);

  % Lower-left faces: (0,5,6,2), (5,10,11,6),
  %                   (2,6,12,7), (6,11,15,12).
  \draw[twocell] ($(v5)!0.3!(v2)$)
    -- node[midway,auto,swap] {$\scriptstyle \bt$} ($(v5)!0.7!(v2)$);
  \draw[twocell] ($(v9)!0.3!(v10)$)
    -- node[midway,auto,sloped] {$\scriptstyle A.\gm$} ($(v9)!0.7!(v10)$);
  \draw[twocell] ($(v6)!0.3!(v7)$)
    -- node[midway,auto] {$\scriptstyle \brg$} ($(v6)!0.7!(v7)$);
  \draw[twocell] ($(v11)!0.3!(v12)$)
    -- node[midway,auto] {$\scriptstyle \al$} ($(v11)!0.7!(v12)$);

  % Right faces: (9,13,16,14), (10,9,14,11),
  %              (14,16,18,17), (11,14,17,15).
  \draw[twocell] ($(v13)!0.4!(v14)$)
    -- node[midway,auto] {$\scriptstyle \gm$} ($(v13)!0.6!(v14)$);
  \draw[twocell] ($(v4)!0.4!(v6)$)
    -- node[midway,auto,swap] {$\scriptstyle \theta$} ($(v4)!0.6!(v6)$);
  \draw[twocell] ($(v16)!0.4!(v17)$)
    -- node[midway,auto] {$\scriptstyle \al$} ($(v16)!0.6!(v17)$);
  \draw[twocell] ($(v14)!0.4!(v15)$)
    -- node[midway,auto] {$\scriptstyle \brg$} ($(v14)!0.6!(v15)$);
\end{tikzpicture} \\
   =\;&\vcenter{\hbox{\begin{tikzpicture}[
  x=0.8cm, y=0.8cm,
  >=stealth,
  line width=0.5pt,
  every node/.style={inner sep=3pt},
  shorten <=1pt, shorten >=1pt,
  twocell/.style={
    double, double distance=1.2pt,
    -implies,
    shorten <=0pt, shorten >=0pt,
    baseline=(v9.base)
  }
]
  \pgfmathsetmacro{\hexh}{sqrt(3)}

  % Top row
  \node (v4)  at (-2,{2*\hexh}) {$A$};
  \node (v8)  at ( 0,{2*\hexh}) {$A^2$};
  \node (v11) at ( 2,{2*\hexh}) {$A$};

  % Upper row
  \node (v1)  at (-3,\hexh) {$A^2$};
  \node (v5)  at (-1,\hexh) {$A^3$};
  \node (v12) at ( 1,\hexh) {$A^3$};
  \node (v15) at ( 3,\hexh) {$A^2$};

  % Middle row: vertex 9 is the origin.
  \node (v0)  at (-4,0) {$A$};
  \node (v2)  at (-2,0) {$A^2$};
  \node (v9)  at ( 0,0) {$A^4$};
  \node (v16) at ( 2,0) {$A^3$};
  \node (v18) at ( 4,0) {$A$};

  % Lower row
  \node (v3)  at (-3,{-\hexh}) {$A^2$};
  \node (v6)  at (-1,{-\hexh}) {$A^3$};
  \node (v13) at ( 1,{-\hexh}) {$A^3$};
  \node (v17) at ( 3,{-\hexh}) {$A^2$};

  % Bottom row
  \node (v7)  at (-2,{-2*\hexh}) {$A$};
  \node (v10) at ( 0,{-2*\hexh}) {$A^2$};
  \node (v14) at ( 2,{-2*\hexh}) {$A$};

  % Single arrows. 
  \draw[->] (v0)  -- node[midway,auto] {$\scriptstyle \D$}  (v1);
  \draw[->] (v0)  -- node[midway,auto,sloped] {$\scriptstyle \D$}  (v2);
  \draw[->] (v0)  -- node[midway,auto,swap] {$\scriptstyle \D$}  (v3);
  \draw[->] (v1)  -- node[midway,auto] {$\scriptstyle m$}  (v4);
  \draw[->] (v1)  -- node[midway,auto,sloped] {$\scriptstyle A.\D$}  (v5);
  \draw[->] (v2)  -- node[midway,auto,sloped] {$\scriptstyle \D.A$}  (v5);
  \draw[->] (v2)  -- node[midway,auto,sloped,swap] {$\scriptstyle A.\D$}  (v6);
  \draw[->] (v3)  -- node[midway,auto,sloped,swap] {$\scriptstyle \D.A$}  (v6);
  \draw[->] (v3)  -- node[midway,auto,swap] {$\scriptstyle m$}  (v7);
  \draw[->] (v4)  -- node[midway,auto,sloped] {$\scriptstyle \D$}  (v8);
  \draw[->] (v5)  -- node[midway,auto,sloped,swap] {$\scriptstyle m.A$}  (v8);
  \draw[->] (v5)  -- node[midway,auto,sloped,swap] {$\scriptstyle A^2.\D$}  (v9);
  \draw[->] (v6)  -- node[midway,auto,sloped] {$\scriptstyle \D.A^2$}  (v9);
  \draw[->] (v6)  -- node[midway,auto,sloped] {$\scriptstyle A.m$}  (v10);
  \draw[->] (v7)  -- node[midway,auto,swap] {$\scriptstyle \D$}  (v10);
  \draw[->] (v8)  -- node[midway,auto,sloped] {$\scriptstyle m$}  (v11);
  \draw[->] (v8)  -- node[midway,auto,sloped] {$\scriptstyle A.\D$}  (v12);
  \draw[->] (v9)  -- node[midway,auto,sloped] {$\scriptstyle m.A^2$}  (v12);
  \draw[->] (v9)  -- node[midway,auto,sloped,swap] {$\scriptstyle A^2.m$}  (v13);
  \draw[->] (v10) -- node[midway,auto,sloped,swap] {$\scriptstyle \D.A$}  (v13);
  \draw[->] (v10) -- node[midway,auto,swap] {$\scriptstyle m$}  (v14);
  \draw[->] (v11) -- node[midway,auto] {$\scriptstyle \D$}  (v15);
  \draw[->] (v12) -- node[midway,auto,sloped] {$\scriptstyle m.A$}  (v15);
  \draw[->] (v9) -- node[midway,auto,sloped] {$\scriptstyle A.m.A$}  (v16);
  \draw[->] (v16) -- node[midway,auto,sloped,swap] {$\scriptstyle m.A$}  (v15);
  \draw[->] (v13) -- node[midway,auto,sloped] {$\scriptstyle A.m$}  (v17);
  \draw[->] (v14) -- node[midway,auto,swap] {$\scriptstyle \D$} (v17);
  \draw[->] (v15) -- node[midway,auto] {$\scriptstyle m$} (v18);
  \draw[->] (v16) -- node[midway,auto,sloped] {$\scriptstyle A.m$} (v17);
  \draw[->] (v17) -- node[midway,auto,swap] {$\scriptstyle m$} (v18);

  % Double arrows
  % A: face (1,4,8,5), down-right.
  \draw[twocell] ($(v4)!0.3!(v5)$)
    -- node[midway,auto] {$\scriptstyle \gm$} ($(v4)!0.7!(v5)$);
  % B: face (8,11,15,12), down-left.
  \draw[twocell] ($(v11)!0.3!(v12)$)
    -- node[midway,auto,swap] {$\scriptstyle \gm$} ($(v11)!0.7!(v12)$);
  % C: face (5,8,12,9), downward.
  \draw[twocell] ($(v8)!0.4!(v9)$)
    -- node[midway,auto] {$\scriptstyle \theta$} ($(v8)!0.6!(v9)$);
  % D: face (0,1,5,2), down-right.
  \draw[twocell] ($(v1)!0.3!(v2)$)
    -- node[midway,auto] {$\scriptstyle \bt$} ($(v1)!0.7!(v2)$);
  % E: face (12,15,18,16), down-left.
  \draw[twocell] ($(v16)!0.3!(v13)$)
    -- node[midway,auto,sloped] {$\scriptstyle A.\al$} ($(v16)!0.7!(v13)$);
  % F: face (2,5,9,6), downward.
  \draw[twocell] ($(v5)!0.4!(v6)$)
    -- node[midway,auto,swap] {$\scriptstyle \theta$} ($(v5)!0.6!(v6)$);
  % G: face (9,12,16,13), downward.
  \draw[twocell] ($(v15)!0.4!(v17)$)
    -- node[midway,auto] {$\scriptstyle \al$} ($(v15)!0.6!(v17)$);
  % H: face (0,2,6,3), down-left.
  \draw[twocell] ($(v2)!0.3!(v3)$)
    -- node[midway,auto,swap] {$\scriptstyle \bt$} ($(v2)!0.7!(v3)$);
  % I: face (13,16,18,17), down-right.
  \draw[twocell] ($(v12)!0.3!(v16)$)
    -- node[midway,auto,sloped] {$\scriptstyle \al.A$} ($(v12)!0.7!(v16)$);
  % J: face (6,9,13,10), downward.
  \draw[twocell] ($(v9)!0.4!(v10)$)
    -- node[midway,auto] {$\scriptstyle \theta$} ($(v9)!0.6!(v10)$);
  % K: face (3,6,10,7), down-left.
  \draw[twocell] ($(v6)!0.3!(v7)$)
    -- node[midway,auto] {$\scriptstyle \brg$} ($(v6)!0.7!(v7)$);
  % L: face (10,13,17,14), down-right.
  \draw[twocell] ($(v13)!0.3!(v14)$)
    -- node[midway,auto] {$\scriptstyle \brg$} ($(v13)!0.7!(v14)$);
\end{tikzpicture}}} \;=\;\vcenter{\hbox{\begin{tikzpicture}[
  x=0.8cm, y=0.8cm,
  >=stealth,
  line width=0.5pt,
  every node/.style={inner sep=3pt},
  shorten <=1pt, shorten >=1pt,
  twocell/.style={
    double, double distance=1.2pt,
    -implies,
    shorten <=0pt, shorten >=0pt,
    baseline=(v0.base)
  }
]
  \pgfmathsetmacro{\hexh}{sqrt(3)}

 % Top row
  \node (v4)  at (-2,{2*\hexh}) {$A$};
  \node (v8)  at ( 0,{2*\hexh}) {$A^2$};
  \node (v11) at ( 2,{2*\hexh}) {$A$};

  % Upper row
  \node (v1)  at (-3,\hexh) {$A^2$};
  \node (v5)  at (-1,\hexh) {$A^3$};
  \node (v12) at ( 1,\hexh) {$A^3$};
  \node (v15) at ( 3,\hexh) {$A^2$};

  % Middle row: vertex 9 is the origin.
  \node (v0)  at (-4,0) {$A$};
  \node (v2)  at (-2,0) {$A^2$};
  \node (v9)  at ( 0,0) {$A^4$};
  \node (v16) at ( 2,0) {$A^2$};
  \node (v18) at ( 4,0) {$A$};

  % Lower row
  \node (v3)  at (-3,{-\hexh}) {$A^2$};
  \node (v6)  at (-1,{-\hexh}) {$A^3$};
  \node (v13) at ( 1,{-\hexh}) {$A^3$};
  \node (v17) at ( 3,{-\hexh}) {$A^2$};

  % Bottom row
  \node (v7)  at (-2,{-2*\hexh}) {$A$};
  \node (v10) at ( 0,{-2*\hexh}) {$A^2$};
  \node (v14) at ( 2,{-2*\hexh}) {$A$};

  % Single arrows.
\draw[->] (v0)  -- node[midway,auto] {$\scriptstyle \D$}  (v1);
  \draw[->] (v0)  -- node[midway,auto,sloped] {$\scriptstyle \D$}  (v2);
  \draw[->] (v0)  -- node[midway,auto,swap] {$\scriptstyle \D$}  (v3);
  \draw[->] (v1)  -- node[midway,auto] {$\scriptstyle m$}  (v4);
  \draw[->] (v1)  -- node[midway,auto,sloped] {$\scriptstyle A.\D$}  (v5);
  \draw[->] (v2)  -- node[midway,auto,sloped] {$\scriptstyle \D.A$}  (v5);
  \draw[->] (v2)  -- node[midway,auto,sloped,swap] {$\scriptstyle A.\D$}  (v6);
  \draw[->] (v3)  -- node[midway,auto,sloped,swap] {$\scriptstyle \D.A$}  (v6);
  \draw[->] (v3)  -- node[midway,auto,swap] {$\scriptstyle m$}  (v7);
  \draw[->] (v4)  -- node[midway,auto,sloped] {$\scriptstyle \D$}  (v8);
  \draw[->] (v5)  -- node[midway,auto,sloped,swap] {$\scriptstyle m.A$}  (v8);
  \draw[->] (v5)  -- node[midway,auto,sloped,swap] {$\scriptstyle A^2.\D$}  (v9);
  \draw[->] (v6)  -- node[midway,auto,sloped] {$\scriptstyle \D.A^2$}  (v9);
  \draw[->] (v6)  -- node[midway,auto,sloped] {$\scriptstyle A.m$}  (v10);
  \draw[->] (v7)  -- node[midway,auto,swap] {$\scriptstyle \D$}  (v10);
  \draw[->] (v8)  -- node[midway,auto,sloped] {$\scriptstyle m$}  (v11);
  \draw[->] (v8)  -- node[midway,auto,sloped] {$\scriptstyle A.\D$}  (v12);
  \draw[->] (v9)  -- node[midway,auto,sloped] {$\scriptstyle m.A^2$}  (v12);
  \draw[->] (v9)  -- node[midway,auto,sloped,swap] {$\scriptstyle A^2.m$}  (v13);
  \draw[->] (v10) -- node[midway,auto,sloped,swap] {$\scriptstyle \D.A$}  (v13);
  \draw[->] (v10) -- node[midway,auto,swap] {$\scriptstyle m$}  (v14);
  \draw[->] (v11) -- node[midway,auto] {$\scriptstyle \D$}  (v15);
  \draw[->] (v12) -- node[midway,auto,sloped] {$\scriptstyle m.A$}  (v15);
  \draw[->] (v12) -- node[midway,auto,sloped,swap] {$\scriptstyle A.m$}  (v16);
  \draw[->] (v13) -- node[midway,auto,sloped] {$\scriptstyle m.A$}  (v16);
  \draw[->] (v13) -- node[midway,auto,sloped] {$\scriptstyle A.m$}  (v17);
  \draw[->] (v14) -- node[midway,auto,swap] {$\scriptstyle \D$} (v17);
  \draw[->] (v15) -- node[midway,auto] {$\scriptstyle m$} (v18);
  \draw[->] (v16) -- node[midway,auto,sloped] {$\scriptstyle m$} (v18);
  \draw[->] (v17) -- node[midway,auto,swap] {$\scriptstyle m$} (v18);

  % Double arrows
  % A: face (1,4,8,5), down-right.
  \draw[twocell] ($(v4)!0.3!(v5)$)
    -- node[midway,auto] {$\scriptstyle \gm$} ($(v4)!0.7!(v5)$);
  % B: face (8,11,15,12), down-left.
  \draw[twocell] ($(v11)!0.3!(v12)$)
    -- node[midway,auto,swap] {$\scriptstyle \gm$} ($(v11)!0.7!(v12)$);
  % C: face (5,8,12,9), downward.
  \draw[twocell] ($(v8)!0.4!(v9)$)
    -- node[midway,auto] {$\scriptstyle \theta$} ($(v8)!0.6!(v9)$);
  % D: face (0,1,5,2), down-right.
  \draw[twocell] ($(v1)!0.3!(v2)$)
    -- node[midway,auto] {$\scriptstyle \bt$} ($(v1)!0.7!(v2)$);
  % E: face (12,15,18,16), down-left.
  \draw[twocell] ($(v15)!0.3!(v16)$)
    -- node[midway,auto,swap] {$\scriptstyle \al$} ($(v15)!0.7!(v16)$);
  % F: face (2,5,9,6), downward.
  \draw[twocell] ($(v5)!0.4!(v6)$)
    -- node[midway,auto,swap] {$\scriptstyle \theta$} ($(v5)!0.6!(v6)$);
  % G: face (9,12,16,13), downward.
  \draw[twocell] ($(v12)!0.4!(v13)$)
    -- node[midway,auto,swap] {$\scriptstyle \theta$} ($(v12)!0.6!(v13)$);
  % H: face (0,2,6,3), down-left.
  \draw[twocell] ($(v2)!0.3!(v3)$)
    -- node[midway,auto,swap] {$\scriptstyle \bt$} ($(v2)!0.7!(v3)$);
  % I: face (13,16,18,17), down-right.
  \draw[twocell] ($(v16)!0.3!(v17)$)
    -- node[midway,auto] {$\scriptstyle \al$} ($(v16)!0.7!(v17)$);
  % J: face (6,9,13,10), downward.
  \draw[twocell] ($(v9)!0.4!(v10)$)
    -- node[midway,auto] {$\scriptstyle \theta$} ($(v9)!0.6!(v10)$);
  % K: face (3,6,10,7), down-left.
  \draw[twocell] ($(v6)!0.3!(v7)$)
    -- node[midway,auto] {$\scriptstyle \brg$} ($(v6)!0.7!(v7)$);
  % L: face (10,13,17,14), down-right.
  \draw[twocell] ($(v13)!0.3!(v14)$)
    -- node[midway,auto] {$\scriptstyle \brg$} ($(v13)!0.7!(v14)$);
\end{tikzpicture}}}  \\
   =\;&\vcenter{\hbox{\begin{tikzpicture}[
  x=0.8cm, y=0.8cm,
  >=stealth,
  line width=0.5pt,
  every node/.style={inner sep=3pt},
  shorten <=1pt, shorten >=1pt,
  twocell/.style={
    double, double distance=1.2pt,
    -implies,
    shorten <=0pt, shorten >=0pt,
    baseline=(v0.base)
  }
]
  \pgfmathsetmacro{\hexh}{sqrt(3)}

  % Top row
  \node (v4)  at (-2,{2*\hexh}) {$A$};
  \node (v8)  at ( 0,{2*\hexh}) {$A^2$};
  \node (v11) at ( 2,{2*\hexh}) {$A$};

  % Upper row
  \node (v1)  at (-3,\hexh) {$A^2$};
  \node (v5)  at (-1,\hexh) {$A^3$};
  \node (v12) at ( 1,\hexh) {$A^3$};
  \node (v15) at ( 3,\hexh) {$A^2$};

  % Middle row: vertex 9 is the origin.
  \node (v0)  at (-4,0) {$A$};
  \node (v2)  at (-2,0) {$A^3$};
  \node (v9)  at ( 0,0) {$A^4$};
  \node (v16) at ( 2,0) {$A^2$};
  \node (v18) at ( 4,0) {$A$};

  % Lower row
  \node (v3)  at (-3,{-\hexh}) {$A^2$};
  \node (v6)  at (-1,{-\hexh}) {$A^3$};
  \node (v13) at ( 1,{-\hexh}) {$A^3$};
  \node (v17) at ( 3,{-\hexh}) {$A^2$};

  % Bottom row
  \node (v7)  at (-2,{-2*\hexh}) {$A$};
  \node (v10) at ( 0,{-2*\hexh}) {$A^2$};
  \node (v14) at ( 2,{-2*\hexh}) {$A$};

\draw[->] (v0)  -- node[midway,auto] {$\scriptstyle \D$}  (v1);
  \draw[->] (v1)  -- node[midway,auto,sloped,swap] {$\scriptstyle A.\D$}  (v2);
  \draw[->] (v0)  -- node[midway,auto,swap] {$\scriptstyle \D$}  (v3);
  \draw[->] (v1)  -- node[midway,auto] {$\scriptstyle m$}  (v4);
  \draw[->] (v1)  -- node[midway,auto,sloped] {$\scriptstyle A.\D$}  (v5);
  \draw[->] (v2)  -- node[midway,auto,sloped] {$\scriptstyle A.\D.A$}  (v9);
  \draw[->] (v3)  -- node[midway,auto,sloped] {$\scriptstyle \D.A$}  (v2);
  \draw[->] (v3)  -- node[midway,auto,sloped,swap] {$\scriptstyle \D.A$}  (v6);
  \draw[->] (v3)  -- node[midway,auto,swap] {$\scriptstyle m$}  (v7);
  \draw[->] (v4)  -- node[midway,auto,sloped] {$\scriptstyle \D$}  (v8);
  \draw[->] (v5)  -- node[midway,auto,sloped,swap] {$\scriptstyle m.A$}  (v8);
  \draw[->] (v5)  -- node[midway,auto,sloped] {$\scriptstyle A^2.\D$}  (v9);
  \draw[->] (v6)  -- node[midway,auto,sloped] {$\scriptstyle \D.A^2$}  (v9);
  \draw[->] (v6)  -- node[midway,auto,sloped] {$\scriptstyle A.m$}  (v10);
  \draw[->] (v7)  -- node[midway,auto,swap] {$\scriptstyle \D$}  (v10);
  \draw[->] (v8)  -- node[midway,auto,sloped] {$\scriptstyle m$}  (v11);
  \draw[->] (v8)  -- node[midway,auto,sloped] {$\scriptstyle A.\D$}  (v12);
  \draw[->] (v9)  -- node[midway,auto,sloped,swap] {$\scriptstyle m.A^2$}  (v12);
  \draw[->] (v9)  -- node[midway,auto,sloped,swap] {$\scriptstyle A^2.m$}  (v13);
  \draw[->] (v10) -- node[midway,auto,sloped,swap] {$\scriptstyle \D.A$}  (v13);
  \draw[->] (v10) -- node[midway,auto,swap] {$\scriptstyle m$}  (v14);
  \draw[->] (v11) -- node[midway,auto] {$\scriptstyle \D$}  (v15);
  \draw[->] (v12) -- node[midway,auto,sloped] {$\scriptstyle m.A$}  (v15);
  \draw[->] (v12) -- node[midway,auto,sloped] {$\scriptstyle A.m$}  (v16);
  \draw[->] (v13) -- node[midway,auto,sloped] {$\scriptstyle m.A$}  (v16);
  \draw[->] (v13) -- node[midway,auto,sloped] {$\scriptstyle A.m$}  (v17);
  \draw[->] (v14) -- node[midway,auto,swap] {$\scriptstyle \D$} (v17);
  \draw[->] (v15) -- node[midway,auto] {$\scriptstyle m$} (v18);
  \draw[->] (v16) -- node[midway,auto,sloped] {$\scriptstyle m$} (v18);
  \draw[->] (v17) -- node[midway,auto,swap] {$\scriptstyle m$} (v18);

  % Double arrows
  % A: face (1,4,8,5), down-right.
  \draw[twocell] ($(v4)!0.3!(v5)$)
    -- node[midway,auto] {$\scriptstyle \gm$} ($(v4)!0.7!(v5)$);
  % B: face (8,11,15,12), down-left.
  \draw[twocell] ($(v11)!0.3!(v12)$)
    -- node[midway,auto,swap] {$\scriptstyle \gm$} ($(v11)!0.7!(v12)$);
  % C: face (5,8,12,9), downward.
  \draw[twocell] ($(v8)!0.4!(v9)$)
    -- node[midway,auto] {$\scriptstyle \theta$} ($(v8)!0.6!(v9)$);
  % D: face (0,1,5,2), down-right.
  \draw[twocell] ($(v2)!0.3!(v6)$)
    -- node[midway,auto, sloped,swap] {$\scriptstyle \bt.A$} ($(v2)!0.7!(v6)$);
  % E: face (12,15,18,16), down-left.
  \draw[twocell] ($(v15)!0.3!(v16)$)
    -- node[midway,auto,swap] {$\scriptstyle \al$} ($(v15)!0.7!(v16)$);
  % F: face (2,5,9,6), downward.
  \draw[twocell] ($(v1)!0.4!(v3)$)
    -- node[midway,auto,swap] {$\scriptstyle \bt$} ($(v1)!0.6!(v3)$);
  % G: face (9,12,16,13), downward.
  \draw[twocell] ($(v12)!0.4!(v13)$)
    -- node[midway,auto,swap] {$\scriptstyle \theta$} ($(v12)!0.6!(v13)$);
  % H: face (0,2,6,3), down-left.
  \draw[twocell] ($(v5)!0.3!(v2)$)
    -- node[midway,auto,sloped] {$\scriptstyle A.\bt$} ($(v5)!0.7!(v2)$);
  % I: face (13,16,18,17), down-right.
  \draw[twocell] ($(v16)!0.3!(v17)$)
    -- node[midway,auto] {$\scriptstyle \al$} ($(v16)!0.7!(v17)$);
  % J: face (6,9,13,10), downward.
  \draw[twocell] ($(v9)!0.4!(v10)$)
    -- node[midway,auto] {$\scriptstyle \theta$} ($(v9)!0.6!(v10)$);
  % K: face (3,6,10,7), down-left.
  \draw[twocell] ($(v6)!0.3!(v7)$)
    -- node[midway,auto] {$\scriptstyle \brg$} ($(v6)!0.7!(v7)$);
  % L: face (10,13,17,14), down-right.
  \draw[twocell] ($(v13)!0.3!(v14)$)
    -- node[midway,auto] {$\scriptstyle \brg$} ($(v13)!0.7!(v14)$);
\end{tikzpicture}}} \;=\;\vcenter{\hbox{\begin{tikzpicture}[
  x=0.8cm, y=0.8cm,
  >=stealth,
  line width=0.5pt,
  every node/.style={inner sep=3pt},
  shorten <=1pt, shorten >=1pt,
  twocell/.style={
    double, double distance=1.2pt,
    -implies,
    shorten <=0pt, shorten >=0pt,
    baseline=(v0.base)
  }
]
  \pgfmathsetmacro{\hexh}{sqrt(3)}

  % Top row
  \node (v4)  at (-2,{2*\hexh}) {$A$};
  \node (v8)  at ( 0,{2*\hexh}) {$A^2$};
  \node (v11) at ( 2,{2*\hexh}) {$A$};

  % Upper row
  \node (v1)  at (-3,\hexh) {$A^2$};
  \node (v5)  at (-1,\hexh) {$A^2$};
  \node (v12) at ( 1,\hexh) {$A^3$};
  \node (v15) at ( 3,\hexh) {$A^2$};

  % Middle row: vertex 9 is the origin.
  \node (v0)  at (-4,0) {$A$};
  \node (v2)  at (-2,0) {$A^3$};
  \node (v9)  at ( 0,0) {$A^4$};
  \node (v16) at ( 2,0) {$A^2$};
  \node (v18) at ( 4,0) {$A$};

  % Lower row
  \node (v3)  at (-3,{-\hexh}) {$A^2$};
  \node (v6)  at (-1,{-\hexh}) {$A^2$};
  \node (v13) at ( 1,{-\hexh}) {$A^3$};
  \node (v17) at ( 3,{-\hexh}) {$A^2$};

  % Bottom row
  \node (v7)  at (-2,{-2*\hexh}) {$A$};
  \node (v10) at ( 0,{-2*\hexh}) {$A^2$};
  \node (v14) at ( 2,{-2*\hexh}) {$A$};

  % Single arrows. 
\draw[->] (v0)  -- node[midway,auto] {$\scriptstyle \D$}  (v1);
  \draw[->] (v1)  -- node[midway,auto,sloped,swap] {$\scriptstyle A.\D$}  (v2);
  \draw[->] (v0)  -- node[midway,auto,swap] {$\scriptstyle \D$}  (v3);
  \draw[->] (v1)  -- node[midway,auto] {$\scriptstyle m$}  (v4);
  \draw[->] (v2)  -- node[midway,auto,sloped] {$\scriptstyle m.A$}  (v5);
  \draw[->] (v2)  -- node[midway,auto,sloped] {$\scriptstyle A.\D.A$}  (v9);
  \draw[->] (v3)  -- node[midway,auto,sloped] {$\scriptstyle \D.A$}  (v2);
  \draw[->] (v2)  -- node[midway,auto,sloped,swap] {$\scriptstyle A.m$}  (v6);
  \draw[->] (v3)  -- node[midway,auto,swap] {$\scriptstyle m$}  (v7);
  \draw[->] (v4)  -- node[midway,auto,sloped] {$\scriptstyle \D$}  (v8);
  \draw[->] (v4)  -- node[midway,auto,sloped,swap] {$\scriptstyle \D$}  (v5);
  \draw[->] (v5)  -- node[midway,auto,sloped] {$\scriptstyle \D.A$}  (v12);
  \draw[->] (v7)  -- node[midway,auto,sloped] {$\scriptstyle \D$}  (v6);
  \draw[->] (v6)  -- node[midway,auto,sloped,swap] {$\scriptstyle A.\D$}  (v13);
  \draw[->] (v7)  -- node[midway,auto,swap] {$\scriptstyle \D$}  (v10);
  \draw[->] (v8)  -- node[midway,auto,sloped] {$\scriptstyle m$}  (v11);
  \draw[->] (v8)  -- node[midway,auto,sloped] {$\scriptstyle A.\D$}  (v12);
  \draw[->] (v9)  -- node[midway,auto,sloped,swap] {$\scriptstyle m.A^2$}  (v12);
  \draw[->] (v9)  -- node[midway,auto,sloped,swap] {$\scriptstyle A^2.m$}  (v13);
  \draw[->] (v10) -- node[midway,auto,sloped,swap] {$\scriptstyle \D.A$}  (v13);
  \draw[->] (v10) -- node[midway,auto,swap] {$\scriptstyle m$}  (v14);
  \draw[->] (v11) -- node[midway,auto] {$\scriptstyle \D$}  (v15);
  \draw[->] (v12) -- node[midway,auto,sloped] {$\scriptstyle m.A$}  (v15);
  \draw[->] (v12) -- node[midway,auto,sloped] {$\scriptstyle A.m$}  (v16);
  \draw[->] (v13) -- node[midway,auto,sloped] {$\scriptstyle m.A$}  (v16);
  \draw[->] (v13) -- node[midway,auto,sloped] {$\scriptstyle A.m$}  (v17);
  \draw[->] (v14) -- node[midway,auto,swap] {$\scriptstyle \D$} (v17);
  \draw[->] (v15) -- node[midway,auto] {$\scriptstyle m$} (v18);
  \draw[->] (v16) -- node[midway,auto,sloped] {$\scriptstyle m$} (v18);
  \draw[->] (v17) -- node[midway,auto,swap] {$\scriptstyle m$} (v18);

  % Double arrows
  % A: face (1,4,8,5), down-right.
  \draw[twocell] ($(v5)!0.3!(v9)$)
    -- node[midway,auto,sloped] {$\scriptstyle \gm.A$} ($(v5)!0.7!(v9)$);
  % B: face (8,11,15,12), down-left.
  \draw[twocell] ($(v11)!0.3!(v12)$)
    -- node[midway,auto,swap] {$\scriptstyle \gm$} ($(v11)!0.7!(v12)$);
  % C: face (5,8,12,9), downward.
  \draw[twocell] ($(v4)!0.4!(v2)$)
    -- node[midway,auto,swap] {$\scriptstyle \gm$} ($(v4)!0.6!(v2)$);
  % D: face (0,1,5,2), down-right.
  \draw[twocell] ($(v6)!0.3!(v10)$)
    -- node[midway,auto,swap] {$\scriptstyle \bt$} ($(v6)!0.7!(v10)$);
  % E: face (12,15,18,16), down-left.
  \draw[twocell] ($(v15)!0.3!(v16)$)
    -- node[midway,auto,swap] {$\scriptstyle \al$} ($(v15)!0.7!(v16)$);
  % F: face (2,5,9,6), downward.
  \draw[twocell] ($(v1)!0.4!(v3)$)
    -- node[midway,auto,swap] {$\scriptstyle \bt$} ($(v1)!0.6!(v3)$);
  % G: face (9,12,16,13), downward.
  \draw[twocell] ($(v12)!0.4!(v13)$)
    -- node[midway,auto,swap] {$\scriptstyle \theta$} ($(v12)!0.6!(v13)$);
  % H: face (0,2,6,3), down-left.
  \draw[twocell] ($(v8)!0.3!(v5)$)
    -- node[midway,auto,swap] {$\scriptstyle \bt$} ($(v8)!0.7!(v5)$);
  % I: face (13,16,18,17), down-right.
  \draw[twocell] ($(v16)!0.3!(v17)$)
    -- node[midway,auto] {$\scriptstyle \al$} ($(v16)!0.7!(v17)$);
  % J: face (6,9,13,10), downward.
  \draw[twocell] ($(v2)!0.4!(v7)$)
    -- node[midway,auto] {$\scriptstyle \brg$} ($(v2)!0.6!(v7)$);
  % K: face (3,6,10,7), down-left.
  \draw[twocell] ($(v9)!0.3!(v6)$)
    -- node[midway,auto,sloped] {$\scriptstyle A.\brg$} ($(v9)!0.7!(v6)$);
  % L: face (10,13,17,14), down-right.
  \draw[twocell] ($(v13)!0.3!(v14)$)
    -- node[midway,auto] {$\scriptstyle \brg$} ($(v13)!0.7!(v14)$);
\end{tikzpicture}}}  \\
   =\;&\vcenter{\hbox{\begin{tikzpicture}[
  x=0.8cm, y=0.8cm,
  >=stealth,
  line width=0.5pt,
  every node/.style={inner sep=3pt},
  shorten <=1pt, shorten >=1pt,
  twocell/.style={
    double, double distance=1.2pt,
    -implies,
    shorten <=0pt, shorten >=0pt,
    baseline=(v0.base)
  }
]
  \pgfmathsetmacro{\hexh}{sqrt(3)}

  % Top row
  \node (v4)  at (-2,{2*\hexh}) {$A$};
  \node (v8)  at ( 0,{2*\hexh}) {$A^2$};
  \node (v11) at ( 2,{2*\hexh}) {$A$};

  % Upper row
  \node (v1)  at (-3,\hexh) {$A^2$};
  \node (v5)  at (-1,\hexh) {$A^2$};
  \node (v12) at ( 1,\hexh) {$A^3$};
  \node (v15) at ( 3,\hexh) {$A^2$};

  % Middle row: vertex 9 is the origin.
  \node (v0)  at (-4,0) {$A$};
  \node (v2)  at (-2,0) {$A^3$};
  \node (v9)  at ( 0,0) {$A$};
  \node (v16) at ( 2,0) {$A^2$};
  \node (v18) at ( 4,0) {$A$};

  % Lower row
  \node (v3)  at (-3,{-\hexh}) {$A^2$};
  \node (v6)  at (-1,{-\hexh}) {$A^2$};
  \node (v13) at ( 1,{-\hexh}) {$A^3$};
  \node (v17) at ( 3,{-\hexh}) {$A^2$};

  % Bottom row
  \node (v7)  at (-2,{-2*\hexh}) {$A$};
  \node (v10) at ( 0,{-2*\hexh}) {$A^2$};
  \node (v14) at ( 2,{-2*\hexh}) {$A$};

  % Single arrows
\draw[->] (v0)  -- node[midway,auto] {$\scriptstyle \D$}  (v1);
  \draw[->] (v1)  -- node[midway,auto,sloped,swap] {$\scriptstyle A.\D$}  (v2);
  \draw[->] (v0)  -- node[midway,auto,swap] {$\scriptstyle \D$}  (v3);
  \draw[->] (v1)  -- node[midway,auto] {$\scriptstyle m$}  (v4);
  \draw[->] (v2)  -- node[midway,auto,sloped] {$\scriptstyle m.A$}  (v5);
  \draw[->] (v5)  -- node[midway,auto,sloped,swap] {$\scriptstyle m$}  (v9);
  \draw[->] (v3)  -- node[midway,auto,sloped] {$\scriptstyle \D.A$}  (v2);
  \draw[->] (v2)  -- node[midway,auto,sloped,swap] {$\scriptstyle A.m$}  (v6);
  \draw[->] (v3)  -- node[midway,auto,swap] {$\scriptstyle m$}  (v7);
  \draw[->] (v4)  -- node[midway,auto,sloped] {$\scriptstyle \D$}  (v8);
  \draw[->] (v4)  -- node[midway,auto,sloped,swap] {$\scriptstyle \D$}  (v5);
  \draw[->] (v5)  -- node[midway,auto,sloped] {$\scriptstyle \D.A$}  (v12);
  \draw[->] (v7)  -- node[midway,auto,sloped] {$\scriptstyle \D$}  (v6);
  \draw[->] (v6)  -- node[midway,auto,sloped,swap] {$\scriptstyle A.\D$}  (v13);
  \draw[->] (v7)  -- node[midway,auto,swap] {$\scriptstyle \D$}  (v10);
  \draw[->] (v8)  -- node[midway,auto,sloped] {$\scriptstyle m$}  (v11);
  \draw[->] (v8)  -- node[midway,auto,sloped] {$\scriptstyle A.\D$}  (v12);
  \draw[->] (v9)  -- node[midway,auto,sloped] {$\scriptstyle \D$}  (v16);
  \draw[->] (v6)  -- node[midway,auto,sloped] {$\scriptstyle m$}  (v9);
  \draw[->] (v10) -- node[midway,auto,sloped,swap] {$\scriptstyle \D.A$}  (v13);
  \draw[->] (v10) -- node[midway,auto,swap] {$\scriptstyle m$}  (v14);
  \draw[->] (v11) -- node[midway,auto] {$\scriptstyle \D$}  (v15);
  \draw[->] (v12) -- node[midway,auto,sloped] {$\scriptstyle m.A$}  (v15);
  \draw[->] (v12) -- node[midway,auto,sloped,swap] {$\scriptstyle A.m$}  (v16);
  \draw[->] (v13) -- node[midway,auto,sloped] {$\scriptstyle m.A$}  (v16);
  \draw[->] (v13) -- node[midway,auto,sloped] {$\scriptstyle A.m$}  (v17);
  \draw[->] (v14) -- node[midway,auto,swap] {$\scriptstyle \D$} (v17);
  \draw[->] (v15) -- node[midway,auto] {$\scriptstyle m$} (v18);
  \draw[->] (v16) -- node[midway,auto,sloped] {$\scriptstyle m$} (v18);
  \draw[->] (v17) -- node[midway,auto,swap] {$\scriptstyle m$} (v18);

  % Double arrows
  % A: face (1,4,8,5), down-right.
  \draw[twocell] ($(v9)!0.3!(v13)$)
    -- node[midway,auto,swap] {$\scriptstyle \gm$} ($(v9)!0.7!(v13)$);
  % B: face (8,11,15,12), down-left.
  \draw[twocell] ($(v11)!0.3!(v12)$)
    -- node[midway,auto,swap] {$\scriptstyle \gm$} ($(v11)!0.7!(v12)$);
  % C: face (5,8,12,9), downward.
  \draw[twocell] ($(v4)!0.4!(v2)$)
    -- node[midway,auto,swap] {$\scriptstyle \gm$} ($(v4)!0.6!(v2)$);
  % D: face (0,1,5,2), down-right.
  \draw[twocell] ($(v6)!0.3!(v10)$)
    -- node[midway,auto,swap] {$\scriptstyle \bt$} ($(v6)!0.7!(v10)$);
  % E: face (12,15,18,16), down-left.
  \draw[twocell] ($(v15)!0.3!(v16)$)
    -- node[midway,auto,swap] {$\scriptstyle \al$} ($(v15)!0.7!(v16)$);
  % F: face (2,5,9,6), downward.
  \draw[twocell] ($(v1)!0.4!(v3)$)
    -- node[midway,auto,swap] {$\scriptstyle \beta$} ($(v1)!0.6!(v3)$);
  % G: face (9,12,16,13), downward.
  \draw[twocell] ($(v5)!0.4!(v6)$)
    -- node[midway,auto,swap] {$\scriptstyle \al$} ($(v5)!0.6!(v6)$);
  % H: face (0,2,6,3), down-left.
  \draw[twocell] ($(v8)!0.3!(v5)$)
    -- node[midway,auto,swap] {$\scriptstyle \bt$} ($(v8)!0.7!(v5)$);
  % I: face (13,16,18,17), down-right.
  \draw[twocell] ($(v16)!0.3!(v17)$)
    -- node[midway,auto] {$\scriptstyle \al$} ($(v16)!0.7!(v17)$);
  % J: face (6,9,13,10), downward.
  \draw[twocell] ($(v2)!0.4!(v7)$)
    -- node[midway,auto,swap] {$\scriptstyle \brg$} ($(v2)!0.6!(v7)$);
  % K: face (3,6,10,7), down-left.
  \draw[twocell] ($(v12)!0.3!(v9)$)
    -- node[midway,auto,swap] {$\scriptstyle \brg$} ($(v12)!0.7!(v9)$);
  % L: face (10,13,17,14), down-right.
  \draw[twocell] ($(v13)!0.3!(v14)$)
    -- node[midway,auto] {$\scriptstyle \brg$} ($(v13)!0.7!(v14)$);
\end{tikzpicture}}} 
 \end{align*}
\fi
\end{proof}

\begin{ex}\label{ex:braid}
Let $C$ be a strict braided monoidal category, and denote its braiding by $b_{X,Y}:X\otimes Y\rightarrow Y\otimes X$ for $X,Y\in C$. 

We consider the semistrict monoidal $2$-category $\cC=\Sigma C$ defined by the delooping of $C$ with the single object $*\in \cC$ (see Subsection~\ref{sec:semi}). Note that the braiding of $C$ gives the coherent $2$-isomorphism $\theta$.

Given any objects $X,Y\in C$, we obtain a Frobenius $2$-algebra $$(*,X,Y,\al=b_{X,X},\bt=b_{Y,Y},\gm=b_{X,Y},\brg=b_{Y,X})$$
in $\cC$. In fact, the eight axioms \eqref{diag:1}--\eqref{diag:8} are precisely the colored Yang-Baxter equations
$$(\id_S \otimes b_{P,Q})(b_{P,S}\otimes \id_Q)(\id_P \otimes b_{Q,S})=(b_{Q,S}\otimes \id_P)(id_Q \otimes b_{P,S})(b_{P,Q}\otimes \id_S)$$
for $P,Q,S\in \{X,Y\}$. 

In this case, the $R$-matrix given by \eqref{eq:R} is the braiding $b_{Z,Z}$ on $Z:=Y\otimes X$ as follows.
\begin{align*}
    R&=(\id_Y \otimes\brg \otimes \id_X) (\bt \otimes \al)(\id_Y \otimes\gm\otimes \id_X) \\
    &=(\id_Y \otimes b_{Y,X} \otimes \id_X) (\id_Y\otimes \id_Y\otimes b_{X,X}) (b_{Y,Y} \otimes \id_X\otimes \id_X)(\id_Y \otimes b_{X,Y}\otimes \id_X) \\
    &=(\id_Y\otimes b_{Z,X})(b_{Z,Y}\otimes \id_X) \\
    &=b_{Z,Z}.
\end{align*}
\end{ex}

\section{Hopf bimodules}\label{sec:hopf bimod}
Let $H=(H,m,\D, 1, \varepsilon, S)$ be a Hopf algebra over a field $\mathbb{K}$ with invertible antipode. In this paper, we adopt Sweedler's notation
$$\D (x)=x_{(1)}\otimes x_{(2)}$$
for $x\in H$, where we omit summation.

If $(V,\lambda)$ is a left $H$-comodule (resp. $(V,\rho)$ is a right $H$-comodule), we use
\begin{align*}
    \lambda(v)=v_{(-1)}\otimes v_{(0)}\in H\otimes V,\quad
    \left(\mathrm{resp.}\; \rho(v)=v_{(0)}\otimes v_{(1)}\in V\otimes H\right).
\end{align*}

For $H$-modules $V$ and $W$, the tensor product $V\otimes W$
carries the $H$-action induced by the coproduct $\Delta$.
Dually, for left $H$-comodules $(V,\lambda)$ and $(W,\lambda')$,
the tensor product $V\otimes W$ carries the left $H$-coaction
$\lambda\star\lambda'$ defined by
$$\left(\lambda\star \lambda^{\prime}\right)(v\otimes w)=v_{(-1)}w_{(-1)}\otimes v_{(0)}\otimes w_{(0)}\in H\otimes V\otimes W,$$
and similarly for right $H$-comodules.
\subsection{Hopf bimodules and Yetter-Drinfeld modules}\label{hopf bimod}

In this subsection, we introduce $H$-Hopf bimodules and Yetter-Drinfeld modules over $H$. Our main references are \cite{AS,FJ,R}.
 
\begin{dfn}
    An $H$-Hopf bimodule is a vector space $V$ equipped simultaneously with $H$-bimodule and $H$-bicomodule structures such that the left coaction $\lambda:V\rightarrow H\otimes V$ and the right coaction $\rho:V\rightarrow V\otimes H$ are $H$-bimodule homomorphisms.
\end{dfn}

Let $\cM(H)$ be the category of $H$-Hopf bimodules (whose morphisms are $H$-bimodule and $H$-bicomodule homomorphisms). This category is monoidal under the relative tensor
product over $H$; the left and right actions are induced by the outer actions, and
the coactions are induced by the tensor product $\star$ of coactions. Woronowicz showed in \cite{W} that the monoidal category $\cM(H)$ is braided.

Next we recall the definition of Yetter-Drinfeld modules over $H$.
\begin{dfn}
    A left Yetter-Drinfeld module over $H$ is a left $H$-module $V$ with a left $H$-comodule structure such that whenever $x\in H$ and $v\in V$,
    $$x_{(1)}v_{(-1)}\otimes x_{(2)}v_{(0)}=\left(x_{(1)}v \right)_{(-1)}x_{(2)}\otimes \left(x_{(1)}v \right)_{(0)}.$$
\end{dfn}
Let $\cY\cD(H)$ be the category of Yetter-Drinfeld modules over $H$ (whose morphisms are left $H$-module and comodule homomorphisms). Together with the usual tensor product of left $H$-module and comodule, the category $\cY\cD(H)$ is monoidal. Moreover, it is braided with the braiding
$$\mathcal{R}_{V,W} :V\otimes W\rightarrow W\otimes V,\quad \mathcal{R}_{V,W}(v\otimes w)=v_{(-1)}w\otimes v_{(0)}.$$

\begin{ex}\label{exA}
    It is known that $H$ is a Yetter-Drinfeld module over itself given by
    $$x\triangleright a:=x_{(1)}aS(x_{(2)}),\quad\lambda(a):=\D(a),$$
    for any $x,a\in H$. Then the standard braiding is given by
    \begin{align}\label{universal R}
        \mathcal{R}=\mathcal{R}_{H,H}:H\otimes H\rightarrow H\otimes H, \quad\mathcal{R}(x\otimes y)=x_{(1)}yS(x_{(2)})\otimes x_{(3)}.
    \end{align}
\end{ex}
Finally, we recall the equivalence between the category of $H$-Hopf bimodules and the category of Yetter-Drinfeld modules over $H$ as follows.
\begin{thm}[\cite{Sch, R}]\label{Rosso}
    Given a Hopf algebra $H$, there is an equivalence of braided monoidal categories $\cM(H)\xrightarrow{\cong} \cY\cD(H)$ sending an $H$-Hopf bimodule $V\in \cM(H)$ to the Yetter-Drinfeld module
    $$V^{coH}:=\{v\in V:\; \rho(v)=v\otimes 1\}\in \cY\cD(H),$$
    where the left $H$-action on $V^{\mathrm{co}H}$ is the adjoint
$
  a\triangleright v=a_{(1)}vS(a_{(2)}),
$
 and its left coaction is the restriction of $V$.
\end{thm}

\subsection{Frobenius $2$-algebras in the Morita $2$-category}
Let $H=(H,m,\D,1,\varepsilon,S)$ be a Hopf algebra over a field $\mathbb{K}$ with invertible antipode. We work in the Morita $2$-category $\cC=\mathbf{Mor}\left(\mathbf{Alg}_{\mathbb{K}}\right)$ defined in Subsection~\ref{sec:semi}. The following data define a Frobenius $2$-algebra
$(A,M,L,\al,\bt,\gm,\brg)$ in $\cC$:
\begin{itemize}
    \item The underlying algebra $A=(H,m,1) \in \cC.$
    \item The $(A^2, A)$-bimodule $M=(A\otimes A)_{\D}$ with the right $A$-action given by the coproduct $\D:A\rightarrow A^2$ and the left regular $A^2$-action. We represent elements of $M$ diagrammatically as  
    \iflatexml
\begin{center}
  \includegraphics{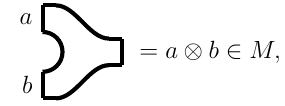}
\end{center}
\else
\begin{equation*}
    \vcenter{\hbox{
\begin{tikzpicture}[scale=0.45, line width=2pt]
 \coordinate (a1) at (0,0);
\coordinate (a2) at (0,1);
\coordinate (b1) at (0,2.5);
\coordinate (b2) at (0,3.5);
\coordinate (c1) at (3,1.25);
\coordinate (c2) at (3,2.25);

\draw (a1) -- (a2);
\draw (b1) -- (b2);
\draw (c1) -- (c2);

%円弧
\draw (0,1) arc[start angle=-90, end angle=90, radius=0.75];

%上の曲線(b2)--(c2)
\draw (b2) -- ($(b2)+(0.5,0)$);
\draw ($(c2)-(0.5,0)$) -- (c2);
\draw ($(b2)+(0.5,0)$) .. controls ($(b2)+(1.25,0)$) and ($(c2)-(1.25,0)$) ..($(c2)-(0.5,0)$);

%下の曲線(a1)--(c1)
\draw (a1) -- ($(a1)+(0.5,0)$);
\draw ($(c1)-(0.5,0)$) -- (c1);
\draw ($(a1)+(0.5,0)$) .. controls ($(a1)+(1.25,0)$) and ($(c1)-(1.25,0)$) ..($(c1)-(0.5,0)$);
\node[left] at (0,3) {$a$};
\node[left] at (0,0.5) {$b$};
\end{tikzpicture}
}} =a\otimes b\in M,
    \end{equation*}
\fi
    and then the right action is visualized by
    \iflatexml
\begin{center}
  \includegraphics{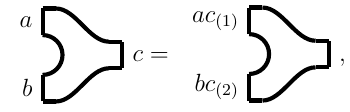}
\end{center}
\else
\begin{equation*}
        \vcenter{\hbox{
\begin{tikzpicture}[scale=0.45, line width=2pt]
 \coordinate (a1) at (0,0);
\coordinate (a2) at (0,1);
\coordinate (b1) at (0,2.5);
\coordinate (b2) at (0,3.5);
\coordinate (c1) at (3,1.25);
\coordinate (c2) at (3,2.25);

\draw (a1) -- (a2);
\draw (b1) -- (b2);
\draw (c1) -- (c2);

%円弧
\draw (0,1) arc[start angle=-90, end angle=90, radius=0.75];

%上の曲線(b2)--(c2)
\draw (b2) -- ($(b2)+(0.5,0)$);
\draw ($(c2)-(0.5,0)$) -- (c2);
\draw ($(b2)+(0.5,0)$) .. controls ($(b2)+(1.25,0)$) and ($(c2)-(1.25,0)$) ..($(c2)-(0.5,0)$);

%下の曲線(a1)--(c1)
\draw (a1) -- ($(a1)+(0.5,0)$);
\draw ($(c1)-(0.5,0)$) -- (c1);
\draw ($(a1)+(0.5,0)$) .. controls ($(a1)+(1.25,0)$) and ($(c1)-(1.25,0)$) ..($(c1)-(0.5,0)$);
\node[left] at (0,3) {$a$};
\node[left] at (0,0.5) {$b$};
\end{tikzpicture}
}} c=\vcenter{\hbox{
\begin{tikzpicture}[scale=0.45, line width=2pt]
 \coordinate (a1) at (0,0);
\coordinate (a2) at (0,1);
\coordinate (b1) at (0,2.5);
\coordinate (b2) at (0,3.5);
\coordinate (c1) at (3,1.25);
\coordinate (c2) at (3,2.25);

\draw (a1) -- (a2);
\draw (b1) -- (b2);
\draw (c1) -- (c2);

%円弧
\draw (0,1) arc[start angle=-90, end angle=90, radius=0.75];

%上の曲線(b2)--(c2)
\draw (b2) -- ($(b2)+(0.5,0)$);
\draw ($(c2)-(0.5,0)$) -- (c2);
\draw ($(b2)+(0.5,0)$) .. controls ($(b2)+(1.25,0)$) and ($(c2)-(1.25,0)$) ..($(c2)-(0.5,0)$);

%下の曲線(a1)--(c1)
\draw (a1) -- ($(a1)+(0.5,0)$);
\draw ($(c1)-(0.5,0)$) -- (c1);
\draw ($(a1)+(0.5,0)$) .. controls ($(a1)+(1.25,0)$) and ($(c1)-(1.25,0)$) ..($(c1)-(0.5,0)$);
\node[left] at (0,3) {$ac_{(1)}$};
\node[left] at (0,0.5) {$bc_{(2)}$};
\end{tikzpicture}
}},
    \end{equation*}
\fi
    where $c\in A$. Similarly, we obtain the $(A,A^2)$-bimodule $L={}_{\D}(A\otimes A)$ and
    \iflatexml
\begin{center}
  \includegraphics{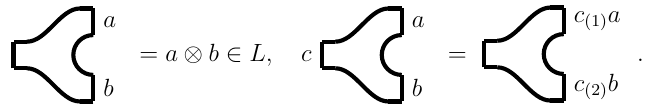}
\end{center}
\else
\begin{equation*}
       \vcenter{\hbox{
\begin{tikzpicture}[scale=0.45, line width=2pt, xscale=-1]
 \coordinate (a1) at (0,0);
\coordinate (a2) at (0,1);
\coordinate (b1) at (0,2.5);
\coordinate (b2) at (0,3.5);
\coordinate (c1) at (3,1.25);
\coordinate (c2) at (3,2.25);

\draw (a1) -- (a2);
\draw (b1) -- (b2);
\draw (c1) -- (c2);

%円弧
\draw (0,1) arc[start angle=-90, end angle=90, radius=0.75];

%上の曲線(b2)--(c2)
\draw (b2) -- ($(b2)+(0.5,0)$);
\draw ($(c2)-(0.5,0)$) -- (c2);
\draw ($(b2)+(0.5,0)$) .. controls ($(b2)+(1.25,0)$) and ($(c2)-(1.25,0)$) ..($(c2)-(0.5,0)$);

%下の曲線(a1)--(c1)
\draw (a1) -- ($(a1)+(0.5,0)$);
\draw ($(c1)-(0.5,0)$) -- (c1);
\draw ($(a1)+(0.5,0)$) .. controls ($(a1)+(1.25,0)$) and ($(c1)-(1.25,0)$) ..($(c1)-(0.5,0)$);
\node[right] at (0,3) {$a$};
\node[right] at (0,0.5) {$b$};
\end{tikzpicture}
}} =a\otimes b\in L,\quad
       c\vcenter{\hbox{
\begin{tikzpicture}[scale=0.45, line width=2pt, xscale=-1]
 \coordinate (a1) at (0,0);
\coordinate (a2) at (0,1);
\coordinate (b1) at (0,2.5);
\coordinate (b2) at (0,3.5);
\coordinate (c1) at (3,1.25);
\coordinate (c2) at (3,2.25);

\draw (a1) -- (a2);
\draw (b1) -- (b2);
\draw (c1) -- (c2);

%円弧
\draw (0,1) arc[start angle=-90, end angle=90, radius=0.75];

%上の曲線(b2)--(c2)
\draw (b2) -- ($(b2)+(0.5,0)$);
\draw ($(c2)-(0.5,0)$) -- (c2);
\draw ($(b2)+(0.5,0)$) .. controls ($(b2)+(1.25,0)$) and ($(c2)-(1.25,0)$) ..($(c2)-(0.5,0)$);

%下の曲線(a1)--(c1)
\draw (a1) -- ($(a1)+(0.5,0)$);
\draw ($(c1)-(0.5,0)$) -- (c1);
\draw ($(a1)+(0.5,0)$) .. controls ($(a1)+(1.25,0)$) and ($(c1)-(1.25,0)$) ..($(c1)-(0.5,0)$);
\node[right] at (0,3) {$a$};
\node[right] at (0,0.5) {$b$};
\end{tikzpicture}
}}=\vcenter{\hbox{
\begin{tikzpicture}[scale=0.45, line width=2pt, xscale=-1]
 \coordinate (a1) at (0,0);
\coordinate (a2) at (0,1);
\coordinate (b1) at (0,2.5);
\coordinate (b2) at (0,3.5);
\coordinate (c1) at (3,1.25);
\coordinate (c2) at (3,2.25);

\draw (a1) -- (a2);
\draw (b1) -- (b2);
\draw (c1) -- (c2);

%円弧
\draw (0,1) arc[start angle=-90, end angle=90, radius=0.75];

%上の曲線(b2)--(c2)
\draw (b2) -- ($(b2)+(0.5,0)$);
\draw ($(c2)-(0.5,0)$) -- (c2);
\draw ($(b2)+(0.5,0)$) .. controls ($(b2)+(1.25,0)$) and ($(c2)-(1.25,0)$) ..($(c2)-(0.5,0)$);

%下の曲線(a1)--(c1)
\draw (a1) -- ($(a1)+(0.5,0)$);
\draw ($(c1)-(0.5,0)$) -- (c1);
\draw ($(a1)+(0.5,0)$) .. controls ($(a1)+(1.25,0)$) and ($(c1)-(1.25,0)$) ..($(c1)-(0.5,0)$);
\node[right] at (0,3) {$c_{(1)}a$};
\node[right] at (0,0.5) {$c_{(2)}b$};
\end{tikzpicture}
}}.
    \end{equation*}
\fi
   \item We have the following bimodule isomorphisms (see Lemma~\ref{lem1}).
   \begin{enumerate}
       \item The $(A^3,A)$-bimodule isomorphism
       $$\al: (M\otimes A)\otimes_{A^2}M\rightarrow (A\otimes M)\otimes_{A^2}M$$
       defined by
       \iflatexml
\begin{center}
  \includegraphics{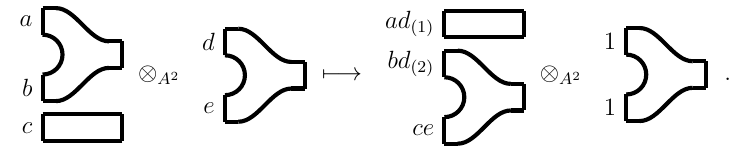}
\end{center}
\else
\begin{equation*}
           \vcenter{\hbox{
\begin{tikzpicture}[scale=0.45, line width=2pt]
\begin{scope}[shift={(0,0.5)}]
    \coordinate (x1) at (0,-1);
\coordinate (x2) at (0,-2);
\coordinate (y1) at (3,-1);
\coordinate (y2) at (3,-2);
\draw (x1) -- (x2);
\draw (y1) -- (y2);
\draw (x1) -- (y1);
\draw (y2) -- (x2);

\end{scope}
 \coordinate (a1) at (0,0);
\coordinate (a2) at (0,1);
\coordinate (b1) at (0,2.5);
\coordinate (b2) at (0,3.5);
\coordinate (c1) at (3,1.25);
\coordinate (c2) at (3,2.25);

\draw (a1) -- (a2);
\draw (b1) -- (b2);
\draw (c1) -- (c2);

%円弧
\draw (0,1) arc[start angle=-90, end angle=90, radius=0.75];

%上の曲線(b2)--(c2)
\draw (b2) -- ($(b2)+(0.5,0)$);
\draw ($(c2)-(0.5,0)$) -- (c2);
\draw ($(b2)+(0.5,0)$) .. controls ($(b2)+(1.25,0)$) and ($(c2)-(1.25,0)$) ..($(c2)-(0.5,0)$);

%下の曲線(a1)--(c1)
\draw (a1) -- ($(a1)+(0.5,0)$);
\draw ($(c1)-(0.5,0)$) -- (c1);
\draw ($(a1)+(0.5,0)$) .. controls ($(a1)+(1.25,0)$) and ($(c1)-(1.25,0)$) ..($(c1)-(0.5,0)$);
\node[left] at (0,3) {$a$};
\node[left] at (0,0.5) {$b$};
\node[left] at (0,-1) {$c$};
\end{tikzpicture}
}}\otimes_{A^2} \vcenter{\hbox{
\begin{tikzpicture}[scale=0.45, line width=2pt]
 \coordinate (a1) at (0,0);
\coordinate (a2) at (0,1);
\coordinate (b1) at (0,2.5);
\coordinate (b2) at (0,3.5);
\coordinate (c1) at (3,1.25);
\coordinate (c2) at (3,2.25);

\draw (a1) -- (a2);
\draw (b1) -- (b2);
\draw (c1) -- (c2);

%円弧
\draw (0,1) arc[start angle=-90, end angle=90, radius=0.75];

%上の曲線(b2)--(c2)
\draw (b2) -- ($(b2)+(0.5,0)$);
\draw ($(c2)-(0.5,0)$) -- (c2);
\draw ($(b2)+(0.5,0)$) .. controls ($(b2)+(1.25,0)$) and ($(c2)-(1.25,0)$) ..($(c2)-(0.5,0)$);

%下の曲線(a1)--(c1)
\draw (a1) -- ($(a1)+(0.5,0)$);
\draw ($(c1)-(0.5,0)$) -- (c1);
\draw ($(a1)+(0.5,0)$) .. controls ($(a1)+(1.25,0)$) and ($(c1)-(1.25,0)$) ..($(c1)-(0.5,0)$);
\node[left] at (0,3) {$d$};
\node[left] at (0,0.5) {$e$};
\end{tikzpicture}
}}
           \longmapsto
           \vcenter{\hbox{
\begin{tikzpicture}[scale=0.45, line width=2pt]
\begin{scope}[shift={(0,6)}]
    \coordinate (x1) at (0,-1);
\coordinate (x2) at (0,-2);
\coordinate (y1) at (3,-1);
\coordinate (y2) at (3,-2);
\draw (x1) -- (x2);
\draw (y1) -- (y2);
\draw (x1) -- (y1);
\draw (y2) -- (x2);

\end{scope}
 \coordinate (a1) at (0,0);
\coordinate (a2) at (0,1);
\coordinate (b1) at (0,2.5);
\coordinate (b2) at (0,3.5);
\coordinate (c1) at (3,1.25);
\coordinate (c2) at (3,2.25);

\draw (a1) -- (a2);
\draw (b1) -- (b2);
\draw (c1) -- (c2);

%円弧
\draw (0,1) arc[start angle=-90, end angle=90, radius=0.75];

%上の曲線(b2)--(c2)
\draw (b2) -- ($(b2)+(0.5,0)$);
\draw ($(c2)-(0.5,0)$) -- (c2);
\draw ($(b2)+(0.5,0)$) .. controls ($(b2)+(1.25,0)$) and ($(c2)-(1.25,0)$) ..($(c2)-(0.5,0)$);

%下の曲線(a1)--(c1)
\draw (a1) -- ($(a1)+(0.5,0)$);
\draw ($(c1)-(0.5,0)$) -- (c1);
\draw ($(a1)+(0.5,0)$) .. controls ($(a1)+(1.25,0)$) and ($(c1)-(1.25,0)$) ..($(c1)-(0.5,0)$);
\node[left] at (0,4.5) {$ad_{(1)}$};
\node[left] at (0,3) {$bd_{(2)}$};
\node[left] at (0,0.5) {$ce$};
\end{tikzpicture}
}}\otimes_{A^2}\vcenter{\hbox{
\begin{tikzpicture}[scale=0.45, line width=2pt]
 \coordinate (a1) at (0,0);
\coordinate (a2) at (0,1);
\coordinate (b1) at (0,2.5);
\coordinate (b2) at (0,3.5);
\coordinate (c1) at (3,1.25);
\coordinate (c2) at (3,2.25);

\draw (a1) -- (a2);
\draw (b1) -- (b2);
\draw (c1) -- (c2);

%円弧
\draw (0,1) arc[start angle=-90, end angle=90, radius=0.75];

%上の曲線(b2)--(c2)
\draw (b2) -- ($(b2)+(0.5,0)$);
\draw ($(c2)-(0.5,0)$) -- (c2);
\draw ($(b2)+(0.5,0)$) .. controls ($(b2)+(1.25,0)$) and ($(c2)-(1.25,0)$) ..($(c2)-(0.5,0)$);

%下の曲線(a1)--(c1)
\draw (a1) -- ($(a1)+(0.5,0)$);
\draw ($(c1)-(0.5,0)$) -- (c1);
\draw ($(a1)+(0.5,0)$) .. controls ($(a1)+(1.25,0)$) and ($(c1)-(1.25,0)$) ..($(c1)-(0.5,0)$);
\node[left] at (0,3) {$1$};
\node[left] at (0,0.5) {$1$};
\end{tikzpicture}

           }}.
       \end{equation*}
\fi
       \item The $(A,A^3)$-bimodule isomorphism 
       $$\bt: L\otimes_{A^2}(A\otimes L)\rightarrow L\otimes_{A^2}(L\otimes A)$$
       defined by
       \iflatexml
\begin{center}
  \includegraphics{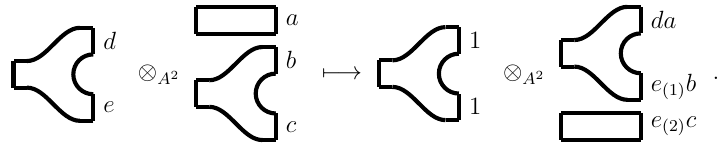}
\end{center}
\else
\begin{equation*}
           \vcenter{\hbox{
\begin{tikzpicture}[scale=0.45, line width=2pt, xscale=-1]
 \coordinate (a1) at (0,0);
\coordinate (a2) at (0,1);
\coordinate (b1) at (0,2.5);
\coordinate (b2) at (0,3.5);
\coordinate (c1) at (3,1.25);
\coordinate (c2) at (3,2.25);

\draw (a1) -- (a2);
\draw (b1) -- (b2);
\draw (c1) -- (c2);

%円弧
\draw (0,1) arc[start angle=-90, end angle=90, radius=0.75];

%上の曲線(b2)--(c2)
\draw (b2) -- ($(b2)+(0.5,0)$);
\draw ($(c2)-(0.5,0)$) -- (c2);
\draw ($(b2)+(0.5,0)$) .. controls ($(b2)+(1.25,0)$) and ($(c2)-(1.25,0)$) ..($(c2)-(0.5,0)$);

%下の曲線(a1)--(c1)
\draw (a1) -- ($(a1)+(0.5,0)$);
\draw ($(c1)-(0.5,0)$) -- (c1);
\draw ($(a1)+(0.5,0)$) .. controls ($(a1)+(1.25,0)$) and ($(c1)-(1.25,0)$) ..($(c1)-(0.5,0)$);
\node[right] at (0,3) {$d$};
\node[right] at (0,0.5) {$e$};
\end{tikzpicture}
}}\otimes_{A^2}\vcenter{\hbox{
\begin{tikzpicture}[scale=0.45, line width=2pt, xscale=-1]
\begin{scope}[shift={(0,6)}]
    \coordinate (x1) at (0,-1);
\coordinate (x2) at (0,-2);
\coordinate (y1) at (3,-1);
\coordinate (y2) at (3,-2);
\draw (x1) -- (x2);
\draw (y1) -- (y2);
\draw (x1) -- (y1);
\draw (y2) -- (x2);

\end{scope}
 \coordinate (a1) at (0,0);
\coordinate (a2) at (0,1);
\coordinate (b1) at (0,2.5);
\coordinate (b2) at (0,3.5);
\coordinate (c1) at (3,1.25);
\coordinate (c2) at (3,2.25);

\draw (a1) -- (a2);
\draw (b1) -- (b2);
\draw (c1) -- (c2);

%円弧
\draw (0,1) arc[start angle=-90, end angle=90, radius=0.75];

%上の曲線(b2)--(c2)
\draw (b2) -- ($(b2)+(0.5,0)$);
\draw ($(c2)-(0.5,0)$) -- (c2);
\draw ($(b2)+(0.5,0)$) .. controls ($(b2)+(1.25,0)$) and ($(c2)-(1.25,0)$) ..($(c2)-(0.5,0)$);

%下の曲線(a1)--(c1)
\draw (a1) -- ($(a1)+(0.5,0)$);
\draw ($(c1)-(0.5,0)$) -- (c1);
\draw ($(a1)+(0.5,0)$) .. controls ($(a1)+(1.25,0)$) and ($(c1)-(1.25,0)$) ..($(c1)-(0.5,0)$);
\node[right] at (0,4.5) {$a$};
\node[right] at (0,3) {$b$};
\node[right] at (0,0.5) {$c$};
\end{tikzpicture}
}}
           \longmapsto \vcenter{\hbox{
\begin{tikzpicture}[scale=0.45, line width=2pt, xscale=-1]
 \coordinate (a1) at (0,0);
\coordinate (a2) at (0,1);
\coordinate (b1) at (0,2.5);
\coordinate (b2) at (0,3.5);
\coordinate (c1) at (3,1.25);
\coordinate (c2) at (3,2.25);

\draw (a1) -- (a2);
\draw (b1) -- (b2);
\draw (c1) -- (c2);

%円弧
\draw (0,1) arc[start angle=-90, end angle=90, radius=0.75];

%上の曲線(b2)--(c2)
\draw (b2) -- ($(b2)+(0.5,0)$);
\draw ($(c2)-(0.5,0)$) -- (c2);
\draw ($(b2)+(0.5,0)$) .. controls ($(b2)+(1.25,0)$) and ($(c2)-(1.25,0)$) ..($(c2)-(0.5,0)$);

%下の曲線(a1)--(c1)
\draw (a1) -- ($(a1)+(0.5,0)$);
\draw ($(c1)-(0.5,0)$) -- (c1);
\draw ($(a1)+(0.5,0)$) .. controls ($(a1)+(1.25,0)$) and ($(c1)-(1.25,0)$) ..($(c1)-(0.5,0)$);
\node[right] at (0,3) {$1$};
\node[right] at (0,0.5) {$1$};
\end{tikzpicture}
}}\otimes_{A^2}\vcenter{\hbox{
\begin{tikzpicture}[scale=0.45, line width=2pt, xscale=-1]
\begin{scope}[shift={(0,0.5)}]
    \coordinate (x1) at (0,-1);
\coordinate (x2) at (0,-2);
\coordinate (y1) at (3,-1);
\coordinate (y2) at (3,-2);
\draw (x1) -- (x2);
\draw (y1) -- (y2);
\draw (x1) -- (y1);
\draw (y2) -- (x2);

\end{scope}
 \coordinate (a1) at (0,0);
\coordinate (a2) at (0,1);
\coordinate (b1) at (0,2.5);
\coordinate (b2) at (0,3.5);
\coordinate (c1) at (3,1.25);
\coordinate (c2) at (3,2.25);

\draw (a1) -- (a2);
\draw (b1) -- (b2);
\draw (c1) -- (c2);

%円弧
\draw (0,1) arc[start angle=-90, end angle=90, radius=0.75];

%上の曲線(b2)--(c2)
\draw (b2) -- ($(b2)+(0.5,0)$);
\draw ($(c2)-(0.5,0)$) -- (c2);
\draw ($(b2)+(0.5,0)$) .. controls ($(b2)+(1.25,0)$) and ($(c2)-(1.25,0)$) ..($(c2)-(0.5,0)$);

%下の曲線(a1)--(c1)
\draw (a1) -- ($(a1)+(0.5,0)$);
\draw ($(c1)-(0.5,0)$) -- (c1);
\draw ($(a1)+(0.5,0)$) .. controls ($(a1)+(1.25,0)$) and ($(c1)-(1.25,0)$) ..($(c1)-(0.5,0)$);
\node[right] at (0,3) {$da$};
\node[right] at (0,0.5) {$e_{(1)}b$};
\node[right] at (0,-1) {$e_{(2)}c$};
\end{tikzpicture}
}}.
       \end{equation*}
\fi
       \item The $(A^2,A^2)$-bimodule isomorphism 
       $$\gm:M\otimes_{A}L\rightarrow (A\otimes L)\otimes_{A^3}(M\otimes A)$$
       defined by
       \iflatexml
\begin{center}
  \includegraphics{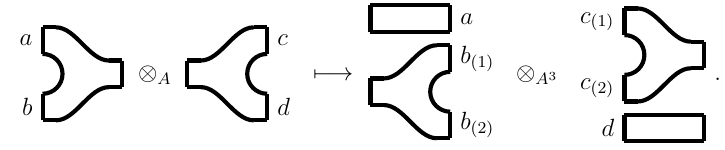}
\end{center}
\else
\begin{equation*}
           \vcenter{\hbox{
\begin{tikzpicture}[scale=0.45, line width=2pt]
 \coordinate (a1) at (0,0);
\coordinate (a2) at (0,1);
\coordinate (b1) at (0,2.5);
\coordinate (b2) at (0,3.5);
\coordinate (c1) at (3,1.25);
\coordinate (c2) at (3,2.25);

\draw (a1) -- (a2);
\draw (b1) -- (b2);
\draw (c1) -- (c2);

%円弧
\draw (0,1) arc[start angle=-90, end angle=90, radius=0.75];

%上の曲線(b2)--(c2)
\draw (b2) -- ($(b2)+(0.5,0)$);
\draw ($(c2)-(0.5,0)$) -- (c2);
\draw ($(b2)+(0.5,0)$) .. controls ($(b2)+(1.25,0)$) and ($(c2)-(1.25,0)$) ..($(c2)-(0.5,0)$);

%下の曲線(a1)--(c1)
\draw (a1) -- ($(a1)+(0.5,0)$);
\draw ($(c1)-(0.5,0)$) -- (c1);
\draw ($(a1)+(0.5,0)$) .. controls ($(a1)+(1.25,0)$) and ($(c1)-(1.25,0)$) ..($(c1)-(0.5,0)$);
\node[left] at (0,3) {$a$};
\node[left] at (0,0.5) {$b$};
\end{tikzpicture}
}}\otimes_A \vcenter{\hbox{
\begin{tikzpicture}[scale=0.45, line width=2pt, xscale=-1]
 \coordinate (a1) at (0,0);
\coordinate (a2) at (0,1);
\coordinate (b1) at (0,2.5);
\coordinate (b2) at (0,3.5);
\coordinate (c1) at (3,1.25);
\coordinate (c2) at (3,2.25);

\draw (a1) -- (a2);
\draw (b1) -- (b2);
\draw (c1) -- (c2);

%円弧
\draw (0,1) arc[start angle=-90, end angle=90, radius=0.75];

%上の曲線(b2)--(c2)
\draw (b2) -- ($(b2)+(0.5,0)$);
\draw ($(c2)-(0.5,0)$) -- (c2);
\draw ($(b2)+(0.5,0)$) .. controls ($(b2)+(1.25,0)$) and ($(c2)-(1.25,0)$) ..($(c2)-(0.5,0)$);

%下の曲線(a1)--(c1)
\draw (a1) -- ($(a1)+(0.5,0)$);
\draw ($(c1)-(0.5,0)$) -- (c1);
\draw ($(a1)+(0.5,0)$) .. controls ($(a1)+(1.25,0)$) and ($(c1)-(1.25,0)$) ..($(c1)-(0.5,0)$);
\node[right] at (0,3) {$c$};
\node[right] at (0,0.5) {$d$};
\end{tikzpicture}
}}
           \longmapsto
           \vcenter{\hbox{
\begin{tikzpicture}[scale=0.45, line width=2pt, xscale=-1]
\begin{scope}[shift={(0,6)}]
    \coordinate (x1) at (0,-1);
\coordinate (x2) at (0,-2);
\coordinate (y1) at (3,-1);
\coordinate (y2) at (3,-2);
\draw (x1) -- (x2);
\draw (y1) -- (y2);
\draw (x1) -- (y1);
\draw (y2) -- (x2);

\end{scope}
 \coordinate (a1) at (0,0);
\coordinate (a2) at (0,1);
\coordinate (b1) at (0,2.5);
\coordinate (b2) at (0,3.5);
\coordinate (c1) at (3,1.25);
\coordinate (c2) at (3,2.25);

\draw (a1) -- (a2);
\draw (b1) -- (b2);
\draw (c1) -- (c2);

%円弧
\draw (0,1) arc[start angle=-90, end angle=90, radius=0.75];

%上の曲線(b2)--(c2)
\draw (b2) -- ($(b2)+(0.5,0)$);
\draw ($(c2)-(0.5,0)$) -- (c2);
\draw ($(b2)+(0.5,0)$) .. controls ($(b2)+(1.25,0)$) and ($(c2)-(1.25,0)$) ..($(c2)-(0.5,0)$);

%下の曲線(a1)--(c1)
\draw (a1) -- ($(a1)+(0.5,0)$);
\draw ($(c1)-(0.5,0)$) -- (c1);
\draw ($(a1)+(0.5,0)$) .. controls ($(a1)+(1.25,0)$) and ($(c1)-(1.25,0)$) ..($(c1)-(0.5,0)$);
\node[right] at (0,4.5) {$a$};
\node[right] at (0,3) {$b_{(1)}$};
\node[right] at (0,0.5) {$b_{(2)}$};
\end{tikzpicture}
}}\otimes_{A^3}\vcenter{\hbox{
\begin{tikzpicture}[scale=0.45, line width=2pt]
\begin{scope}[shift={(0,0.5)}]
    \coordinate (x1) at (0,-1);
\coordinate (x2) at (0,-2);
\coordinate (y1) at (3,-1);
\coordinate (y2) at (3,-2);
\draw (x1) -- (x2);
\draw (y1) -- (y2);
\draw (x1) -- (y1);
\draw (y2) -- (x2);

\end{scope}
 \coordinate (a1) at (0,0);
\coordinate (a2) at (0,1);
\coordinate (b1) at (0,2.5);
\coordinate (b2) at (0,3.5);
\coordinate (c1) at (3,1.25);
\coordinate (c2) at (3,2.25);

\draw (a1) -- (a2);
\draw (b1) -- (b2);
\draw (c1) -- (c2);

%円弧
\draw (0,1) arc[start angle=-90, end angle=90, radius=0.75];

%上の曲線(b2)--(c2)
\draw (b2) -- ($(b2)+(0.5,0)$);
\draw ($(c2)-(0.5,0)$) -- (c2);
\draw ($(b2)+(0.5,0)$) .. controls ($(b2)+(1.25,0)$) and ($(c2)-(1.25,0)$) ..($(c2)-(0.5,0)$);

%下の曲線(a1)--(c1)
\draw (a1) -- ($(a1)+(0.5,0)$);
\draw ($(c1)-(0.5,0)$) -- (c1);
\draw ($(a1)+(0.5,0)$) .. controls ($(a1)+(1.25,0)$) and ($(c1)-(1.25,0)$) ..($(c1)-(0.5,0)$);
\node[left] at (0,3) {$c_{(1)}$};
\node[left] at (0,0.5) {$c_{(2)}$};
\node[left] at (0,-1) {$d$};
\end{tikzpicture}
}}.
       \end{equation*}
\fi
       \item The $(A^2,A^2)$-bimodule isomorphism
       $$\brg:(L\otimes A)\otimes_{A^3}(A\otimes M)\rightarrow M\otimes_{A}L$$
       defined by
       \iflatexml
\begin{center}
  \includegraphics{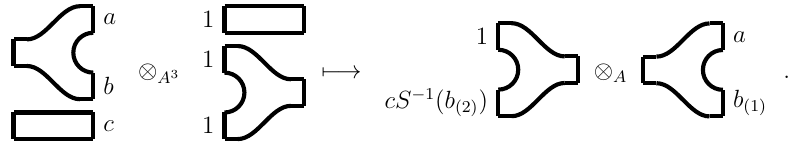}
\end{center}
\else
\begin{equation*}
           \vcenter{\hbox{
\begin{tikzpicture}[scale=0.45, line width=2pt, xscale=-1]
\begin{scope}[shift={(0,0.5)}]
    \coordinate (x1) at (0,-1);
\coordinate (x2) at (0,-2);
\coordinate (y1) at (3,-1);
\coordinate (y2) at (3,-2);
\draw (x1) -- (x2);
\draw (y1) -- (y2);
\draw (x1) -- (y1);
\draw (y2) -- (x2);

\end{scope}
 \coordinate (a1) at (0,0);
\coordinate (a2) at (0,1);
\coordinate (b1) at (0,2.5);
\coordinate (b2) at (0,3.5);
\coordinate (c1) at (3,1.25);
\coordinate (c2) at (3,2.25);

\draw (a1) -- (a2);
\draw (b1) -- (b2);
\draw (c1) -- (c2);

%円弧
\draw (0,1) arc[start angle=-90, end angle=90, radius=0.75];

%上の曲線(b2)--(c2)
\draw (b2) -- ($(b2)+(0.5,0)$);
\draw ($(c2)-(0.5,0)$) -- (c2);
\draw ($(b2)+(0.5,0)$) .. controls ($(b2)+(1.25,0)$) and ($(c2)-(1.25,0)$) ..($(c2)-(0.5,0)$);

%下の曲線(a1)--(c1)
\draw (a1) -- ($(a1)+(0.5,0)$);
\draw ($(c1)-(0.5,0)$) -- (c1);
\draw ($(a1)+(0.5,0)$) .. controls ($(a1)+(1.25,0)$) and ($(c1)-(1.25,0)$) ..($(c1)-(0.5,0)$);
\node[right] at (0,3) {$a$};
\node[right] at (0,0.5) {$b$};
\node[right] at (0,-1) {$c$};
\end{tikzpicture}
}}\otimes_{A^3}\vcenter{\hbox{
\begin{tikzpicture}[scale=0.45, line width=2pt]
\begin{scope}[shift={(0,6)}]
    \coordinate (x1) at (0,-1);
\coordinate (x2) at (0,-2);
\coordinate (y1) at (3,-1);
\coordinate (y2) at (3,-2);
\draw (x1) -- (x2);
\draw (y1) -- (y2);
\draw (x1) -- (y1);
\draw (y2) -- (x2);

\end{scope}
 \coordinate (a1) at (0,0);
\coordinate (a2) at (0,1);
\coordinate (b1) at (0,2.5);
\coordinate (b2) at (0,3.5);
\coordinate (c1) at (3,1.25);
\coordinate (c2) at (3,2.25);

\draw (a1) -- (a2);
\draw (b1) -- (b2);
\draw (c1) -- (c2);

%円弧
\draw (0,1) arc[start angle=-90, end angle=90, radius=0.75];

%上の曲線(b2)--(c2)
\draw (b2) -- ($(b2)+(0.5,0)$);
\draw ($(c2)-(0.5,0)$) -- (c2);
\draw ($(b2)+(0.5,0)$) .. controls ($(b2)+(1.25,0)$) and ($(c2)-(1.25,0)$) ..($(c2)-(0.5,0)$);

%下の曲線(a1)--(c1)
\draw (a1) -- ($(a1)+(0.5,0)$);
\draw ($(c1)-(0.5,0)$) -- (c1);
\draw ($(a1)+(0.5,0)$) .. controls ($(a1)+(1.25,0)$) and ($(c1)-(1.25,0)$) ..($(c1)-(0.5,0)$);
\node[left] at (0,4.5) {$1$};
\node[left] at (0,3) {$1$};
\node[left] at (0,0.5) {$1$};
\end{tikzpicture}
}}
           \longmapsto \vcenter{\hbox{
\begin{tikzpicture}[scale=0.45, line width=2pt]
 \coordinate (a1) at (0,0);
\coordinate (a2) at (0,1);
\coordinate (b1) at (0,2.5);
\coordinate (b2) at (0,3.5);
\coordinate (c1) at (3,1.25);
\coordinate (c2) at (3,2.25);

\draw (a1) -- (a2);
\draw (b1) -- (b2);
\draw (c1) -- (c2);

%円弧
\draw (0,1) arc[start angle=-90, end angle=90, radius=0.75];

%上の曲線(b2)--(c2)
\draw (b2) -- ($(b2)+(0.5,0)$);
\draw ($(c2)-(0.5,0)$) -- (c2);
\draw ($(b2)+(0.5,0)$) .. controls ($(b2)+(1.25,0)$) and ($(c2)-(1.25,0)$) ..($(c2)-(0.5,0)$);

%下の曲線(a1)--(c1)
\draw (a1) -- ($(a1)+(0.5,0)$);
\draw ($(c1)-(0.5,0)$) -- (c1);
\draw ($(a1)+(0.5,0)$) .. controls ($(a1)+(1.25,0)$) and ($(c1)-(1.25,0)$) ..($(c1)-(0.5,0)$);
\node[left] at (0,3) {$1$};
\node[left] at (0,0.5) {$cS^{-1}(b_{(2)})$};
\end{tikzpicture}
}}\otimes_{A}\vcenter{\hbox{
\begin{tikzpicture}[scale=0.45, line width=2pt, xscale=-1]
 \coordinate (a1) at (0,0);
\coordinate (a2) at (0,1);
\coordinate (b1) at (0,2.5);
\coordinate (b2) at (0,3.5);
\coordinate (c1) at (3,1.25);
\coordinate (c2) at (3,2.25);

\draw (a1) -- (a2);
\draw (b1) -- (b2);
\draw (c1) -- (c2);

%円弧
\draw (0,1) arc[start angle=-90, end angle=90, radius=0.75];

%上の曲線(b2)--(c2)
\draw (b2) -- ($(b2)+(0.5,0)$);
\draw ($(c2)-(0.5,0)$) -- (c2);
\draw ($(b2)+(0.5,0)$) .. controls ($(b2)+(1.25,0)$) and ($(c2)-(1.25,0)$) ..($(c2)-(0.5,0)$);

%下の曲線(a1)--(c1)
\draw (a1) -- ($(a1)+(0.5,0)$);
\draw ($(c1)-(0.5,0)$) -- (c1);
\draw ($(a1)+(0.5,0)$) .. controls ($(a1)+(1.25,0)$) and ($(c1)-(1.25,0)$) ..($(c1)-(0.5,0)$);
\node[right] at (0,3) {$a$};
\node[right] at (0,0.5) {$b_{(1)}$};
\end{tikzpicture}
}}.
       \end{equation*}
\fi
   \end{enumerate}
\end{itemize}
In fact, these isomorphisms satisfy the eight axioms \eqref{diag:1}--\eqref{diag:8} as shown in Lemma~\ref{lem0}. Theorem~\ref{main} therefore yields the $R$-matrix on
$X=L\otimes_{A^2}M$ defined by \eqref{eq:R}. Its inverse is given by
the following diagram, using the inverse isomorphisms from the proof of Lemma~\ref{lem1}.

%begin novalidate
\iflatexml
\begin{center}
  \includegraphics{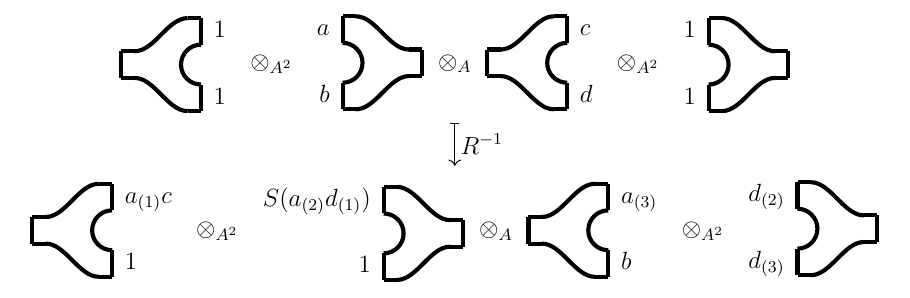}
\end{center}
\else
\begin{equation*}
\vcenter{\hbox{
    \begin{tikzcd}
   \vcenter{\hbox{
\begin{tikzpicture}[scale=0.45, line width=2pt, xscale=-1]
 \coordinate (a1) at (0,0);
\coordinate (a2) at (0,1);
\coordinate (b1) at (0,2.5);
\coordinate (b2) at (0,3.5);
\coordinate (c1) at (3,1.25);
\coordinate (c2) at (3,2.25);

\draw (a1) -- (a2);
\draw (b1) -- (b2);
\draw (c1) -- (c2);

%円弧
\draw (0,1) arc[start angle=-90, end angle=90, radius=0.75];

%上の曲線(b2)--(c2)
\draw (b2) -- ($(b2)+(0.5,0)$);
\draw ($(c2)-(0.5,0)$) -- (c2);
\draw ($(b2)+(0.5,0)$) .. controls ($(b2)+(1.25,0)$) and ($(c2)-(1.25,0)$) ..($(c2)-(0.5,0)$);

%下の曲線(a1)--(c1)
\draw (a1) -- ($(a1)+(0.5,0)$);
\draw ($(c1)-(0.5,0)$) -- (c1);
\draw ($(a1)+(0.5,0)$) .. controls ($(a1)+(1.25,0)$) and ($(c1)-(1.25,0)$) ..($(c1)-(0.5,0)$);
\node[right] at (0,3) {$1$};
\node[right] at (0,0.5) {$1$};
\end{tikzpicture}
}}\otimes_{A^2} \vcenter{\hbox{
\begin{tikzpicture}[scale=0.45, line width=2pt]
 \coordinate (a1) at (0,0);
\coordinate (a2) at (0,1);
\coordinate (b1) at (0,2.5);
\coordinate (b2) at (0,3.5);
\coordinate (c1) at (3,1.25);
\coordinate (c2) at (3,2.25);

\draw (a1) -- (a2);
\draw (b1) -- (b2);
\draw (c1) -- (c2);

%円弧
\draw (0,1) arc[start angle=-90, end angle=90, radius=0.75];

%上の曲線(b2)--(c2)
\draw (b2) -- ($(b2)+(0.5,0)$);
\draw ($(c2)-(0.5,0)$) -- (c2);
\draw ($(b2)+(0.5,0)$) .. controls ($(b2)+(1.25,0)$) and ($(c2)-(1.25,0)$) ..($(c2)-(0.5,0)$);

%下の曲線(a1)--(c1)
\draw (a1) -- ($(a1)+(0.5,0)$);
\draw ($(c1)-(0.5,0)$) -- (c1);
\draw ($(a1)+(0.5,0)$) .. controls ($(a1)+(1.25,0)$) and ($(c1)-(1.25,0)$) ..($(c1)-(0.5,0)$);
\node[left] at (0,3) {$a$};
\node[left] at (0,0.5) {$b$};
\end{tikzpicture}
}}\otimes_A \vcenter{\hbox{
\begin{tikzpicture}[scale=0.45, line width=2pt, xscale=-1]
 \coordinate (a1) at (0,0);
\coordinate (a2) at (0,1);
\coordinate (b1) at (0,2.5);
\coordinate (b2) at (0,3.5);
\coordinate (c1) at (3,1.25);
\coordinate (c2) at (3,2.25);

\draw (a1) -- (a2);
\draw (b1) -- (b2);
\draw (c1) -- (c2);

%円弧
\draw (0,1) arc[start angle=-90, end angle=90, radius=0.75];

%上の曲線(b2)--(c2)
\draw (b2) -- ($(b2)+(0.5,0)$);
\draw ($(c2)-(0.5,0)$) -- (c2);
\draw ($(b2)+(0.5,0)$) .. controls ($(b2)+(1.25,0)$) and ($(c2)-(1.25,0)$) ..($(c2)-(0.5,0)$);

%下の曲線(a1)--(c1)
\draw (a1) -- ($(a1)+(0.5,0)$);
\draw ($(c1)-(0.5,0)$) -- (c1);
\draw ($(a1)+(0.5,0)$) .. controls ($(a1)+(1.25,0)$) and ($(c1)-(1.25,0)$) ..($(c1)-(0.5,0)$);
\node[right] at (0,3) {$c$};
\node[right] at (0,0.5) {$d$};
\end{tikzpicture}
}}\otimes_{A^2} \vcenter{\hbox{
\begin{tikzpicture}[scale=0.45, line width=2pt]
 \coordinate (a1) at (0,0);
\coordinate (a2) at (0,1);
\coordinate (b1) at (0,2.5);
\coordinate (b2) at (0,3.5);
\coordinate (c1) at (3,1.25);
\coordinate (c2) at (3,2.25);

\draw (a1) -- (a2);
\draw (b1) -- (b2);
\draw (c1) -- (c2);

%円弧
\draw (0,1) arc[start angle=-90, end angle=90, radius=0.75];

%上の曲線(b2)--(c2)
\draw (b2) -- ($(b2)+(0.5,0)$);
\draw ($(c2)-(0.5,0)$) -- (c2);
\draw ($(b2)+(0.5,0)$) .. controls ($(b2)+(1.25,0)$) and ($(c2)-(1.25,0)$) ..($(c2)-(0.5,0)$);

%下の曲線(a1)--(c1)
\draw (a1) -- ($(a1)+(0.5,0)$);
\draw ($(c1)-(0.5,0)$) -- (c1);
\draw ($(a1)+(0.5,0)$) .. controls ($(a1)+(1.25,0)$) and ($(c1)-(1.25,0)$) ..($(c1)-(0.5,0)$);
\node[left] at (0,3) {$1$};
\node[left] at (0,0.5) {$1$};
\end{tikzpicture}
}} \arrow[d, mapsto, "{\displaystyle R^{-1}}"] \\
       \vcenter{\hbox{
\begin{tikzpicture}[scale=0.45, line width=2pt, xscale=-1]
 \coordinate (a1) at (0,0);
\coordinate (a2) at (0,1);
\coordinate (b1) at (0,2.5);
\coordinate (b2) at (0,3.5);
\coordinate (c1) at (3,1.25);
\coordinate (c2) at (3,2.25);

\draw (a1) -- (a2);
\draw (b1) -- (b2);
\draw (c1) -- (c2);

%円弧
\draw (0,1) arc[start angle=-90, end angle=90, radius=0.75];

%上の曲線(b2)--(c2)
\draw (b2) -- ($(b2)+(0.5,0)$);
\draw ($(c2)-(0.5,0)$) -- (c2);
\draw ($(b2)+(0.5,0)$) .. controls ($(b2)+(1.25,0)$) and ($(c2)-(1.25,0)$) ..($(c2)-(0.5,0)$);

%下の曲線(a1)--(c1)
\draw (a1) -- ($(a1)+(0.5,0)$);
\draw ($(c1)-(0.5,0)$) -- (c1);
\draw ($(a1)+(0.5,0)$) .. controls ($(a1)+(1.25,0)$) and ($(c1)-(1.25,0)$) ..($(c1)-(0.5,0)$);
\node[right] at (0,3) {$a_{(1)}c$};
\node[right] at (0,0.5) {$1$};
\end{tikzpicture}
}}\otimes_{A^2} \vcenter{\hbox{
\begin{tikzpicture}[scale=0.45, line width=2pt]
 \coordinate (a1) at (0,0);
\coordinate (a2) at (0,1);
\coordinate (b1) at (0,2.5);
\coordinate (b2) at (0,3.5);
\coordinate (c1) at (3,1.25);
\coordinate (c2) at (3,2.25);

\draw (a1) -- (a2);
\draw (b1) -- (b2);
\draw (c1) -- (c2);

%円弧
\draw (0,1) arc[start angle=-90, end angle=90, radius=0.75];

%上の曲線(b2)--(c2)
\draw (b2) -- ($(b2)+(0.5,0)$);
\draw ($(c2)-(0.5,0)$) -- (c2);
\draw ($(b2)+(0.5,0)$) .. controls ($(b2)+(1.25,0)$) and ($(c2)-(1.25,0)$) ..($(c2)-(0.5,0)$);

%下の曲線(a1)--(c1)
\draw (a1) -- ($(a1)+(0.5,0)$);
\draw ($(c1)-(0.5,0)$) -- (c1);
\draw ($(a1)+(0.5,0)$) .. controls ($(a1)+(1.25,0)$) and ($(c1)-(1.25,0)$) ..($(c1)-(0.5,0)$);
\node[left] at (0,3) {$S(a_{(2)}d_{(1)})$};
\node[left] at (0,0.5) {$1$};
\end{tikzpicture}
}}\otimes_A \vcenter{\hbox{
\begin{tikzpicture}[scale=0.45, line width=2pt, xscale=-1]
 \coordinate (a1) at (0,0);
\coordinate (a2) at (0,1);
\coordinate (b1) at (0,2.5);
\coordinate (b2) at (0,3.5);
\coordinate (c1) at (3,1.25);
\coordinate (c2) at (3,2.25);

\draw (a1) -- (a2);
\draw (b1) -- (b2);
\draw (c1) -- (c2);

%円弧
\draw (0,1) arc[start angle=-90, end angle=90, radius=0.75];

%上の曲線(b2)--(c2)
\draw (b2) -- ($(b2)+(0.5,0)$);
\draw ($(c2)-(0.5,0)$) -- (c2);
\draw ($(b2)+(0.5,0)$) .. controls ($(b2)+(1.25,0)$) and ($(c2)-(1.25,0)$) ..($(c2)-(0.5,0)$);

%下の曲線(a1)--(c1)
\draw (a1) -- ($(a1)+(0.5,0)$);
\draw ($(c1)-(0.5,0)$) -- (c1);
\draw ($(a1)+(0.5,0)$) .. controls ($(a1)+(1.25,0)$) and ($(c1)-(1.25,0)$) ..($(c1)-(0.5,0)$);
\node[right] at (0,3) {$a_{(3)}$};
\node[right] at (0,0.5) {$b$};
\end{tikzpicture}
}}\otimes_{A^2} \vcenter{\hbox{
\begin{tikzpicture}[scale=0.45, line width=2pt]
 \coordinate (a1) at (0,0);
\coordinate (a2) at (0,1);
\coordinate (b1) at (0,2.5);
\coordinate (b2) at (0,3.5);
\coordinate (c1) at (3,1.25);
\coordinate (c2) at (3,2.25);

\draw (a1) -- (a2);
\draw (b1) -- (b2);
\draw (c1) -- (c2);

%円弧
\draw (0,1) arc[start angle=-90, end angle=90, radius=0.75];

%上の曲線(b2)--(c2)
\draw (b2) -- ($(b2)+(0.5,0)$);
\draw ($(c2)-(0.5,0)$) -- (c2);
\draw ($(b2)+(0.5,0)$) .. controls ($(b2)+(1.25,0)$) and ($(c2)-(1.25,0)$) ..($(c2)-(0.5,0)$);

%下の曲線(a1)--(c1)
\draw (a1) -- ($(a1)+(0.5,0)$);
\draw ($(c1)-(0.5,0)$) -- (c1);
\draw ($(a1)+(0.5,0)$) .. controls ($(a1)+(1.25,0)$) and ($(c1)-(1.25,0)$) ..($(c1)-(0.5,0)$);
\node[left] at (0,3) {$d_{(2)}$};
\node[left] at (0,0.5) {$d_{(3)}$};
\end{tikzpicture}
}}
\end{tikzcd}
}}
\end{equation*}
\fi
%end novalidate

Now, the $(A,A)$-bimodule $X$ carries a canonical $H$-Hopf bimodule
structure, with left and right coactions
\[
  \lambda=\Delta\otimes 1:X\longrightarrow H\otimes X,
  \qquad
  \rho=1\otimes\Delta:X\longrightarrow X\otimes H,
\]
respectively.
By Lemma~\ref{lem2}, $R^{-1}$ is an isomorphism of $H$-Hopf bimodules.

Under the equivalence of Theorem~\ref{Rosso}, the $H$-Hopf bimodule $X$ corresponds to the
Yetter-Drinfeld module of right coinvariants
\[
  X^{\mathrm{co}H}:=\{x\in X:\rho(x)=x\otimes 1\}.
\]
By Lemma~\ref{lem3}, the map
\[
  \varphi:H\longrightarrow X^{\mathrm{co}H},
  \qquad
  \varphi(a)=a\otimes 1,
\]
is an isomorphism of Yetter-Drinfeld modules, where $H$ is equipped
with the structure in Example~\ref{exA}.

Consequently, the following assertion shows that $R^{-1}$ corresponds to
the standard braiding on $H$ under this identification.
\begin{prop}\label{ex:bimod}
    The following diagram commutes.
\iflatexml
\begin{center}
  \includegraphics{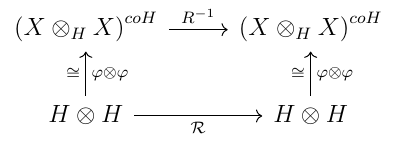}
\end{center}
\else
\[
\begin{tikzcd}
  \left(X\otimes_H X\right)^{coH} \arrow[r, "R^{-1}"] & \left(X\otimes_H X\right)^{coH}  \\
  H\otimes H \arrow[r, "{\mathcal{R}}"']  \arrow[u,"\cong", "\varphi\otimes\varphi"']                    & H\otimes H \arrow[u,"\cong", "\varphi\otimes\varphi"']
\end{tikzcd}
\]
\fi
    where the standard braiding $\mathcal{R}(x\otimes y)=x_{(1)}yS(x_{(2)})\otimes x_{(3)}$ is given by \eqref{universal R}.
\end{prop}
\begin{proof}
  If $b=d=1$ for $R^{-1}$, then we have
  \iflatexml
\begin{center}
  \includegraphics{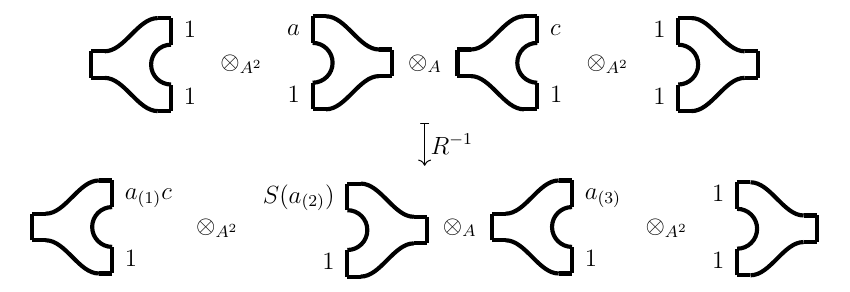}
\end{center}
\else
\begin{equation*}
\vcenter{\hbox{
    \begin{tikzcd}
   \vcenter{\hbox{
\begin{tikzpicture}[scale=0.45, line width=2pt, xscale=-1]
 \coordinate (a1) at (0,0);
\coordinate (a2) at (0,1);
\coordinate (b1) at (0,2.5);
\coordinate (b2) at (0,3.5);
\coordinate (c1) at (3,1.25);
\coordinate (c2) at (3,2.25);

\draw (a1) -- (a2);
\draw (b1) -- (b2);
\draw (c1) -- (c2);

%円弧
\draw (0,1) arc[start angle=-90, end angle=90, radius=0.75];

%上の曲線(b2)--(c2)
\draw (b2) -- ($(b2)+(0.5,0)$);
\draw ($(c2)-(0.5,0)$) -- (c2);
\draw ($(b2)+(0.5,0)$) .. controls ($(b2)+(1.25,0)$) and ($(c2)-(1.25,0)$) ..($(c2)-(0.5,0)$);

%下の曲線(a1)--(c1)
\draw (a1) -- ($(a1)+(0.5,0)$);
\draw ($(c1)-(0.5,0)$) -- (c1);
\draw ($(a1)+(0.5,0)$) .. controls ($(a1)+(1.25,0)$) and ($(c1)-(1.25,0)$) ..($(c1)-(0.5,0)$);
\node[right] at (0,3) {$1$};
\node[right] at (0,0.5) {$1$};
\end{tikzpicture}
}}\otimes_{A^2} \vcenter{\hbox{
\begin{tikzpicture}[scale=0.45, line width=2pt]
 \coordinate (a1) at (0,0);
\coordinate (a2) at (0,1);
\coordinate (b1) at (0,2.5);
\coordinate (b2) at (0,3.5);
\coordinate (c1) at (3,1.25);
\coordinate (c2) at (3,2.25);

\draw (a1) -- (a2);
\draw (b1) -- (b2);
\draw (c1) -- (c2);

%円弧
\draw (0,1) arc[start angle=-90, end angle=90, radius=0.75];

%上の曲線(b2)--(c2)
\draw (b2) -- ($(b2)+(0.5,0)$);
\draw ($(c2)-(0.5,0)$) -- (c2);
\draw ($(b2)+(0.5,0)$) .. controls ($(b2)+(1.25,0)$) and ($(c2)-(1.25,0)$) ..($(c2)-(0.5,0)$);

%下の曲線(a1)--(c1)
\draw (a1) -- ($(a1)+(0.5,0)$);
\draw ($(c1)-(0.5,0)$) -- (c1);
\draw ($(a1)+(0.5,0)$) .. controls ($(a1)+(1.25,0)$) and ($(c1)-(1.25,0)$) ..($(c1)-(0.5,0)$);
\node[left] at (0,3) {$a$};
\node[left] at (0,0.5) {$1$};
\end{tikzpicture}
}}\otimes_A \vcenter{\hbox{
\begin{tikzpicture}[scale=0.45, line width=2pt, xscale=-1]
 \coordinate (a1) at (0,0);
\coordinate (a2) at (0,1);
\coordinate (b1) at (0,2.5);
\coordinate (b2) at (0,3.5);
\coordinate (c1) at (3,1.25);
\coordinate (c2) at (3,2.25);

\draw (a1) -- (a2);
\draw (b1) -- (b2);
\draw (c1) -- (c2);

%円弧
\draw (0,1) arc[start angle=-90, end angle=90, radius=0.75];

%上の曲線(b2)--(c2)
\draw (b2) -- ($(b2)+(0.5,0)$);
\draw ($(c2)-(0.5,0)$) -- (c2);
\draw ($(b2)+(0.5,0)$) .. controls ($(b2)+(1.25,0)$) and ($(c2)-(1.25,0)$) ..($(c2)-(0.5,0)$);

%下の曲線(a1)--(c1)
\draw (a1) -- ($(a1)+(0.5,0)$);
\draw ($(c1)-(0.5,0)$) -- (c1);
\draw ($(a1)+(0.5,0)$) .. controls ($(a1)+(1.25,0)$) and ($(c1)-(1.25,0)$) ..($(c1)-(0.5,0)$);
\node[right] at (0,3) {$c$};
\node[right] at (0,0.5) {$1$};
\end{tikzpicture}
}}\otimes_{A^2} \vcenter{\hbox{
\begin{tikzpicture}[scale=0.45, line width=2pt]
 \coordinate (a1) at (0,0);
\coordinate (a2) at (0,1);
\coordinate (b1) at (0,2.5);
\coordinate (b2) at (0,3.5);
\coordinate (c1) at (3,1.25);
\coordinate (c2) at (3,2.25);

\draw (a1) -- (a2);
\draw (b1) -- (b2);
\draw (c1) -- (c2);

%円弧
\draw (0,1) arc[start angle=-90, end angle=90, radius=0.75];

%上の曲線(b2)--(c2)
\draw (b2) -- ($(b2)+(0.5,0)$);
\draw ($(c2)-(0.5,0)$) -- (c2);
\draw ($(b2)+(0.5,0)$) .. controls ($(b2)+(1.25,0)$) and ($(c2)-(1.25,0)$) ..($(c2)-(0.5,0)$);

%下の曲線(a1)--(c1)
\draw (a1) -- ($(a1)+(0.5,0)$);
\draw ($(c1)-(0.5,0)$) -- (c1);
\draw ($(a1)+(0.5,0)$) .. controls ($(a1)+(1.25,0)$) and ($(c1)-(1.25,0)$) ..($(c1)-(0.5,0)$);
\node[left] at (0,3) {$1$};
\node[left] at (0,0.5) {$1$};
\end{tikzpicture}
}} \arrow[d, mapsto, "{\displaystyle R^{-1}}"] \\
       \vcenter{\hbox{
\begin{tikzpicture}[scale=0.45, line width=2pt, xscale=-1]
 \coordinate (a1) at (0,0);
\coordinate (a2) at (0,1);
\coordinate (b1) at (0,2.5);
\coordinate (b2) at (0,3.5);
\coordinate (c1) at (3,1.25);
\coordinate (c2) at (3,2.25);

\draw (a1) -- (a2);
\draw (b1) -- (b2);
\draw (c1) -- (c2);

%円弧
\draw (0,1) arc[start angle=-90, end angle=90, radius=0.75];

%上の曲線(b2)--(c2)
\draw (b2) -- ($(b2)+(0.5,0)$);
\draw ($(c2)-(0.5,0)$) -- (c2);
\draw ($(b2)+(0.5,0)$) .. controls ($(b2)+(1.25,0)$) and ($(c2)-(1.25,0)$) ..($(c2)-(0.5,0)$);

%下の曲線(a1)--(c1)
\draw (a1) -- ($(a1)+(0.5,0)$);
\draw ($(c1)-(0.5,0)$) -- (c1);
\draw ($(a1)+(0.5,0)$) .. controls ($(a1)+(1.25,0)$) and ($(c1)-(1.25,0)$) ..($(c1)-(0.5,0)$);
\node[right] at (0,3) {$a_{(1)}c$};
\node[right] at (0,0.5) {$1$};
\end{tikzpicture}
}}\otimes_{A^2} \vcenter{\hbox{
\begin{tikzpicture}[scale=0.45, line width=2pt]
 \coordinate (a1) at (0,0);
\coordinate (a2) at (0,1);
\coordinate (b1) at (0,2.5);
\coordinate (b2) at (0,3.5);
\coordinate (c1) at (3,1.25);
\coordinate (c2) at (3,2.25);

\draw (a1) -- (a2);
\draw (b1) -- (b2);
\draw (c1) -- (c2);

%円弧
\draw (0,1) arc[start angle=-90, end angle=90, radius=0.75];

%上の曲線(b2)--(c2)
\draw (b2) -- ($(b2)+(0.5,0)$);
\draw ($(c2)-(0.5,0)$) -- (c2);
\draw ($(b2)+(0.5,0)$) .. controls ($(b2)+(1.25,0)$) and ($(c2)-(1.25,0)$) ..($(c2)-(0.5,0)$);

%下の曲線(a1)--(c1)
\draw (a1) -- ($(a1)+(0.5,0)$);
\draw ($(c1)-(0.5,0)$) -- (c1);
\draw ($(a1)+(0.5,0)$) .. controls ($(a1)+(1.25,0)$) and ($(c1)-(1.25,0)$) ..($(c1)-(0.5,0)$);
\node[left] at (0,3) {$S(a_{(2)})$};
\node[left] at (0,0.5) {$1$};
\end{tikzpicture}
}}\otimes_A \vcenter{\hbox{
\begin{tikzpicture}[scale=0.45, line width=2pt, xscale=-1]
 \coordinate (a1) at (0,0);
\coordinate (a2) at (0,1);
\coordinate (b1) at (0,2.5);
\coordinate (b2) at (0,3.5);
\coordinate (c1) at (3,1.25);
\coordinate (c2) at (3,2.25);

\draw (a1) -- (a2);
\draw (b1) -- (b2);
\draw (c1) -- (c2);

%円弧
\draw (0,1) arc[start angle=-90, end angle=90, radius=0.75];

%上の曲線(b2)--(c2)
\draw (b2) -- ($(b2)+(0.5,0)$);
\draw ($(c2)-(0.5,0)$) -- (c2);
\draw ($(b2)+(0.5,0)$) .. controls ($(b2)+(1.25,0)$) and ($(c2)-(1.25,0)$) ..($(c2)-(0.5,0)$);

%下の曲線(a1)--(c1)
\draw (a1) -- ($(a1)+(0.5,0)$);
\draw ($(c1)-(0.5,0)$) -- (c1);
\draw ($(a1)+(0.5,0)$) .. controls ($(a1)+(1.25,0)$) and ($(c1)-(1.25,0)$) ..($(c1)-(0.5,0)$);
\node[right] at (0,3) {$a_{(3)}$};
\node[right] at (0,0.5) {$1$};
\end{tikzpicture}
}}\otimes_{A^2} \vcenter{\hbox{
\begin{tikzpicture}[scale=0.45, line width=2pt]
 \coordinate (a1) at (0,0);
\coordinate (a2) at (0,1);
\coordinate (b1) at (0,2.5);
\coordinate (b2) at (0,3.5);
\coordinate (c1) at (3,1.25);
\coordinate (c2) at (3,2.25);

\draw (a1) -- (a2);
\draw (b1) -- (b2);
\draw (c1) -- (c2);

%円弧
\draw (0,1) arc[start angle=-90, end angle=90, radius=0.75];

%上の曲線(b2)--(c2)
\draw (b2) -- ($(b2)+(0.5,0)$);
\draw ($(c2)-(0.5,0)$) -- (c2);
\draw ($(b2)+(0.5,0)$) .. controls ($(b2)+(1.25,0)$) and ($(c2)-(1.25,0)$) ..($(c2)-(0.5,0)$);

%下の曲線(a1)--(c1)
\draw (a1) -- ($(a1)+(0.5,0)$);
\draw ($(c1)-(0.5,0)$) -- (c1);
\draw ($(a1)+(0.5,0)$) .. controls ($(a1)+(1.25,0)$) and ($(c1)-(1.25,0)$) ..($(c1)-(0.5,0)$);
\node[left] at (0,3) {$1$};
\node[left] at (0,0.5) {$1$};
\end{tikzpicture}
}}
\end{tikzcd}
}}
\end{equation*}
\fi
which implies the claim.
\end{proof}
\begin{lem}\label{lem1}
    The coherent maps $\al$, $\bt$, $\gm$ and $\brg$ are bimodule isomorphisms.
\end{lem}
\begin{proof}
  It is clear for $\al$ and $\bt$. The map $\gm$ is clearly an $(A^2,A^2)$-bimodule homomorphism. Its inverse is given by
  \iflatexml
\begin{center}
  \includegraphics{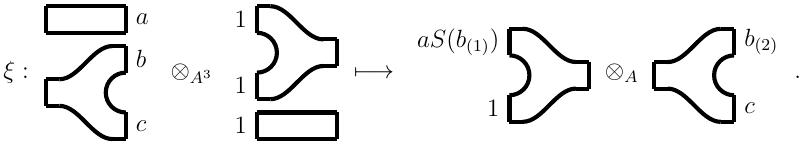}
\end{center}
\else
\begin{equation*}\xi:
  \vcenter{\hbox{
\begin{tikzpicture}[scale=0.45, line width=2pt, xscale=-1]
\begin{scope}[shift={(0,6)}]
    \coordinate (x1) at (0,-1);
\coordinate (x2) at (0,-2);
\coordinate (y1) at (3,-1);
\coordinate (y2) at (3,-2);
\draw (x1) -- (x2);
\draw (y1) -- (y2);
\draw (x1) -- (y1);
\draw (y2) -- (x2);

\end{scope}
 \coordinate (a1) at (0,0);
\coordinate (a2) at (0,1);
\coordinate (b1) at (0,2.5);
\coordinate (b2) at (0,3.5);
\coordinate (c1) at (3,1.25);
\coordinate (c2) at (3,2.25);

\draw (a1) -- (a2);
\draw (b1) -- (b2);
\draw (c1) -- (c2);

%円弧
\draw (0,1) arc[start angle=-90, end angle=90, radius=0.75];

%上の曲線(b2)--(c2)
\draw (b2) -- ($(b2)+(0.5,0)$);
\draw ($(c2)-(0.5,0)$) -- (c2);
\draw ($(b2)+(0.5,0)$) .. controls ($(b2)+(1.25,0)$) and ($(c2)-(1.25,0)$) ..($(c2)-(0.5,0)$);

%下の曲線(a1)--(c1)
\draw (a1) -- ($(a1)+(0.5,0)$);
\draw ($(c1)-(0.5,0)$) -- (c1);
\draw ($(a1)+(0.5,0)$) .. controls ($(a1)+(1.25,0)$) and ($(c1)-(1.25,0)$) ..($(c1)-(0.5,0)$);
\node[right] at (0,4.5) {$a$};
\node[right] at (0,3) {$b$};
\node[right] at (0,0.5) {$c$};
\end{tikzpicture}
}}\otimes_{A^3}\vcenter{\hbox{
\begin{tikzpicture}[scale=0.45, line width=2pt]
\begin{scope}[shift={(0,0.5)}]
    \coordinate (x1) at (0,-1);
\coordinate (x2) at (0,-2);
\coordinate (y1) at (3,-1);
\coordinate (y2) at (3,-2);
\draw (x1) -- (x2);
\draw (y1) -- (y2);
\draw (x1) -- (y1);
\draw (y2) -- (x2);

\end{scope}
 \coordinate (a1) at (0,0);
\coordinate (a2) at (0,1);
\coordinate (b1) at (0,2.5);
\coordinate (b2) at (0,3.5);
\coordinate (c1) at (3,1.25);
\coordinate (c2) at (3,2.25);

\draw (a1) -- (a2);
\draw (b1) -- (b2);
\draw (c1) -- (c2);

%円弧
\draw (0,1) arc[start angle=-90, end angle=90, radius=0.75];

%上の曲線(b2)--(c2)
\draw (b2) -- ($(b2)+(0.5,0)$);
\draw ($(c2)-(0.5,0)$) -- (c2);
\draw ($(b2)+(0.5,0)$) .. controls ($(b2)+(1.25,0)$) and ($(c2)-(1.25,0)$) ..($(c2)-(0.5,0)$);

%下の曲線(a1)--(c1)
\draw (a1) -- ($(a1)+(0.5,0)$);
\draw ($(c1)-(0.5,0)$) -- (c1);
\draw ($(a1)+(0.5,0)$) .. controls ($(a1)+(1.25,0)$) and ($(c1)-(1.25,0)$) ..($(c1)-(0.5,0)$);
\node[left] at (0,3) {$1$};
\node[left] at (0,0.5) {$1$};
\node[left] at (0,-1) {$1$};
\end{tikzpicture}
}}
  \longmapsto
           \vcenter{\hbox{
\begin{tikzpicture}[scale=0.45, line width=2pt]
 \coordinate (a1) at (0,0);
\coordinate (a2) at (0,1);
\coordinate (b1) at (0,2.5);
\coordinate (b2) at (0,3.5);
\coordinate (c1) at (3,1.25);
\coordinate (c2) at (3,2.25);

\draw (a1) -- (a2);
\draw (b1) -- (b2);
\draw (c1) -- (c2);

%円弧
\draw (0,1) arc[start angle=-90, end angle=90, radius=0.75];

%上の曲線(b2)--(c2)
\draw (b2) -- ($(b2)+(0.5,0)$);
\draw ($(c2)-(0.5,0)$) -- (c2);
\draw ($(b2)+(0.5,0)$) .. controls ($(b2)+(1.25,0)$) and ($(c2)-(1.25,0)$) ..($(c2)-(0.5,0)$);

%下の曲線(a1)--(c1)
\draw (a1) -- ($(a1)+(0.5,0)$);
\draw ($(c1)-(0.5,0)$) -- (c1);
\draw ($(a1)+(0.5,0)$) .. controls ($(a1)+(1.25,0)$) and ($(c1)-(1.25,0)$) ..($(c1)-(0.5,0)$);
\node[left] at (0,3) {$aS(b_{(1)})$};
\node[left] at (0,0.5) {$1$};
\end{tikzpicture}
}}\otimes_A \vcenter{\hbox{
\begin{tikzpicture}[scale=0.45, line width=2pt, xscale=-1]
 \coordinate (a1) at (0,0);
\coordinate (a2) at (0,1);
\coordinate (b1) at (0,2.5);
\coordinate (b2) at (0,3.5);
\coordinate (c1) at (3,1.25);
\coordinate (c2) at (3,2.25);

\draw (a1) -- (a2);
\draw (b1) -- (b2);
\draw (c1) -- (c2);

%円弧
\draw (0,1) arc[start angle=-90, end angle=90, radius=0.75];

%上の曲線(b2)--(c2)
\draw (b2) -- ($(b2)+(0.5,0)$);
\draw ($(c2)-(0.5,0)$) -- (c2);
\draw ($(b2)+(0.5,0)$) .. controls ($(b2)+(1.25,0)$) and ($(c2)-(1.25,0)$) ..($(c2)-(0.5,0)$);

%下の曲線(a1)--(c1)
\draw (a1) -- ($(a1)+(0.5,0)$);
\draw ($(c1)-(0.5,0)$) -- (c1);
\draw ($(a1)+(0.5,0)$) .. controls ($(a1)+(1.25,0)$) and ($(c1)-(1.25,0)$) ..($(c1)-(0.5,0)$);
\node[right] at (0,3) {$b_{(2)}$};
\node[right] at (0,0.5) {$c$};
\end{tikzpicture}
}}.
       \end{equation*}
\fi
       Indeed, the following calculation shows that $\gm^{-1}=\xi$:
       \iflatexml
\begin{center}
  \includegraphics{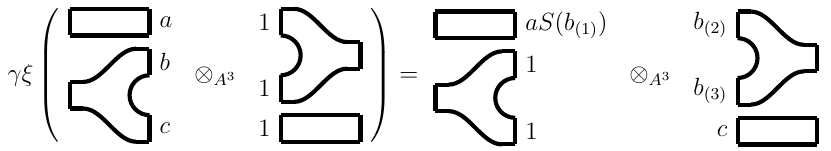}\par
  \includegraphics{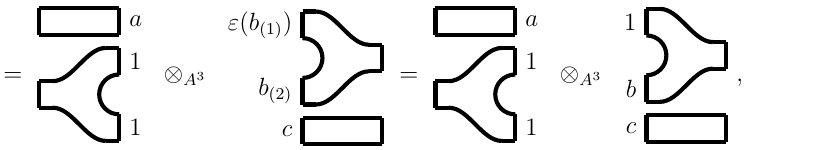}
\end{center}
\else
\begin{align*}
          \gm\xi\left(\vcenter{\hbox{
\begin{tikzpicture}[scale=0.45, line width=2pt, xscale=-1]
\begin{scope}[shift={(0,6)}]
    \coordinate (x1) at (0,-1);
\coordinate (x2) at (0,-2);
\coordinate (y1) at (3,-1);
\coordinate (y2) at (3,-2);
\draw (x1) -- (x2);
\draw (y1) -- (y2);
\draw (x1) -- (y1);
\draw (y2) -- (x2);

\end{scope}
 \coordinate (a1) at (0,0);
\coordinate (a2) at (0,1);
\coordinate (b1) at (0,2.5);
\coordinate (b2) at (0,3.5);
\coordinate (c1) at (3,1.25);
\coordinate (c2) at (3,2.25);

\draw (a1) -- (a2);
\draw (b1) -- (b2);
\draw (c1) -- (c2);

%円弧
\draw (0,1) arc[start angle=-90, end angle=90, radius=0.75];

%上の曲線(b2)--(c2)
\draw (b2) -- ($(b2)+(0.5,0)$);
\draw ($(c2)-(0.5,0)$) -- (c2);
\draw ($(b2)+(0.5,0)$) .. controls ($(b2)+(1.25,0)$) and ($(c2)-(1.25,0)$) ..($(c2)-(0.5,0)$);

%下の曲線(a1)--(c1)
\draw (a1) -- ($(a1)+(0.5,0)$);
\draw ($(c1)-(0.5,0)$) -- (c1);
\draw ($(a1)+(0.5,0)$) .. controls ($(a1)+(1.25,0)$) and ($(c1)-(1.25,0)$) ..($(c1)-(0.5,0)$);
\node[right] at (0,4.5) {$a$};
\node[right] at (0,3) {$b$};
\node[right] at (0,0.5) {$c$};
\end{tikzpicture}
}}\otimes_{A^3}\vcenter{\hbox{
\begin{tikzpicture}[scale=0.45, line width=2pt]
\begin{scope}[shift={(0,0.5)}]
    \coordinate (x1) at (0,-1);
\coordinate (x2) at (0,-2);
\coordinate (y1) at (3,-1);
\coordinate (y2) at (3,-2);
\draw (x1) -- (x2);
\draw (y1) -- (y2);
\draw (x1) -- (y1);
\draw (y2) -- (x2);

\end{scope}
 \coordinate (a1) at (0,0);
\coordinate (a2) at (0,1);
\coordinate (b1) at (0,2.5);
\coordinate (b2) at (0,3.5);
\coordinate (c1) at (3,1.25);
\coordinate (c2) at (3,2.25);

\draw (a1) -- (a2);
\draw (b1) -- (b2);
\draw (c1) -- (c2);

%円弧
\draw (0,1) arc[start angle=-90, end angle=90, radius=0.75];

%上の曲線(b2)--(c2)
\draw (b2) -- ($(b2)+(0.5,0)$);
\draw ($(c2)-(0.5,0)$) -- (c2);
\draw ($(b2)+(0.5,0)$) .. controls ($(b2)+(1.25,0)$) and ($(c2)-(1.25,0)$) ..($(c2)-(0.5,0)$);

%下の曲線(a1)--(c1)
\draw (a1) -- ($(a1)+(0.5,0)$);
\draw ($(c1)-(0.5,0)$) -- (c1);
\draw ($(a1)+(0.5,0)$) .. controls ($(a1)+(1.25,0)$) and ($(c1)-(1.25,0)$) ..($(c1)-(0.5,0)$);
\node[left] at (0,3) {$1$};
\node[left] at (0,0.5) {$1$};
\node[left] at (0,-1) {$1$};
\end{tikzpicture}
}} \right)
          &=\vcenter{\hbox{
\begin{tikzpicture}[scale=0.45, line width=2pt, xscale=-1]
\begin{scope}[shift={(0,6)}]
    \coordinate (x1) at (0,-1);
\coordinate (x2) at (0,-2);
\coordinate (y1) at (3,-1);
\coordinate (y2) at (3,-2);
\draw (x1) -- (x2);
\draw (y1) -- (y2);
\draw (x1) -- (y1);
\draw (y2) -- (x2);

\end{scope}
 \coordinate (a1) at (0,0);
\coordinate (a2) at (0,1);
\coordinate (b1) at (0,2.5);
\coordinate (b2) at (0,3.5);
\coordinate (c1) at (3,1.25);
\coordinate (c2) at (3,2.25);

\draw (a1) -- (a2);
\draw (b1) -- (b2);
\draw (c1) -- (c2);

%円弧
\draw (0,1) arc[start angle=-90, end angle=90, radius=0.75];

%上の曲線(b2)--(c2)
\draw (b2) -- ($(b2)+(0.5,0)$);
\draw ($(c2)-(0.5,0)$) -- (c2);
\draw ($(b2)+(0.5,0)$) .. controls ($(b2)+(1.25,0)$) and ($(c2)-(1.25,0)$) ..($(c2)-(0.5,0)$);

%下の曲線(a1)--(c1)
\draw (a1) -- ($(a1)+(0.5,0)$);
\draw ($(c1)-(0.5,0)$) -- (c1);
\draw ($(a1)+(0.5,0)$) .. controls ($(a1)+(1.25,0)$) and ($(c1)-(1.25,0)$) ..($(c1)-(0.5,0)$);
\node[right] at (0,4.5) {$aS(b_{(1)})$};
\node[right] at (0,3) {$1$};
\node[right] at (0,0.5) {$1$};
\end{tikzpicture}
}}\otimes_{A^3}\vcenter{\hbox{
\begin{tikzpicture}[scale=0.45, line width=2pt]
\begin{scope}[shift={(0,0.5)}]
    \coordinate (x1) at (0,-1);
\coordinate (x2) at (0,-2);
\coordinate (y1) at (3,-1);
\coordinate (y2) at (3,-2);
\draw (x1) -- (x2);
\draw (y1) -- (y2);
\draw (x1) -- (y1);
\draw (y2) -- (x2);

\end{scope}
 \coordinate (a1) at (0,0);
\coordinate (a2) at (0,1);
\coordinate (b1) at (0,2.5);
\coordinate (b2) at (0,3.5);
\coordinate (c1) at (3,1.25);
\coordinate (c2) at (3,2.25);

\draw (a1) -- (a2);
\draw (b1) -- (b2);
\draw (c1) -- (c2);

%円弧
\draw (0,1) arc[start angle=-90, end angle=90, radius=0.75];

%上の曲線(b2)--(c2)
\draw (b2) -- ($(b2)+(0.5,0)$);
\draw ($(c2)-(0.5,0)$) -- (c2);
\draw ($(b2)+(0.5,0)$) .. controls ($(b2)+(1.25,0)$) and ($(c2)-(1.25,0)$) ..($(c2)-(0.5,0)$);

%下の曲線(a1)--(c1)
\draw (a1) -- ($(a1)+(0.5,0)$);
\draw ($(c1)-(0.5,0)$) -- (c1);
\draw ($(a1)+(0.5,0)$) .. controls ($(a1)+(1.25,0)$) and ($(c1)-(1.25,0)$) ..($(c1)-(0.5,0)$);
\node[left] at (0,3) {$b_{(2)}$};
\node[left] at (0,0.5) {$b_{(3)}$};
\node[left] at (0,-1) {$c$};
\end{tikzpicture}
}} \\
          =\vcenter{\hbox{
\begin{tikzpicture}[scale=0.45, line width=2pt, xscale=-1]
\begin{scope}[shift={(0,6)}]
    \coordinate (x1) at (0,-1);
\coordinate (x2) at (0,-2);
\coordinate (y1) at (3,-1);
\coordinate (y2) at (3,-2);
\draw (x1) -- (x2);
\draw (y1) -- (y2);
\draw (x1) -- (y1);
\draw (y2) -- (x2);

\end{scope}
 \coordinate (a1) at (0,0);
\coordinate (a2) at (0,1);
\coordinate (b1) at (0,2.5);
\coordinate (b2) at (0,3.5);
\coordinate (c1) at (3,1.25);
\coordinate (c2) at (3,2.25);

\draw (a1) -- (a2);
\draw (b1) -- (b2);
\draw (c1) -- (c2);

%円弧
\draw (0,1) arc[start angle=-90, end angle=90, radius=0.75];

%上の曲線(b2)--(c2)
\draw (b2) -- ($(b2)+(0.5,0)$);
\draw ($(c2)-(0.5,0)$) -- (c2);
\draw ($(b2)+(0.5,0)$) .. controls ($(b2)+(1.25,0)$) and ($(c2)-(1.25,0)$) ..($(c2)-(0.5,0)$);

%下の曲線(a1)--(c1)
\draw (a1) -- ($(a1)+(0.5,0)$);
\draw ($(c1)-(0.5,0)$) -- (c1);
\draw ($(a1)+(0.5,0)$) .. controls ($(a1)+(1.25,0)$) and ($(c1)-(1.25,0)$) ..($(c1)-(0.5,0)$);
\node[right] at (0,4.5) {$a$};
\node[right] at (0,3) {$1$};
\node[right] at (0,0.5) {$1$};
\end{tikzpicture}
}}\otimes_{A^3}\vcenter{\hbox{
\begin{tikzpicture}[scale=0.45, line width=2pt]
\begin{scope}[shift={(0,0.5)}]
    \coordinate (x1) at (0,-1);
\coordinate (x2) at (0,-2);
\coordinate (y1) at (3,-1);
\coordinate (y2) at (3,-2);
\draw (x1) -- (x2);
\draw (y1) -- (y2);
\draw (x1) -- (y1);
\draw (y2) -- (x2);

\end{scope}
 \coordinate (a1) at (0,0);
\coordinate (a2) at (0,1);
\coordinate (b1) at (0,2.5);
\coordinate (b2) at (0,3.5);
\coordinate (c1) at (3,1.25);
\coordinate (c2) at (3,2.25);

\draw (a1) -- (a2);
\draw (b1) -- (b2);
\draw (c1) -- (c2);

%円弧
\draw (0,1) arc[start angle=-90, end angle=90, radius=0.75];

%上の曲線(b2)--(c2)
\draw (b2) -- ($(b2)+(0.5,0)$);
\draw ($(c2)-(0.5,0)$) -- (c2);
\draw ($(b2)+(0.5,0)$) .. controls ($(b2)+(1.25,0)$) and ($(c2)-(1.25,0)$) ..($(c2)-(0.5,0)$);

%下の曲線(a1)--(c1)
\draw (a1) -- ($(a1)+(0.5,0)$);
\draw ($(c1)-(0.5,0)$) -- (c1);
\draw ($(a1)+(0.5,0)$) .. controls ($(a1)+(1.25,0)$) and ($(c1)-(1.25,0)$) ..($(c1)-(0.5,0)$);
\node[left] at (0,3) {$\varepsilon(b_{(1)})$};
\node[left] at (0,0.5) {$b_{(2)}$};
\node[left] at (0,-1) {$c$};
\end{tikzpicture}
}} 
          &=\vcenter{\hbox{
\begin{tikzpicture}[scale=0.45, line width=2pt, xscale=-1]
\begin{scope}[shift={(0,6)}]
    \coordinate (x1) at (0,-1);
\coordinate (x2) at (0,-2);
\coordinate (y1) at (3,-1);
\coordinate (y2) at (3,-2);
\draw (x1) -- (x2);
\draw (y1) -- (y2);
\draw (x1) -- (y1);
\draw (y2) -- (x2);

\end{scope}
 \coordinate (a1) at (0,0);
\coordinate (a2) at (0,1);
\coordinate (b1) at (0,2.5);
\coordinate (b2) at (0,3.5);
\coordinate (c1) at (3,1.25);
\coordinate (c2) at (3,2.25);

\draw (a1) -- (a2);
\draw (b1) -- (b2);
\draw (c1) -- (c2);

%円弧
\draw (0,1) arc[start angle=-90, end angle=90, radius=0.75];

%上の曲線(b2)--(c2)
\draw (b2) -- ($(b2)+(0.5,0)$);
\draw ($(c2)-(0.5,0)$) -- (c2);
\draw ($(b2)+(0.5,0)$) .. controls ($(b2)+(1.25,0)$) and ($(c2)-(1.25,0)$) ..($(c2)-(0.5,0)$);

%下の曲線(a1)--(c1)
\draw (a1) -- ($(a1)+(0.5,0)$);
\draw ($(c1)-(0.5,0)$) -- (c1);
\draw ($(a1)+(0.5,0)$) .. controls ($(a1)+(1.25,0)$) and ($(c1)-(1.25,0)$) ..($(c1)-(0.5,0)$);
\node[right] at (0,4.5) {$a$};
\node[right] at (0,3) {$1$};
\node[right] at (0,0.5) {$1$};
\end{tikzpicture}
}}\otimes_{A^3}\vcenter{\hbox{
\begin{tikzpicture}[scale=0.45, line width=2pt]
\begin{scope}[shift={(0,0.5)}]
    \coordinate (x1) at (0,-1);
\coordinate (x2) at (0,-2);
\coordinate (y1) at (3,-1);
\coordinate (y2) at (3,-2);
\draw (x1) -- (x2);
\draw (y1) -- (y2);
\draw (x1) -- (y1);
\draw (y2) -- (x2);

\end{scope}
 \coordinate (a1) at (0,0);
\coordinate (a2) at (0,1);
\coordinate (b1) at (0,2.5);
\coordinate (b2) at (0,3.5);
\coordinate (c1) at (3,1.25);
\coordinate (c2) at (3,2.25);

\draw (a1) -- (a2);
\draw (b1) -- (b2);
\draw (c1) -- (c2);

%円弧
\draw (0,1) arc[start angle=-90, end angle=90, radius=0.75];

%上の曲線(b2)--(c2)
\draw (b2) -- ($(b2)+(0.5,0)$);
\draw ($(c2)-(0.5,0)$) -- (c2);
\draw ($(b2)+(0.5,0)$) .. controls ($(b2)+(1.25,0)$) and ($(c2)-(1.25,0)$) ..($(c2)-(0.5,0)$);

%下の曲線(a1)--(c1)
\draw (a1) -- ($(a1)+(0.5,0)$);
\draw ($(c1)-(0.5,0)$) -- (c1);
\draw ($(a1)+(0.5,0)$) .. controls ($(a1)+(1.25,0)$) and ($(c1)-(1.25,0)$) ..($(c1)-(0.5,0)$);
\node[left] at (0,3) {$1$};
\node[left] at (0,0.5) {$b$};
\node[left] at (0,-1) {$c$};
\end{tikzpicture}
}},
       \end{align*}
\fi
       \iflatexml
\begin{center}
  \includegraphics{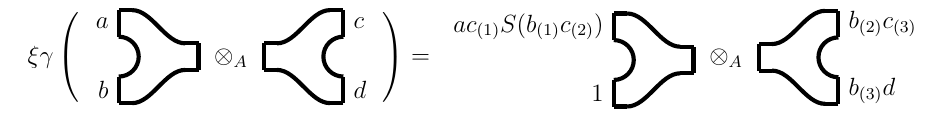}\par
  \includegraphics{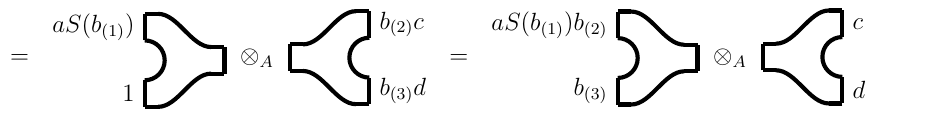}\par
  \includegraphics{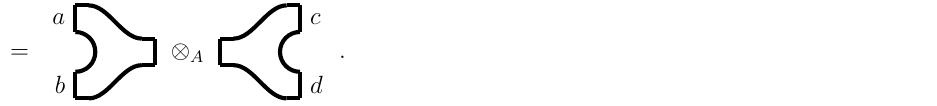}
\end{center}
\else
\begin{align*}
         &\quad\xi\gm\left(\vcenter{\hbox{
\begin{tikzpicture}[scale=0.45, line width=2pt]
 \coordinate (a1) at (0,0);
\coordinate (a2) at (0,1);
\coordinate (b1) at (0,2.5);
\coordinate (b2) at (0,3.5);
\coordinate (c1) at (3,1.25);
\coordinate (c2) at (3,2.25);

\draw (a1) -- (a2);
\draw (b1) -- (b2);
\draw (c1) -- (c2);

%円弧
\draw (0,1) arc[start angle=-90, end angle=90, radius=0.75];

%上の曲線(b2)--(c2)
\draw (b2) -- ($(b2)+(0.5,0)$);
\draw ($(c2)-(0.5,0)$) -- (c2);
\draw ($(b2)+(0.5,0)$) .. controls ($(b2)+(1.25,0)$) and ($(c2)-(1.25,0)$) ..($(c2)-(0.5,0)$);

%下の曲線(a1)--(c1)
\draw (a1) -- ($(a1)+(0.5,0)$);
\draw ($(c1)-(0.5,0)$) -- (c1);
\draw ($(a1)+(0.5,0)$) .. controls ($(a1)+(1.25,0)$) and ($(c1)-(1.25,0)$) ..($(c1)-(0.5,0)$);
\node[left] at (0,3) {$a$};
\node[left] at (0,0.5) {$b$};
\end{tikzpicture}
}}\otimes_{A}\vcenter{\hbox{
\begin{tikzpicture}[scale=0.45, line width=2pt, xscale=-1]
 \coordinate (a1) at (0,0);
\coordinate (a2) at (0,1);
\coordinate (b1) at (0,2.5);
\coordinate (b2) at (0,3.5);
\coordinate (c1) at (3,1.25);
\coordinate (c2) at (3,2.25);

\draw (a1) -- (a2);
\draw (b1) -- (b2);
\draw (c1) -- (c2);

%円弧
\draw (0,1) arc[start angle=-90, end angle=90, radius=0.75];

%上の曲線(b2)--(c2)
\draw (b2) -- ($(b2)+(0.5,0)$);
\draw ($(c2)-(0.5,0)$) -- (c2);
\draw ($(b2)+(0.5,0)$) .. controls ($(b2)+(1.25,0)$) and ($(c2)-(1.25,0)$) ..($(c2)-(0.5,0)$);

%下の曲線(a1)--(c1)
\draw (a1) -- ($(a1)+(0.5,0)$);
\draw ($(c1)-(0.5,0)$) -- (c1);
\draw ($(a1)+(0.5,0)$) .. controls ($(a1)+(1.25,0)$) and ($(c1)-(1.25,0)$) ..($(c1)-(0.5,0)$);
\node[right] at (0,3) {$c$};
\node[right] at (0,0.5) {$d$};
\end{tikzpicture}
}}\right)
         = \vcenter{\hbox{
\begin{tikzpicture}[scale=0.45, line width=2pt]
 \coordinate (a1) at (0,0);
\coordinate (a2) at (0,1);
\coordinate (b1) at (0,2.5);
\coordinate (b2) at (0,3.5);
\coordinate (c1) at (3,1.25);
\coordinate (c2) at (3,2.25);

\draw (a1) -- (a2);
\draw (b1) -- (b2);
\draw (c1) -- (c2);

%円弧
\draw (0,1) arc[start angle=-90, end angle=90, radius=0.75];

%上の曲線(b2)--(c2)
\draw (b2) -- ($(b2)+(0.5,0)$);
\draw ($(c2)-(0.5,0)$) -- (c2);
\draw ($(b2)+(0.5,0)$) .. controls ($(b2)+(1.25,0)$) and ($(c2)-(1.25,0)$) ..($(c2)-(0.5,0)$);

%下の曲線(a1)--(c1)
\draw (a1) -- ($(a1)+(0.5,0)$);
\draw ($(c1)-(0.5,0)$) -- (c1);
\draw ($(a1)+(0.5,0)$) .. controls ($(a1)+(1.25,0)$) and ($(c1)-(1.25,0)$) ..($(c1)-(0.5,0)$);
\node[left] at (0,3) {$ac_{(1)}S(b_{(1)}c_{(2)})$};
\node[left] at (0,0.5) {$1$};
\end{tikzpicture}
}}\otimes_{A}\vcenter{\hbox{
\begin{tikzpicture}[scale=0.45, line width=2pt, xscale=-1]
 \coordinate (a1) at (0,0);
\coordinate (a2) at (0,1);
\coordinate (b1) at (0,2.5);
\coordinate (b2) at (0,3.5);
\coordinate (c1) at (3,1.25);
\coordinate (c2) at (3,2.25);

\draw (a1) -- (a2);
\draw (b1) -- (b2);
\draw (c1) -- (c2);

%円弧
\draw (0,1) arc[start angle=-90, end angle=90, radius=0.75];

%上の曲線(b2)--(c2)
\draw (b2) -- ($(b2)+(0.5,0)$);
\draw ($(c2)-(0.5,0)$) -- (c2);
\draw ($(b2)+(0.5,0)$) .. controls ($(b2)+(1.25,0)$) and ($(c2)-(1.25,0)$) ..($(c2)-(0.5,0)$);

%下の曲線(a1)--(c1)
\draw (a1) -- ($(a1)+(0.5,0)$);
\draw ($(c1)-(0.5,0)$) -- (c1);
\draw ($(a1)+(0.5,0)$) .. controls ($(a1)+(1.25,0)$) and ($(c1)-(1.25,0)$) ..($(c1)-(0.5,0)$);
\node[right] at (0,3) {$b_{(2)}c_{(3)}$};
\node[right] at (0,0.5) {$b_{(3)}d$};
\end{tikzpicture}
}} \\
        &= \vcenter{\hbox{
\begin{tikzpicture}[scale=0.45, line width=2pt]
 \coordinate (a1) at (0,0);
\coordinate (a2) at (0,1);
\coordinate (b1) at (0,2.5);
\coordinate (b2) at (0,3.5);
\coordinate (c1) at (3,1.25);
\coordinate (c2) at (3,2.25);

\draw (a1) -- (a2);
\draw (b1) -- (b2);
\draw (c1) -- (c2);

%円弧
\draw (0,1) arc[start angle=-90, end angle=90, radius=0.75];

%上の曲線(b2)--(c2)
\draw (b2) -- ($(b2)+(0.5,0)$);
\draw ($(c2)-(0.5,0)$) -- (c2);
\draw ($(b2)+(0.5,0)$) .. controls ($(b2)+(1.25,0)$) and ($(c2)-(1.25,0)$) ..($(c2)-(0.5,0)$);

%下の曲線(a1)--(c1)
\draw (a1) -- ($(a1)+(0.5,0)$);
\draw ($(c1)-(0.5,0)$) -- (c1);
\draw ($(a1)+(0.5,0)$) .. controls ($(a1)+(1.25,0)$) and ($(c1)-(1.25,0)$) ..($(c1)-(0.5,0)$);
\node[left] at (0,3) {$aS(b_{(1)})$};
\node[left] at (0,0.5) {$1$};
\end{tikzpicture}
}}\otimes_{A}\vcenter{\hbox{
\begin{tikzpicture}[scale=0.45, line width=2pt, xscale=-1]
 \coordinate (a1) at (0,0);
\coordinate (a2) at (0,1);
\coordinate (b1) at (0,2.5);
\coordinate (b2) at (0,3.5);
\coordinate (c1) at (3,1.25);
\coordinate (c2) at (3,2.25);

\draw (a1) -- (a2);
\draw (b1) -- (b2);
\draw (c1) -- (c2);

%円弧
\draw (0,1) arc[start angle=-90, end angle=90, radius=0.75];

%上の曲線(b2)--(c2)
\draw (b2) -- ($(b2)+(0.5,0)$);
\draw ($(c2)-(0.5,0)$) -- (c2);
\draw ($(b2)+(0.5,0)$) .. controls ($(b2)+(1.25,0)$) and ($(c2)-(1.25,0)$) ..($(c2)-(0.5,0)$);

%下の曲線(a1)--(c1)
\draw (a1) -- ($(a1)+(0.5,0)$);
\draw ($(c1)-(0.5,0)$) -- (c1);
\draw ($(a1)+(0.5,0)$) .. controls ($(a1)+(1.25,0)$) and ($(c1)-(1.25,0)$) ..($(c1)-(0.5,0)$);
\node[right] at (0,3) {$b_{(2)}c$};
\node[right] at (0,0.5) {$b_{(3)}d$};
\end{tikzpicture}
}}
         =\vcenter{\hbox{
\begin{tikzpicture}[scale=0.45, line width=2pt]
 \coordinate (a1) at (0,0);
\coordinate (a2) at (0,1);
\coordinate (b1) at (0,2.5);
\coordinate (b2) at (0,3.5);
\coordinate (c1) at (3,1.25);
\coordinate (c2) at (3,2.25);

\draw (a1) -- (a2);
\draw (b1) -- (b2);
\draw (c1) -- (c2);

%円弧
\draw (0,1) arc[start angle=-90, end angle=90, radius=0.75];

%上の曲線(b2)--(c2)
\draw (b2) -- ($(b2)+(0.5,0)$);
\draw ($(c2)-(0.5,0)$) -- (c2);
\draw ($(b2)+(0.5,0)$) .. controls ($(b2)+(1.25,0)$) and ($(c2)-(1.25,0)$) ..($(c2)-(0.5,0)$);

%下の曲線(a1)--(c1)
\draw (a1) -- ($(a1)+(0.5,0)$);
\draw ($(c1)-(0.5,0)$) -- (c1);
\draw ($(a1)+(0.5,0)$) .. controls ($(a1)+(1.25,0)$) and ($(c1)-(1.25,0)$) ..($(c1)-(0.5,0)$);
\node[left] at (0,3) {$aS(b_{(1)})b_{(2)}$};
\node[left] at (0,0.5) {$b_{(3)}$};
\end{tikzpicture}
}}\otimes_{A}\vcenter{\hbox{
\begin{tikzpicture}[scale=0.45, line width=2pt, xscale=-1]
 \coordinate (a1) at (0,0);
\coordinate (a2) at (0,1);
\coordinate (b1) at (0,2.5);
\coordinate (b2) at (0,3.5);
\coordinate (c1) at (3,1.25);
\coordinate (c2) at (3,2.25);

\draw (a1) -- (a2);
\draw (b1) -- (b2);
\draw (c1) -- (c2);

%円弧
\draw (0,1) arc[start angle=-90, end angle=90, radius=0.75];

%上の曲線(b2)--(c2)
\draw (b2) -- ($(b2)+(0.5,0)$);
\draw ($(c2)-(0.5,0)$) -- (c2);
\draw ($(b2)+(0.5,0)$) .. controls ($(b2)+(1.25,0)$) and ($(c2)-(1.25,0)$) ..($(c2)-(0.5,0)$);

%下の曲線(a1)--(c1)
\draw (a1) -- ($(a1)+(0.5,0)$);
\draw ($(c1)-(0.5,0)$) -- (c1);
\draw ($(a1)+(0.5,0)$) .. controls ($(a1)+(1.25,0)$) and ($(c1)-(1.25,0)$) ..($(c1)-(0.5,0)$);
\node[right] at (0,3) {$c$};
\node[right] at (0,0.5) {$d$};
\end{tikzpicture}
}} \\
         &=\vcenter{\hbox{
\begin{tikzpicture}[scale=0.45, line width=2pt]
 \coordinate (a1) at (0,0);
\coordinate (a2) at (0,1);
\coordinate (b1) at (0,2.5);
\coordinate (b2) at (0,3.5);
\coordinate (c1) at (3,1.25);
\coordinate (c2) at (3,2.25);

\draw (a1) -- (a2);
\draw (b1) -- (b2);
\draw (c1) -- (c2);

%円弧
\draw (0,1) arc[start angle=-90, end angle=90, radius=0.75];

%上の曲線(b2)--(c2)
\draw (b2) -- ($(b2)+(0.5,0)$);
\draw ($(c2)-(0.5,0)$) -- (c2);
\draw ($(b2)+(0.5,0)$) .. controls ($(b2)+(1.25,0)$) and ($(c2)-(1.25,0)$) ..($(c2)-(0.5,0)$);

%下の曲線(a1)--(c1)
\draw (a1) -- ($(a1)+(0.5,0)$);
\draw ($(c1)-(0.5,0)$) -- (c1);
\draw ($(a1)+(0.5,0)$) .. controls ($(a1)+(1.25,0)$) and ($(c1)-(1.25,0)$) ..($(c1)-(0.5,0)$);
\node[left] at (0,3) {$a$};
\node[left] at (0,0.5) {$b$};
\end{tikzpicture}
}}\otimes_{A}\vcenter{\hbox{
\begin{tikzpicture}[scale=0.45, line width=2pt, xscale=-1]
 \coordinate (a1) at (0,0);
\coordinate (a2) at (0,1);
\coordinate (b1) at (0,2.5);
\coordinate (b2) at (0,3.5);
\coordinate (c1) at (3,1.25);
\coordinate (c2) at (3,2.25);

\draw (a1) -- (a2);
\draw (b1) -- (b2);
\draw (c1) -- (c2);

%円弧
\draw (0,1) arc[start angle=-90, end angle=90, radius=0.75];

%上の曲線(b2)--(c2)
\draw (b2) -- ($(b2)+(0.5,0)$);
\draw ($(c2)-(0.5,0)$) -- (c2);
\draw ($(b2)+(0.5,0)$) .. controls ($(b2)+(1.25,0)$) and ($(c2)-(1.25,0)$) ..($(c2)-(0.5,0)$);

%下の曲線(a1)--(c1)
\draw (a1) -- ($(a1)+(0.5,0)$);
\draw ($(c1)-(0.5,0)$) -- (c1);
\draw ($(a1)+(0.5,0)$) .. controls ($(a1)+(1.25,0)$) and ($(c1)-(1.25,0)$) ..($(c1)-(0.5,0)$);
\node[right] at (0,3) {$c$};
\node[right] at (0,0.5) {$d$};
\end{tikzpicture}
}}. 
       \end{align*}
\fi
  Similarly, $\brg$ is invertible, with inverse given by 
  \iflatexml
\begin{center}
  \includegraphics{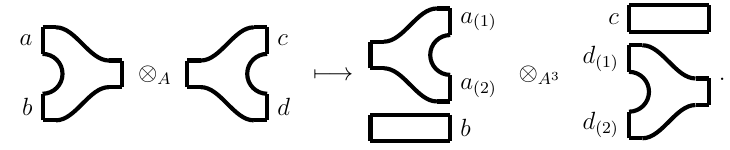}
\end{center}
\else
\begin{equation*}
  \vcenter{\hbox{
\begin{tikzpicture}[scale=0.45, line width=2pt]
 \coordinate (a1) at (0,0);
\coordinate (a2) at (0,1);
\coordinate (b1) at (0,2.5);
\coordinate (b2) at (0,3.5);
\coordinate (c1) at (3,1.25);
\coordinate (c2) at (3,2.25);

\draw (a1) -- (a2);
\draw (b1) -- (b2);
\draw (c1) -- (c2);

%円弧
\draw (0,1) arc[start angle=-90, end angle=90, radius=0.75];

%上の曲線(b2)--(c2)
\draw (b2) -- ($(b2)+(0.5,0)$);
\draw ($(c2)-(0.5,0)$) -- (c2);
\draw ($(b2)+(0.5,0)$) .. controls ($(b2)+(1.25,0)$) and ($(c2)-(1.25,0)$) ..($(c2)-(0.5,0)$);

%下の曲線(a1)--(c1)
\draw (a1) -- ($(a1)+(0.5,0)$);
\draw ($(c1)-(0.5,0)$) -- (c1);
\draw ($(a1)+(0.5,0)$) .. controls ($(a1)+(1.25,0)$) and ($(c1)-(1.25,0)$) ..($(c1)-(0.5,0)$);
\node[left] at (0,3) {$a$};
\node[left] at (0,0.5) {$b$};
\end{tikzpicture}
}}\otimes_{A}\vcenter{\hbox{
\begin{tikzpicture}[scale=0.45, line width=2pt, xscale=-1]
 \coordinate (a1) at (0,0);
\coordinate (a2) at (0,1);
\coordinate (b1) at (0,2.5);
\coordinate (b2) at (0,3.5);
\coordinate (c1) at (3,1.25);
\coordinate (c2) at (3,2.25);

\draw (a1) -- (a2);
\draw (b1) -- (b2);
\draw (c1) -- (c2);

%円弧
\draw (0,1) arc[start angle=-90, end angle=90, radius=0.75];

%上の曲線(b2)--(c2)
\draw (b2) -- ($(b2)+(0.5,0)$);
\draw ($(c2)-(0.5,0)$) -- (c2);
\draw ($(b2)+(0.5,0)$) .. controls ($(b2)+(1.25,0)$) and ($(c2)-(1.25,0)$) ..($(c2)-(0.5,0)$);

%下の曲線(a1)--(c1)
\draw (a1) -- ($(a1)+(0.5,0)$);
\draw ($(c1)-(0.5,0)$) -- (c1);
\draw ($(a1)+(0.5,0)$) .. controls ($(a1)+(1.25,0)$) and ($(c1)-(1.25,0)$) ..($(c1)-(0.5,0)$);
\node[right] at (0,3) {$c$};
\node[right] at (0,0.5) {$d$};
\end{tikzpicture}
}}
  \longmapsto 
    \vcenter{\hbox{
\begin{tikzpicture}[scale=0.45, line width=2pt, xscale=-1]
\begin{scope}[shift={(0,0.5)}]
    \coordinate (x1) at (0,-1);
\coordinate (x2) at (0,-2);
\coordinate (y1) at (3,-1);
\coordinate (y2) at (3,-2);
\draw (x1) -- (x2);
\draw (y1) -- (y2);
\draw (x1) -- (y1);
\draw (y2) -- (x2);

\end{scope}
 \coordinate (a1) at (0,0);
\coordinate (a2) at (0,1);
\coordinate (b1) at (0,2.5);
\coordinate (b2) at (0,3.5);
\coordinate (c1) at (3,1.25);
\coordinate (c2) at (3,2.25);

\draw (a1) -- (a2);
\draw (b1) -- (b2);
\draw (c1) -- (c2);

%円弧
\draw (0,1) arc[start angle=-90, end angle=90, radius=0.75];

%上の曲線(b2)--(c2)
\draw (b2) -- ($(b2)+(0.5,0)$);
\draw ($(c2)-(0.5,0)$) -- (c2);
\draw ($(b2)+(0.5,0)$) .. controls ($(b2)+(1.25,0)$) and ($(c2)-(1.25,0)$) ..($(c2)-(0.5,0)$);

%下の曲線(a1)--(c1)
\draw (a1) -- ($(a1)+(0.5,0)$);
\draw ($(c1)-(0.5,0)$) -- (c1);
\draw ($(a1)+(0.5,0)$) .. controls ($(a1)+(1.25,0)$) and ($(c1)-(1.25,0)$) ..($(c1)-(0.5,0)$);
\node[right] at (0,3) {$a_{(1)}$};
\node[right] at (0,0.5) {$a_{(2)}$};
\node[right] at (0,-1) {$b$};
\end{tikzpicture}
}}\otimes_{A^3}\vcenter{\hbox{
\begin{tikzpicture}[scale=0.45, line width=2pt]
\begin{scope}[shift={(0,6)}]
    \coordinate (x1) at (0,-1);
\coordinate (x2) at (0,-2);
\coordinate (y1) at (3,-1);
\coordinate (y2) at (3,-2);
\draw (x1) -- (x2);
\draw (y1) -- (y2);
\draw (x1) -- (y1);
\draw (y2) -- (x2);

\end{scope}
 \coordinate (a1) at (0,0);
\coordinate (a2) at (0,1);
\coordinate (b1) at (0,2.5);
\coordinate (b2) at (0,3.5);
\coordinate (c1) at (3,1.25);
\coordinate (c2) at (3,2.25);

\draw (a1) -- (a2);
\draw (b1) -- (b2);
\draw (c1) -- (c2);

%円弧
\draw (0,1) arc[start angle=-90, end angle=90, radius=0.75];

%上の曲線(b2)--(c2)
\draw (b2) -- ($(b2)+(0.5,0)$);
\draw ($(c2)-(0.5,0)$) -- (c2);
\draw ($(b2)+(0.5,0)$) .. controls ($(b2)+(1.25,0)$) and ($(c2)-(1.25,0)$) ..($(c2)-(0.5,0)$);

%下の曲線(a1)--(c1)
\draw (a1) -- ($(a1)+(0.5,0)$);
\draw ($(c1)-(0.5,0)$) -- (c1);
\draw ($(a1)+(0.5,0)$) .. controls ($(a1)+(1.25,0)$) and ($(c1)-(1.25,0)$) ..($(c1)-(0.5,0)$);
\node[left] at (0,4.5) {$c$};
\node[left] at (0,3) {$d_{(1)}$};
\node[left] at (0,0.5) {$d_{(2)}$};
\end{tikzpicture}
}}.
  \end{equation*}
\fi
Indeed, the remaining identities follow from similar calculations
using
$$x_{(1)}\otimes S^{-1}\left(x_{(3)}\right)x_{(2)}=x_{(1)}\otimes x_{(3)} S^{-1}\left(x_{(2)}\right)=x\otimes 1$$
  for $x\in H$.
  This inverse is clearly an $(A^2,A^2)$-bimodule homomorphism, and so is $\brg$.
\end{proof}

\begin{lem}\label{lem2}
    The mapping $R^{-1}:X\otimes_A X\rightarrow X\otimes_A X$ is compatible with both the left and right $H$-coactions $\lambda\star \lambda$ and $\rho\star \rho$.
\end{lem}
\begin{proof}
  Recall that \(R^{-1}\) is given by
  $$R^{-1}:X\otimes_A X\rightarrow X\otimes_A X, \quad R^{-1}(a\otimes b\otimes c\otimes d)=a_{(1)}cS(a_{(2)}d_{(1)})\otimes 1\otimes a_{(3)}d_{(2)}\otimes bd_{(3)}.$$
  By definition, the left and right $H$-coactions are given by
  $$(\lambda\star\lambda)(a\otimes b\otimes c\otimes d)=a_{(1)}c_{(1)}\otimes a_{(2)}\otimes b\otimes c_{(2)}\otimes d,$$
  $$(\rho\star\rho)(a\otimes b\otimes c\otimes d)=a\otimes b_{(1)}\otimes c\otimes d_{(1)}\otimes b_{(2)}d_{(2)}.$$
  We first show that $R^{-1}$ is compatible with the right $H$-coaction $\rho\star\rho$ by the following direct calculation.
  \begin{align*}
    (\rho\star\rho)R^{-1}(a\otimes b\otimes c\otimes d)
    &=a_{(1)}cS(a_{(2)}d_{(1)})\otimes 1\otimes a_{(3)}d_{(2)}\otimes b_{(1)}d_{(3)}\otimes b_{(2)}d_{(4)} \\
    &=(R^{-1}\otimes \id_H)(\rho\star\rho)(a\otimes b\otimes c\otimes d).
  \end{align*}
  Next we verify compatibility with the left $H$-coaction $\lambda\star\lambda$. Using $\D S=(S\otimes S)\D^{op}$ and
  $$
  (\D\otimes\D\otimes\D)(a_{(1)}c\otimes a_{(2)}d_{(1)}\otimes a_{(3)}d_{(2)})=a_{(1)}c_{(1)}\otimes a_{(2)}c_{(2)}\otimes a_{(3)}d_{(1)}\otimes a_{(4)}d_{(2)}\otimes a_{(5)}d_{(3)}\otimes a_{(6)}d_{(4)},
  $$
  we obtain
  \begin{align*}
    &(\lambda\star\lambda)R^{-1}(a\otimes b\otimes c\otimes d) \\
    &\;=\left(a_{(1)}c\right)_{(1)}\left(S\left(a_{(2)}d_{(1)}\right)\right)_{(1)}\left(a_{(3)}d_{(2)}\right)_{(1)}\otimes \left(a_{(1)}c\right)_{(2)}\left(S\left(a_{(2)}d_{(1)}\right)\right)_{(2)}  \otimes 1\otimes \left(a_{(3)}d_{(2)}\right)_{(2)}\otimes bd_{(3)} \\
    &\;=a_{(1)}c_{(1)}S\left(a_{(4)}d_{(2)}\right)a_{(5)}d_{(3)}\otimes a_{(2)}c_{(2)}S\left(a_{(3)}d_{(1)}\right)\otimes 1\otimes a_{(6)}d_{(4)}\otimes bd_{(5)} \\
    &\;=a_{(1)}c_{(1)}\varepsilon\left(a_{(4)}d_{(2)}\right)\otimes a_{(2)}c_{(2)}S\left(a_{(3)}d_{(1)}\right)\otimes 1\otimes a_{(5)}d_{(3)}\otimes bd_{(4)} \\
    &\;=a_{(1)}c_{(1)}\otimes a_{(2)}c_{(2)}S\left(a_{(3)}d_{(1)}\right)\otimes 1\otimes a_{(4)}d_{(2)}\otimes bd_{(3)} \\
    &\;=(\id_H \otimes R^{-1})(\lambda\star\lambda)(a\otimes b\otimes c\otimes d).
  \end{align*}
\end{proof}

\begin{lem}\label{lem3}
    The Yetter-Drinfeld module $X^{coH}$ is isomorphic to the standard module $H$ given by Example~\ref{exA}. 
\end{lem}
\begin{proof}
Define a linear mapping
$$\varphi: H\rightarrow X^{coH},\quad \varphi(a):=a\otimes 1.$$ 
Clearly, it preserves the left $H$-coactions. Moreover, it is a left $H$-module homomorphism since it follows from $\D S=(S\otimes S)\D ^{op}$ that
\begin{align*}
  b\triangleright\varphi(a)&=\D(b_{(1)})(a\otimes 1)\D(S(b_{(2)})) 
  =b_{(1)}aS(b_{(4)})\otimes b_{(2)}S(b_{(3)}) \\
  &=b_{(1)}aS(b_{(3)})\otimes \varepsilon(b_{(2)}) 
  =b_{(1)}aS(b_{(2)})\otimes 1=\varphi(b\triangleright a).
\end{align*}
It remains to show that $\varphi$ is invertible.
Define
 $$
 \psi:X^{coH}\rightarrow H,\quad \psi(a\otimes b):=\varepsilon(b)a.
 $$
Clearly, $\psi\varphi=\mathrm{id}_H$.
For any $z=\sum_i a_i\otimes b_i\in X^{\mathrm{co}H}$,
the coinvariance condition gives
\[
  \sum_i a_i\otimes \D(b_i)
  =\sum_i a_i\otimes b_i\otimes 1.
\]
Applying $\mathrm{id}_H\otimes\varepsilon\otimes\mathrm{id}_H$
to both sides, we obtain
\[
  z=\sum_i a_i\varepsilon(b_i)\otimes 1=\varphi\psi(z).
\]
% So it remains to show that $\varphi$ is invertible.
% To do that, we set 
% $$
% \psi:X^{coH}\rightarrow H,\quad \psi(a\otimes b):=\varepsilon(b)a.
% $$
% Then clearly $\psi\varphi(a)=a$ by $\varepsilon(1)=1$. To prove that $\varphi\psi(a\otimes b)=a\otimes b$ for $a\otimes b\in X^{coH}$, we may assume that $a\neq 0$. Then, by the definition of $X^{coH}$, we have
% $$a\otimes b\otimes 1=\rho(a\otimes b)=a\otimes \D(b),$$
% which shows that $\D(b)=b\otimes 1$ by $a\neq 0$. The counitarity therefore induces that $b=\varepsilon(b).$
% Thus we obtain $\varphi\psi(a\otimes b)=\varepsilon(b)a\otimes 1=a\otimes b.$
\end{proof}

\begin{lem}\label{lem0}
    The bimodule isomorphisms $\al$, $\bt$, $\gm$ and $\brg$ satisfy the eight axioms \eqref{diag:1}--\eqref{diag:8}.
\end{lem}
\begin{proof}
  Axioms \eqref{diag:1} and \eqref{diag:5} are immediate. We show the others as follows, where we omit relative tensor product symbols $\otimes_{A^n}$ and $1\in A$.
\iflatexml
\begin{center}
  \includegraphics{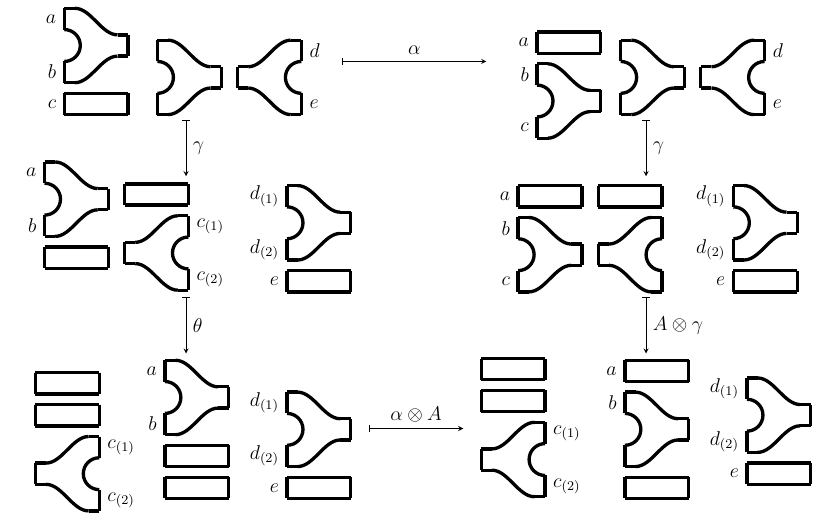}
\end{center}
\else
\[\scalebox{0.8}{%
  \DiagramRectangle[
    AB=\gamma,
    BC=\theta,
    AD=\alpha,
    DE=\gamma,
    EF=A\otimes \gamma,
    CF=\alpha\otimes A
  ]
  {
\begin{tikzpicture}[scale=0.45, line width=2pt]
\begin{scope}[shift={(0,0.5)}]
    \coordinate (x1) at (0,-1);
\coordinate (x2) at (0,-2);
\coordinate (y1) at (3,-1);
\coordinate (y2) at (3,-2);
\draw (x1) -- (x2);
\draw (y1) -- (y2);
\draw (x1) -- (y1);
\draw (y2) -- (x2);

\end{scope}
 \coordinate (a1) at (0,0);
\coordinate (a2) at (0,1);
\coordinate (b1) at (0,2.5);
\coordinate (b2) at (0,3.5);
\coordinate (c1) at (3,1.25);
\coordinate (c2) at (3,2.25);

\draw (a1) -- (a2);
\draw (b1) -- (b2);
\draw (c1) -- (c2);

%円弧
\draw (0,1) arc[start angle=-90, end angle=90, radius=0.75];

%上の曲線(b2)--(c2)
\draw (b2) -- ($(b2)+(0.5,0)$);
\draw ($(c2)-(0.5,0)$) -- (c2);
\draw ($(b2)+(0.5,0)$) .. controls ($(b2)+(1.25,0)$) and ($(c2)-(1.25,0)$) ..($(c2)-(0.5,0)$);

%下の曲線(a1)--(c1)
\draw (a1) -- ($(a1)+(0.5,0)$);
\draw ($(c1)-(0.5,0)$) -- (c1);
\draw ($(a1)+(0.5,0)$) .. controls ($(a1)+(1.25,0)$) and ($(c1)-(1.25,0)$) ..($(c1)-(0.5,0)$);
\node[left] at (0,3) {$a$};
\node[left] at (0,0.5) {$b$};
\node[left] at (0,-1) {$c$};
\end{tikzpicture}

\begin{tikzpicture}[scale=0.45, line width=2pt]
 \coordinate (a1) at (0,0);
\coordinate (a2) at (0,1);
\coordinate (b1) at (0,2.5);
\coordinate (b2) at (0,3.5);
\coordinate (c1) at (3,1.25);
\coordinate (c2) at (3,2.25);

\draw (a1) -- (a2);
\draw (b1) -- (b2);
\draw (c1) -- (c2);

%円弧
\draw (0,1) arc[start angle=-90, end angle=90, radius=0.75];

%上の曲線(b2)--(c2)
\draw (b2) -- ($(b2)+(0.5,0)$);
\draw ($(c2)-(0.5,0)$) -- (c2);
\draw ($(b2)+(0.5,0)$) .. controls ($(b2)+(1.25,0)$) and ($(c2)-(1.25,0)$) ..($(c2)-(0.5,0)$);

%下の曲線(a1)--(c1)
\draw (a1) -- ($(a1)+(0.5,0)$);
\draw ($(c1)-(0.5,0)$) -- (c1);
\draw ($(a1)+(0.5,0)$) .. controls ($(a1)+(1.25,0)$) and ($(c1)-(1.25,0)$) ..($(c1)-(0.5,0)$);
\node[left] at (0,3) {$$};
\node[left] at (0,0.5) {$$};
\end{tikzpicture}

\begin{tikzpicture}[scale=0.45, line width=2pt, xscale=-1]
 \coordinate (a1) at (0,0);
\coordinate (a2) at (0,1);
\coordinate (b1) at (0,2.5);
\coordinate (b2) at (0,3.5);
\coordinate (c1) at (3,1.25);
\coordinate (c2) at (3,2.25);

\draw (a1) -- (a2);
\draw (b1) -- (b2);
\draw (c1) -- (c2);

%円弧
\draw (0,1) arc[start angle=-90, end angle=90, radius=0.75];

%上の曲線(b2)--(c2)
\draw (b2) -- ($(b2)+(0.5,0)$);
\draw ($(c2)-(0.5,0)$) -- (c2);
\draw ($(b2)+(0.5,0)$) .. controls ($(b2)+(1.25,0)$) and ($(c2)-(1.25,0)$) ..($(c2)-(0.5,0)$);

%下の曲線(a1)--(c1)
\draw (a1) -- ($(a1)+(0.5,0)$);
\draw ($(c1)-(0.5,0)$) -- (c1);
\draw ($(a1)+(0.5,0)$) .. controls ($(a1)+(1.25,0)$) and ($(c1)-(1.25,0)$) ..($(c1)-(0.5,0)$);
\node[right] at (0,3) {$d$};
\node[right] at (0,0.5) {$e$};
\end{tikzpicture}

  }{
  \shiftfig{0mm}{5mm}{
\begin{tikzpicture}[scale=0.45, line width=2pt]
\begin{scope}[shift={(0,0.5)}]
    \coordinate (x1) at (0,-1);
\coordinate (x2) at (0,-2);
\coordinate (y1) at (3,-1);
\coordinate (y2) at (3,-2);
\draw (x1) -- (x2);
\draw (y1) -- (y2);
\draw (x1) -- (y1);
\draw (y2) -- (x2);

\end{scope}
 \coordinate (a1) at (0,0);
\coordinate (a2) at (0,1);
\coordinate (b1) at (0,2.5);
\coordinate (b2) at (0,3.5);
\coordinate (c1) at (3,1.25);
\coordinate (c2) at (3,2.25);

\draw (a1) -- (a2);
\draw (b1) -- (b2);
\draw (c1) -- (c2);

%円弧
\draw (0,1) arc[start angle=-90, end angle=90, radius=0.75];

%上の曲線(b2)--(c2)
\draw (b2) -- ($(b2)+(0.5,0)$);
\draw ($(c2)-(0.5,0)$) -- (c2);
\draw ($(b2)+(0.5,0)$) .. controls ($(b2)+(1.25,0)$) and ($(c2)-(1.25,0)$) ..($(c2)-(0.5,0)$);

%下の曲線(a1)--(c1)
\draw (a1) -- ($(a1)+(0.5,0)$);
\draw ($(c1)-(0.5,0)$) -- (c1);
\draw ($(a1)+(0.5,0)$) .. controls ($(a1)+(1.25,0)$) and ($(c1)-(1.25,0)$) ..($(c1)-(0.5,0)$);
\node[left] at (0,3) {$a$};
\node[left] at (0,0.5) {$b$};
\node[left] at (0,-1) {$$};
\end{tikzpicture}
}
\begin{tikzpicture}[scale=0.45, line width=2pt, xscale=-1]
\begin{scope}[shift={(0,6)}]
    \coordinate (x1) at (0,-1);
\coordinate (x2) at (0,-2);
\coordinate (y1) at (3,-1);
\coordinate (y2) at (3,-2);
\draw (x1) -- (x2);
\draw (y1) -- (y2);
\draw (x1) -- (y1);
\draw (y2) -- (x2);

\end{scope}
 \coordinate (a1) at (0,0);
\coordinate (a2) at (0,1);
\coordinate (b1) at (0,2.5);
\coordinate (b2) at (0,3.5);
\coordinate (c1) at (3,1.25);
\coordinate (c2) at (3,2.25);

\draw (a1) -- (a2);
\draw (b1) -- (b2);
\draw (c1) -- (c2);

%円弧
\draw (0,1) arc[start angle=-90, end angle=90, radius=0.75];

%上の曲線(b2)--(c2)
\draw (b2) -- ($(b2)+(0.5,0)$);
\draw ($(c2)-(0.5,0)$) -- (c2);
\draw ($(b2)+(0.5,0)$) .. controls ($(b2)+(1.25,0)$) and ($(c2)-(1.25,0)$) ..($(c2)-(0.5,0)$);

%下の曲線(a1)--(c1)
\draw (a1) -- ($(a1)+(0.5,0)$);
\draw ($(c1)-(0.5,0)$) -- (c1);
\draw ($(a1)+(0.5,0)$) .. controls ($(a1)+(1.25,0)$) and ($(c1)-(1.25,0)$) ..($(c1)-(0.5,0)$);
\node[right] at (0,4.5) {$$};
\node[right] at (0,3) {$c_{(1)}$};
\node[right] at (0,0.5) {$c_{(2)}$};
\end{tikzpicture}

\begin{tikzpicture}[scale=0.45, line width=2pt]
\begin{scope}[shift={(0,0.5)}]
    \coordinate (x1) at (0,-1);
\coordinate (x2) at (0,-2);
\coordinate (y1) at (3,-1);
\coordinate (y2) at (3,-2);
\draw (x1) -- (x2);
\draw (y1) -- (y2);
\draw (x1) -- (y1);
\draw (y2) -- (x2);

\end{scope}
 \coordinate (a1) at (0,0);
\coordinate (a2) at (0,1);
\coordinate (b1) at (0,2.5);
\coordinate (b2) at (0,3.5);
\coordinate (c1) at (3,1.25);
\coordinate (c2) at (3,2.25);

\draw (a1) -- (a2);
\draw (b1) -- (b2);
\draw (c1) -- (c2);

%円弧
\draw (0,1) arc[start angle=-90, end angle=90, radius=0.75];

%上の曲線(b2)--(c2)
\draw (b2) -- ($(b2)+(0.5,0)$);
\draw ($(c2)-(0.5,0)$) -- (c2);
\draw ($(b2)+(0.5,0)$) .. controls ($(b2)+(1.25,0)$) and ($(c2)-(1.25,0)$) ..($(c2)-(0.5,0)$);

%下の曲線(a1)--(c1)
\draw (a1) -- ($(a1)+(0.5,0)$);
\draw ($(c1)-(0.5,0)$) -- (c1);
\draw ($(a1)+(0.5,0)$) .. controls ($(a1)+(1.25,0)$) and ($(c1)-(1.25,0)$) ..($(c1)-(0.5,0)$);
\node[left] at (0,3) {$d_{(1)}$};
\node[left] at (0,0.5) {$d_{(2)}$};
\node[left] at (0,-1) {$e$};
\end{tikzpicture}

  }{
  \shiftfig{3mm}{-3mm}{
\begin{tikzpicture}[scale=0.45, line width=2pt, xscale=-1]
\begin{scope}[shift={(0,7.5)}]
    \coordinate (x1) at (0,-1);
\coordinate (x2) at (0,-2);
\coordinate (y1) at (3,-1);
\coordinate (y2) at (3,-2);
\draw (x1) -- (x2);
\draw (y1) -- (y2);
\draw (x1) -- (y1);
\draw (y2) -- (x2);

\end{scope}
\begin{scope}[shift={(0,6)}]
    \coordinate (x1) at (0,-1);
\coordinate (x2) at (0,-2);
\coordinate (y1) at (3,-1);
\coordinate (y2) at (3,-2);
\draw (x1) -- (x2);
\draw (y1) -- (y2);
\draw (x1) -- (y1);
\draw (y2) -- (x2);

\end{scope}
 \coordinate (a1) at (0,0);
\coordinate (a2) at (0,1);
\coordinate (b1) at (0,2.5);
\coordinate (b2) at (0,3.5);
\coordinate (c1) at (3,1.25);
\coordinate (c2) at (3,2.25);

\draw (a1) -- (a2);
\draw (b1) -- (b2);
\draw (c1) -- (c2);

%円弧
\draw (0,1) arc[start angle=-90, end angle=90, radius=0.75];

%上の曲線(b2)--(c2)
\draw (b2) -- ($(b2)+(0.5,0)$);
\draw ($(c2)-(0.5,0)$) -- (c2);
\draw ($(b2)+(0.5,0)$) .. controls ($(b2)+(1.25,0)$) and ($(c2)-(1.25,0)$) ..($(c2)-(0.5,0)$);

%下の曲線(a1)--(c1)
\draw (a1) -- ($(a1)+(0.5,0)$);
\draw ($(c1)-(0.5,0)$) -- (c1);
\draw ($(a1)+(0.5,0)$) .. controls ($(a1)+(1.25,0)$) and ($(c1)-(1.25,0)$) ..($(c1)-(0.5,0)$);
\node[right] at (0,6) {$$};
\node[right] at (0,4.5) {$$};
\node[right] at (0,3) {$c_{(1)}$};
\node[right] at (0,0.5) {$c_{(2)}$};
\end{tikzpicture}
}
\begin{tikzpicture}[scale=0.45, line width=2pt]
\begin{scope}[shift={(0,0.5)}]
    \coordinate (x1) at (0,-1);
\coordinate (x2) at (0,-2);
\coordinate (y1) at (3,-1);
\coordinate (y2) at (3,-2);
\draw (x1) -- (x2);
\draw (y1) -- (y2);
\draw (x1) -- (y1);
\draw (y2) -- (x2);

\end{scope}
\begin{scope}[shift={(0,-1)}]
    \coordinate (x1) at (0,-1);
\coordinate (x2) at (0,-2);
\coordinate (y1) at (3,-1);
\coordinate (y2) at (3,-2);
\draw (x1) -- (x2);
\draw (y1) -- (y2);
\draw (x1) -- (y1);
\draw (y2) -- (x2);

\end{scope}
 \coordinate (a1) at (0,0);
\coordinate (a2) at (0,1);
\coordinate (b1) at (0,2.5);
\coordinate (b2) at (0,3.5);
\coordinate (c1) at (3,1.25);
\coordinate (c2) at (3,2.25);

\draw (a1) -- (a2);
\draw (b1) -- (b2);
\draw (c1) -- (c2);

%円弧
\draw (0,1) arc[start angle=-90, end angle=90, radius=0.75];

%上の曲線(b2)--(c2)
\draw (b2) -- ($(b2)+(0.5,0)$);
\draw ($(c2)-(0.5,0)$) -- (c2);
\draw ($(b2)+(0.5,0)$) .. controls ($(b2)+(1.25,0)$) and ($(c2)-(1.25,0)$) ..($(c2)-(0.5,0)$);

%下の曲線(a1)--(c1)
\draw (a1) -- ($(a1)+(0.5,0)$);
\draw ($(c1)-(0.5,0)$) -- (c1);
\draw ($(a1)+(0.5,0)$) .. controls ($(a1)+(1.25,0)$) and ($(c1)-(1.25,0)$) ..($(c1)-(0.5,0)$);
\node[left] at (0,3) {$a$};
\node[left] at (0,0.5) {$b$};
\node[left] at (0,-1) {$$};
\node[left] at (0,-2.5) {$$};
\end{tikzpicture}

\begin{tikzpicture}[scale=0.45, line width=2pt]
\begin{scope}[shift={(0,0.5)}]
    \coordinate (x1) at (0,-1);
\coordinate (x2) at (0,-2);
\coordinate (y1) at (3,-1);
\coordinate (y2) at (3,-2);
\draw (x1) -- (x2);
\draw (y1) -- (y2);
\draw (x1) -- (y1);
\draw (y2) -- (x2);

\end{scope}
 \coordinate (a1) at (0,0);
\coordinate (a2) at (0,1);
\coordinate (b1) at (0,2.5);
\coordinate (b2) at (0,3.5);
\coordinate (c1) at (3,1.25);
\coordinate (c2) at (3,2.25);

\draw (a1) -- (a2);
\draw (b1) -- (b2);
\draw (c1) -- (c2);

%円弧
\draw (0,1) arc[start angle=-90, end angle=90, radius=0.75];

%上の曲線(b2)--(c2)
\draw (b2) -- ($(b2)+(0.5,0)$);
\draw ($(c2)-(0.5,0)$) -- (c2);
\draw ($(b2)+(0.5,0)$) .. controls ($(b2)+(1.25,0)$) and ($(c2)-(1.25,0)$) ..($(c2)-(0.5,0)$);

%下の曲線(a1)--(c1)
\draw (a1) -- ($(a1)+(0.5,0)$);
\draw ($(c1)-(0.5,0)$) -- (c1);
\draw ($(a1)+(0.5,0)$) .. controls ($(a1)+(1.25,0)$) and ($(c1)-(1.25,0)$) ..($(c1)-(0.5,0)$);
\node[left] at (0,3) {$d_{(1)}$};
\node[left] at (0,0.5) {$d_{(2)}$};
\node[left] at (0,-1) {$e$};
\end{tikzpicture}

  }{
  \shiftfig{2mm}{-5mm}{
\begin{tikzpicture}[scale=0.45, line width=2pt]
\begin{scope}[shift={(0,6)}]
    \coordinate (x1) at (0,-1);
\coordinate (x2) at (0,-2);
\coordinate (y1) at (3,-1);
\coordinate (y2) at (3,-2);
\draw (x1) -- (x2);
\draw (y1) -- (y2);
\draw (x1) -- (y1);
\draw (y2) -- (x2);

\end{scope}
 \coordinate (a1) at (0,0);
\coordinate (a2) at (0,1);
\coordinate (b1) at (0,2.5);
\coordinate (b2) at (0,3.5);
\coordinate (c1) at (3,1.25);
\coordinate (c2) at (3,2.25);

\draw (a1) -- (a2);
\draw (b1) -- (b2);
\draw (c1) -- (c2);

%円弧
\draw (0,1) arc[start angle=-90, end angle=90, radius=0.75];

%上の曲線(b2)--(c2)
\draw (b2) -- ($(b2)+(0.5,0)$);
\draw ($(c2)-(0.5,0)$) -- (c2);
\draw ($(b2)+(0.5,0)$) .. controls ($(b2)+(1.25,0)$) and ($(c2)-(1.25,0)$) ..($(c2)-(0.5,0)$);

%下の曲線(a1)--(c1)
\draw (a1) -- ($(a1)+(0.5,0)$);
\draw ($(c1)-(0.5,0)$) -- (c1);
\draw ($(a1)+(0.5,0)$) .. controls ($(a1)+(1.25,0)$) and ($(c1)-(1.25,0)$) ..($(c1)-(0.5,0)$);
\node[left] at (0,4.5) {$a$};
\node[left] at (0,3) {$b$};
\node[left] at (0,0.5) {$c$};
\end{tikzpicture}
}
\begin{tikzpicture}[scale=0.45, line width=2pt]
 \coordinate (a1) at (0,0);
\coordinate (a2) at (0,1);
\coordinate (b1) at (0,2.5);
\coordinate (b2) at (0,3.5);
\coordinate (c1) at (3,1.25);
\coordinate (c2) at (3,2.25);

\draw (a1) -- (a2);
\draw (b1) -- (b2);
\draw (c1) -- (c2);

%円弧
\draw (0,1) arc[start angle=-90, end angle=90, radius=0.75];

%上の曲線(b2)--(c2)
\draw (b2) -- ($(b2)+(0.5,0)$);
\draw ($(c2)-(0.5,0)$) -- (c2);
\draw ($(b2)+(0.5,0)$) .. controls ($(b2)+(1.25,0)$) and ($(c2)-(1.25,0)$) ..($(c2)-(0.5,0)$);

%下の曲線(a1)--(c1)
\draw (a1) -- ($(a1)+(0.5,0)$);
\draw ($(c1)-(0.5,0)$) -- (c1);
\draw ($(a1)+(0.5,0)$) .. controls ($(a1)+(1.25,0)$) and ($(c1)-(1.25,0)$) ..($(c1)-(0.5,0)$);
\node[left] at (0,3) {$$};
\node[left] at (0,0.5) {$$};
\end{tikzpicture}

\begin{tikzpicture}[scale=0.45, line width=2pt, xscale=-1]
 \coordinate (a1) at (0,0);
\coordinate (a2) at (0,1);
\coordinate (b1) at (0,2.5);
\coordinate (b2) at (0,3.5);
\coordinate (c1) at (3,1.25);
\coordinate (c2) at (3,2.25);

\draw (a1) -- (a2);
\draw (b1) -- (b2);
\draw (c1) -- (c2);

%円弧
\draw (0,1) arc[start angle=-90, end angle=90, radius=0.75];

%上の曲線(b2)--(c2)
\draw (b2) -- ($(b2)+(0.5,0)$);
\draw ($(c2)-(0.5,0)$) -- (c2);
\draw ($(b2)+(0.5,0)$) .. controls ($(b2)+(1.25,0)$) and ($(c2)-(1.25,0)$) ..($(c2)-(0.5,0)$);

%下の曲線(a1)--(c1)
\draw (a1) -- ($(a1)+(0.5,0)$);
\draw ($(c1)-(0.5,0)$) -- (c1);
\draw ($(a1)+(0.5,0)$) .. controls ($(a1)+(1.25,0)$) and ($(c1)-(1.25,0)$) ..($(c1)-(0.5,0)$);
\node[right] at (0,3) {$d$};
\node[right] at (0,0.5) {$e$};
\end{tikzpicture}

  }{
  
\begin{tikzpicture}[scale=0.45, line width=2pt]
\begin{scope}[shift={(0,6)}]
    \coordinate (x1) at (0,-1);
\coordinate (x2) at (0,-2);
\coordinate (y1) at (3,-1);
\coordinate (y2) at (3,-2);
\draw (x1) -- (x2);
\draw (y1) -- (y2);
\draw (x1) -- (y1);
\draw (y2) -- (x2);

\end{scope}
 \coordinate (a1) at (0,0);
\coordinate (a2) at (0,1);
\coordinate (b1) at (0,2.5);
\coordinate (b2) at (0,3.5);
\coordinate (c1) at (3,1.25);
\coordinate (c2) at (3,2.25);

\draw (a1) -- (a2);
\draw (b1) -- (b2);
\draw (c1) -- (c2);

%円弧
\draw (0,1) arc[start angle=-90, end angle=90, radius=0.75];

%上の曲線(b2)--(c2)
\draw (b2) -- ($(b2)+(0.5,0)$);
\draw ($(c2)-(0.5,0)$) -- (c2);
\draw ($(b2)+(0.5,0)$) .. controls ($(b2)+(1.25,0)$) and ($(c2)-(1.25,0)$) ..($(c2)-(0.5,0)$);

%下の曲線(a1)--(c1)
\draw (a1) -- ($(a1)+(0.5,0)$);
\draw ($(c1)-(0.5,0)$) -- (c1);
\draw ($(a1)+(0.5,0)$) .. controls ($(a1)+(1.25,0)$) and ($(c1)-(1.25,0)$) ..($(c1)-(0.5,0)$);
\node[left] at (0,4.5) {$a$};
\node[left] at (0,3) {$b$};
\node[left] at (0,0.5) {$c$};
\end{tikzpicture}

\begin{tikzpicture}[scale=0.45, line width=2pt, xscale=-1]
\begin{scope}[shift={(0,6)}]
    \coordinate (x1) at (0,-1);
\coordinate (x2) at (0,-2);
\coordinate (y1) at (3,-1);
\coordinate (y2) at (3,-2);
\draw (x1) -- (x2);
\draw (y1) -- (y2);
\draw (x1) -- (y1);
\draw (y2) -- (x2);

\end{scope}
 \coordinate (a1) at (0,0);
\coordinate (a2) at (0,1);
\coordinate (b1) at (0,2.5);
\coordinate (b2) at (0,3.5);
\coordinate (c1) at (3,1.25);
\coordinate (c2) at (3,2.25);

\draw (a1) -- (a2);
\draw (b1) -- (b2);
\draw (c1) -- (c2);

%円弧
\draw (0,1) arc[start angle=-90, end angle=90, radius=0.75];

%上の曲線(b2)--(c2)
\draw (b2) -- ($(b2)+(0.5,0)$);
\draw ($(c2)-(0.5,0)$) -- (c2);
\draw ($(b2)+(0.5,0)$) .. controls ($(b2)+(1.25,0)$) and ($(c2)-(1.25,0)$) ..($(c2)-(0.5,0)$);

%下の曲線(a1)--(c1)
\draw (a1) -- ($(a1)+(0.5,0)$);
\draw ($(c1)-(0.5,0)$) -- (c1);
\draw ($(a1)+(0.5,0)$) .. controls ($(a1)+(1.25,0)$) and ($(c1)-(1.25,0)$) ..($(c1)-(0.5,0)$);
\node[right] at (0,4.5) {$$};
\node[right] at (0,3) {$$};
\node[right] at (0,0.5) {$$};
\end{tikzpicture}

\begin{tikzpicture}[scale=0.45, line width=2pt]
\begin{scope}[shift={(0,0.5)}]
    \coordinate (x1) at (0,-1);
\coordinate (x2) at (0,-2);
\coordinate (y1) at (3,-1);
\coordinate (y2) at (3,-2);
\draw (x1) -- (x2);
\draw (y1) -- (y2);
\draw (x1) -- (y1);
\draw (y2) -- (x2);

\end{scope}
 \coordinate (a1) at (0,0);
\coordinate (a2) at (0,1);
\coordinate (b1) at (0,2.5);
\coordinate (b2) at (0,3.5);
\coordinate (c1) at (3,1.25);
\coordinate (c2) at (3,2.25);

\draw (a1) -- (a2);
\draw (b1) -- (b2);
\draw (c1) -- (c2);

%円弧
\draw (0,1) arc[start angle=-90, end angle=90, radius=0.75];

%上の曲線(b2)--(c2)
\draw (b2) -- ($(b2)+(0.5,0)$);
\draw ($(c2)-(0.5,0)$) -- (c2);
\draw ($(b2)+(0.5,0)$) .. controls ($(b2)+(1.25,0)$) and ($(c2)-(1.25,0)$) ..($(c2)-(0.5,0)$);

%下の曲線(a1)--(c1)
\draw (a1) -- ($(a1)+(0.5,0)$);
\draw ($(c1)-(0.5,0)$) -- (c1);
\draw ($(a1)+(0.5,0)$) .. controls ($(a1)+(1.25,0)$) and ($(c1)-(1.25,0)$) ..($(c1)-(0.5,0)$);
\node[left] at (0,3) {$d_{(1)}$};
\node[left] at (0,0.5) {$d_{(2)}$};
\node[left] at (0,-1) {$e$};
\end{tikzpicture}

  }{
  
\begin{tikzpicture}[scale=0.45, line width=2pt, xscale=-1]
\begin{scope}[shift={(0,7.5)}]
    \coordinate (x1) at (0,-1);
\coordinate (x2) at (0,-2);
\coordinate (y1) at (3,-1);
\coordinate (y2) at (3,-2);
\draw (x1) -- (x2);
\draw (y1) -- (y2);
\draw (x1) -- (y1);
\draw (y2) -- (x2);

\end{scope}
\begin{scope}[shift={(0,6)}]
    \coordinate (x1) at (0,-1);
\coordinate (x2) at (0,-2);
\coordinate (y1) at (3,-1);
\coordinate (y2) at (3,-2);
\draw (x1) -- (x2);
\draw (y1) -- (y2);
\draw (x1) -- (y1);
\draw (y2) -- (x2);

\end{scope}
 \coordinate (a1) at (0,0);
\coordinate (a2) at (0,1);
\coordinate (b1) at (0,2.5);
\coordinate (b2) at (0,3.5);
\coordinate (c1) at (3,1.25);
\coordinate (c2) at (3,2.25);

\draw (a1) -- (a2);
\draw (b1) -- (b2);
\draw (c1) -- (c2);

%円弧
\draw (0,1) arc[start angle=-90, end angle=90, radius=0.75];

%上の曲線(b2)--(c2)
\draw (b2) -- ($(b2)+(0.5,0)$);
\draw ($(c2)-(0.5,0)$) -- (c2);
\draw ($(b2)+(0.5,0)$) .. controls ($(b2)+(1.25,0)$) and ($(c2)-(1.25,0)$) ..($(c2)-(0.5,0)$);

%下の曲線(a1)--(c1)
\draw (a1) -- ($(a1)+(0.5,0)$);
\draw ($(c1)-(0.5,0)$) -- (c1);
\draw ($(a1)+(0.5,0)$) .. controls ($(a1)+(1.25,0)$) and ($(c1)-(1.25,0)$) ..($(c1)-(0.5,0)$);
\node[right] at (0,6) {$$};
\node[right] at (0,4.5) {$$};
\node[right] at (0,3) {$c_{(1)}$};
\node[right] at (0,0.5) {$c_{(2)}$};
\end{tikzpicture}

\begin{tikzpicture}[scale=0.45, line width=2pt]
\begin{scope}[shift={(0,6)}]
    \coordinate (x1) at (0,-1);
\coordinate (x2) at (0,-2);
\coordinate (y1) at (3,-1);
\coordinate (y2) at (3,-2);
\draw (x1) -- (x2);
\draw (y1) -- (y2);
\draw (x1) -- (y1);
\draw (y2) -- (x2);

\end{scope}
\begin{scope}[shift={(0,0.5)}]
    \coordinate (x1) at (0,-1);
\coordinate (x2) at (0,-2);
\coordinate (y1) at (3,-1);
\coordinate (y2) at (3,-2);
\draw (x1) -- (x2);
\draw (y1) -- (y2);
\draw (x1) -- (y1);
\draw (y2) -- (x2);

\end{scope}
 \coordinate (a1) at (0,0);
\coordinate (a2) at (0,1);
\coordinate (b1) at (0,2.5);
\coordinate (b2) at (0,3.5);
\coordinate (c1) at (3,1.25);
\coordinate (c2) at (3,2.25);

\draw (a1) -- (a2);
\draw (b1) -- (b2);
\draw (c1) -- (c2);

%円弧
\draw (0,1) arc[start angle=-90, end angle=90, radius=0.75];

%上の曲線(b2)--(c2)
\draw (b2) -- ($(b2)+(0.5,0)$);
\draw ($(c2)-(0.5,0)$) -- (c2);
\draw ($(b2)+(0.5,0)$) .. controls ($(b2)+(1.25,0)$) and ($(c2)-(1.25,0)$) ..($(c2)-(0.5,0)$);

%下の曲線(a1)--(c1)
\draw (a1) -- ($(a1)+(0.5,0)$);
\draw ($(c1)-(0.5,0)$) -- (c1);
\draw ($(a1)+(0.5,0)$) .. controls ($(a1)+(1.25,0)$) and ($(c1)-(1.25,0)$) ..($(c1)-(0.5,0)$);
\node[left] at (0,4.5) {$a$};
\node[left] at (0,3) {$b$};
\node[left] at (0,0.5) {$$};
\node[left] at (0,-1) {$$};
\end{tikzpicture}
\shiftfig{0mm}{3mm}{
\begin{tikzpicture}[scale=0.45, line width=2pt]
\begin{scope}[shift={(0,0.5)}]
    \coordinate (x1) at (0,-1);
\coordinate (x2) at (0,-2);
\coordinate (y1) at (3,-1);
\coordinate (y2) at (3,-2);
\draw (x1) -- (x2);
\draw (y1) -- (y2);
\draw (x1) -- (y1);
\draw (y2) -- (x2);

\end{scope}
 \coordinate (a1) at (0,0);
\coordinate (a2) at (0,1);
\coordinate (b1) at (0,2.5);
\coordinate (b2) at (0,3.5);
\coordinate (c1) at (3,1.25);
\coordinate (c2) at (3,2.25);

\draw (a1) -- (a2);
\draw (b1) -- (b2);
\draw (c1) -- (c2);

%円弧
\draw (0,1) arc[start angle=-90, end angle=90, radius=0.75];

%上の曲線(b2)--(c2)
\draw (b2) -- ($(b2)+(0.5,0)$);
\draw ($(c2)-(0.5,0)$) -- (c2);
\draw ($(b2)+(0.5,0)$) .. controls ($(b2)+(1.25,0)$) and ($(c2)-(1.25,0)$) ..($(c2)-(0.5,0)$);

%下の曲線(a1)--(c1)
\draw (a1) -- ($(a1)+(0.5,0)$);
\draw ($(c1)-(0.5,0)$) -- (c1);
\draw ($(a1)+(0.5,0)$) .. controls ($(a1)+(1.25,0)$) and ($(c1)-(1.25,0)$) ..($(c1)-(0.5,0)$);
\node[left] at (0,3) {$d_{(1)}$};
\node[left] at (0,0.5) {$d_{(2)}$};
\node[left] at (0,-1) {$e$};
\end{tikzpicture}
}
  }
  }
\]
\fi

\iflatexml
\begin{center}
  \includegraphics{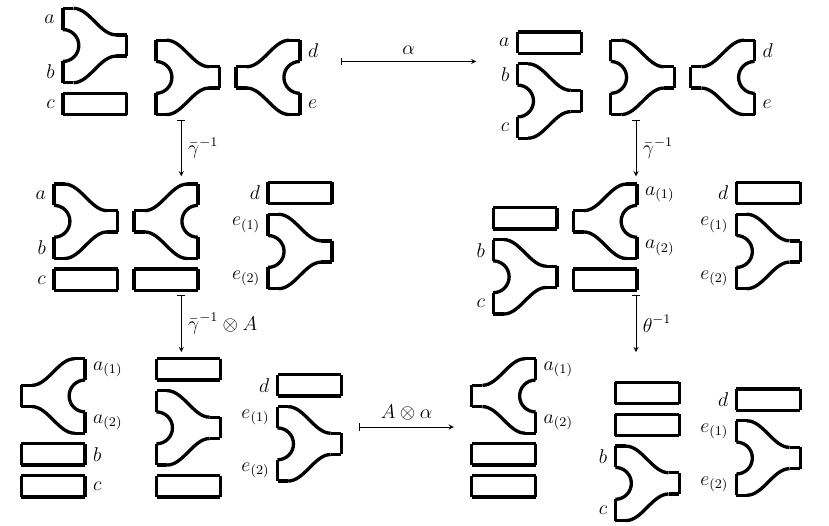}
\end{center}
\else
\[\scalebox{0.8}{%
  \DiagramRectangle[
    AB=\brg^{-1},
    BC=\brg^{-1}\otimes A,
    AD=\alpha,
    DE=\brg^{-1},
    EF=\theta^{-1},
    CF=A\otimes\alpha
  ]
    {
    
\begin{tikzpicture}[scale=0.45, line width=2pt]
\begin{scope}[shift={(0,0.5)}]
    \coordinate (x1) at (0,-1);
\coordinate (x2) at (0,-2);
\coordinate (y1) at (3,-1);
\coordinate (y2) at (3,-2);
\draw (x1) -- (x2);
\draw (y1) -- (y2);
\draw (x1) -- (y1);
\draw (y2) -- (x2);

\end{scope}
 \coordinate (a1) at (0,0);
\coordinate (a2) at (0,1);
\coordinate (b1) at (0,2.5);
\coordinate (b2) at (0,3.5);
\coordinate (c1) at (3,1.25);
\coordinate (c2) at (3,2.25);

\draw (a1) -- (a2);
\draw (b1) -- (b2);
\draw (c1) -- (c2);

%円弧
\draw (0,1) arc[start angle=-90, end angle=90, radius=0.75];

%上の曲線(b2)--(c2)
\draw (b2) -- ($(b2)+(0.5,0)$);
\draw ($(c2)-(0.5,0)$) -- (c2);
\draw ($(b2)+(0.5,0)$) .. controls ($(b2)+(1.25,0)$) and ($(c2)-(1.25,0)$) ..($(c2)-(0.5,0)$);

%下の曲線(a1)--(c1)
\draw (a1) -- ($(a1)+(0.5,0)$);
\draw ($(c1)-(0.5,0)$) -- (c1);
\draw ($(a1)+(0.5,0)$) .. controls ($(a1)+(1.25,0)$) and ($(c1)-(1.25,0)$) ..($(c1)-(0.5,0)$);
\node[left] at (0,3) {$a$};
\node[left] at (0,0.5) {$b$};
\node[left] at (0,-1) {$c$};
\end{tikzpicture}

\begin{tikzpicture}[scale=0.45, line width=2pt]
 \coordinate (a1) at (0,0);
\coordinate (a2) at (0,1);
\coordinate (b1) at (0,2.5);
\coordinate (b2) at (0,3.5);
\coordinate (c1) at (3,1.25);
\coordinate (c2) at (3,2.25);

\draw (a1) -- (a2);
\draw (b1) -- (b2);
\draw (c1) -- (c2);

%円弧
\draw (0,1) arc[start angle=-90, end angle=90, radius=0.75];

%上の曲線(b2)--(c2)
\draw (b2) -- ($(b2)+(0.5,0)$);
\draw ($(c2)-(0.5,0)$) -- (c2);
\draw ($(b2)+(0.5,0)$) .. controls ($(b2)+(1.25,0)$) and ($(c2)-(1.25,0)$) ..($(c2)-(0.5,0)$);

%下の曲線(a1)--(c1)
\draw (a1) -- ($(a1)+(0.5,0)$);
\draw ($(c1)-(0.5,0)$) -- (c1);
\draw ($(a1)+(0.5,0)$) .. controls ($(a1)+(1.25,0)$) and ($(c1)-(1.25,0)$) ..($(c1)-(0.5,0)$);
\node[left] at (0,3) {$$};
\node[left] at (0,0.5) {$$};
\end{tikzpicture}

\begin{tikzpicture}[scale=0.45, line width=2pt, xscale=-1]
 \coordinate (a1) at (0,0);
\coordinate (a2) at (0,1);
\coordinate (b1) at (0,2.5);
\coordinate (b2) at (0,3.5);
\coordinate (c1) at (3,1.25);
\coordinate (c2) at (3,2.25);

\draw (a1) -- (a2);
\draw (b1) -- (b2);
\draw (c1) -- (c2);

%円弧
\draw (0,1) arc[start angle=-90, end angle=90, radius=0.75];

%上の曲線(b2)--(c2)
\draw (b2) -- ($(b2)+(0.5,0)$);
\draw ($(c2)-(0.5,0)$) -- (c2);
\draw ($(b2)+(0.5,0)$) .. controls ($(b2)+(1.25,0)$) and ($(c2)-(1.25,0)$) ..($(c2)-(0.5,0)$);

%下の曲線(a1)--(c1)
\draw (a1) -- ($(a1)+(0.5,0)$);
\draw ($(c1)-(0.5,0)$) -- (c1);
\draw ($(a1)+(0.5,0)$) .. controls ($(a1)+(1.25,0)$) and ($(c1)-(1.25,0)$) ..($(c1)-(0.5,0)$);
\node[right] at (0,3) {$d$};
\node[right] at (0,0.5) {$e$};
\end{tikzpicture}

    }{
    
\begin{tikzpicture}[scale=0.45, line width=2pt]
\begin{scope}[shift={(0,0.5)}]
    \coordinate (x1) at (0,-1);
\coordinate (x2) at (0,-2);
\coordinate (y1) at (3,-1);
\coordinate (y2) at (3,-2);
\draw (x1) -- (x2);
\draw (y1) -- (y2);
\draw (x1) -- (y1);
\draw (y2) -- (x2);

\end{scope}
 \coordinate (a1) at (0,0);
\coordinate (a2) at (0,1);
\coordinate (b1) at (0,2.5);
\coordinate (b2) at (0,3.5);
\coordinate (c1) at (3,1.25);
\coordinate (c2) at (3,2.25);

\draw (a1) -- (a2);
\draw (b1) -- (b2);
\draw (c1) -- (c2);

%円弧
\draw (0,1) arc[start angle=-90, end angle=90, radius=0.75];

%上の曲線(b2)--(c2)
\draw (b2) -- ($(b2)+(0.5,0)$);
\draw ($(c2)-(0.5,0)$) -- (c2);
\draw ($(b2)+(0.5,0)$) .. controls ($(b2)+(1.25,0)$) and ($(c2)-(1.25,0)$) ..($(c2)-(0.5,0)$);

%下の曲線(a1)--(c1)
\draw (a1) -- ($(a1)+(0.5,0)$);
\draw ($(c1)-(0.5,0)$) -- (c1);
\draw ($(a1)+(0.5,0)$) .. controls ($(a1)+(1.25,0)$) and ($(c1)-(1.25,0)$) ..($(c1)-(0.5,0)$);
\node[left] at (0,3) {$a$};
\node[left] at (0,0.5) {$b$};
\node[left] at (0,-1) {$c$};
\end{tikzpicture}

\begin{tikzpicture}[scale=0.45, line width=2pt, xscale=-1]
\begin{scope}[shift={(0,0.5)}]
    \coordinate (x1) at (0,-1);
\coordinate (x2) at (0,-2);
\coordinate (y1) at (3,-1);
\coordinate (y2) at (3,-2);
\draw (x1) -- (x2);
\draw (y1) -- (y2);
\draw (x1) -- (y1);
\draw (y2) -- (x2);

\end{scope}
 \coordinate (a1) at (0,0);
\coordinate (a2) at (0,1);
\coordinate (b1) at (0,2.5);
\coordinate (b2) at (0,3.5);
\coordinate (c1) at (3,1.25);
\coordinate (c2) at (3,2.25);

\draw (a1) -- (a2);
\draw (b1) -- (b2);
\draw (c1) -- (c2);

%円弧
\draw (0,1) arc[start angle=-90, end angle=90, radius=0.75];

%上の曲線(b2)--(c2)
\draw (b2) -- ($(b2)+(0.5,0)$);
\draw ($(c2)-(0.5,0)$) -- (c2);
\draw ($(b2)+(0.5,0)$) .. controls ($(b2)+(1.25,0)$) and ($(c2)-(1.25,0)$) ..($(c2)-(0.5,0)$);

%下の曲線(a1)--(c1)
\draw (a1) -- ($(a1)+(0.5,0)$);
\draw ($(c1)-(0.5,0)$) -- (c1);
\draw ($(a1)+(0.5,0)$) .. controls ($(a1)+(1.25,0)$) and ($(c1)-(1.25,0)$) ..($(c1)-(0.5,0)$);
\node[right] at (0,3) {$$};
\node[right] at (0,0.5) {$$};
\node[right] at (0,-1) {$$};
\end{tikzpicture}

\begin{tikzpicture}[scale=0.45, line width=2pt]
\begin{scope}[shift={(0,6)}]
    \coordinate (x1) at (0,-1);
\coordinate (x2) at (0,-2);
\coordinate (y1) at (3,-1);
\coordinate (y2) at (3,-2);
\draw (x1) -- (x2);
\draw (y1) -- (y2);
\draw (x1) -- (y1);
\draw (y2) -- (x2);

\end{scope}
 \coordinate (a1) at (0,0);
\coordinate (a2) at (0,1);
\coordinate (b1) at (0,2.5);
\coordinate (b2) at (0,3.5);
\coordinate (c1) at (3,1.25);
\coordinate (c2) at (3,2.25);

\draw (a1) -- (a2);
\draw (b1) -- (b2);
\draw (c1) -- (c2);

%円弧
\draw (0,1) arc[start angle=-90, end angle=90, radius=0.75];

%上の曲線(b2)--(c2)
\draw (b2) -- ($(b2)+(0.5,0)$);
\draw ($(c2)-(0.5,0)$) -- (c2);
\draw ($(b2)+(0.5,0)$) .. controls ($(b2)+(1.25,0)$) and ($(c2)-(1.25,0)$) ..($(c2)-(0.5,0)$);

%下の曲線(a1)--(c1)
\draw (a1) -- ($(a1)+(0.5,0)$);
\draw ($(c1)-(0.5,0)$) -- (c1);
\draw ($(a1)+(0.5,0)$) .. controls ($(a1)+(1.25,0)$) and ($(c1)-(1.25,0)$) ..($(c1)-(0.5,0)$);
\node[left] at (0,4.5) {$d$};
\node[left] at (0,3) {$e_{(1)}$};
\node[left] at (0,0.5) {$e_{(2)}$};
\end{tikzpicture}

    }{
    
\begin{tikzpicture}[scale=0.45, line width=2pt, xscale=-1]
\begin{scope}[shift={(0,0.5)}]
    \coordinate (x1) at (0,-1);
\coordinate (x2) at (0,-2);
\coordinate (y1) at (3,-1);
\coordinate (y2) at (3,-2);
\draw (x1) -- (x2);
\draw (y1) -- (y2);
\draw (x1) -- (y1);
\draw (y2) -- (x2);

\end{scope}
\begin{scope}[shift={(0,-1)}]
    \coordinate (x1) at (0,-1);
\coordinate (x2) at (0,-2);
\coordinate (y1) at (3,-1);
\coordinate (y2) at (3,-2);
\draw (x1) -- (x2);
\draw (y1) -- (y2);
\draw (x1) -- (y1);
\draw (y2) -- (x2);

\end{scope}
 \coordinate (a1) at (0,0);
\coordinate (a2) at (0,1);
\coordinate (b1) at (0,2.5);
\coordinate (b2) at (0,3.5);
\coordinate (c1) at (3,1.25);
\coordinate (c2) at (3,2.25);

\draw (a1) -- (a2);
\draw (b1) -- (b2);
\draw (c1) -- (c2);

%円弧
\draw (0,1) arc[start angle=-90, end angle=90, radius=0.75];

%上の曲線(b2)--(c2)
\draw (b2) -- ($(b2)+(0.5,0)$);
\draw ($(c2)-(0.5,0)$) -- (c2);
\draw ($(b2)+(0.5,0)$) .. controls ($(b2)+(1.25,0)$) and ($(c2)-(1.25,0)$) ..($(c2)-(0.5,0)$);

%下の曲線(a1)--(c1)
\draw (a1) -- ($(a1)+(0.5,0)$);
\draw ($(c1)-(0.5,0)$) -- (c1);
\draw ($(a1)+(0.5,0)$) .. controls ($(a1)+(1.25,0)$) and ($(c1)-(1.25,0)$) ..($(c1)-(0.5,0)$);
\node[right] at (0,3) {$a_{(1)}$};
\node[right] at (0,0.5) {$a_{(2)}$};
\node[right] at (0,-1) {$b$};
\node[right] at (0,-2.5) {$c$};
\end{tikzpicture}

\begin{tikzpicture}[scale=0.45, line width=2pt]
\begin{scope}[shift={(0,6)}]
    \coordinate (x1) at (0,-1);
\coordinate (x2) at (0,-2);
\coordinate (y1) at (3,-1);
\coordinate (y2) at (3,-2);
\draw (x1) -- (x2);
\draw (y1) -- (y2);
\draw (x1) -- (y1);
\draw (y2) -- (x2);

\end{scope}
\begin{scope}[shift={(0,0.5)}]
    \coordinate (x1) at (0,-1);
\coordinate (x2) at (0,-2);
\coordinate (y1) at (3,-1);
\coordinate (y2) at (3,-2);
\draw (x1) -- (x2);
\draw (y1) -- (y2);
\draw (x1) -- (y1);
\draw (y2) -- (x2);

\end{scope}
 \coordinate (a1) at (0,0);
\coordinate (a2) at (0,1);
\coordinate (b1) at (0,2.5);
\coordinate (b2) at (0,3.5);
\coordinate (c1) at (3,1.25);
\coordinate (c2) at (3,2.25);

\draw (a1) -- (a2);
\draw (b1) -- (b2);
\draw (c1) -- (c2);

%円弧
\draw (0,1) arc[start angle=-90, end angle=90, radius=0.75];

%上の曲線(b2)--(c2)
\draw (b2) -- ($(b2)+(0.5,0)$);
\draw ($(c2)-(0.5,0)$) -- (c2);
\draw ($(b2)+(0.5,0)$) .. controls ($(b2)+(1.25,0)$) and ($(c2)-(1.25,0)$) ..($(c2)-(0.5,0)$);

%下の曲線(a1)--(c1)
\draw (a1) -- ($(a1)+(0.5,0)$);
\draw ($(c1)-(0.5,0)$) -- (c1);
\draw ($(a1)+(0.5,0)$) .. controls ($(a1)+(1.25,0)$) and ($(c1)-(1.25,0)$) ..($(c1)-(0.5,0)$);
\node[left] at (0,4.5) {$$};
\node[left] at (0,3) {$$};
\node[left] at (0,0.5) {$$};
\node[left] at (0,-1) {$$};
\end{tikzpicture}
\shiftfig{0mm}{3mm}{
\begin{tikzpicture}[scale=0.45, line width=2pt]
\begin{scope}[shift={(0,6)}]
    \coordinate (x1) at (0,-1);
\coordinate (x2) at (0,-2);
\coordinate (y1) at (3,-1);
\coordinate (y2) at (3,-2);
\draw (x1) -- (x2);
\draw (y1) -- (y2);
\draw (x1) -- (y1);
\draw (y2) -- (x2);

\end{scope}
 \coordinate (a1) at (0,0);
\coordinate (a2) at (0,1);
\coordinate (b1) at (0,2.5);
\coordinate (b2) at (0,3.5);
\coordinate (c1) at (3,1.25);
\coordinate (c2) at (3,2.25);

\draw (a1) -- (a2);
\draw (b1) -- (b2);
\draw (c1) -- (c2);

%円弧
\draw (0,1) arc[start angle=-90, end angle=90, radius=0.75];

%上の曲線(b2)--(c2)
\draw (b2) -- ($(b2)+(0.5,0)$);
\draw ($(c2)-(0.5,0)$) -- (c2);
\draw ($(b2)+(0.5,0)$) .. controls ($(b2)+(1.25,0)$) and ($(c2)-(1.25,0)$) ..($(c2)-(0.5,0)$);

%下の曲線(a1)--(c1)
\draw (a1) -- ($(a1)+(0.5,0)$);
\draw ($(c1)-(0.5,0)$) -- (c1);
\draw ($(a1)+(0.5,0)$) .. controls ($(a1)+(1.25,0)$) and ($(c1)-(1.25,0)$) ..($(c1)-(0.5,0)$);
\node[left] at (0,4.5) {$d$};
\node[left] at (0,3) {$e_{(1)}$};
\node[left] at (0,0.5) {$e_{(2)}$};
\end{tikzpicture}
}
    }{
    \shiftfig{0mm}{-5mm}{
\begin{tikzpicture}[scale=0.45, line width=2pt]
\begin{scope}[shift={(0,6)}]
    \coordinate (x1) at (0,-1);
\coordinate (x2) at (0,-2);
\coordinate (y1) at (3,-1);
\coordinate (y2) at (3,-2);
\draw (x1) -- (x2);
\draw (y1) -- (y2);
\draw (x1) -- (y1);
\draw (y2) -- (x2);

\end{scope}
 \coordinate (a1) at (0,0);
\coordinate (a2) at (0,1);
\coordinate (b1) at (0,2.5);
\coordinate (b2) at (0,3.5);
\coordinate (c1) at (3,1.25);
\coordinate (c2) at (3,2.25);

\draw (a1) -- (a2);
\draw (b1) -- (b2);
\draw (c1) -- (c2);

%円弧
\draw (0,1) arc[start angle=-90, end angle=90, radius=0.75];

%上の曲線(b2)--(c2)
\draw (b2) -- ($(b2)+(0.5,0)$);
\draw ($(c2)-(0.5,0)$) -- (c2);
\draw ($(b2)+(0.5,0)$) .. controls ($(b2)+(1.25,0)$) and ($(c2)-(1.25,0)$) ..($(c2)-(0.5,0)$);

%下の曲線(a1)--(c1)
\draw (a1) -- ($(a1)+(0.5,0)$);
\draw ($(c1)-(0.5,0)$) -- (c1);
\draw ($(a1)+(0.5,0)$) .. controls ($(a1)+(1.25,0)$) and ($(c1)-(1.25,0)$) ..($(c1)-(0.5,0)$);
\node[left] at (0,4.5) {$a$};
\node[left] at (0,3) {$b$};
\node[left] at (0,0.5) {$c$};
\end{tikzpicture}
}
\begin{tikzpicture}[scale=0.45, line width=2pt]
 \coordinate (a1) at (0,0);
\coordinate (a2) at (0,1);
\coordinate (b1) at (0,2.5);
\coordinate (b2) at (0,3.5);
\coordinate (c1) at (3,1.25);
\coordinate (c2) at (3,2.25);

\draw (a1) -- (a2);
\draw (b1) -- (b2);
\draw (c1) -- (c2);

%円弧
\draw (0,1) arc[start angle=-90, end angle=90, radius=0.75];

%上の曲線(b2)--(c2)
\draw (b2) -- ($(b2)+(0.5,0)$);
\draw ($(c2)-(0.5,0)$) -- (c2);
\draw ($(b2)+(0.5,0)$) .. controls ($(b2)+(1.25,0)$) and ($(c2)-(1.25,0)$) ..($(c2)-(0.5,0)$);

%下の曲線(a1)--(c1)
\draw (a1) -- ($(a1)+(0.5,0)$);
\draw ($(c1)-(0.5,0)$) -- (c1);
\draw ($(a1)+(0.5,0)$) .. controls ($(a1)+(1.25,0)$) and ($(c1)-(1.25,0)$) ..($(c1)-(0.5,0)$);
\node[left] at (0,3) {$$};
\node[left] at (0,0.5) {$$};
\end{tikzpicture}

\begin{tikzpicture}[scale=0.45, line width=2pt, xscale=-1]
 \coordinate (a1) at (0,0);
\coordinate (a2) at (0,1);
\coordinate (b1) at (0,2.5);
\coordinate (b2) at (0,3.5);
\coordinate (c1) at (3,1.25);
\coordinate (c2) at (3,2.25);

\draw (a1) -- (a2);
\draw (b1) -- (b2);
\draw (c1) -- (c2);

%円弧
\draw (0,1) arc[start angle=-90, end angle=90, radius=0.75];

%上の曲線(b2)--(c2)
\draw (b2) -- ($(b2)+(0.5,0)$);
\draw ($(c2)-(0.5,0)$) -- (c2);
\draw ($(b2)+(0.5,0)$) .. controls ($(b2)+(1.25,0)$) and ($(c2)-(1.25,0)$) ..($(c2)-(0.5,0)$);

%下の曲線(a1)--(c1)
\draw (a1) -- ($(a1)+(0.5,0)$);
\draw ($(c1)-(0.5,0)$) -- (c1);
\draw ($(a1)+(0.5,0)$) .. controls ($(a1)+(1.25,0)$) and ($(c1)-(1.25,0)$) ..($(c1)-(0.5,0)$);
\node[right] at (0,3) {$d$};
\node[right] at (0,0.5) {$e$};
\end{tikzpicture}

    }{
    \shiftfig{0mm}{-5mm}{
\begin{tikzpicture}[scale=0.45, line width=2pt]
\begin{scope}[shift={(0,6)}]
    \coordinate (x1) at (0,-1);
\coordinate (x2) at (0,-2);
\coordinate (y1) at (3,-1);
\coordinate (y2) at (3,-2);
\draw (x1) -- (x2);
\draw (y1) -- (y2);
\draw (x1) -- (y1);
\draw (y2) -- (x2);

\end{scope}
 \coordinate (a1) at (0,0);
\coordinate (a2) at (0,1);
\coordinate (b1) at (0,2.5);
\coordinate (b2) at (0,3.5);
\coordinate (c1) at (3,1.25);
\coordinate (c2) at (3,2.25);

\draw (a1) -- (a2);
\draw (b1) -- (b2);
\draw (c1) -- (c2);

%円弧
\draw (0,1) arc[start angle=-90, end angle=90, radius=0.75];

%上の曲線(b2)--(c2)
\draw (b2) -- ($(b2)+(0.5,0)$);
\draw ($(c2)-(0.5,0)$) -- (c2);
\draw ($(b2)+(0.5,0)$) .. controls ($(b2)+(1.25,0)$) and ($(c2)-(1.25,0)$) ..($(c2)-(0.5,0)$);

%下の曲線(a1)--(c1)
\draw (a1) -- ($(a1)+(0.5,0)$);
\draw ($(c1)-(0.5,0)$) -- (c1);
\draw ($(a1)+(0.5,0)$) .. controls ($(a1)+(1.25,0)$) and ($(c1)-(1.25,0)$) ..($(c1)-(0.5,0)$);
\node[left] at (0,4.5) {$$};
\node[left] at (0,3) {$b$};
\node[left] at (0,0.5) {$c$};
\end{tikzpicture}
}
\begin{tikzpicture}[scale=0.45, line width=2pt, xscale=-1]
\begin{scope}[shift={(0,0.5)}]
    \coordinate (x1) at (0,-1);
\coordinate (x2) at (0,-2);
\coordinate (y1) at (3,-1);
\coordinate (y2) at (3,-2);
\draw (x1) -- (x2);
\draw (y1) -- (y2);
\draw (x1) -- (y1);
\draw (y2) -- (x2);

\end{scope}
 \coordinate (a1) at (0,0);
\coordinate (a2) at (0,1);
\coordinate (b1) at (0,2.5);
\coordinate (b2) at (0,3.5);
\coordinate (c1) at (3,1.25);
\coordinate (c2) at (3,2.25);

\draw (a1) -- (a2);
\draw (b1) -- (b2);
\draw (c1) -- (c2);

%円弧
\draw (0,1) arc[start angle=-90, end angle=90, radius=0.75];

%上の曲線(b2)--(c2)
\draw (b2) -- ($(b2)+(0.5,0)$);
\draw ($(c2)-(0.5,0)$) -- (c2);
\draw ($(b2)+(0.5,0)$) .. controls ($(b2)+(1.25,0)$) and ($(c2)-(1.25,0)$) ..($(c2)-(0.5,0)$);

%下の曲線(a1)--(c1)
\draw (a1) -- ($(a1)+(0.5,0)$);
\draw ($(c1)-(0.5,0)$) -- (c1);
\draw ($(a1)+(0.5,0)$) .. controls ($(a1)+(1.25,0)$) and ($(c1)-(1.25,0)$) ..($(c1)-(0.5,0)$);
\node[right] at (0,3) {$a_{(1)}$};
\node[right] at (0,0.5) {$a_{(2)}$};
\node[right] at (0,-1) {$$};
\end{tikzpicture}

\begin{tikzpicture}[scale=0.45, line width=2pt]
\begin{scope}[shift={(0,6)}]
    \coordinate (x1) at (0,-1);
\coordinate (x2) at (0,-2);
\coordinate (y1) at (3,-1);
\coordinate (y2) at (3,-2);
\draw (x1) -- (x2);
\draw (y1) -- (y2);
\draw (x1) -- (y1);
\draw (y2) -- (x2);

\end{scope}
 \coordinate (a1) at (0,0);
\coordinate (a2) at (0,1);
\coordinate (b1) at (0,2.5);
\coordinate (b2) at (0,3.5);
\coordinate (c1) at (3,1.25);
\coordinate (c2) at (3,2.25);

\draw (a1) -- (a2);
\draw (b1) -- (b2);
\draw (c1) -- (c2);

%円弧
\draw (0,1) arc[start angle=-90, end angle=90, radius=0.75];

%上の曲線(b2)--(c2)
\draw (b2) -- ($(b2)+(0.5,0)$);
\draw ($(c2)-(0.5,0)$) -- (c2);
\draw ($(b2)+(0.5,0)$) .. controls ($(b2)+(1.25,0)$) and ($(c2)-(1.25,0)$) ..($(c2)-(0.5,0)$);

%下の曲線(a1)--(c1)
\draw (a1) -- ($(a1)+(0.5,0)$);
\draw ($(c1)-(0.5,0)$) -- (c1);
\draw ($(a1)+(0.5,0)$) .. controls ($(a1)+(1.25,0)$) and ($(c1)-(1.25,0)$) ..($(c1)-(0.5,0)$);
\node[left] at (0,4.5) {$d$};
\node[left] at (0,3) {$e_{(1)}$};
\node[left] at (0,0.5) {$e_{(2)}$};
\end{tikzpicture}

    }{
    
\begin{tikzpicture}[scale=0.45, line width=2pt, xscale=-1]
\begin{scope}[shift={(0,0.5)}]
    \coordinate (x1) at (0,-1);
\coordinate (x2) at (0,-2);
\coordinate (y1) at (3,-1);
\coordinate (y2) at (3,-2);
\draw (x1) -- (x2);
\draw (y1) -- (y2);
\draw (x1) -- (y1);
\draw (y2) -- (x2);

\end{scope}
\begin{scope}[shift={(0,-1)}]
    \coordinate (x1) at (0,-1);
\coordinate (x2) at (0,-2);
\coordinate (y1) at (3,-1);
\coordinate (y2) at (3,-2);
\draw (x1) -- (x2);
\draw (y1) -- (y2);
\draw (x1) -- (y1);
\draw (y2) -- (x2);

\end{scope}
 \coordinate (a1) at (0,0);
\coordinate (a2) at (0,1);
\coordinate (b1) at (0,2.5);
\coordinate (b2) at (0,3.5);
\coordinate (c1) at (3,1.25);
\coordinate (c2) at (3,2.25);

\draw (a1) -- (a2);
\draw (b1) -- (b2);
\draw (c1) -- (c2);

%円弧
\draw (0,1) arc[start angle=-90, end angle=90, radius=0.75];

%上の曲線(b2)--(c2)
\draw (b2) -- ($(b2)+(0.5,0)$);
\draw ($(c2)-(0.5,0)$) -- (c2);
\draw ($(b2)+(0.5,0)$) .. controls ($(b2)+(1.25,0)$) and ($(c2)-(1.25,0)$) ..($(c2)-(0.5,0)$);

%下の曲線(a1)--(c1)
\draw (a1) -- ($(a1)+(0.5,0)$);
\draw ($(c1)-(0.5,0)$) -- (c1);
\draw ($(a1)+(0.5,0)$) .. controls ($(a1)+(1.25,0)$) and ($(c1)-(1.25,0)$) ..($(c1)-(0.5,0)$);
\node[right] at (0,3) {$a_{(1)}$};
\node[right] at (0,0.5) {$a_{(2)}$};
\node[right] at (0,-1) {$$};
\node[right] at (0,-2.5) {$$};
\end{tikzpicture}
\shiftfig{0mm}{-5mm}{
\begin{tikzpicture}[scale=0.45, line width=2pt]
\begin{scope}[shift={(0,7.5)}]
    \coordinate (x1) at (0,-1);
\coordinate (x2) at (0,-2);
\coordinate (y1) at (3,-1);
\coordinate (y2) at (3,-2);
\draw (x1) -- (x2);
\draw (y1) -- (y2);
\draw (x1) -- (y1);
\draw (y2) -- (x2);

\end{scope}
\begin{scope}[shift={(0,6)}]
    \coordinate (x1) at (0,-1);
\coordinate (x2) at (0,-2);
\coordinate (y1) at (3,-1);
\coordinate (y2) at (3,-2);
\draw (x1) -- (x2);
\draw (y1) -- (y2);
\draw (x1) -- (y1);
\draw (y2) -- (x2);

\end{scope}
 \coordinate (a1) at (0,0);
\coordinate (a2) at (0,1);
\coordinate (b1) at (0,2.5);
\coordinate (b2) at (0,3.5);
\coordinate (c1) at (3,1.25);
\coordinate (c2) at (3,2.25);

\draw (a1) -- (a2);
\draw (b1) -- (b2);
\draw (c1) -- (c2);

%円弧
\draw (0,1) arc[start angle=-90, end angle=90, radius=0.75];

%上の曲線(b2)--(c2)
\draw (b2) -- ($(b2)+(0.5,0)$);
\draw ($(c2)-(0.5,0)$) -- (c2);
\draw ($(b2)+(0.5,0)$) .. controls ($(b2)+(1.25,0)$) and ($(c2)-(1.25,0)$) ..($(c2)-(0.5,0)$);

%下の曲線(a1)--(c1)
\draw (a1) -- ($(a1)+(0.5,0)$);
\draw ($(c1)-(0.5,0)$) -- (c1);
\draw ($(a1)+(0.5,0)$) .. controls ($(a1)+(1.25,0)$) and ($(c1)-(1.25,0)$) ..($(c1)-(0.5,0)$);
\node[left] at (0,6) {$$};
\node[left] at (0,4.5) {$$};
\node[left] at (0,3) {$b$};
\node[left] at (0,0.5) {$c$};
\end{tikzpicture}
}
\begin{tikzpicture}[scale=0.45, line width=2pt]
\begin{scope}[shift={(0,6)}]
    \coordinate (x1) at (0,-1);
\coordinate (x2) at (0,-2);
\coordinate (y1) at (3,-1);
\coordinate (y2) at (3,-2);
\draw (x1) -- (x2);
\draw (y1) -- (y2);
\draw (x1) -- (y1);
\draw (y2) -- (x2);

\end{scope}
 \coordinate (a1) at (0,0);
\coordinate (a2) at (0,1);
\coordinate (b1) at (0,2.5);
\coordinate (b2) at (0,3.5);
\coordinate (c1) at (3,1.25);
\coordinate (c2) at (3,2.25);

\draw (a1) -- (a2);
\draw (b1) -- (b2);
\draw (c1) -- (c2);

%円弧
\draw (0,1) arc[start angle=-90, end angle=90, radius=0.75];

%上の曲線(b2)--(c2)
\draw (b2) -- ($(b2)+(0.5,0)$);
\draw ($(c2)-(0.5,0)$) -- (c2);
\draw ($(b2)+(0.5,0)$) .. controls ($(b2)+(1.25,0)$) and ($(c2)-(1.25,0)$) ..($(c2)-(0.5,0)$);

%下の曲線(a1)--(c1)
\draw (a1) -- ($(a1)+(0.5,0)$);
\draw ($(c1)-(0.5,0)$) -- (c1);
\draw ($(a1)+(0.5,0)$) .. controls ($(a1)+(1.25,0)$) and ($(c1)-(1.25,0)$) ..($(c1)-(0.5,0)$);
\node[left] at (0,4.5) {$d$};
\node[left] at (0,3) {$e_{(1)}$};
\node[left] at (0,0.5) {$e_{(2)}$};
\end{tikzpicture}

    }
    }
\]
\fi

\iflatexml
\begin{center}
  \includegraphics{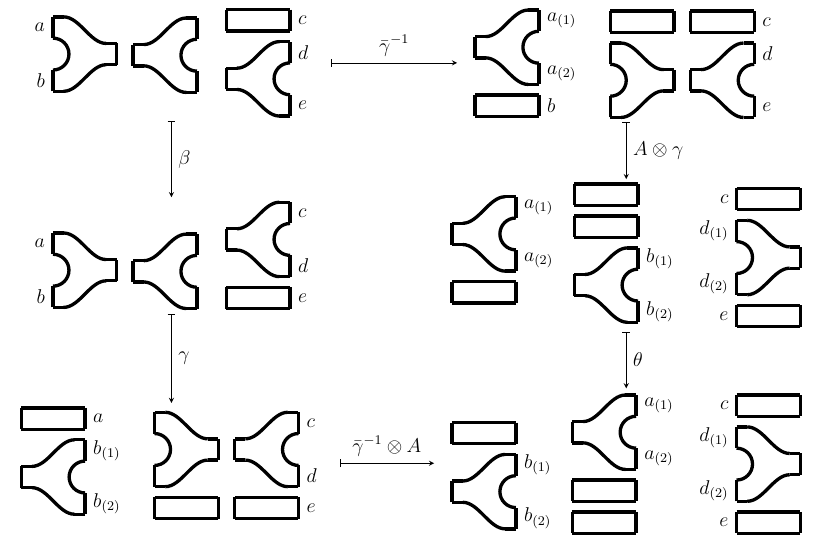}
\end{center}
\else
\[\scalebox{0.8}{%
  \DiagramRectangle[
    AB=\beta,
    BC=\gamma,
    AD=\brg^{-1},
    DE=A\otimes \gamma,
    EF=\theta,
    CF=\brg^{-1}\otimes A
  ]
    {
    \shiftfig{0mm}{5mm}{
\begin{tikzpicture}[scale=0.45, line width=2pt]
 \coordinate (a1) at (0,0);
\coordinate (a2) at (0,1);
\coordinate (b1) at (0,2.5);
\coordinate (b2) at (0,3.5);
\coordinate (c1) at (3,1.25);
\coordinate (c2) at (3,2.25);

\draw (a1) -- (a2);
\draw (b1) -- (b2);
\draw (c1) -- (c2);

%円弧
\draw (0,1) arc[start angle=-90, end angle=90, radius=0.75];

%上の曲線(b2)--(c2)
\draw (b2) -- ($(b2)+(0.5,0)$);
\draw ($(c2)-(0.5,0)$) -- (c2);
\draw ($(b2)+(0.5,0)$) .. controls ($(b2)+(1.25,0)$) and ($(c2)-(1.25,0)$) ..($(c2)-(0.5,0)$);

%下の曲線(a1)--(c1)
\draw (a1) -- ($(a1)+(0.5,0)$);
\draw ($(c1)-(0.5,0)$) -- (c1);
\draw ($(a1)+(0.5,0)$) .. controls ($(a1)+(1.25,0)$) and ($(c1)-(1.25,0)$) ..($(c1)-(0.5,0)$);
\node[left] at (0,3) {$a$};
\node[left] at (0,0.5) {$b$};
\end{tikzpicture}

\begin{tikzpicture}[scale=0.45, line width=2pt, xscale=-1]
 \coordinate (a1) at (0,0);
\coordinate (a2) at (0,1);
\coordinate (b1) at (0,2.5);
\coordinate (b2) at (0,3.5);
\coordinate (c1) at (3,1.25);
\coordinate (c2) at (3,2.25);

\draw (a1) -- (a2);
\draw (b1) -- (b2);
\draw (c1) -- (c2);

%円弧
\draw (0,1) arc[start angle=-90, end angle=90, radius=0.75];

%上の曲線(b2)--(c2)
\draw (b2) -- ($(b2)+(0.5,0)$);
\draw ($(c2)-(0.5,0)$) -- (c2);
\draw ($(b2)+(0.5,0)$) .. controls ($(b2)+(1.25,0)$) and ($(c2)-(1.25,0)$) ..($(c2)-(0.5,0)$);

%下の曲線(a1)--(c1)
\draw (a1) -- ($(a1)+(0.5,0)$);
\draw ($(c1)-(0.5,0)$) -- (c1);
\draw ($(a1)+(0.5,0)$) .. controls ($(a1)+(1.25,0)$) and ($(c1)-(1.25,0)$) ..($(c1)-(0.5,0)$);
\node[right] at (0,3) {$$};
\node[right] at (0,0.5) {$$};
\end{tikzpicture}
}
\begin{tikzpicture}[scale=0.45, line width=2pt, xscale=-1]
\begin{scope}[shift={(0,6)}]
    \coordinate (x1) at (0,-1);
\coordinate (x2) at (0,-2);
\coordinate (y1) at (3,-1);
\coordinate (y2) at (3,-2);
\draw (x1) -- (x2);
\draw (y1) -- (y2);
\draw (x1) -- (y1);
\draw (y2) -- (x2);

\end{scope}
 \coordinate (a1) at (0,0);
\coordinate (a2) at (0,1);
\coordinate (b1) at (0,2.5);
\coordinate (b2) at (0,3.5);
\coordinate (c1) at (3,1.25);
\coordinate (c2) at (3,2.25);

\draw (a1) -- (a2);
\draw (b1) -- (b2);
\draw (c1) -- (c2);

%円弧
\draw (0,1) arc[start angle=-90, end angle=90, radius=0.75];

%上の曲線(b2)--(c2)
\draw (b2) -- ($(b2)+(0.5,0)$);
\draw ($(c2)-(0.5,0)$) -- (c2);
\draw ($(b2)+(0.5,0)$) .. controls ($(b2)+(1.25,0)$) and ($(c2)-(1.25,0)$) ..($(c2)-(0.5,0)$);

%下の曲線(a1)--(c1)
\draw (a1) -- ($(a1)+(0.5,0)$);
\draw ($(c1)-(0.5,0)$) -- (c1);
\draw ($(a1)+(0.5,0)$) .. controls ($(a1)+(1.25,0)$) and ($(c1)-(1.25,0)$) ..($(c1)-(0.5,0)$);
\node[right] at (0,4.5) {$c$};
\node[right] at (0,3) {$d$};
\node[right] at (0,0.5) {$e$};
\end{tikzpicture}

    }{
    
\begin{tikzpicture}[scale=0.45, line width=2pt]
 \coordinate (a1) at (0,0);
\coordinate (a2) at (0,1);
\coordinate (b1) at (0,2.5);
\coordinate (b2) at (0,3.5);
\coordinate (c1) at (3,1.25);
\coordinate (c2) at (3,2.25);

\draw (a1) -- (a2);
\draw (b1) -- (b2);
\draw (c1) -- (c2);

%円弧
\draw (0,1) arc[start angle=-90, end angle=90, radius=0.75];

%上の曲線(b2)--(c2)
\draw (b2) -- ($(b2)+(0.5,0)$);
\draw ($(c2)-(0.5,0)$) -- (c2);
\draw ($(b2)+(0.5,0)$) .. controls ($(b2)+(1.25,0)$) and ($(c2)-(1.25,0)$) ..($(c2)-(0.5,0)$);

%下の曲線(a1)--(c1)
\draw (a1) -- ($(a1)+(0.5,0)$);
\draw ($(c1)-(0.5,0)$) -- (c1);
\draw ($(a1)+(0.5,0)$) .. controls ($(a1)+(1.25,0)$) and ($(c1)-(1.25,0)$) ..($(c1)-(0.5,0)$);
\node[left] at (0,3) {$a$};
\node[left] at (0,0.5) {$b$};
\end{tikzpicture}

\begin{tikzpicture}[scale=0.45, line width=2pt, xscale=-1]
 \coordinate (a1) at (0,0);
\coordinate (a2) at (0,1);
\coordinate (b1) at (0,2.5);
\coordinate (b2) at (0,3.5);
\coordinate (c1) at (3,1.25);
\coordinate (c2) at (3,2.25);

\draw (a1) -- (a2);
\draw (b1) -- (b2);
\draw (c1) -- (c2);

%円弧
\draw (0,1) arc[start angle=-90, end angle=90, radius=0.75];

%上の曲線(b2)--(c2)
\draw (b2) -- ($(b2)+(0.5,0)$);
\draw ($(c2)-(0.5,0)$) -- (c2);
\draw ($(b2)+(0.5,0)$) .. controls ($(b2)+(1.25,0)$) and ($(c2)-(1.25,0)$) ..($(c2)-(0.5,0)$);

%下の曲線(a1)--(c1)
\draw (a1) -- ($(a1)+(0.5,0)$);
\draw ($(c1)-(0.5,0)$) -- (c1);
\draw ($(a1)+(0.5,0)$) .. controls ($(a1)+(1.25,0)$) and ($(c1)-(1.25,0)$) ..($(c1)-(0.5,0)$);
\node[right] at (0,3) {$$};
\node[right] at (0,0.5) {$$};
\end{tikzpicture}

\begin{tikzpicture}[scale=0.45, line width=2pt, xscale=-1]
\begin{scope}[shift={(0,0.5)}]
    \coordinate (x1) at (0,-1);
\coordinate (x2) at (0,-2);
\coordinate (y1) at (3,-1);
\coordinate (y2) at (3,-2);
\draw (x1) -- (x2);
\draw (y1) -- (y2);
\draw (x1) -- (y1);
\draw (y2) -- (x2);

\end{scope}
 \coordinate (a1) at (0,0);
\coordinate (a2) at (0,1);
\coordinate (b1) at (0,2.5);
\coordinate (b2) at (0,3.5);
\coordinate (c1) at (3,1.25);
\coordinate (c2) at (3,2.25);

\draw (a1) -- (a2);
\draw (b1) -- (b2);
\draw (c1) -- (c2);

%円弧
\draw (0,1) arc[start angle=-90, end angle=90, radius=0.75];

%上の曲線(b2)--(c2)
\draw (b2) -- ($(b2)+(0.5,0)$);
\draw ($(c2)-(0.5,0)$) -- (c2);
\draw ($(b2)+(0.5,0)$) .. controls ($(b2)+(1.25,0)$) and ($(c2)-(1.25,0)$) ..($(c2)-(0.5,0)$);

%下の曲線(a1)--(c1)
\draw (a1) -- ($(a1)+(0.5,0)$);
\draw ($(c1)-(0.5,0)$) -- (c1);
\draw ($(a1)+(0.5,0)$) .. controls ($(a1)+(1.25,0)$) and ($(c1)-(1.25,0)$) ..($(c1)-(0.5,0)$);
\node[right] at (0,3) {$c$};
\node[right] at (0,0.5) {$d$};
\node[right] at (0,-1) {$e$};
\end{tikzpicture}

    }{
    
\begin{tikzpicture}[scale=0.45, line width=2pt, xscale=-1]
\begin{scope}[shift={(0,6)}]
    \coordinate (x1) at (0,-1);
\coordinate (x2) at (0,-2);
\coordinate (y1) at (3,-1);
\coordinate (y2) at (3,-2);
\draw (x1) -- (x2);
\draw (y1) -- (y2);
\draw (x1) -- (y1);
\draw (y2) -- (x2);

\end{scope}
 \coordinate (a1) at (0,0);
\coordinate (a2) at (0,1);
\coordinate (b1) at (0,2.5);
\coordinate (b2) at (0,3.5);
\coordinate (c1) at (3,1.25);
\coordinate (c2) at (3,2.25);

\draw (a1) -- (a2);
\draw (b1) -- (b2);
\draw (c1) -- (c2);

%円弧
\draw (0,1) arc[start angle=-90, end angle=90, radius=0.75];

%上の曲線(b2)--(c2)
\draw (b2) -- ($(b2)+(0.5,0)$);
\draw ($(c2)-(0.5,0)$) -- (c2);
\draw ($(b2)+(0.5,0)$) .. controls ($(b2)+(1.25,0)$) and ($(c2)-(1.25,0)$) ..($(c2)-(0.5,0)$);

%下の曲線(a1)--(c1)
\draw (a1) -- ($(a1)+(0.5,0)$);
\draw ($(c1)-(0.5,0)$) -- (c1);
\draw ($(a1)+(0.5,0)$) .. controls ($(a1)+(1.25,0)$) and ($(c1)-(1.25,0)$) ..($(c1)-(0.5,0)$);
\node[right] at (0,4.5) {$a$};
\node[right] at (0,3) {$b_{(1)}$};
\node[right] at (0,0.5) {$b_{(2)}$};
\end{tikzpicture}

\begin{tikzpicture}[scale=0.45, line width=2pt]
\begin{scope}[shift={(0,0.5)}]
    \coordinate (x1) at (0,-1);
\coordinate (x2) at (0,-2);
\coordinate (y1) at (3,-1);
\coordinate (y2) at (3,-2);
\draw (x1) -- (x2);
\draw (y1) -- (y2);
\draw (x1) -- (y1);
\draw (y2) -- (x2);

\end{scope}
 \coordinate (a1) at (0,0);
\coordinate (a2) at (0,1);
\coordinate (b1) at (0,2.5);
\coordinate (b2) at (0,3.5);
\coordinate (c1) at (3,1.25);
\coordinate (c2) at (3,2.25);

\draw (a1) -- (a2);
\draw (b1) -- (b2);
\draw (c1) -- (c2);

%円弧
\draw (0,1) arc[start angle=-90, end angle=90, radius=0.75];

%上の曲線(b2)--(c2)
\draw (b2) -- ($(b2)+(0.5,0)$);
\draw ($(c2)-(0.5,0)$) -- (c2);
\draw ($(b2)+(0.5,0)$) .. controls ($(b2)+(1.25,0)$) and ($(c2)-(1.25,0)$) ..($(c2)-(0.5,0)$);

%下の曲線(a1)--(c1)
\draw (a1) -- ($(a1)+(0.5,0)$);
\draw ($(c1)-(0.5,0)$) -- (c1);
\draw ($(a1)+(0.5,0)$) .. controls ($(a1)+(1.25,0)$) and ($(c1)-(1.25,0)$) ..($(c1)-(0.5,0)$);
\node[left] at (0,3) {$$};
\node[left] at (0,0.5) {$$};
\node[left] at (0,-1) {$$};
\end{tikzpicture}

\begin{tikzpicture}[scale=0.45, line width=2pt, xscale=-1]
\begin{scope}[shift={(0,0.5)}]
    \coordinate (x1) at (0,-1);
\coordinate (x2) at (0,-2);
\coordinate (y1) at (3,-1);
\coordinate (y2) at (3,-2);
\draw (x1) -- (x2);
\draw (y1) -- (y2);
\draw (x1) -- (y1);
\draw (y2) -- (x2);

\end{scope}
 \coordinate (a1) at (0,0);
\coordinate (a2) at (0,1);
\coordinate (b1) at (0,2.5);
\coordinate (b2) at (0,3.5);
\coordinate (c1) at (3,1.25);
\coordinate (c2) at (3,2.25);

\draw (a1) -- (a2);
\draw (b1) -- (b2);
\draw (c1) -- (c2);

%円弧
\draw (0,1) arc[start angle=-90, end angle=90, radius=0.75];

%上の曲線(b2)--(c2)
\draw (b2) -- ($(b2)+(0.5,0)$);
\draw ($(c2)-(0.5,0)$) -- (c2);
\draw ($(b2)+(0.5,0)$) .. controls ($(b2)+(1.25,0)$) and ($(c2)-(1.25,0)$) ..($(c2)-(0.5,0)$);

%下の曲線(a1)--(c1)
\draw (a1) -- ($(a1)+(0.5,0)$);
\draw ($(c1)-(0.5,0)$) -- (c1);
\draw ($(a1)+(0.5,0)$) .. controls ($(a1)+(1.25,0)$) and ($(c1)-(1.25,0)$) ..($(c1)-(0.5,0)$);
\node[right] at (0,3) {$c$};
\node[right] at (0,0.5) {$d$};
\node[right] at (0,-1) {$e$};
\end{tikzpicture}

    }{
     
\begin{tikzpicture}[scale=0.45, line width=2pt, xscale=-1]
\begin{scope}[shift={(0,0.5)}]
    \coordinate (x1) at (0,-1);
\coordinate (x2) at (0,-2);
\coordinate (y1) at (3,-1);
\coordinate (y2) at (3,-2);
\draw (x1) -- (x2);
\draw (y1) -- (y2);
\draw (x1) -- (y1);
\draw (y2) -- (x2);

\end{scope}
 \coordinate (a1) at (0,0);
\coordinate (a2) at (0,1);
\coordinate (b1) at (0,2.5);
\coordinate (b2) at (0,3.5);
\coordinate (c1) at (3,1.25);
\coordinate (c2) at (3,2.25);

\draw (a1) -- (a2);
\draw (b1) -- (b2);
\draw (c1) -- (c2);

%円弧
\draw (0,1) arc[start angle=-90, end angle=90, radius=0.75];

%上の曲線(b2)--(c2)
\draw (b2) -- ($(b2)+(0.5,0)$);
\draw ($(c2)-(0.5,0)$) -- (c2);
\draw ($(b2)+(0.5,0)$) .. controls ($(b2)+(1.25,0)$) and ($(c2)-(1.25,0)$) ..($(c2)-(0.5,0)$);

%下の曲線(a1)--(c1)
\draw (a1) -- ($(a1)+(0.5,0)$);
\draw ($(c1)-(0.5,0)$) -- (c1);
\draw ($(a1)+(0.5,0)$) .. controls ($(a1)+(1.25,0)$) and ($(c1)-(1.25,0)$) ..($(c1)-(0.5,0)$);
\node[right] at (0,3) {$a_{(1)}$};
\node[right] at (0,0.5) {$a_{(2)}$};
\node[right] at (0,-1) {$b$};
\end{tikzpicture}

\begin{tikzpicture}[scale=0.45, line width=2pt]
\begin{scope}[shift={(0,6)}]
    \coordinate (x1) at (0,-1);
\coordinate (x2) at (0,-2);
\coordinate (y1) at (3,-1);
\coordinate (y2) at (3,-2);
\draw (x1) -- (x2);
\draw (y1) -- (y2);
\draw (x1) -- (y1);
\draw (y2) -- (x2);

\end{scope}
 \coordinate (a1) at (0,0);
\coordinate (a2) at (0,1);
\coordinate (b1) at (0,2.5);
\coordinate (b2) at (0,3.5);
\coordinate (c1) at (3,1.25);
\coordinate (c2) at (3,2.25);

\draw (a1) -- (a2);
\draw (b1) -- (b2);
\draw (c1) -- (c2);

%円弧
\draw (0,1) arc[start angle=-90, end angle=90, radius=0.75];

%上の曲線(b2)--(c2)
\draw (b2) -- ($(b2)+(0.5,0)$);
\draw ($(c2)-(0.5,0)$) -- (c2);
\draw ($(b2)+(0.5,0)$) .. controls ($(b2)+(1.25,0)$) and ($(c2)-(1.25,0)$) ..($(c2)-(0.5,0)$);

%下の曲線(a1)--(c1)
\draw (a1) -- ($(a1)+(0.5,0)$);
\draw ($(c1)-(0.5,0)$) -- (c1);
\draw ($(a1)+(0.5,0)$) .. controls ($(a1)+(1.25,0)$) and ($(c1)-(1.25,0)$) ..($(c1)-(0.5,0)$);
\node[left] at (0,4.5) {$$};
\node[left] at (0,3) {$$};
\node[left] at (0,0.5) {$$};
\end{tikzpicture}

\begin{tikzpicture}[scale=0.45, line width=2pt, xscale=-1]
\begin{scope}[shift={(0,6)}]
    \coordinate (x1) at (0,-1);
\coordinate (x2) at (0,-2);
\coordinate (y1) at (3,-1);
\coordinate (y2) at (3,-2);
\draw (x1) -- (x2);
\draw (y1) -- (y2);
\draw (x1) -- (y1);
\draw (y2) -- (x2);

\end{scope}
 \coordinate (a1) at (0,0);
\coordinate (a2) at (0,1);
\coordinate (b1) at (0,2.5);
\coordinate (b2) at (0,3.5);
\coordinate (c1) at (3,1.25);
\coordinate (c2) at (3,2.25);

\draw (a1) -- (a2);
\draw (b1) -- (b2);
\draw (c1) -- (c2);

%円弧
\draw (0,1) arc[start angle=-90, end angle=90, radius=0.75];

%上の曲線(b2)--(c2)
\draw (b2) -- ($(b2)+(0.5,0)$);
\draw ($(c2)-(0.5,0)$) -- (c2);
\draw ($(b2)+(0.5,0)$) .. controls ($(b2)+(1.25,0)$) and ($(c2)-(1.25,0)$) ..($(c2)-(0.5,0)$);

%下の曲線(a1)--(c1)
\draw (a1) -- ($(a1)+(0.5,0)$);
\draw ($(c1)-(0.5,0)$) -- (c1);
\draw ($(a1)+(0.5,0)$) .. controls ($(a1)+(1.25,0)$) and ($(c1)-(1.25,0)$) ..($(c1)-(0.5,0)$);
\node[right] at (0,4.5) {$c$};
\node[right] at (0,3) {$d$};
\node[right] at (0,0.5) {$e$};
\end{tikzpicture}

    }{
   \shiftfig{0mm}{5mm}{
\begin{tikzpicture}[scale=0.45, line width=2pt, xscale=-1]
\begin{scope}[shift={(0,0.5)}]
    \coordinate (x1) at (0,-1);
\coordinate (x2) at (0,-2);
\coordinate (y1) at (3,-1);
\coordinate (y2) at (3,-2);
\draw (x1) -- (x2);
\draw (y1) -- (y2);
\draw (x1) -- (y1);
\draw (y2) -- (x2);

\end{scope}
 \coordinate (a1) at (0,0);
\coordinate (a2) at (0,1);
\coordinate (b1) at (0,2.5);
\coordinate (b2) at (0,3.5);
\coordinate (c1) at (3,1.25);
\coordinate (c2) at (3,2.25);

\draw (a1) -- (a2);
\draw (b1) -- (b2);
\draw (c1) -- (c2);

%円弧
\draw (0,1) arc[start angle=-90, end angle=90, radius=0.75];

%上の曲線(b2)--(c2)
\draw (b2) -- ($(b2)+(0.5,0)$);
\draw ($(c2)-(0.5,0)$) -- (c2);
\draw ($(b2)+(0.5,0)$) .. controls ($(b2)+(1.25,0)$) and ($(c2)-(1.25,0)$) ..($(c2)-(0.5,0)$);

%下の曲線(a1)--(c1)
\draw (a1) -- ($(a1)+(0.5,0)$);
\draw ($(c1)-(0.5,0)$) -- (c1);
\draw ($(a1)+(0.5,0)$) .. controls ($(a1)+(1.25,0)$) and ($(c1)-(1.25,0)$) ..($(c1)-(0.5,0)$);
\node[right] at (0,3) {$a_{(1)}$};
\node[right] at (0,0.5) {$a_{(2)}$};
\node[right] at (0,-1) {$$};
\end{tikzpicture}
}
\begin{tikzpicture}[scale=0.45, line width=2pt, xscale=-1]
\begin{scope}[shift={(0,7.5)}]
    \coordinate (x1) at (0,-1);
\coordinate (x2) at (0,-2);
\coordinate (y1) at (3,-1);
\coordinate (y2) at (3,-2);
\draw (x1) -- (x2);
\draw (y1) -- (y2);
\draw (x1) -- (y1);
\draw (y2) -- (x2);

\end{scope}
\begin{scope}[shift={(0,6)}]
    \coordinate (x1) at (0,-1);
\coordinate (x2) at (0,-2);
\coordinate (y1) at (3,-1);
\coordinate (y2) at (3,-2);
\draw (x1) -- (x2);
\draw (y1) -- (y2);
\draw (x1) -- (y1);
\draw (y2) -- (x2);

\end{scope}
 \coordinate (a1) at (0,0);
\coordinate (a2) at (0,1);
\coordinate (b1) at (0,2.5);
\coordinate (b2) at (0,3.5);
\coordinate (c1) at (3,1.25);
\coordinate (c2) at (3,2.25);

\draw (a1) -- (a2);
\draw (b1) -- (b2);
\draw (c1) -- (c2);

%円弧
\draw (0,1) arc[start angle=-90, end angle=90, radius=0.75];

%上の曲線(b2)--(c2)
\draw (b2) -- ($(b2)+(0.5,0)$);
\draw ($(c2)-(0.5,0)$) -- (c2);
\draw ($(b2)+(0.5,0)$) .. controls ($(b2)+(1.25,0)$) and ($(c2)-(1.25,0)$) ..($(c2)-(0.5,0)$);

%下の曲線(a1)--(c1)
\draw (a1) -- ($(a1)+(0.5,0)$);
\draw ($(c1)-(0.5,0)$) -- (c1);
\draw ($(a1)+(0.5,0)$) .. controls ($(a1)+(1.25,0)$) and ($(c1)-(1.25,0)$) ..($(c1)-(0.5,0)$);
\node[right] at (0,6) {$$};
\node[right] at (0,4.5) {$$};
\node[right] at (0,3) {$b_{(1)}$};
\node[right] at (0,0.5) {$b_{(2)}$};
\end{tikzpicture}

\begin{tikzpicture}[scale=0.45, line width=2pt]
\begin{scope}[shift={(0,6)}]
    \coordinate (x1) at (0,-1);
\coordinate (x2) at (0,-2);
\coordinate (y1) at (3,-1);
\coordinate (y2) at (3,-2);
\draw (x1) -- (x2);
\draw (y1) -- (y2);
\draw (x1) -- (y1);
\draw (y2) -- (x2);

\end{scope}
\begin{scope}[shift={(0,0.5)}]
    \coordinate (x1) at (0,-1);
\coordinate (x2) at (0,-2);
\coordinate (y1) at (3,-1);
\coordinate (y2) at (3,-2);
\draw (x1) -- (x2);
\draw (y1) -- (y2);
\draw (x1) -- (y1);
\draw (y2) -- (x2);

\end{scope}
 \coordinate (a1) at (0,0);
\coordinate (a2) at (0,1);
\coordinate (b1) at (0,2.5);
\coordinate (b2) at (0,3.5);
\coordinate (c1) at (3,1.25);
\coordinate (c2) at (3,2.25);

\draw (a1) -- (a2);
\draw (b1) -- (b2);
\draw (c1) -- (c2);

%円弧
\draw (0,1) arc[start angle=-90, end angle=90, radius=0.75];

%上の曲線(b2)--(c2)
\draw (b2) -- ($(b2)+(0.5,0)$);
\draw ($(c2)-(0.5,0)$) -- (c2);
\draw ($(b2)+(0.5,0)$) .. controls ($(b2)+(1.25,0)$) and ($(c2)-(1.25,0)$) ..($(c2)-(0.5,0)$);

%下の曲線(a1)--(c1)
\draw (a1) -- ($(a1)+(0.5,0)$);
\draw ($(c1)-(0.5,0)$) -- (c1);
\draw ($(a1)+(0.5,0)$) .. controls ($(a1)+(1.25,0)$) and ($(c1)-(1.25,0)$) ..($(c1)-(0.5,0)$);
\node[left] at (0,4.5) {$c$};
\node[left] at (0,3) {$d_{(1)}$};
\node[left] at (0,0.5) {$d_{(2)}$};
\node[left] at (0,-1) {$e$};
\end{tikzpicture}

    }{
    
\begin{tikzpicture}[scale=0.45, line width=2pt, xscale=-1]
\begin{scope}[shift={(0,6)}]
    \coordinate (x1) at (0,-1);
\coordinate (x2) at (0,-2);
\coordinate (y1) at (3,-1);
\coordinate (y2) at (3,-2);
\draw (x1) -- (x2);
\draw (y1) -- (y2);
\draw (x1) -- (y1);
\draw (y2) -- (x2);

\end{scope}
 \coordinate (a1) at (0,0);
\coordinate (a2) at (0,1);
\coordinate (b1) at (0,2.5);
\coordinate (b2) at (0,3.5);
\coordinate (c1) at (3,1.25);
\coordinate (c2) at (3,2.25);

\draw (a1) -- (a2);
\draw (b1) -- (b2);
\draw (c1) -- (c2);

%円弧
\draw (0,1) arc[start angle=-90, end angle=90, radius=0.75];

%上の曲線(b2)--(c2)
\draw (b2) -- ($(b2)+(0.5,0)$);
\draw ($(c2)-(0.5,0)$) -- (c2);
\draw ($(b2)+(0.5,0)$) .. controls ($(b2)+(1.25,0)$) and ($(c2)-(1.25,0)$) ..($(c2)-(0.5,0)$);

%下の曲線(a1)--(c1)
\draw (a1) -- ($(a1)+(0.5,0)$);
\draw ($(c1)-(0.5,0)$) -- (c1);
\draw ($(a1)+(0.5,0)$) .. controls ($(a1)+(1.25,0)$) and ($(c1)-(1.25,0)$) ..($(c1)-(0.5,0)$);
\node[right] at (0,4.5) {$$};
\node[right] at (0,3) {$b_{(1)}$};
\node[right] at (0,0.5) {$b_{(2)}$};
\end{tikzpicture}

\begin{tikzpicture}[scale=0.45, line width=2pt, xscale=-1]
\begin{scope}[shift={(0,0.5)}]
    \coordinate (x1) at (0,-1);
\coordinate (x2) at (0,-2);
\coordinate (y1) at (3,-1);
\coordinate (y2) at (3,-2);
\draw (x1) -- (x2);
\draw (y1) -- (y2);
\draw (x1) -- (y1);
\draw (y2) -- (x2);

\end{scope}
\begin{scope}[shift={(0,-1)}]
    \coordinate (x1) at (0,-1);
\coordinate (x2) at (0,-2);
\coordinate (y1) at (3,-1);
\coordinate (y2) at (3,-2);
\draw (x1) -- (x2);
\draw (y1) -- (y2);
\draw (x1) -- (y1);
\draw (y2) -- (x2);

\end{scope}
 \coordinate (a1) at (0,0);
\coordinate (a2) at (0,1);
\coordinate (b1) at (0,2.5);
\coordinate (b2) at (0,3.5);
\coordinate (c1) at (3,1.25);
\coordinate (c2) at (3,2.25);

\draw (a1) -- (a2);
\draw (b1) -- (b2);
\draw (c1) -- (c2);

%円弧
\draw (0,1) arc[start angle=-90, end angle=90, radius=0.75];

%上の曲線(b2)--(c2)
\draw (b2) -- ($(b2)+(0.5,0)$);
\draw ($(c2)-(0.5,0)$) -- (c2);
\draw ($(b2)+(0.5,0)$) .. controls ($(b2)+(1.25,0)$) and ($(c2)-(1.25,0)$) ..($(c2)-(0.5,0)$);

%下の曲線(a1)--(c1)
\draw (a1) -- ($(a1)+(0.5,0)$);
\draw ($(c1)-(0.5,0)$) -- (c1);
\draw ($(a1)+(0.5,0)$) .. controls ($(a1)+(1.25,0)$) and ($(c1)-(1.25,0)$) ..($(c1)-(0.5,0)$);
\node[right] at (0,3) {$a_{(1)}$};
\node[right] at (0,0.5) {$a_{(2)}$};
\node[right] at (0,-1) {$$};
\node[right] at (0,-2.5) {$$};
\end{tikzpicture}

\begin{tikzpicture}[scale=0.45, line width=2pt]
\begin{scope}[shift={(0,6)}]
    \coordinate (x1) at (0,-1);
\coordinate (x2) at (0,-2);
\coordinate (y1) at (3,-1);
\coordinate (y2) at (3,-2);
\draw (x1) -- (x2);
\draw (y1) -- (y2);
\draw (x1) -- (y1);
\draw (y2) -- (x2);

\end{scope}
\begin{scope}[shift={(0,0.5)}]
    \coordinate (x1) at (0,-1);
\coordinate (x2) at (0,-2);
\coordinate (y1) at (3,-1);
\coordinate (y2) at (3,-2);
\draw (x1) -- (x2);
\draw (y1) -- (y2);
\draw (x1) -- (y1);
\draw (y2) -- (x2);

\end{scope}
 \coordinate (a1) at (0,0);
\coordinate (a2) at (0,1);
\coordinate (b1) at (0,2.5);
\coordinate (b2) at (0,3.5);
\coordinate (c1) at (3,1.25);
\coordinate (c2) at (3,2.25);

\draw (a1) -- (a2);
\draw (b1) -- (b2);
\draw (c1) -- (c2);

%円弧
\draw (0,1) arc[start angle=-90, end angle=90, radius=0.75];

%上の曲線(b2)--(c2)
\draw (b2) -- ($(b2)+(0.5,0)$);
\draw ($(c2)-(0.5,0)$) -- (c2);
\draw ($(b2)+(0.5,0)$) .. controls ($(b2)+(1.25,0)$) and ($(c2)-(1.25,0)$) ..($(c2)-(0.5,0)$);

%下の曲線(a1)--(c1)
\draw (a1) -- ($(a1)+(0.5,0)$);
\draw ($(c1)-(0.5,0)$) -- (c1);
\draw ($(a1)+(0.5,0)$) .. controls ($(a1)+(1.25,0)$) and ($(c1)-(1.25,0)$) ..($(c1)-(0.5,0)$);
\node[left] at (0,4.5) {$c$};
\node[left] at (0,3) {$d_{(1)}$};
\node[left] at (0,0.5) {$d_{(2)}$};
\node[left] at (0,-1) {$e$};
\end{tikzpicture}

    }
    }
\]
\fi

\iflatexml
\begin{center}
  \includegraphics{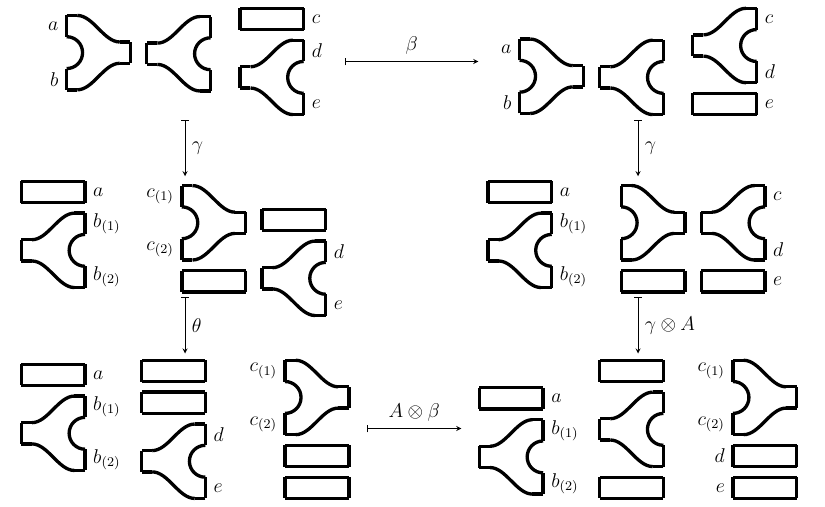}
\end{center}
\else
\[\scalebox{0.8}{%
  \DiagramRectangle[
    AB=\gamma,
    BC=\theta,
    AD=\beta,
    DE=\gamma,
    EF=\gamma\otimes A,
    CF=A\otimes \beta
  ]
   {
   \shiftfig{0mm}{5mm}{
\begin{tikzpicture}[scale=0.45, line width=2pt]
 \coordinate (a1) at (0,0);
\coordinate (a2) at (0,1);
\coordinate (b1) at (0,2.5);
\coordinate (b2) at (0,3.5);
\coordinate (c1) at (3,1.25);
\coordinate (c2) at (3,2.25);

\draw (a1) -- (a2);
\draw (b1) -- (b2);
\draw (c1) -- (c2);

%円弧
\draw (0,1) arc[start angle=-90, end angle=90, radius=0.75];

%上の曲線(b2)--(c2)
\draw (b2) -- ($(b2)+(0.5,0)$);
\draw ($(c2)-(0.5,0)$) -- (c2);
\draw ($(b2)+(0.5,0)$) .. controls ($(b2)+(1.25,0)$) and ($(c2)-(1.25,0)$) ..($(c2)-(0.5,0)$);

%下の曲線(a1)--(c1)
\draw (a1) -- ($(a1)+(0.5,0)$);
\draw ($(c1)-(0.5,0)$) -- (c1);
\draw ($(a1)+(0.5,0)$) .. controls ($(a1)+(1.25,0)$) and ($(c1)-(1.25,0)$) ..($(c1)-(0.5,0)$);
\node[left] at (0,3) {$a$};
\node[left] at (0,0.5) {$b$};
\end{tikzpicture}

\begin{tikzpicture}[scale=0.45, line width=2pt, xscale=-1]
 \coordinate (a1) at (0,0);
\coordinate (a2) at (0,1);
\coordinate (b1) at (0,2.5);
\coordinate (b2) at (0,3.5);
\coordinate (c1) at (3,1.25);
\coordinate (c2) at (3,2.25);

\draw (a1) -- (a2);
\draw (b1) -- (b2);
\draw (c1) -- (c2);

%円弧
\draw (0,1) arc[start angle=-90, end angle=90, radius=0.75];

%上の曲線(b2)--(c2)
\draw (b2) -- ($(b2)+(0.5,0)$);
\draw ($(c2)-(0.5,0)$) -- (c2);
\draw ($(b2)+(0.5,0)$) .. controls ($(b2)+(1.25,0)$) and ($(c2)-(1.25,0)$) ..($(c2)-(0.5,0)$);

%下の曲線(a1)--(c1)
\draw (a1) -- ($(a1)+(0.5,0)$);
\draw ($(c1)-(0.5,0)$) -- (c1);
\draw ($(a1)+(0.5,0)$) .. controls ($(a1)+(1.25,0)$) and ($(c1)-(1.25,0)$) ..($(c1)-(0.5,0)$);
\node[right] at (0,3) {$$};
\node[right] at (0,0.5) {$$};
\end{tikzpicture}
}
\begin{tikzpicture}[scale=0.45, line width=2pt, xscale=-1]
\begin{scope}[shift={(0,6)}]
    \coordinate (x1) at (0,-1);
\coordinate (x2) at (0,-2);
\coordinate (y1) at (3,-1);
\coordinate (y2) at (3,-2);
\draw (x1) -- (x2);
\draw (y1) -- (y2);
\draw (x1) -- (y1);
\draw (y2) -- (x2);

\end{scope}
 \coordinate (a1) at (0,0);
\coordinate (a2) at (0,1);
\coordinate (b1) at (0,2.5);
\coordinate (b2) at (0,3.5);
\coordinate (c1) at (3,1.25);
\coordinate (c2) at (3,2.25);

\draw (a1) -- (a2);
\draw (b1) -- (b2);
\draw (c1) -- (c2);

%円弧
\draw (0,1) arc[start angle=-90, end angle=90, radius=0.75];

%上の曲線(b2)--(c2)
\draw (b2) -- ($(b2)+(0.5,0)$);
\draw ($(c2)-(0.5,0)$) -- (c2);
\draw ($(b2)+(0.5,0)$) .. controls ($(b2)+(1.25,0)$) and ($(c2)-(1.25,0)$) ..($(c2)-(0.5,0)$);

%下の曲線(a1)--(c1)
\draw (a1) -- ($(a1)+(0.5,0)$);
\draw ($(c1)-(0.5,0)$) -- (c1);
\draw ($(a1)+(0.5,0)$) .. controls ($(a1)+(1.25,0)$) and ($(c1)-(1.25,0)$) ..($(c1)-(0.5,0)$);
\node[right] at (0,4.5) {$c$};
\node[right] at (0,3) {$d$};
\node[right] at (0,0.5) {$e$};
\end{tikzpicture}

   }{
   
\begin{tikzpicture}[scale=0.45, line width=2pt, xscale=-1]
\begin{scope}[shift={(0,6)}]
    \coordinate (x1) at (0,-1);
\coordinate (x2) at (0,-2);
\coordinate (y1) at (3,-1);
\coordinate (y2) at (3,-2);
\draw (x1) -- (x2);
\draw (y1) -- (y2);
\draw (x1) -- (y1);
\draw (y2) -- (x2);

\end{scope}
 \coordinate (a1) at (0,0);
\coordinate (a2) at (0,1);
\coordinate (b1) at (0,2.5);
\coordinate (b2) at (0,3.5);
\coordinate (c1) at (3,1.25);
\coordinate (c2) at (3,2.25);

\draw (a1) -- (a2);
\draw (b1) -- (b2);
\draw (c1) -- (c2);

%円弧
\draw (0,1) arc[start angle=-90, end angle=90, radius=0.75];

%上の曲線(b2)--(c2)
\draw (b2) -- ($(b2)+(0.5,0)$);
\draw ($(c2)-(0.5,0)$) -- (c2);
\draw ($(b2)+(0.5,0)$) .. controls ($(b2)+(1.25,0)$) and ($(c2)-(1.25,0)$) ..($(c2)-(0.5,0)$);

%下の曲線(a1)--(c1)
\draw (a1) -- ($(a1)+(0.5,0)$);
\draw ($(c1)-(0.5,0)$) -- (c1);
\draw ($(a1)+(0.5,0)$) .. controls ($(a1)+(1.25,0)$) and ($(c1)-(1.25,0)$) ..($(c1)-(0.5,0)$);
\node[right] at (0,4.5) {$a$};
\node[right] at (0,3) {$b_{(1)}$};
\node[right] at (0,0.5) {$b_{(2)}$};
\end{tikzpicture}

\begin{tikzpicture}[scale=0.45, line width=2pt]
\begin{scope}[shift={(0,0.5)}]
    \coordinate (x1) at (0,-1);
\coordinate (x2) at (0,-2);
\coordinate (y1) at (3,-1);
\coordinate (y2) at (3,-2);
\draw (x1) -- (x2);
\draw (y1) -- (y2);
\draw (x1) -- (y1);
\draw (y2) -- (x2);

\end{scope}
 \coordinate (a1) at (0,0);
\coordinate (a2) at (0,1);
\coordinate (b1) at (0,2.5);
\coordinate (b2) at (0,3.5);
\coordinate (c1) at (3,1.25);
\coordinate (c2) at (3,2.25);

\draw (a1) -- (a2);
\draw (b1) -- (b2);
\draw (c1) -- (c2);

%円弧
\draw (0,1) arc[start angle=-90, end angle=90, radius=0.75];

%上の曲線(b2)--(c2)
\draw (b2) -- ($(b2)+(0.5,0)$);
\draw ($(c2)-(0.5,0)$) -- (c2);
\draw ($(b2)+(0.5,0)$) .. controls ($(b2)+(1.25,0)$) and ($(c2)-(1.25,0)$) ..($(c2)-(0.5,0)$);

%下の曲線(a1)--(c1)
\draw (a1) -- ($(a1)+(0.5,0)$);
\draw ($(c1)-(0.5,0)$) -- (c1);
\draw ($(a1)+(0.5,0)$) .. controls ($(a1)+(1.25,0)$) and ($(c1)-(1.25,0)$) ..($(c1)-(0.5,0)$);
\node[left] at (0,3) {$c_{(1)}$};
\node[left] at (0,0.5) {$c_{(2)}$};
\node[left] at (0,-1) {$$};
\end{tikzpicture}
\shiftfig{0mm}{-5mm}{
\begin{tikzpicture}[scale=0.45, line width=2pt, xscale=-1]
\begin{scope}[shift={(0,6)}]
    \coordinate (x1) at (0,-1);
\coordinate (x2) at (0,-2);
\coordinate (y1) at (3,-1);
\coordinate (y2) at (3,-2);
\draw (x1) -- (x2);
\draw (y1) -- (y2);
\draw (x1) -- (y1);
\draw (y2) -- (x2);

\end{scope}
 \coordinate (a1) at (0,0);
\coordinate (a2) at (0,1);
\coordinate (b1) at (0,2.5);
\coordinate (b2) at (0,3.5);
\coordinate (c1) at (3,1.25);
\coordinate (c2) at (3,2.25);

\draw (a1) -- (a2);
\draw (b1) -- (b2);
\draw (c1) -- (c2);

%円弧
\draw (0,1) arc[start angle=-90, end angle=90, radius=0.75];

%上の曲線(b2)--(c2)
\draw (b2) -- ($(b2)+(0.5,0)$);
\draw ($(c2)-(0.5,0)$) -- (c2);
\draw ($(b2)+(0.5,0)$) .. controls ($(b2)+(1.25,0)$) and ($(c2)-(1.25,0)$) ..($(c2)-(0.5,0)$);

%下の曲線(a1)--(c1)
\draw (a1) -- ($(a1)+(0.5,0)$);
\draw ($(c1)-(0.5,0)$) -- (c1);
\draw ($(a1)+(0.5,0)$) .. controls ($(a1)+(1.25,0)$) and ($(c1)-(1.25,0)$) ..($(c1)-(0.5,0)$);
\node[right] at (0,4.5) {$$};
\node[right] at (0,3) {$d$};
\node[right] at (0,0.5) {$e$};
\end{tikzpicture}
}
   }{
   \shiftfig{0mm}{5mm}{
\begin{tikzpicture}[scale=0.45, line width=2pt, xscale=-1]
\begin{scope}[shift={(0,6)}]
    \coordinate (x1) at (0,-1);
\coordinate (x2) at (0,-2);
\coordinate (y1) at (3,-1);
\coordinate (y2) at (3,-2);
\draw (x1) -- (x2);
\draw (y1) -- (y2);
\draw (x1) -- (y1);
\draw (y2) -- (x2);

\end{scope}
 \coordinate (a1) at (0,0);
\coordinate (a2) at (0,1);
\coordinate (b1) at (0,2.5);
\coordinate (b2) at (0,3.5);
\coordinate (c1) at (3,1.25);
\coordinate (c2) at (3,2.25);

\draw (a1) -- (a2);
\draw (b1) -- (b2);
\draw (c1) -- (c2);

%円弧
\draw (0,1) arc[start angle=-90, end angle=90, radius=0.75];

%上の曲線(b2)--(c2)
\draw (b2) -- ($(b2)+(0.5,0)$);
\draw ($(c2)-(0.5,0)$) -- (c2);
\draw ($(b2)+(0.5,0)$) .. controls ($(b2)+(1.25,0)$) and ($(c2)-(1.25,0)$) ..($(c2)-(0.5,0)$);

%下の曲線(a1)--(c1)
\draw (a1) -- ($(a1)+(0.5,0)$);
\draw ($(c1)-(0.5,0)$) -- (c1);
\draw ($(a1)+(0.5,0)$) .. controls ($(a1)+(1.25,0)$) and ($(c1)-(1.25,0)$) ..($(c1)-(0.5,0)$);
\node[right] at (0,4.5) {$a$};
\node[right] at (0,3) {$b_{(1)}$};
\node[right] at (0,0.5) {$b_{(2)}$};
\end{tikzpicture}
}
\begin{tikzpicture}[scale=0.45, line width=2pt, xscale=-1]
\begin{scope}[shift={(0,7.5)}]
    \coordinate (x1) at (0,-1);
\coordinate (x2) at (0,-2);
\coordinate (y1) at (3,-1);
\coordinate (y2) at (3,-2);
\draw (x1) -- (x2);
\draw (y1) -- (y2);
\draw (x1) -- (y1);
\draw (y2) -- (x2);

\end{scope}
\begin{scope}[shift={(0,6)}]
    \coordinate (x1) at (0,-1);
\coordinate (x2) at (0,-2);
\coordinate (y1) at (3,-1);
\coordinate (y2) at (3,-2);
\draw (x1) -- (x2);
\draw (y1) -- (y2);
\draw (x1) -- (y1);
\draw (y2) -- (x2);

\end{scope}
 \coordinate (a1) at (0,0);
\coordinate (a2) at (0,1);
\coordinate (b1) at (0,2.5);
\coordinate (b2) at (0,3.5);
\coordinate (c1) at (3,1.25);
\coordinate (c2) at (3,2.25);

\draw (a1) -- (a2);
\draw (b1) -- (b2);
\draw (c1) -- (c2);

%円弧
\draw (0,1) arc[start angle=-90, end angle=90, radius=0.75];

%上の曲線(b2)--(c2)
\draw (b2) -- ($(b2)+(0.5,0)$);
\draw ($(c2)-(0.5,0)$) -- (c2);
\draw ($(b2)+(0.5,0)$) .. controls ($(b2)+(1.25,0)$) and ($(c2)-(1.25,0)$) ..($(c2)-(0.5,0)$);

%下の曲線(a1)--(c1)
\draw (a1) -- ($(a1)+(0.5,0)$);
\draw ($(c1)-(0.5,0)$) -- (c1);
\draw ($(a1)+(0.5,0)$) .. controls ($(a1)+(1.25,0)$) and ($(c1)-(1.25,0)$) ..($(c1)-(0.5,0)$);
\node[right] at (0,6) {$$};
\node[right] at (0,4.5) {$$};
\node[right] at (0,3) {$d$};
\node[right] at (0,0.5) {$e$};
\end{tikzpicture}

\begin{tikzpicture}[scale=0.45, line width=2pt]
\begin{scope}[shift={(0,0.5)}]
    \coordinate (x1) at (0,-1);
\coordinate (x2) at (0,-2);
\coordinate (y1) at (3,-1);
\coordinate (y2) at (3,-2);
\draw (x1) -- (x2);
\draw (y1) -- (y2);
\draw (x1) -- (y1);
\draw (y2) -- (x2);

\end{scope}
\begin{scope}[shift={(0,-1)}]
    \coordinate (x1) at (0,-1);
\coordinate (x2) at (0,-2);
\coordinate (y1) at (3,-1);
\coordinate (y2) at (3,-2);
\draw (x1) -- (x2);
\draw (y1) -- (y2);
\draw (x1) -- (y1);
\draw (y2) -- (x2);

\end{scope}
 \coordinate (a1) at (0,0);
\coordinate (a2) at (0,1);
\coordinate (b1) at (0,2.5);
\coordinate (b2) at (0,3.5);
\coordinate (c1) at (3,1.25);
\coordinate (c2) at (3,2.25);

\draw (a1) -- (a2);
\draw (b1) -- (b2);
\draw (c1) -- (c2);

%円弧
\draw (0,1) arc[start angle=-90, end angle=90, radius=0.75];

%上の曲線(b2)--(c2)
\draw (b2) -- ($(b2)+(0.5,0)$);
\draw ($(c2)-(0.5,0)$) -- (c2);
\draw ($(b2)+(0.5,0)$) .. controls ($(b2)+(1.25,0)$) and ($(c2)-(1.25,0)$) ..($(c2)-(0.5,0)$);

%下の曲線(a1)--(c1)
\draw (a1) -- ($(a1)+(0.5,0)$);
\draw ($(c1)-(0.5,0)$) -- (c1);
\draw ($(a1)+(0.5,0)$) .. controls ($(a1)+(1.25,0)$) and ($(c1)-(1.25,0)$) ..($(c1)-(0.5,0)$);
\node[left] at (0,3) {$c_{(1)}$};
\node[left] at (0,0.5) {$c_{(2)}$};
\node[left] at (0,-1) {$$};
\node[left] at (0,-2.5) {$$};
\end{tikzpicture}

   }{
   
\begin{tikzpicture}[scale=0.45, line width=2pt]
 \coordinate (a1) at (0,0);
\coordinate (a2) at (0,1);
\coordinate (b1) at (0,2.5);
\coordinate (b2) at (0,3.5);
\coordinate (c1) at (3,1.25);
\coordinate (c2) at (3,2.25);

\draw (a1) -- (a2);
\draw (b1) -- (b2);
\draw (c1) -- (c2);

%円弧
\draw (0,1) arc[start angle=-90, end angle=90, radius=0.75];

%上の曲線(b2)--(c2)
\draw (b2) -- ($(b2)+(0.5,0)$);
\draw ($(c2)-(0.5,0)$) -- (c2);
\draw ($(b2)+(0.5,0)$) .. controls ($(b2)+(1.25,0)$) and ($(c2)-(1.25,0)$) ..($(c2)-(0.5,0)$);

%下の曲線(a1)--(c1)
\draw (a1) -- ($(a1)+(0.5,0)$);
\draw ($(c1)-(0.5,0)$) -- (c1);
\draw ($(a1)+(0.5,0)$) .. controls ($(a1)+(1.25,0)$) and ($(c1)-(1.25,0)$) ..($(c1)-(0.5,0)$);
\node[left] at (0,3) {$a$};
\node[left] at (0,0.5) {$b$};
\end{tikzpicture}

\begin{tikzpicture}[scale=0.45, line width=2pt, xscale=-1]
 \coordinate (a1) at (0,0);
\coordinate (a2) at (0,1);
\coordinate (b1) at (0,2.5);
\coordinate (b2) at (0,3.5);
\coordinate (c1) at (3,1.25);
\coordinate (c2) at (3,2.25);

\draw (a1) -- (a2);
\draw (b1) -- (b2);
\draw (c1) -- (c2);

%円弧
\draw (0,1) arc[start angle=-90, end angle=90, radius=0.75];

%上の曲線(b2)--(c2)
\draw (b2) -- ($(b2)+(0.5,0)$);
\draw ($(c2)-(0.5,0)$) -- (c2);
\draw ($(b2)+(0.5,0)$) .. controls ($(b2)+(1.25,0)$) and ($(c2)-(1.25,0)$) ..($(c2)-(0.5,0)$);

%下の曲線(a1)--(c1)
\draw (a1) -- ($(a1)+(0.5,0)$);
\draw ($(c1)-(0.5,0)$) -- (c1);
\draw ($(a1)+(0.5,0)$) .. controls ($(a1)+(1.25,0)$) and ($(c1)-(1.25,0)$) ..($(c1)-(0.5,0)$);
\node[right] at (0,3) {$$};
\node[right] at (0,0.5) {$$};
\end{tikzpicture}

\begin{tikzpicture}[scale=0.45, line width=2pt, xscale=-1]
\begin{scope}[shift={(0,0.5)}]
    \coordinate (x1) at (0,-1);
\coordinate (x2) at (0,-2);
\coordinate (y1) at (3,-1);
\coordinate (y2) at (3,-2);
\draw (x1) -- (x2);
\draw (y1) -- (y2);
\draw (x1) -- (y1);
\draw (y2) -- (x2);

\end{scope}
 \coordinate (a1) at (0,0);
\coordinate (a2) at (0,1);
\coordinate (b1) at (0,2.5);
\coordinate (b2) at (0,3.5);
\coordinate (c1) at (3,1.25);
\coordinate (c2) at (3,2.25);

\draw (a1) -- (a2);
\draw (b1) -- (b2);
\draw (c1) -- (c2);

%円弧
\draw (0,1) arc[start angle=-90, end angle=90, radius=0.75];

%上の曲線(b2)--(c2)
\draw (b2) -- ($(b2)+(0.5,0)$);
\draw ($(c2)-(0.5,0)$) -- (c2);
\draw ($(b2)+(0.5,0)$) .. controls ($(b2)+(1.25,0)$) and ($(c2)-(1.25,0)$) ..($(c2)-(0.5,0)$);

%下の曲線(a1)--(c1)
\draw (a1) -- ($(a1)+(0.5,0)$);
\draw ($(c1)-(0.5,0)$) -- (c1);
\draw ($(a1)+(0.5,0)$) .. controls ($(a1)+(1.25,0)$) and ($(c1)-(1.25,0)$) ..($(c1)-(0.5,0)$);
\node[right] at (0,3) {$c$};
\node[right] at (0,0.5) {$d$};
\node[right] at (0,-1) {$e$};
\end{tikzpicture}

   }{
   
\begin{tikzpicture}[scale=0.45, line width=2pt, xscale=-1]
\begin{scope}[shift={(0,6)}]
    \coordinate (x1) at (0,-1);
\coordinate (x2) at (0,-2);
\coordinate (y1) at (3,-1);
\coordinate (y2) at (3,-2);
\draw (x1) -- (x2);
\draw (y1) -- (y2);
\draw (x1) -- (y1);
\draw (y2) -- (x2);

\end{scope}
 \coordinate (a1) at (0,0);
\coordinate (a2) at (0,1);
\coordinate (b1) at (0,2.5);
\coordinate (b2) at (0,3.5);
\coordinate (c1) at (3,1.25);
\coordinate (c2) at (3,2.25);

\draw (a1) -- (a2);
\draw (b1) -- (b2);
\draw (c1) -- (c2);

%円弧
\draw (0,1) arc[start angle=-90, end angle=90, radius=0.75];

%上の曲線(b2)--(c2)
\draw (b2) -- ($(b2)+(0.5,0)$);
\draw ($(c2)-(0.5,0)$) -- (c2);
\draw ($(b2)+(0.5,0)$) .. controls ($(b2)+(1.25,0)$) and ($(c2)-(1.25,0)$) ..($(c2)-(0.5,0)$);

%下の曲線(a1)--(c1)
\draw (a1) -- ($(a1)+(0.5,0)$);
\draw ($(c1)-(0.5,0)$) -- (c1);
\draw ($(a1)+(0.5,0)$) .. controls ($(a1)+(1.25,0)$) and ($(c1)-(1.25,0)$) ..($(c1)-(0.5,0)$);
\node[right] at (0,4.5) {$a$};
\node[right] at (0,3) {$b_{(1)}$};
\node[right] at (0,0.5) {$b_{(2)}$};
\end{tikzpicture}

\begin{tikzpicture}[scale=0.45, line width=2pt]
\begin{scope}[shift={(0,0.5)}]
    \coordinate (x1) at (0,-1);
\coordinate (x2) at (0,-2);
\coordinate (y1) at (3,-1);
\coordinate (y2) at (3,-2);
\draw (x1) -- (x2);
\draw (y1) -- (y2);
\draw (x1) -- (y1);
\draw (y2) -- (x2);

\end{scope}
 \coordinate (a1) at (0,0);
\coordinate (a2) at (0,1);
\coordinate (b1) at (0,2.5);
\coordinate (b2) at (0,3.5);
\coordinate (c1) at (3,1.25);
\coordinate (c2) at (3,2.25);

\draw (a1) -- (a2);
\draw (b1) -- (b2);
\draw (c1) -- (c2);

%円弧
\draw (0,1) arc[start angle=-90, end angle=90, radius=0.75];

%上の曲線(b2)--(c2)
\draw (b2) -- ($(b2)+(0.5,0)$);
\draw ($(c2)-(0.5,0)$) -- (c2);
\draw ($(b2)+(0.5,0)$) .. controls ($(b2)+(1.25,0)$) and ($(c2)-(1.25,0)$) ..($(c2)-(0.5,0)$);

%下の曲線(a1)--(c1)
\draw (a1) -- ($(a1)+(0.5,0)$);
\draw ($(c1)-(0.5,0)$) -- (c1);
\draw ($(a1)+(0.5,0)$) .. controls ($(a1)+(1.25,0)$) and ($(c1)-(1.25,0)$) ..($(c1)-(0.5,0)$);
\node[left] at (0,3) {$$};
\node[left] at (0,0.5) {$$};
\node[left] at (0,-1) {$$};
\end{tikzpicture}

\begin{tikzpicture}[scale=0.45, line width=2pt, xscale=-1]
\begin{scope}[shift={(0,0.5)}]
    \coordinate (x1) at (0,-1);
\coordinate (x2) at (0,-2);
\coordinate (y1) at (3,-1);
\coordinate (y2) at (3,-2);
\draw (x1) -- (x2);
\draw (y1) -- (y2);
\draw (x1) -- (y1);
\draw (y2) -- (x2);

\end{scope}
 \coordinate (a1) at (0,0);
\coordinate (a2) at (0,1);
\coordinate (b1) at (0,2.5);
\coordinate (b2) at (0,3.5);
\coordinate (c1) at (3,1.25);
\coordinate (c2) at (3,2.25);

\draw (a1) -- (a2);
\draw (b1) -- (b2);
\draw (c1) -- (c2);

%円弧
\draw (0,1) arc[start angle=-90, end angle=90, radius=0.75];

%上の曲線(b2)--(c2)
\draw (b2) -- ($(b2)+(0.5,0)$);
\draw ($(c2)-(0.5,0)$) -- (c2);
\draw ($(b2)+(0.5,0)$) .. controls ($(b2)+(1.25,0)$) and ($(c2)-(1.25,0)$) ..($(c2)-(0.5,0)$);

%下の曲線(a1)--(c1)
\draw (a1) -- ($(a1)+(0.5,0)$);
\draw ($(c1)-(0.5,0)$) -- (c1);
\draw ($(a1)+(0.5,0)$) .. controls ($(a1)+(1.25,0)$) and ($(c1)-(1.25,0)$) ..($(c1)-(0.5,0)$);
\node[right] at (0,3) {$c$};
\node[right] at (0,0.5) {$d$};
\node[right] at (0,-1) {$e$};
\end{tikzpicture}

   }{
   
\begin{tikzpicture}[scale=0.45, line width=2pt, xscale=-1]
\begin{scope}[shift={(0,6)}]
    \coordinate (x1) at (0,-1);
\coordinate (x2) at (0,-2);
\coordinate (y1) at (3,-1);
\coordinate (y2) at (3,-2);
\draw (x1) -- (x2);
\draw (y1) -- (y2);
\draw (x1) -- (y1);
\draw (y2) -- (x2);

\end{scope}
 \coordinate (a1) at (0,0);
\coordinate (a2) at (0,1);
\coordinate (b1) at (0,2.5);
\coordinate (b2) at (0,3.5);
\coordinate (c1) at (3,1.25);
\coordinate (c2) at (3,2.25);

\draw (a1) -- (a2);
\draw (b1) -- (b2);
\draw (c1) -- (c2);

%円弧
\draw (0,1) arc[start angle=-90, end angle=90, radius=0.75];

%上の曲線(b2)--(c2)
\draw (b2) -- ($(b2)+(0.5,0)$);
\draw ($(c2)-(0.5,0)$) -- (c2);
\draw ($(b2)+(0.5,0)$) .. controls ($(b2)+(1.25,0)$) and ($(c2)-(1.25,0)$) ..($(c2)-(0.5,0)$);

%下の曲線(a1)--(c1)
\draw (a1) -- ($(a1)+(0.5,0)$);
\draw ($(c1)-(0.5,0)$) -- (c1);
\draw ($(a1)+(0.5,0)$) .. controls ($(a1)+(1.25,0)$) and ($(c1)-(1.25,0)$) ..($(c1)-(0.5,0)$);
\node[right] at (0,4.5) {$a$};
\node[right] at (0,3) {$b_{(1)}$};
\node[right] at (0,0.5) {$b_{(2)}$};
\end{tikzpicture}

\begin{tikzpicture}[scale=0.45, line width=2pt, xscale=-1]
\begin{scope}[shift={(0,6)}]
    \coordinate (x1) at (0,-1);
\coordinate (x2) at (0,-2);
\coordinate (y1) at (3,-1);
\coordinate (y2) at (3,-2);
\draw (x1) -- (x2);
\draw (y1) -- (y2);
\draw (x1) -- (y1);
\draw (y2) -- (x2);

\end{scope}
\begin{scope}[shift={(0,0.5)}]
    \coordinate (x1) at (0,-1);
\coordinate (x2) at (0,-2);
\coordinate (y1) at (3,-1);
\coordinate (y2) at (3,-2);
\draw (x1) -- (x2);
\draw (y1) -- (y2);
\draw (x1) -- (y1);
\draw (y2) -- (x2);

\end{scope}
 \coordinate (a1) at (0,0);
\coordinate (a2) at (0,1);
\coordinate (b1) at (0,2.5);
\coordinate (b2) at (0,3.5);
\coordinate (c1) at (3,1.25);
\coordinate (c2) at (3,2.25);

\draw (a1) -- (a2);
\draw (b1) -- (b2);
\draw (c1) -- (c2);

%円弧
\draw (0,1) arc[start angle=-90, end angle=90, radius=0.75];

%上の曲線(b2)--(c2)
\draw (b2) -- ($(b2)+(0.5,0)$);
\draw ($(c2)-(0.5,0)$) -- (c2);
\draw ($(b2)+(0.5,0)$) .. controls ($(b2)+(1.25,0)$) and ($(c2)-(1.25,0)$) ..($(c2)-(0.5,0)$);

%下の曲線(a1)--(c1)
\draw (a1) -- ($(a1)+(0.5,0)$);
\draw ($(c1)-(0.5,0)$) -- (c1);
\draw ($(a1)+(0.5,0)$) .. controls ($(a1)+(1.25,0)$) and ($(c1)-(1.25,0)$) ..($(c1)-(0.5,0)$);
\node[right] at (0,4.5) {$$};
\node[right] at (0,3) {$$};
\node[right] at (0,0.5) {$$};
\node[right] at (0,-1) {$$};
\end{tikzpicture}

\begin{tikzpicture}[scale=0.45, line width=2pt]
\begin{scope}[shift={(0,0.5)}]
    \coordinate (x1) at (0,-1);
\coordinate (x2) at (0,-2);
\coordinate (y1) at (3,-1);
\coordinate (y2) at (3,-2);
\draw (x1) -- (x2);
\draw (y1) -- (y2);
\draw (x1) -- (y1);
\draw (y2) -- (x2);

\end{scope}
\begin{scope}[shift={(0,-1)}]
    \coordinate (x1) at (0,-1);
\coordinate (x2) at (0,-2);
\coordinate (y1) at (3,-1);
\coordinate (y2) at (3,-2);
\draw (x1) -- (x2);
\draw (y1) -- (y2);
\draw (x1) -- (y1);
\draw (y2) -- (x2);

\end{scope}
 \coordinate (a1) at (0,0);
\coordinate (a2) at (0,1);
\coordinate (b1) at (0,2.5);
\coordinate (b2) at (0,3.5);
\coordinate (c1) at (3,1.25);
\coordinate (c2) at (3,2.25);

\draw (a1) -- (a2);
\draw (b1) -- (b2);
\draw (c1) -- (c2);

%円弧
\draw (0,1) arc[start angle=-90, end angle=90, radius=0.75];

%上の曲線(b2)--(c2)
\draw (b2) -- ($(b2)+(0.5,0)$);
\draw ($(c2)-(0.5,0)$) -- (c2);
\draw ($(b2)+(0.5,0)$) .. controls ($(b2)+(1.25,0)$) and ($(c2)-(1.25,0)$) ..($(c2)-(0.5,0)$);

%下の曲線(a1)--(c1)
\draw (a1) -- ($(a1)+(0.5,0)$);
\draw ($(c1)-(0.5,0)$) -- (c1);
\draw ($(a1)+(0.5,0)$) .. controls ($(a1)+(1.25,0)$) and ($(c1)-(1.25,0)$) ..($(c1)-(0.5,0)$);
\node[left] at (0,3) {$c_{(1)}$};
\node[left] at (0,0.5) {$c_{(2)}$};
\node[left] at (0,-1) {$d$};
\node[left] at (0,-2.5) {$e$};
\end{tikzpicture}

   }
   }
\]
\fi

\iflatexml
\begin{center}
  \includegraphics{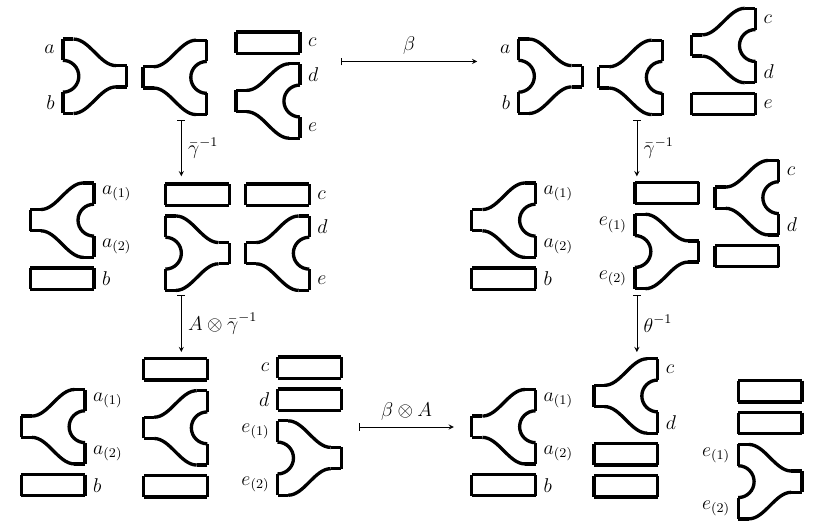}
\end{center}
\else
\[\scalebox{0.8}{%
  \DiagramRectangle[
    AB=\brg^{-1},
    BC=A\otimes \brg^{-1},
    AD=\beta,
    DE=\brg^{-1},
    EF=\theta^{-1},
    CF=\beta\otimes A
  ]
    {
    
\begin{tikzpicture}[scale=0.45, line width=2pt]
 \coordinate (a1) at (0,0);
\coordinate (a2) at (0,1);
\coordinate (b1) at (0,2.5);
\coordinate (b2) at (0,3.5);
\coordinate (c1) at (3,1.25);
\coordinate (c2) at (3,2.25);

\draw (a1) -- (a2);
\draw (b1) -- (b2);
\draw (c1) -- (c2);

%円弧
\draw (0,1) arc[start angle=-90, end angle=90, radius=0.75];

%上の曲線(b2)--(c2)
\draw (b2) -- ($(b2)+(0.5,0)$);
\draw ($(c2)-(0.5,0)$) -- (c2);
\draw ($(b2)+(0.5,0)$) .. controls ($(b2)+(1.25,0)$) and ($(c2)-(1.25,0)$) ..($(c2)-(0.5,0)$);

%下の曲線(a1)--(c1)
\draw (a1) -- ($(a1)+(0.5,0)$);
\draw ($(c1)-(0.5,0)$) -- (c1);
\draw ($(a1)+(0.5,0)$) .. controls ($(a1)+(1.25,0)$) and ($(c1)-(1.25,0)$) ..($(c1)-(0.5,0)$);
\node[left] at (0,3) {$a$};
\node[left] at (0,0.5) {$b$};
\end{tikzpicture}

\begin{tikzpicture}[scale=0.45, line width=2pt, xscale=-1]
 \coordinate (a1) at (0,0);
\coordinate (a2) at (0,1);
\coordinate (b1) at (0,2.5);
\coordinate (b2) at (0,3.5);
\coordinate (c1) at (3,1.25);
\coordinate (c2) at (3,2.25);

\draw (a1) -- (a2);
\draw (b1) -- (b2);
\draw (c1) -- (c2);

%円弧
\draw (0,1) arc[start angle=-90, end angle=90, radius=0.75];

%上の曲線(b2)--(c2)
\draw (b2) -- ($(b2)+(0.5,0)$);
\draw ($(c2)-(0.5,0)$) -- (c2);
\draw ($(b2)+(0.5,0)$) .. controls ($(b2)+(1.25,0)$) and ($(c2)-(1.25,0)$) ..($(c2)-(0.5,0)$);

%下の曲線(a1)--(c1)
\draw (a1) -- ($(a1)+(0.5,0)$);
\draw ($(c1)-(0.5,0)$) -- (c1);
\draw ($(a1)+(0.5,0)$) .. controls ($(a1)+(1.25,0)$) and ($(c1)-(1.25,0)$) ..($(c1)-(0.5,0)$);
\node[right] at (0,3) {$$};
\node[right] at (0,0.5) {$$};
\end{tikzpicture}
\shiftfig{0mm}{-5mm}{
\begin{tikzpicture}[scale=0.45, line width=2pt, xscale=-1]
\begin{scope}[shift={(0,6)}]
    \coordinate (x1) at (0,-1);
\coordinate (x2) at (0,-2);
\coordinate (y1) at (3,-1);
\coordinate (y2) at (3,-2);
\draw (x1) -- (x2);
\draw (y1) -- (y2);
\draw (x1) -- (y1);
\draw (y2) -- (x2);

\end{scope}
 \coordinate (a1) at (0,0);
\coordinate (a2) at (0,1);
\coordinate (b1) at (0,2.5);
\coordinate (b2) at (0,3.5);
\coordinate (c1) at (3,1.25);
\coordinate (c2) at (3,2.25);

\draw (a1) -- (a2);
\draw (b1) -- (b2);
\draw (c1) -- (c2);

%円弧
\draw (0,1) arc[start angle=-90, end angle=90, radius=0.75];

%上の曲線(b2)--(c2)
\draw (b2) -- ($(b2)+(0.5,0)$);
\draw ($(c2)-(0.5,0)$) -- (c2);
\draw ($(b2)+(0.5,0)$) .. controls ($(b2)+(1.25,0)$) and ($(c2)-(1.25,0)$) ..($(c2)-(0.5,0)$);

%下の曲線(a1)--(c1)
\draw (a1) -- ($(a1)+(0.5,0)$);
\draw ($(c1)-(0.5,0)$) -- (c1);
\draw ($(a1)+(0.5,0)$) .. controls ($(a1)+(1.25,0)$) and ($(c1)-(1.25,0)$) ..($(c1)-(0.5,0)$);
\node[right] at (0,4.5) {$c$};
\node[right] at (0,3) {$d$};
\node[right] at (0,0.5) {$e$};
\end{tikzpicture}
}
    }{
    
\begin{tikzpicture}[scale=0.45, line width=2pt, xscale=-1]
\begin{scope}[shift={(0,0.5)}]
    \coordinate (x1) at (0,-1);
\coordinate (x2) at (0,-2);
\coordinate (y1) at (3,-1);
\coordinate (y2) at (3,-2);
\draw (x1) -- (x2);
\draw (y1) -- (y2);
\draw (x1) -- (y1);
\draw (y2) -- (x2);

\end{scope}
 \coordinate (a1) at (0,0);
\coordinate (a2) at (0,1);
\coordinate (b1) at (0,2.5);
\coordinate (b2) at (0,3.5);
\coordinate (c1) at (3,1.25);
\coordinate (c2) at (3,2.25);

\draw (a1) -- (a2);
\draw (b1) -- (b2);
\draw (c1) -- (c2);

%円弧
\draw (0,1) arc[start angle=-90, end angle=90, radius=0.75];

%上の曲線(b2)--(c2)
\draw (b2) -- ($(b2)+(0.5,0)$);
\draw ($(c2)-(0.5,0)$) -- (c2);
\draw ($(b2)+(0.5,0)$) .. controls ($(b2)+(1.25,0)$) and ($(c2)-(1.25,0)$) ..($(c2)-(0.5,0)$);

%下の曲線(a1)--(c1)
\draw (a1) -- ($(a1)+(0.5,0)$);
\draw ($(c1)-(0.5,0)$) -- (c1);
\draw ($(a1)+(0.5,0)$) .. controls ($(a1)+(1.25,0)$) and ($(c1)-(1.25,0)$) ..($(c1)-(0.5,0)$);
\node[right] at (0,3) {$a_{(1)}$};
\node[right] at (0,0.5) {$a_{(2)}$};
\node[right] at (0,-1) {$b$};
\end{tikzpicture}

\begin{tikzpicture}[scale=0.45, line width=2pt]
\begin{scope}[shift={(0,6)}]
    \coordinate (x1) at (0,-1);
\coordinate (x2) at (0,-2);
\coordinate (y1) at (3,-1);
\coordinate (y2) at (3,-2);
\draw (x1) -- (x2);
\draw (y1) -- (y2);
\draw (x1) -- (y1);
\draw (y2) -- (x2);

\end{scope}
 \coordinate (a1) at (0,0);
\coordinate (a2) at (0,1);
\coordinate (b1) at (0,2.5);
\coordinate (b2) at (0,3.5);
\coordinate (c1) at (3,1.25);
\coordinate (c2) at (3,2.25);

\draw (a1) -- (a2);
\draw (b1) -- (b2);
\draw (c1) -- (c2);

%円弧
\draw (0,1) arc[start angle=-90, end angle=90, radius=0.75];

%上の曲線(b2)--(c2)
\draw (b2) -- ($(b2)+(0.5,0)$);
\draw ($(c2)-(0.5,0)$) -- (c2);
\draw ($(b2)+(0.5,0)$) .. controls ($(b2)+(1.25,0)$) and ($(c2)-(1.25,0)$) ..($(c2)-(0.5,0)$);

%下の曲線(a1)--(c1)
\draw (a1) -- ($(a1)+(0.5,0)$);
\draw ($(c1)-(0.5,0)$) -- (c1);
\draw ($(a1)+(0.5,0)$) .. controls ($(a1)+(1.25,0)$) and ($(c1)-(1.25,0)$) ..($(c1)-(0.5,0)$);
\node[left] at (0,4.5) {$$};
\node[left] at (0,3) {$$};
\node[left] at (0,0.5) {$$};
\end{tikzpicture}

\begin{tikzpicture}[scale=0.45, line width=2pt, xscale=-1]
\begin{scope}[shift={(0,6)}]
    \coordinate (x1) at (0,-1);
\coordinate (x2) at (0,-2);
\coordinate (y1) at (3,-1);
\coordinate (y2) at (3,-2);
\draw (x1) -- (x2);
\draw (y1) -- (y2);
\draw (x1) -- (y1);
\draw (y2) -- (x2);

\end{scope}
 \coordinate (a1) at (0,0);
\coordinate (a2) at (0,1);
\coordinate (b1) at (0,2.5);
\coordinate (b2) at (0,3.5);
\coordinate (c1) at (3,1.25);
\coordinate (c2) at (3,2.25);

\draw (a1) -- (a2);
\draw (b1) -- (b2);
\draw (c1) -- (c2);

%円弧
\draw (0,1) arc[start angle=-90, end angle=90, radius=0.75];

%上の曲線(b2)--(c2)
\draw (b2) -- ($(b2)+(0.5,0)$);
\draw ($(c2)-(0.5,0)$) -- (c2);
\draw ($(b2)+(0.5,0)$) .. controls ($(b2)+(1.25,0)$) and ($(c2)-(1.25,0)$) ..($(c2)-(0.5,0)$);

%下の曲線(a1)--(c1)
\draw (a1) -- ($(a1)+(0.5,0)$);
\draw ($(c1)-(0.5,0)$) -- (c1);
\draw ($(a1)+(0.5,0)$) .. controls ($(a1)+(1.25,0)$) and ($(c1)-(1.25,0)$) ..($(c1)-(0.5,0)$);
\node[right] at (0,4.5) {$c$};
\node[right] at (0,3) {$d$};
\node[right] at (0,0.5) {$e$};
\end{tikzpicture}

    }{
    
\begin{tikzpicture}[scale=0.45, line width=2pt, xscale=-1]
\begin{scope}[shift={(0,0.5)}]
    \coordinate (x1) at (0,-1);
\coordinate (x2) at (0,-2);
\coordinate (y1) at (3,-1);
\coordinate (y2) at (3,-2);
\draw (x1) -- (x2);
\draw (y1) -- (y2);
\draw (x1) -- (y1);
\draw (y2) -- (x2);

\end{scope}
 \coordinate (a1) at (0,0);
\coordinate (a2) at (0,1);
\coordinate (b1) at (0,2.5);
\coordinate (b2) at (0,3.5);
\coordinate (c1) at (3,1.25);
\coordinate (c2) at (3,2.25);

\draw (a1) -- (a2);
\draw (b1) -- (b2);
\draw (c1) -- (c2);

%円弧
\draw (0,1) arc[start angle=-90, end angle=90, radius=0.75];

%上の曲線(b2)--(c2)
\draw (b2) -- ($(b2)+(0.5,0)$);
\draw ($(c2)-(0.5,0)$) -- (c2);
\draw ($(b2)+(0.5,0)$) .. controls ($(b2)+(1.25,0)$) and ($(c2)-(1.25,0)$) ..($(c2)-(0.5,0)$);

%下の曲線(a1)--(c1)
\draw (a1) -- ($(a1)+(0.5,0)$);
\draw ($(c1)-(0.5,0)$) -- (c1);
\draw ($(a1)+(0.5,0)$) .. controls ($(a1)+(1.25,0)$) and ($(c1)-(1.25,0)$) ..($(c1)-(0.5,0)$);
\node[right] at (0,3) {$a_{(1)}$};
\node[right] at (0,0.5) {$a_{(2)}$};
\node[right] at (0,-1) {$b$};
\end{tikzpicture}

\begin{tikzpicture}[scale=0.45, line width=2pt, xscale=-1]
\begin{scope}[shift={(0,6)}]
    \coordinate (x1) at (0,-1);
\coordinate (x2) at (0,-2);
\coordinate (y1) at (3,-1);
\coordinate (y2) at (3,-2);
\draw (x1) -- (x2);
\draw (y1) -- (y2);
\draw (x1) -- (y1);
\draw (y2) -- (x2);

\end{scope}
\begin{scope}[shift={(0,0.5)}]
    \coordinate (x1) at (0,-1);
\coordinate (x2) at (0,-2);
\coordinate (y1) at (3,-1);
\coordinate (y2) at (3,-2);
\draw (x1) -- (x2);
\draw (y1) -- (y2);
\draw (x1) -- (y1);
\draw (y2) -- (x2);

\end{scope}
 \coordinate (a1) at (0,0);
\coordinate (a2) at (0,1);
\coordinate (b1) at (0,2.5);
\coordinate (b2) at (0,3.5);
\coordinate (c1) at (3,1.25);
\coordinate (c2) at (3,2.25);

\draw (a1) -- (a2);
\draw (b1) -- (b2);
\draw (c1) -- (c2);

%円弧
\draw (0,1) arc[start angle=-90, end angle=90, radius=0.75];

%上の曲線(b2)--(c2)
\draw (b2) -- ($(b2)+(0.5,0)$);
\draw ($(c2)-(0.5,0)$) -- (c2);
\draw ($(b2)+(0.5,0)$) .. controls ($(b2)+(1.25,0)$) and ($(c2)-(1.25,0)$) ..($(c2)-(0.5,0)$);

%下の曲線(a1)--(c1)
\draw (a1) -- ($(a1)+(0.5,0)$);
\draw ($(c1)-(0.5,0)$) -- (c1);
\draw ($(a1)+(0.5,0)$) .. controls ($(a1)+(1.25,0)$) and ($(c1)-(1.25,0)$) ..($(c1)-(0.5,0)$);
\node[right] at (0,4.5) {$$};
\node[right] at (0,3) {$$};
\node[right] at (0,0.5) {$$};
\node[right] at (0,-1) {$$};
\end{tikzpicture}

\begin{tikzpicture}[scale=0.45, line width=2pt]
\begin{scope}[shift={(0,7.5)}]
    \coordinate (x1) at (0,-1);
\coordinate (x2) at (0,-2);
\coordinate (y1) at (3,-1);
\coordinate (y2) at (3,-2);
\draw (x1) -- (x2);
\draw (y1) -- (y2);
\draw (x1) -- (y1);
\draw (y2) -- (x2);

\end{scope}
\begin{scope}[shift={(0,6)}]
    \coordinate (x1) at (0,-1);
\coordinate (x2) at (0,-2);
\coordinate (y1) at (3,-1);
\coordinate (y2) at (3,-2);
\draw (x1) -- (x2);
\draw (y1) -- (y2);
\draw (x1) -- (y1);
\draw (y2) -- (x2);

\end{scope}
 \coordinate (a1) at (0,0);
\coordinate (a2) at (0,1);
\coordinate (b1) at (0,2.5);
\coordinate (b2) at (0,3.5);
\coordinate (c1) at (3,1.25);
\coordinate (c2) at (3,2.25);

\draw (a1) -- (a2);
\draw (b1) -- (b2);
\draw (c1) -- (c2);

%円弧
\draw (0,1) arc[start angle=-90, end angle=90, radius=0.75];

%上の曲線(b2)--(c2)
\draw (b2) -- ($(b2)+(0.5,0)$);
\draw ($(c2)-(0.5,0)$) -- (c2);
\draw ($(b2)+(0.5,0)$) .. controls ($(b2)+(1.25,0)$) and ($(c2)-(1.25,0)$) ..($(c2)-(0.5,0)$);

%下の曲線(a1)--(c1)
\draw (a1) -- ($(a1)+(0.5,0)$);
\draw ($(c1)-(0.5,0)$) -- (c1);
\draw ($(a1)+(0.5,0)$) .. controls ($(a1)+(1.25,0)$) and ($(c1)-(1.25,0)$) ..($(c1)-(0.5,0)$);
\node[left] at (0,6) {$c$};
\node[left] at (0,4.5) {$d$};
\node[left] at (0,3) {$e_{(1)}$};
\node[left] at (0,0.5) {$e_{(2)}$};
\end{tikzpicture}
    }{

\begin{tikzpicture}[scale=0.45, line width=2pt]
 \coordinate (a1) at (0,0);
\coordinate (a2) at (0,1);
\coordinate (b1) at (0,2.5);
\coordinate (b2) at (0,3.5);
\coordinate (c1) at (3,1.25);
\coordinate (c2) at (3,2.25);

\draw (a1) -- (a2);
\draw (b1) -- (b2);
\draw (c1) -- (c2);

%円弧
\draw (0,1) arc[start angle=-90, end angle=90, radius=0.75];

%上の曲線(b2)--(c2)
\draw (b2) -- ($(b2)+(0.5,0)$);
\draw ($(c2)-(0.5,0)$) -- (c2);
\draw ($(b2)+(0.5,0)$) .. controls ($(b2)+(1.25,0)$) and ($(c2)-(1.25,0)$) ..($(c2)-(0.5,0)$);

%下の曲線(a1)--(c1)
\draw (a1) -- ($(a1)+(0.5,0)$);
\draw ($(c1)-(0.5,0)$) -- (c1);
\draw ($(a1)+(0.5,0)$) .. controls ($(a1)+(1.25,0)$) and ($(c1)-(1.25,0)$) ..($(c1)-(0.5,0)$);
\node[left] at (0,3) {$a$};
\node[left] at (0,0.5) {$b$};
\end{tikzpicture}

\begin{tikzpicture}[scale=0.45, line width=2pt, xscale=-1]
 \coordinate (a1) at (0,0);
\coordinate (a2) at (0,1);
\coordinate (b1) at (0,2.5);
\coordinate (b2) at (0,3.5);
\coordinate (c1) at (3,1.25);
\coordinate (c2) at (3,2.25);

\draw (a1) -- (a2);
\draw (b1) -- (b2);
\draw (c1) -- (c2);

%円弧
\draw (0,1) arc[start angle=-90, end angle=90, radius=0.75];

%上の曲線(b2)--(c2)
\draw (b2) -- ($(b2)+(0.5,0)$);
\draw ($(c2)-(0.5,0)$) -- (c2);
\draw ($(b2)+(0.5,0)$) .. controls ($(b2)+(1.25,0)$) and ($(c2)-(1.25,0)$) ..($(c2)-(0.5,0)$);

%下の曲線(a1)--(c1)
\draw (a1) -- ($(a1)+(0.5,0)$);
\draw ($(c1)-(0.5,0)$) -- (c1);
\draw ($(a1)+(0.5,0)$) .. controls ($(a1)+(1.25,0)$) and ($(c1)-(1.25,0)$) ..($(c1)-(0.5,0)$);
\node[right] at (0,3) {$$};
\node[right] at (0,0.5) {$$};
\end{tikzpicture}

\begin{tikzpicture}[scale=0.45, line width=2pt, xscale=-1]
\begin{scope}[shift={(0,0.5)}]
    \coordinate (x1) at (0,-1);
\coordinate (x2) at (0,-2);
\coordinate (y1) at (3,-1);
\coordinate (y2) at (3,-2);
\draw (x1) -- (x2);
\draw (y1) -- (y2);
\draw (x1) -- (y1);
\draw (y2) -- (x2);

\end{scope}
 \coordinate (a1) at (0,0);
\coordinate (a2) at (0,1);
\coordinate (b1) at (0,2.5);
\coordinate (b2) at (0,3.5);
\coordinate (c1) at (3,1.25);
\coordinate (c2) at (3,2.25);

\draw (a1) -- (a2);
\draw (b1) -- (b2);
\draw (c1) -- (c2);

%円弧
\draw (0,1) arc[start angle=-90, end angle=90, radius=0.75];

%上の曲線(b2)--(c2)
\draw (b2) -- ($(b2)+(0.5,0)$);
\draw ($(c2)-(0.5,0)$) -- (c2);
\draw ($(b2)+(0.5,0)$) .. controls ($(b2)+(1.25,0)$) and ($(c2)-(1.25,0)$) ..($(c2)-(0.5,0)$);

%下の曲線(a1)--(c1)
\draw (a1) -- ($(a1)+(0.5,0)$);
\draw ($(c1)-(0.5,0)$) -- (c1);
\draw ($(a1)+(0.5,0)$) .. controls ($(a1)+(1.25,0)$) and ($(c1)-(1.25,0)$) ..($(c1)-(0.5,0)$);
\node[right] at (0,3) {$c$};
\node[right] at (0,0.5) {$d$};
\node[right] at (0,-1) {$e$};
\end{tikzpicture}

}{

\begin{tikzpicture}[scale=0.45, line width=2pt, xscale=-1]
\begin{scope}[shift={(0,0.5)}]
    \coordinate (x1) at (0,-1);
\coordinate (x2) at (0,-2);
\coordinate (y1) at (3,-1);
\coordinate (y2) at (3,-2);
\draw (x1) -- (x2);
\draw (y1) -- (y2);
\draw (x1) -- (y1);
\draw (y2) -- (x2);

\end{scope}
 \coordinate (a1) at (0,0);
\coordinate (a2) at (0,1);
\coordinate (b1) at (0,2.5);
\coordinate (b2) at (0,3.5);
\coordinate (c1) at (3,1.25);
\coordinate (c2) at (3,2.25);

\draw (a1) -- (a2);
\draw (b1) -- (b2);
\draw (c1) -- (c2);

%円弧
\draw (0,1) arc[start angle=-90, end angle=90, radius=0.75];

%上の曲線(b2)--(c2)
\draw (b2) -- ($(b2)+(0.5,0)$);
\draw ($(c2)-(0.5,0)$) -- (c2);
\draw ($(b2)+(0.5,0)$) .. controls ($(b2)+(1.25,0)$) and ($(c2)-(1.25,0)$) ..($(c2)-(0.5,0)$);

%下の曲線(a1)--(c1)
\draw (a1) -- ($(a1)+(0.5,0)$);
\draw ($(c1)-(0.5,0)$) -- (c1);
\draw ($(a1)+(0.5,0)$) .. controls ($(a1)+(1.25,0)$) and ($(c1)-(1.25,0)$) ..($(c1)-(0.5,0)$);
\node[right] at (0,3) {$a_{(1)}$};
\node[right] at (0,0.5) {$a_{(2)}$};
\node[right] at (0,-1) {$b$};
\end{tikzpicture}

\begin{tikzpicture}[scale=0.45, line width=2pt]
\begin{scope}[shift={(0,6)}]
    \coordinate (x1) at (0,-1);
\coordinate (x2) at (0,-2);
\coordinate (y1) at (3,-1);
\coordinate (y2) at (3,-2);
\draw (x1) -- (x2);
\draw (y1) -- (y2);
\draw (x1) -- (y1);
\draw (y2) -- (x2);

\end{scope}
 \coordinate (a1) at (0,0);
\coordinate (a2) at (0,1);
\coordinate (b1) at (0,2.5);
\coordinate (b2) at (0,3.5);
\coordinate (c1) at (3,1.25);
\coordinate (c2) at (3,2.25);

\draw (a1) -- (a2);
\draw (b1) -- (b2);
\draw (c1) -- (c2);

%円弧
\draw (0,1) arc[start angle=-90, end angle=90, radius=0.75];

%上の曲線(b2)--(c2)
\draw (b2) -- ($(b2)+(0.5,0)$);
\draw ($(c2)-(0.5,0)$) -- (c2);
\draw ($(b2)+(0.5,0)$) .. controls ($(b2)+(1.25,0)$) and ($(c2)-(1.25,0)$) ..($(c2)-(0.5,0)$);

%下の曲線(a1)--(c1)
\draw (a1) -- ($(a1)+(0.5,0)$);
\draw ($(c1)-(0.5,0)$) -- (c1);
\draw ($(a1)+(0.5,0)$) .. controls ($(a1)+(1.25,0)$) and ($(c1)-(1.25,0)$) ..($(c1)-(0.5,0)$);
\node[left] at (0,4.5) {$$};
\node[left] at (0,3) {$e_{(1)}$};
\node[left] at (0,0.5) {$e_{(2)}$};
\end{tikzpicture}
\shiftfig{0mm}{5mm}{
\begin{tikzpicture}[scale=0.45, line width=2pt, xscale=-1]
\begin{scope}[shift={(0,0.5)}]
    \coordinate (x1) at (0,-1);
\coordinate (x2) at (0,-2);
\coordinate (y1) at (3,-1);
\coordinate (y2) at (3,-2);
\draw (x1) -- (x2);
\draw (y1) -- (y2);
\draw (x1) -- (y1);
\draw (y2) -- (x2);

\end{scope}
 \coordinate (a1) at (0,0);
\coordinate (a2) at (0,1);
\coordinate (b1) at (0,2.5);
\coordinate (b2) at (0,3.5);
\coordinate (c1) at (3,1.25);
\coordinate (c2) at (3,2.25);

\draw (a1) -- (a2);
\draw (b1) -- (b2);
\draw (c1) -- (c2);

%円弧
\draw (0,1) arc[start angle=-90, end angle=90, radius=0.75];

%上の曲線(b2)--(c2)
\draw (b2) -- ($(b2)+(0.5,0)$);
\draw ($(c2)-(0.5,0)$) -- (c2);
\draw ($(b2)+(0.5,0)$) .. controls ($(b2)+(1.25,0)$) and ($(c2)-(1.25,0)$) ..($(c2)-(0.5,0)$);

%下の曲線(a1)--(c1)
\draw (a1) -- ($(a1)+(0.5,0)$);
\draw ($(c1)-(0.5,0)$) -- (c1);
\draw ($(a1)+(0.5,0)$) .. controls ($(a1)+(1.25,0)$) and ($(c1)-(1.25,0)$) ..($(c1)-(0.5,0)$);
\node[right] at (0,3) {$c$};
\node[right] at (0,0.5) {$d$};
\node[right] at (0,-1) {$$};
\end{tikzpicture}

}}{

\begin{tikzpicture}[scale=0.45, line width=2pt, xscale=-1]
\begin{scope}[shift={(0,0.5)}]
    \coordinate (x1) at (0,-1);
\coordinate (x2) at (0,-2);
\coordinate (y1) at (3,-1);
\coordinate (y2) at (3,-2);
\draw (x1) -- (x2);
\draw (y1) -- (y2);
\draw (x1) -- (y1);
\draw (y2) -- (x2);

\end{scope}
 \coordinate (a1) at (0,0);
\coordinate (a2) at (0,1);
\coordinate (b1) at (0,2.5);
\coordinate (b2) at (0,3.5);
\coordinate (c1) at (3,1.25);
\coordinate (c2) at (3,2.25);

\draw (a1) -- (a2);
\draw (b1) -- (b2);
\draw (c1) -- (c2);

%円弧
\draw (0,1) arc[start angle=-90, end angle=90, radius=0.75];

%上の曲線(b2)--(c2)
\draw (b2) -- ($(b2)+(0.5,0)$);
\draw ($(c2)-(0.5,0)$) -- (c2);
\draw ($(b2)+(0.5,0)$) .. controls ($(b2)+(1.25,0)$) and ($(c2)-(1.25,0)$) ..($(c2)-(0.5,0)$);

%下の曲線(a1)--(c1)
\draw (a1) -- ($(a1)+(0.5,0)$);
\draw ($(c1)-(0.5,0)$) -- (c1);
\draw ($(a1)+(0.5,0)$) .. controls ($(a1)+(1.25,0)$) and ($(c1)-(1.25,0)$) ..($(c1)-(0.5,0)$);
\node[right] at (0,3) {$a_{(1)}$};
\node[right] at (0,0.5) {$a_{(2)}$};
\node[right] at (0,-1) {$b$};
\end{tikzpicture}

\begin{tikzpicture}[scale=0.45, line width=2pt, xscale=-1]
\begin{scope}[shift={(0,0.5)}]
    \coordinate (x1) at (0,-1);
\coordinate (x2) at (0,-2);
\coordinate (y1) at (3,-1);
\coordinate (y2) at (3,-2);
\draw (x1) -- (x2);
\draw (y1) -- (y2);
\draw (x1) -- (y1);
\draw (y2) -- (x2);

\end{scope}
\begin{scope}[shift={(0,-1)}]
    \coordinate (x1) at (0,-1);
\coordinate (x2) at (0,-2);
\coordinate (y1) at (3,-1);
\coordinate (y2) at (3,-2);
\draw (x1) -- (x2);
\draw (y1) -- (y2);
\draw (x1) -- (y1);
\draw (y2) -- (x2);

\end{scope}
 \coordinate (a1) at (0,0);
\coordinate (a2) at (0,1);
\coordinate (b1) at (0,2.5);
\coordinate (b2) at (0,3.5);
\coordinate (c1) at (3,1.25);
\coordinate (c2) at (3,2.25);

\draw (a1) -- (a2);
\draw (b1) -- (b2);
\draw (c1) -- (c2);

%円弧
\draw (0,1) arc[start angle=-90, end angle=90, radius=0.75];

%上の曲線(b2)--(c2)
\draw (b2) -- ($(b2)+(0.5,0)$);
\draw ($(c2)-(0.5,0)$) -- (c2);
\draw ($(b2)+(0.5,0)$) .. controls ($(b2)+(1.25,0)$) and ($(c2)-(1.25,0)$) ..($(c2)-(0.5,0)$);

%下の曲線(a1)--(c1)
\draw (a1) -- ($(a1)+(0.5,0)$);
\draw ($(c1)-(0.5,0)$) -- (c1);
\draw ($(a1)+(0.5,0)$) .. controls ($(a1)+(1.25,0)$) and ($(c1)-(1.25,0)$) ..($(c1)-(0.5,0)$);
\node[right] at (0,3) {$c$};
\node[right] at (0,0.5) {$d$};
\node[right] at (0,-1) {$$};
\node[right] at (0,-2.5) {$$};
\end{tikzpicture}
\shiftfig{0mm}{-5mm}{
\begin{tikzpicture}[scale=0.45, line width=2pt]
\begin{scope}[shift={(0,7.5)}]
    \coordinate (x1) at (0,-1);
\coordinate (x2) at (0,-2);
\coordinate (y1) at (3,-1);
\coordinate (y2) at (3,-2);
\draw (x1) -- (x2);
\draw (y1) -- (y2);
\draw (x1) -- (y1);
\draw (y2) -- (x2);

\end{scope}
\begin{scope}[shift={(0,6)}]
    \coordinate (x1) at (0,-1);
\coordinate (x2) at (0,-2);
\coordinate (y1) at (3,-1);
\coordinate (y2) at (3,-2);
\draw (x1) -- (x2);
\draw (y1) -- (y2);
\draw (x1) -- (y1);
\draw (y2) -- (x2);

\end{scope}
 \coordinate (a1) at (0,0);
\coordinate (a2) at (0,1);
\coordinate (b1) at (0,2.5);
\coordinate (b2) at (0,3.5);
\coordinate (c1) at (3,1.25);
\coordinate (c2) at (3,2.25);

\draw (a1) -- (a2);
\draw (b1) -- (b2);
\draw (c1) -- (c2);

%円弧
\draw (0,1) arc[start angle=-90, end angle=90, radius=0.75];

%上の曲線(b2)--(c2)
\draw (b2) -- ($(b2)+(0.5,0)$);
\draw ($(c2)-(0.5,0)$) -- (c2);
\draw ($(b2)+(0.5,0)$) .. controls ($(b2)+(1.25,0)$) and ($(c2)-(1.25,0)$) ..($(c2)-(0.5,0)$);

%下の曲線(a1)--(c1)
\draw (a1) -- ($(a1)+(0.5,0)$);
\draw ($(c1)-(0.5,0)$) -- (c1);
\draw ($(a1)+(0.5,0)$) .. controls ($(a1)+(1.25,0)$) and ($(c1)-(1.25,0)$) ..($(c1)-(0.5,0)$);
\node[left] at (0,6) {$$};
\node[left] at (0,4.5) {$$};
\node[left] at (0,3) {$e_{(1)}$};
\node[left] at (0,0.5) {$e_{(2)}$};
\end{tikzpicture}
}
}
}
\]
\fi

\iflatexml
\begin{center}
  \includegraphics{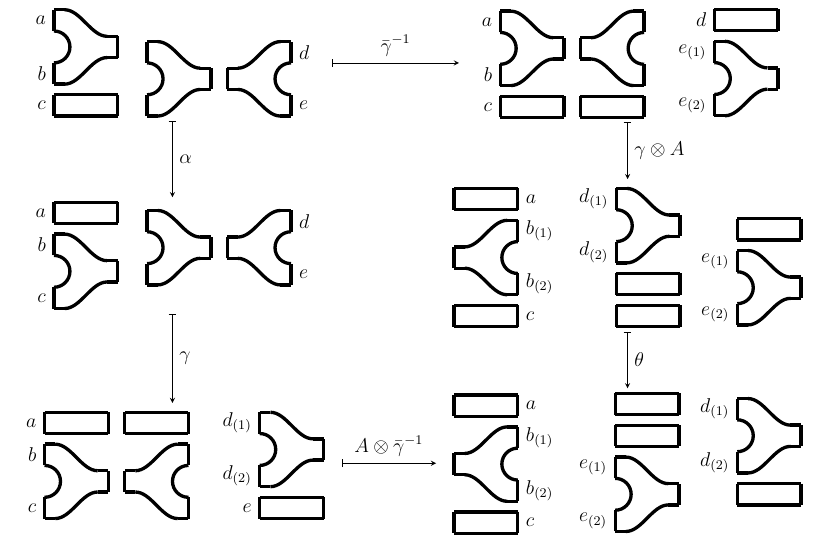}
\end{center}
\else
\[\scalebox{0.8}{%
  \DiagramRectangle[
    AB=\alpha,
    BC=\gamma,
    AD=\brg^{-1},
    DE=\gamma\otimes A,
    EF=\theta,
    CF=A\otimes \brg^{-1}
  ]
    {
    
\begin{tikzpicture}[scale=0.45, line width=2pt]
\begin{scope}[shift={(0,0.5)}]
    \coordinate (x1) at (0,-1);
\coordinate (x2) at (0,-2);
\coordinate (y1) at (3,-1);
\coordinate (y2) at (3,-2);
\draw (x1) -- (x2);
\draw (y1) -- (y2);
\draw (x1) -- (y1);
\draw (y2) -- (x2);

\end{scope}
 \coordinate (a1) at (0,0);
\coordinate (a2) at (0,1);
\coordinate (b1) at (0,2.5);
\coordinate (b2) at (0,3.5);
\coordinate (c1) at (3,1.25);
\coordinate (c2) at (3,2.25);

\draw (a1) -- (a2);
\draw (b1) -- (b2);
\draw (c1) -- (c2);

%円弧
\draw (0,1) arc[start angle=-90, end angle=90, radius=0.75];

%上の曲線(b2)--(c2)
\draw (b2) -- ($(b2)+(0.5,0)$);
\draw ($(c2)-(0.5,0)$) -- (c2);
\draw ($(b2)+(0.5,0)$) .. controls ($(b2)+(1.25,0)$) and ($(c2)-(1.25,0)$) ..($(c2)-(0.5,0)$);

%下の曲線(a1)--(c1)
\draw (a1) -- ($(a1)+(0.5,0)$);
\draw ($(c1)-(0.5,0)$) -- (c1);
\draw ($(a1)+(0.5,0)$) .. controls ($(a1)+(1.25,0)$) and ($(c1)-(1.25,0)$) ..($(c1)-(0.5,0)$);
\node[left] at (0,3) {$a$};
\node[left] at (0,0.5) {$b$};
\node[left] at (0,-1) {$c$};
\end{tikzpicture}

\begin{tikzpicture}[scale=0.45, line width=2pt]
 \coordinate (a1) at (0,0);
\coordinate (a2) at (0,1);
\coordinate (b1) at (0,2.5);
\coordinate (b2) at (0,3.5);
\coordinate (c1) at (3,1.25);
\coordinate (c2) at (3,2.25);

\draw (a1) -- (a2);
\draw (b1) -- (b2);
\draw (c1) -- (c2);

%円弧
\draw (0,1) arc[start angle=-90, end angle=90, radius=0.75];

%上の曲線(b2)--(c2)
\draw (b2) -- ($(b2)+(0.5,0)$);
\draw ($(c2)-(0.5,0)$) -- (c2);
\draw ($(b2)+(0.5,0)$) .. controls ($(b2)+(1.25,0)$) and ($(c2)-(1.25,0)$) ..($(c2)-(0.5,0)$);

%下の曲線(a1)--(c1)
\draw (a1) -- ($(a1)+(0.5,0)$);
\draw ($(c1)-(0.5,0)$) -- (c1);
\draw ($(a1)+(0.5,0)$) .. controls ($(a1)+(1.25,0)$) and ($(c1)-(1.25,0)$) ..($(c1)-(0.5,0)$);
\node[left] at (0,3) {$$};
\node[left] at (0,0.5) {$$};
\end{tikzpicture}

\begin{tikzpicture}[scale=0.45, line width=2pt, xscale=-1]
 \coordinate (a1) at (0,0);
\coordinate (a2) at (0,1);
\coordinate (b1) at (0,2.5);
\coordinate (b2) at (0,3.5);
\coordinate (c1) at (3,1.25);
\coordinate (c2) at (3,2.25);

\draw (a1) -- (a2);
\draw (b1) -- (b2);
\draw (c1) -- (c2);

%円弧
\draw (0,1) arc[start angle=-90, end angle=90, radius=0.75];

%上の曲線(b2)--(c2)
\draw (b2) -- ($(b2)+(0.5,0)$);
\draw ($(c2)-(0.5,0)$) -- (c2);
\draw ($(b2)+(0.5,0)$) .. controls ($(b2)+(1.25,0)$) and ($(c2)-(1.25,0)$) ..($(c2)-(0.5,0)$);

%下の曲線(a1)--(c1)
\draw (a1) -- ($(a1)+(0.5,0)$);
\draw ($(c1)-(0.5,0)$) -- (c1);
\draw ($(a1)+(0.5,0)$) .. controls ($(a1)+(1.25,0)$) and ($(c1)-(1.25,0)$) ..($(c1)-(0.5,0)$);
\node[right] at (0,3) {$d$};
\node[right] at (0,0.5) {$e$};
\end{tikzpicture}

    }{
    
\begin{tikzpicture}[scale=0.45, line width=2pt]
\begin{scope}[shift={(0,6)}]
    \coordinate (x1) at (0,-1);
\coordinate (x2) at (0,-2);
\coordinate (y1) at (3,-1);
\coordinate (y2) at (3,-2);
\draw (x1) -- (x2);
\draw (y1) -- (y2);
\draw (x1) -- (y1);
\draw (y2) -- (x2);

\end{scope}
 \coordinate (a1) at (0,0);
\coordinate (a2) at (0,1);
\coordinate (b1) at (0,2.5);
\coordinate (b2) at (0,3.5);
\coordinate (c1) at (3,1.25);
\coordinate (c2) at (3,2.25);

\draw (a1) -- (a2);
\draw (b1) -- (b2);
\draw (c1) -- (c2);

%円弧
\draw (0,1) arc[start angle=-90, end angle=90, radius=0.75];

%上の曲線(b2)--(c2)
\draw (b2) -- ($(b2)+(0.5,0)$);
\draw ($(c2)-(0.5,0)$) -- (c2);
\draw ($(b2)+(0.5,0)$) .. controls ($(b2)+(1.25,0)$) and ($(c2)-(1.25,0)$) ..($(c2)-(0.5,0)$);

%下の曲線(a1)--(c1)
\draw (a1) -- ($(a1)+(0.5,0)$);
\draw ($(c1)-(0.5,0)$) -- (c1);
\draw ($(a1)+(0.5,0)$) .. controls ($(a1)+(1.25,0)$) and ($(c1)-(1.25,0)$) ..($(c1)-(0.5,0)$);
\node[left] at (0,4.5) {$a$};
\node[left] at (0,3) {$b$};
\node[left] at (0,0.5) {$c$};
\end{tikzpicture}
\shiftfig{0mm}{5mm}{
\begin{tikzpicture}[scale=0.45, line width=2pt]
 \coordinate (a1) at (0,0);
\coordinate (a2) at (0,1);
\coordinate (b1) at (0,2.5);
\coordinate (b2) at (0,3.5);
\coordinate (c1) at (3,1.25);
\coordinate (c2) at (3,2.25);

\draw (a1) -- (a2);
\draw (b1) -- (b2);
\draw (c1) -- (c2);

%円弧
\draw (0,1) arc[start angle=-90, end angle=90, radius=0.75];

%上の曲線(b2)--(c2)
\draw (b2) -- ($(b2)+(0.5,0)$);
\draw ($(c2)-(0.5,0)$) -- (c2);
\draw ($(b2)+(0.5,0)$) .. controls ($(b2)+(1.25,0)$) and ($(c2)-(1.25,0)$) ..($(c2)-(0.5,0)$);

%下の曲線(a1)--(c1)
\draw (a1) -- ($(a1)+(0.5,0)$);
\draw ($(c1)-(0.5,0)$) -- (c1);
\draw ($(a1)+(0.5,0)$) .. controls ($(a1)+(1.25,0)$) and ($(c1)-(1.25,0)$) ..($(c1)-(0.5,0)$);
\node[left] at (0,3) {$$};
\node[left] at (0,0.5) {$$};
\end{tikzpicture}

\begin{tikzpicture}[scale=0.45, line width=2pt, xscale=-1]
 \coordinate (a1) at (0,0);
\coordinate (a2) at (0,1);
\coordinate (b1) at (0,2.5);
\coordinate (b2) at (0,3.5);
\coordinate (c1) at (3,1.25);
\coordinate (c2) at (3,2.25);

\draw (a1) -- (a2);
\draw (b1) -- (b2);
\draw (c1) -- (c2);

%円弧
\draw (0,1) arc[start angle=-90, end angle=90, radius=0.75];

%上の曲線(b2)--(c2)
\draw (b2) -- ($(b2)+(0.5,0)$);
\draw ($(c2)-(0.5,0)$) -- (c2);
\draw ($(b2)+(0.5,0)$) .. controls ($(b2)+(1.25,0)$) and ($(c2)-(1.25,0)$) ..($(c2)-(0.5,0)$);

%下の曲線(a1)--(c1)
\draw (a1) -- ($(a1)+(0.5,0)$);
\draw ($(c1)-(0.5,0)$) -- (c1);
\draw ($(a1)+(0.5,0)$) .. controls ($(a1)+(1.25,0)$) and ($(c1)-(1.25,0)$) ..($(c1)-(0.5,0)$);
\node[right] at (0,3) {$d$};
\node[right] at (0,0.5) {$e$};
\end{tikzpicture}

    }}{
    
\begin{tikzpicture}[scale=0.45, line width=2pt]
\begin{scope}[shift={(0,6)}]
    \coordinate (x1) at (0,-1);
\coordinate (x2) at (0,-2);
\coordinate (y1) at (3,-1);
\coordinate (y2) at (3,-2);
\draw (x1) -- (x2);
\draw (y1) -- (y2);
\draw (x1) -- (y1);
\draw (y2) -- (x2);

\end{scope}
 \coordinate (a1) at (0,0);
\coordinate (a2) at (0,1);
\coordinate (b1) at (0,2.5);
\coordinate (b2) at (0,3.5);
\coordinate (c1) at (3,1.25);
\coordinate (c2) at (3,2.25);

\draw (a1) -- (a2);
\draw (b1) -- (b2);
\draw (c1) -- (c2);

%円弧
\draw (0,1) arc[start angle=-90, end angle=90, radius=0.75];

%上の曲線(b2)--(c2)
\draw (b2) -- ($(b2)+(0.5,0)$);
\draw ($(c2)-(0.5,0)$) -- (c2);
\draw ($(b2)+(0.5,0)$) .. controls ($(b2)+(1.25,0)$) and ($(c2)-(1.25,0)$) ..($(c2)-(0.5,0)$);

%下の曲線(a1)--(c1)
\draw (a1) -- ($(a1)+(0.5,0)$);
\draw ($(c1)-(0.5,0)$) -- (c1);
\draw ($(a1)+(0.5,0)$) .. controls ($(a1)+(1.25,0)$) and ($(c1)-(1.25,0)$) ..($(c1)-(0.5,0)$);
\node[left] at (0,4.5) {$a$};
\node[left] at (0,3) {$b$};
\node[left] at (0,0.5) {$c$};
\end{tikzpicture}

\begin{tikzpicture}[scale=0.45, line width=2pt, xscale=-1]
\begin{scope}[shift={(0,6)}]
    \coordinate (x1) at (0,-1);
\coordinate (x2) at (0,-2);
\coordinate (y1) at (3,-1);
\coordinate (y2) at (3,-2);
\draw (x1) -- (x2);
\draw (y1) -- (y2);
\draw (x1) -- (y1);
\draw (y2) -- (x2);

\end{scope}
 \coordinate (a1) at (0,0);
\coordinate (a2) at (0,1);
\coordinate (b1) at (0,2.5);
\coordinate (b2) at (0,3.5);
\coordinate (c1) at (3,1.25);
\coordinate (c2) at (3,2.25);

\draw (a1) -- (a2);
\draw (b1) -- (b2);
\draw (c1) -- (c2);

%円弧
\draw (0,1) arc[start angle=-90, end angle=90, radius=0.75];

%上の曲線(b2)--(c2)
\draw (b2) -- ($(b2)+(0.5,0)$);
\draw ($(c2)-(0.5,0)$) -- (c2);
\draw ($(b2)+(0.5,0)$) .. controls ($(b2)+(1.25,0)$) and ($(c2)-(1.25,0)$) ..($(c2)-(0.5,0)$);

%下の曲線(a1)--(c1)
\draw (a1) -- ($(a1)+(0.5,0)$);
\draw ($(c1)-(0.5,0)$) -- (c1);
\draw ($(a1)+(0.5,0)$) .. controls ($(a1)+(1.25,0)$) and ($(c1)-(1.25,0)$) ..($(c1)-(0.5,0)$);
\node[right] at (0,4.5) {$$};
\node[right] at (0,3) {$$};
\node[right] at (0,0.5) {$$};
\end{tikzpicture}

\begin{tikzpicture}[scale=0.45, line width=2pt]
\begin{scope}[shift={(0,0.5)}]
    \coordinate (x1) at (0,-1);
\coordinate (x2) at (0,-2);
\coordinate (y1) at (3,-1);
\coordinate (y2) at (3,-2);
\draw (x1) -- (x2);
\draw (y1) -- (y2);
\draw (x1) -- (y1);
\draw (y2) -- (x2);

\end{scope}
 \coordinate (a1) at (0,0);
\coordinate (a2) at (0,1);
\coordinate (b1) at (0,2.5);
\coordinate (b2) at (0,3.5);
\coordinate (c1) at (3,1.25);
\coordinate (c2) at (3,2.25);

\draw (a1) -- (a2);
\draw (b1) -- (b2);
\draw (c1) -- (c2);

%円弧
\draw (0,1) arc[start angle=-90, end angle=90, radius=0.75];

%上の曲線(b2)--(c2)
\draw (b2) -- ($(b2)+(0.5,0)$);
\draw ($(c2)-(0.5,0)$) -- (c2);
\draw ($(b2)+(0.5,0)$) .. controls ($(b2)+(1.25,0)$) and ($(c2)-(1.25,0)$) ..($(c2)-(0.5,0)$);

%下の曲線(a1)--(c1)
\draw (a1) -- ($(a1)+(0.5,0)$);
\draw ($(c1)-(0.5,0)$) -- (c1);
\draw ($(a1)+(0.5,0)$) .. controls ($(a1)+(1.25,0)$) and ($(c1)-(1.25,0)$) ..($(c1)-(0.5,0)$);
\node[left] at (0,3) {$d_{(1)}$};
\node[left] at (0,0.5) {$d_{(2)}$};
\node[left] at (0,-1) {$e$};
\end{tikzpicture}

    }{
    
\begin{tikzpicture}[scale=0.45, line width=2pt]
\begin{scope}[shift={(0,0.5)}]
    \coordinate (x1) at (0,-1);
\coordinate (x2) at (0,-2);
\coordinate (y1) at (3,-1);
\coordinate (y2) at (3,-2);
\draw (x1) -- (x2);
\draw (y1) -- (y2);
\draw (x1) -- (y1);
\draw (y2) -- (x2);

\end{scope}
 \coordinate (a1) at (0,0);
\coordinate (a2) at (0,1);
\coordinate (b1) at (0,2.5);
\coordinate (b2) at (0,3.5);
\coordinate (c1) at (3,1.25);
\coordinate (c2) at (3,2.25);

\draw (a1) -- (a2);
\draw (b1) -- (b2);
\draw (c1) -- (c2);

%円弧
\draw (0,1) arc[start angle=-90, end angle=90, radius=0.75];

%上の曲線(b2)--(c2)
\draw (b2) -- ($(b2)+(0.5,0)$);
\draw ($(c2)-(0.5,0)$) -- (c2);
\draw ($(b2)+(0.5,0)$) .. controls ($(b2)+(1.25,0)$) and ($(c2)-(1.25,0)$) ..($(c2)-(0.5,0)$);

%下の曲線(a1)--(c1)
\draw (a1) -- ($(a1)+(0.5,0)$);
\draw ($(c1)-(0.5,0)$) -- (c1);
\draw ($(a1)+(0.5,0)$) .. controls ($(a1)+(1.25,0)$) and ($(c1)-(1.25,0)$) ..($(c1)-(0.5,0)$);
\node[left] at (0,3) {$a$};
\node[left] at (0,0.5) {$b$};
\node[left] at (0,-1) {$c$};
\end{tikzpicture}

\begin{tikzpicture}[scale=0.45, line width=2pt, xscale=-1]
\begin{scope}[shift={(0,0.5)}]
    \coordinate (x1) at (0,-1);
\coordinate (x2) at (0,-2);
\coordinate (y1) at (3,-1);
\coordinate (y2) at (3,-2);
\draw (x1) -- (x2);
\draw (y1) -- (y2);
\draw (x1) -- (y1);
\draw (y2) -- (x2);

\end{scope}
 \coordinate (a1) at (0,0);
\coordinate (a2) at (0,1);
\coordinate (b1) at (0,2.5);
\coordinate (b2) at (0,3.5);
\coordinate (c1) at (3,1.25);
\coordinate (c2) at (3,2.25);

\draw (a1) -- (a2);
\draw (b1) -- (b2);
\draw (c1) -- (c2);

%円弧
\draw (0,1) arc[start angle=-90, end angle=90, radius=0.75];

%上の曲線(b2)--(c2)
\draw (b2) -- ($(b2)+(0.5,0)$);
\draw ($(c2)-(0.5,0)$) -- (c2);
\draw ($(b2)+(0.5,0)$) .. controls ($(b2)+(1.25,0)$) and ($(c2)-(1.25,0)$) ..($(c2)-(0.5,0)$);

%下の曲線(a1)--(c1)
\draw (a1) -- ($(a1)+(0.5,0)$);
\draw ($(c1)-(0.5,0)$) -- (c1);
\draw ($(a1)+(0.5,0)$) .. controls ($(a1)+(1.25,0)$) and ($(c1)-(1.25,0)$) ..($(c1)-(0.5,0)$);
\node[right] at (0,3) {$$};
\node[right] at (0,0.5) {$$};
\node[right] at (0,-1) {$$};
\end{tikzpicture}

\begin{tikzpicture}[scale=0.45, line width=2pt]
\begin{scope}[shift={(0,6)}]
    \coordinate (x1) at (0,-1);
\coordinate (x2) at (0,-2);
\coordinate (y1) at (3,-1);
\coordinate (y2) at (3,-2);
\draw (x1) -- (x2);
\draw (y1) -- (y2);
\draw (x1) -- (y1);
\draw (y2) -- (x2);

\end{scope}
 \coordinate (a1) at (0,0);
\coordinate (a2) at (0,1);
\coordinate (b1) at (0,2.5);
\coordinate (b2) at (0,3.5);
\coordinate (c1) at (3,1.25);
\coordinate (c2) at (3,2.25);

\draw (a1) -- (a2);
\draw (b1) -- (b2);
\draw (c1) -- (c2);

%円弧
\draw (0,1) arc[start angle=-90, end angle=90, radius=0.75];

%上の曲線(b2)--(c2)
\draw (b2) -- ($(b2)+(0.5,0)$);
\draw ($(c2)-(0.5,0)$) -- (c2);
\draw ($(b2)+(0.5,0)$) .. controls ($(b2)+(1.25,0)$) and ($(c2)-(1.25,0)$) ..($(c2)-(0.5,0)$);

%下の曲線(a1)--(c1)
\draw (a1) -- ($(a1)+(0.5,0)$);
\draw ($(c1)-(0.5,0)$) -- (c1);
\draw ($(a1)+(0.5,0)$) .. controls ($(a1)+(1.25,0)$) and ($(c1)-(1.25,0)$) ..($(c1)-(0.5,0)$);
\node[left] at (0,4.5) {$d$};
\node[left] at (0,3) {$e_{(1)}$};
\node[left] at (0,0.5) {$e_{(2)}$};
\end{tikzpicture}

    }{
    
\begin{tikzpicture}[scale=0.45, line width=2pt, xscale=-1]
\begin{scope}[shift={(0,6)}]
    \coordinate (x1) at (0,-1);
\coordinate (x2) at (0,-2);
\coordinate (y1) at (3,-1);
\coordinate (y2) at (3,-2);
\draw (x1) -- (x2);
\draw (y1) -- (y2);
\draw (x1) -- (y1);
\draw (y2) -- (x2);

\end{scope}
\begin{scope}[shift={(0,0.5)}]
    \coordinate (x1) at (0,-1);
\coordinate (x2) at (0,-2);
\coordinate (y1) at (3,-1);
\coordinate (y2) at (3,-2);
\draw (x1) -- (x2);
\draw (y1) -- (y2);
\draw (x1) -- (y1);
\draw (y2) -- (x2);

\end{scope}
 \coordinate (a1) at (0,0);
\coordinate (a2) at (0,1);
\coordinate (b1) at (0,2.5);
\coordinate (b2) at (0,3.5);
\coordinate (c1) at (3,1.25);
\coordinate (c2) at (3,2.25);

\draw (a1) -- (a2);
\draw (b1) -- (b2);
\draw (c1) -- (c2);

%円弧
\draw (0,1) arc[start angle=-90, end angle=90, radius=0.75];

%上の曲線(b2)--(c2)
\draw (b2) -- ($(b2)+(0.5,0)$);
\draw ($(c2)-(0.5,0)$) -- (c2);
\draw ($(b2)+(0.5,0)$) .. controls ($(b2)+(1.25,0)$) and ($(c2)-(1.25,0)$) ..($(c2)-(0.5,0)$);

%下の曲線(a1)--(c1)
\draw (a1) -- ($(a1)+(0.5,0)$);
\draw ($(c1)-(0.5,0)$) -- (c1);
\draw ($(a1)+(0.5,0)$) .. controls ($(a1)+(1.25,0)$) and ($(c1)-(1.25,0)$) ..($(c1)-(0.5,0)$);
\node[right] at (0,4.5) {$a$};
\node[right] at (0,3) {$b_{(1)}$};
\node[right] at (0,0.5) {$b_{(2)}$};
\node[right] at (0,-1) {$c$};
\end{tikzpicture}

\begin{tikzpicture}[scale=0.45, line width=2pt]
\begin{scope}[shift={(0,0.5)}]
    \coordinate (x1) at (0,-1);
\coordinate (x2) at (0,-2);
\coordinate (y1) at (3,-1);
\coordinate (y2) at (3,-2);
\draw (x1) -- (x2);
\draw (y1) -- (y2);
\draw (x1) -- (y1);
\draw (y2) -- (x2);

\end{scope}
\begin{scope}[shift={(0,-1)}]
    \coordinate (x1) at (0,-1);
\coordinate (x2) at (0,-2);
\coordinate (y1) at (3,-1);
\coordinate (y2) at (3,-2);
\draw (x1) -- (x2);
\draw (y1) -- (y2);
\draw (x1) -- (y1);
\draw (y2) -- (x2);

\end{scope}
 \coordinate (a1) at (0,0);
\coordinate (a2) at (0,1);
\coordinate (b1) at (0,2.5);
\coordinate (b2) at (0,3.5);
\coordinate (c1) at (3,1.25);
\coordinate (c2) at (3,2.25);

\draw (a1) -- (a2);
\draw (b1) -- (b2);
\draw (c1) -- (c2);

%円弧
\draw (0,1) arc[start angle=-90, end angle=90, radius=0.75];

%上の曲線(b2)--(c2)
\draw (b2) -- ($(b2)+(0.5,0)$);
\draw ($(c2)-(0.5,0)$) -- (c2);
\draw ($(b2)+(0.5,0)$) .. controls ($(b2)+(1.25,0)$) and ($(c2)-(1.25,0)$) ..($(c2)-(0.5,0)$);

%下の曲線(a1)--(c1)
\draw (a1) -- ($(a1)+(0.5,0)$);
\draw ($(c1)-(0.5,0)$) -- (c1);
\draw ($(a1)+(0.5,0)$) .. controls ($(a1)+(1.25,0)$) and ($(c1)-(1.25,0)$) ..($(c1)-(0.5,0)$);
\node[left] at (0,3) {$d_{(1)}$};
\node[left] at (0,0.5) {$d_{(2)}$};
\node[left] at (0,-1) {$$};
\node[left] at (0,-2.5) {$$};
\end{tikzpicture}

\begin{tikzpicture}[scale=0.45, line width=2pt]
\begin{scope}[shift={(0,6)}]
    \coordinate (x1) at (0,-1);
\coordinate (x2) at (0,-2);
\coordinate (y1) at (3,-1);
\coordinate (y2) at (3,-2);
\draw (x1) -- (x2);
\draw (y1) -- (y2);
\draw (x1) -- (y1);
\draw (y2) -- (x2);

\end{scope}
 \coordinate (a1) at (0,0);
\coordinate (a2) at (0,1);
\coordinate (b1) at (0,2.5);
\coordinate (b2) at (0,3.5);
\coordinate (c1) at (3,1.25);
\coordinate (c2) at (3,2.25);

\draw (a1) -- (a2);
\draw (b1) -- (b2);
\draw (c1) -- (c2);

%円弧
\draw (0,1) arc[start angle=-90, end angle=90, radius=0.75];

%上の曲線(b2)--(c2)
\draw (b2) -- ($(b2)+(0.5,0)$);
\draw ($(c2)-(0.5,0)$) -- (c2);
\draw ($(b2)+(0.5,0)$) .. controls ($(b2)+(1.25,0)$) and ($(c2)-(1.25,0)$) ..($(c2)-(0.5,0)$);

%下の曲線(a1)--(c1)
\draw (a1) -- ($(a1)+(0.5,0)$);
\draw ($(c1)-(0.5,0)$) -- (c1);
\draw ($(a1)+(0.5,0)$) .. controls ($(a1)+(1.25,0)$) and ($(c1)-(1.25,0)$) ..($(c1)-(0.5,0)$);
\node[left] at (0,4.5) {$$};
\node[left] at (0,3) {$e_{(1)}$};
\node[left] at (0,0.5) {$e_{(2)}$};
\end{tikzpicture}

    }{
    
\begin{tikzpicture}[scale=0.45, line width=2pt, xscale=-1]
\begin{scope}[shift={(0,6)}]
    \coordinate (x1) at (0,-1);
\coordinate (x2) at (0,-2);
\coordinate (y1) at (3,-1);
\coordinate (y2) at (3,-2);
\draw (x1) -- (x2);
\draw (y1) -- (y2);
\draw (x1) -- (y1);
\draw (y2) -- (x2);

\end{scope}
\begin{scope}[shift={(0,0.5)}]
    \coordinate (x1) at (0,-1);
\coordinate (x2) at (0,-2);
\coordinate (y1) at (3,-1);
\coordinate (y2) at (3,-2);
\draw (x1) -- (x2);
\draw (y1) -- (y2);
\draw (x1) -- (y1);
\draw (y2) -- (x2);

\end{scope}
 \coordinate (a1) at (0,0);
\coordinate (a2) at (0,1);
\coordinate (b1) at (0,2.5);
\coordinate (b2) at (0,3.5);
\coordinate (c1) at (3,1.25);
\coordinate (c2) at (3,2.25);

\draw (a1) -- (a2);
\draw (b1) -- (b2);
\draw (c1) -- (c2);

%円弧
\draw (0,1) arc[start angle=-90, end angle=90, radius=0.75];

%上の曲線(b2)--(c2)
\draw (b2) -- ($(b2)+(0.5,0)$);
\draw ($(c2)-(0.5,0)$) -- (c2);
\draw ($(b2)+(0.5,0)$) .. controls ($(b2)+(1.25,0)$) and ($(c2)-(1.25,0)$) ..($(c2)-(0.5,0)$);

%下の曲線(a1)--(c1)
\draw (a1) -- ($(a1)+(0.5,0)$);
\draw ($(c1)-(0.5,0)$) -- (c1);
\draw ($(a1)+(0.5,0)$) .. controls ($(a1)+(1.25,0)$) and ($(c1)-(1.25,0)$) ..($(c1)-(0.5,0)$);
\node[right] at (0,4.5) {$a$};
\node[right] at (0,3) {$b_{(1)}$};
\node[right] at (0,0.5) {$b_{(2)}$};
\node[right] at (0,-1) {$c$};
\end{tikzpicture}

\begin{tikzpicture}[scale=0.45, line width=2pt]
\begin{scope}[shift={(0,7.5)}]
    \coordinate (x1) at (0,-1);
\coordinate (x2) at (0,-2);
\coordinate (y1) at (3,-1);
\coordinate (y2) at (3,-2);
\draw (x1) -- (x2);
\draw (y1) -- (y2);
\draw (x1) -- (y1);
\draw (y2) -- (x2);

\end{scope}
\begin{scope}[shift={(0,6)}]
    \coordinate (x1) at (0,-1);
\coordinate (x2) at (0,-2);
\coordinate (y1) at (3,-1);
\coordinate (y2) at (3,-2);
\draw (x1) -- (x2);
\draw (y1) -- (y2);
\draw (x1) -- (y1);
\draw (y2) -- (x2);

\end{scope}
 \coordinate (a1) at (0,0);
\coordinate (a2) at (0,1);
\coordinate (b1) at (0,2.5);
\coordinate (b2) at (0,3.5);
\coordinate (c1) at (3,1.25);
\coordinate (c2) at (3,2.25);

\draw (a1) -- (a2);
\draw (b1) -- (b2);
\draw (c1) -- (c2);

%円弧
\draw (0,1) arc[start angle=-90, end angle=90, radius=0.75];

%上の曲線(b2)--(c2)
\draw (b2) -- ($(b2)+(0.5,0)$);
\draw ($(c2)-(0.5,0)$) -- (c2);
\draw ($(b2)+(0.5,0)$) .. controls ($(b2)+(1.25,0)$) and ($(c2)-(1.25,0)$) ..($(c2)-(0.5,0)$);

%下の曲線(a1)--(c1)
\draw (a1) -- ($(a1)+(0.5,0)$);
\draw ($(c1)-(0.5,0)$) -- (c1);
\draw ($(a1)+(0.5,0)$) .. controls ($(a1)+(1.25,0)$) and ($(c1)-(1.25,0)$) ..($(c1)-(0.5,0)$);
\node[left] at (0,6) {$$};
\node[left] at (0,4.5) {$$};
\node[left] at (0,3) {$e_{(1)}$};
\node[left] at (0,0.5) {$e_{(2)}$};
\end{tikzpicture}
\shiftfig{0mm}{6mm}{
\begin{tikzpicture}[scale=0.45, line width=2pt]
\begin{scope}[shift={(0,0.5)}]
    \coordinate (x1) at (0,-1);
\coordinate (x2) at (0,-2);
\coordinate (y1) at (3,-1);
\coordinate (y2) at (3,-2);
\draw (x1) -- (x2);
\draw (y1) -- (y2);
\draw (x1) -- (y1);
\draw (y2) -- (x2);

\end{scope}
 \coordinate (a1) at (0,0);
\coordinate (a2) at (0,1);
\coordinate (b1) at (0,2.5);
\coordinate (b2) at (0,3.5);
\coordinate (c1) at (3,1.25);
\coordinate (c2) at (3,2.25);

\draw (a1) -- (a2);
\draw (b1) -- (b2);
\draw (c1) -- (c2);

%円弧
\draw (0,1) arc[start angle=-90, end angle=90, radius=0.75];

%上の曲線(b2)--(c2)
\draw (b2) -- ($(b2)+(0.5,0)$);
\draw ($(c2)-(0.5,0)$) -- (c2);
\draw ($(b2)+(0.5,0)$) .. controls ($(b2)+(1.25,0)$) and ($(c2)-(1.25,0)$) ..($(c2)-(0.5,0)$);

%下の曲線(a1)--(c1)
\draw (a1) -- ($(a1)+(0.5,0)$);
\draw ($(c1)-(0.5,0)$) -- (c1);
\draw ($(a1)+(0.5,0)$) .. controls ($(a1)+(1.25,0)$) and ($(c1)-(1.25,0)$) ..($(c1)-(0.5,0)$);
\node[left] at (0,3) {$d_{(1)}$};
\node[left] at (0,0.5) {$d_{(2)}$};
\node[left] at (0,-1) {$$};
\end{tikzpicture}
}
    }
    }
\]
\fi

\end{proof}

\section{Heisenberg doubles}\label{sec:hsnbrg}
\subsection{Canonical elements and pentagon relations}
In this subsection, we mainly follow \cite{K, Sz}. Let $H=(H,m,\D, 1, \varepsilon, S)$ be a finite-dimensional Hopf algebra over a field $\mathbb{K}$ with invertible antipode. 

We use the left action of $H$ on $H^*$ defined by $(a\rightharpoonup f)(x)=f(xa)$, for $a,x\in H$ and $f\in H^*$. The \emph{Heisenberg double} 
$$\cH(H)=H^*\otimes H$$
of $H$ is a $\mathbb{K}$-algebra with unit $\varepsilon\otimes 1$ and multiplication given by
$$(f\otimes a)(g\otimes b)=f(a_{(1)}\rightharpoonup g)\otimes a_{(2)}b,$$
for $f\otimes a, g\otimes b\in \cH(H)$.

Let $\{e_i\}$ be a basis of $H$ and $\{e^i\}$ its dual basis. We set $\tilde{e}_i =S(e_i)$ and $\tilde{e}^i =(S^{-1})^*(e^i)$. The \emph{canonical elements} are given by
\begin{align}
    T&=\sum_i (\varepsilon\otimes e_i)\otimes (e^i\otimes 1)\in \cH(H)\otimes\cH(H), \label{can} \\
    \tilde{T}&=\sum_i (\varepsilon\otimes \tilde{e}_i)\otimes (\tilde{e}^i\otimes 1)\in \cH(H)^{op}\otimes\cH(H)^{op}, \\
    T^{\prime}&=\sum_i (\varepsilon\otimes \tilde{e}_i)\otimes (e^i\otimes 1)\in \cH(H)^{op}\otimes\cH(H), \\
    T^{\prime\prime}&=\sum_i (\varepsilon\otimes e_i)\otimes (\tilde{e}^i\otimes 1)\in \cH(H)\otimes\cH(H)^{op}. \label{can'}
\end{align}
Their inverses are given by
\begin{align*}
    T^{-1}&=\sum_i (\varepsilon\otimes S(e_i))\otimes (e^i\otimes 1)\in \cH(H)\otimes\cH(H), \\
    \tilde{T}^{-1}&=\sum_i (\varepsilon\otimes S(\tilde{e}_i))\otimes (\tilde{e}^i\otimes 1)\in \cH(H)^{op}\otimes\cH(H)^{op}, \\
    (T^{\prime})^{-1}&=\sum_i (\varepsilon\otimes S(\tilde{e}_i))\otimes (e^i\otimes 1)\in \cH(H)^{op}\otimes\cH(H), \\
    (T^{\prime\prime})^{-1}&=\sum_i (\varepsilon\otimes S(e_i))\otimes (\tilde{e}^i\otimes 1)\in \cH(H)\otimes\cH(H)^{op}.
\end{align*}

Consider the Drinfeld double $D(H)$ of $H$ \cite{D}. Then there is an algebra homomorphism \cite{K, Sz}
$$\phi: D(H)\rightarrow \cH(H)\otimes \cH(H)^{op}$$
defined by
$$\phi(f\otimes x)=f_{(1)}(x_{(3)})f_{(3)}\otimes x_{(1)}\otimes (S^{-1})^*(f_{(2)})\otimes S(x_{(2)}).$$
Let $\mathcal{R}\in D(H)\otimes D(H)$ be the universal $R$-matrix and let $\hat{\mathcal{R}}=(\phi\otimes\phi)\mathcal{R}$. Since $\phi\otimes\phi$ is also an algebra homomorphism, the element $\hat{\mathcal{R}}$ satisfies the Yang-Baxter equation:
\begin{align}\label{eq:KashR}
\hat{\mathcal{R}}_{12}\hat{\mathcal{R}}_{13}\hat{\mathcal{R}}_{23}=\hat{\mathcal{R}}_{23}\hat{\mathcal{R}}_{13}\hat{\mathcal{R}}_{12}
\end{align}

Kashaev showed in \cite{K} that
\begin{align}\label{kash}
\hat{\mathcal{R}}=T^{\prime\prime}_{14}T_{13}\tilde{T}_{24}T^{\prime}_{23}.
\end{align}

In fact, this factorization \eqref{kash} means that the Yang-Baxter equation \eqref{eq:KashR} is a consequence of the following eight mixed-pentagon relations:
\begin{alignat}{2}
    T_{12}T_{13}T_{23}&=T_{23}T_{12},&\qquad \tilde{T}_{12}\tilde{T}_{23}&=\tilde{T}_{23}\tilde{T}_{13}\tilde{T}_{12},\label{ptg} \\
    T^{\prime}_{12}T^{\prime}_{13}T_{23}&=T_{23}T^{\prime}_{12}, &\qquad 
   \tilde{T}_{12}T^{\prime}_{23}&=T^{\prime}_{23}T^{\prime}_{13}\tilde{T}_{12},\\
    T_{12}T^{\prime\prime}_{13}T^{\prime\prime}_{23}&=T^{\prime\prime}_{23}T_{12}, &\qquad 
   T^{\prime\prime}_{12}\tilde{T}_{23}&=\tilde{T}_{23}T^{\prime\prime}_{13}T^{\prime\prime}_{12},\\
    T^{\prime}_{12}\tilde{T}_{13}T^{\prime\prime}_{23}&=T^{\prime\prime}_{23}T^{\prime}_{12}, &\qquad 
   T^{\prime\prime}_{12}T^{\prime}_{23}&=T^{\prime}_{23}T_{13}T^{\prime\prime}_{12}. \label{ptg'}
\end{alignat}

We now recall the definition of \emph{the Fock representation} $\cH(H)\rightarrow \mathrm{End}(H^*)$ given in \cite{MST} as 
$$(f\otimes a)g=f(a\rightharpoonup g)$$
for $(f\otimes a)\in \cH(H)$ and $g\in H^*$. In addition, we have \emph{the opposite-Fock representation} $\cH(H)^{op}\rightarrow \mathrm{End}(H)$ given by
$$(f\otimes a)x=f(x_{(1)})x_{(2)}a$$
for $(f\otimes a)\in \cH(H)^{op}$ and $x\in H.$ 

Applying these representations to the four canonical elements, we obtain the following operators 
\begin{alignat*}{2}
    T&:H^*\otimes H^*\rightarrow H^*\otimes H^*, &\qquad \tilde{T}&:H\otimes H\rightarrow H\otimes H, \\
    T^{\prime}&: H\otimes H^*\rightarrow H\otimes H^*\cong \mathrm{End}(H),&\qquad T^{\prime\prime}&:H^*\otimes H\rightarrow H^*\otimes H \cong \mathrm{End}(H) 
\end{alignat*}
which are explicitly described by
\begin{alignat*}{2}
    T&(f\otimes g)=f_{(1)}\otimes f_{(2)}g, &\qquad \tilde{T}&(x\otimes y)=xy_{(1)}\otimes y_{(2)},\\
    T^{\prime}&(x\otimes f) :y\mapsto f(y_{(2)})xS(y_{(1)}),&\qquad T^{\prime\prime}&(f\otimes x) :y\mapsto f(yS^{-1}(x_{(1)}))x_{(2)}
\end{alignat*}
for $x, y\in H$ and $f, g\in H^*$. 

In particular, the operator $\tilde{T}(x\otimes y)=xy_{(1)}\otimes y_{(2)}$ coincides with the Galois map (also called the canonical map) of the $H$-Hopf-Galois extension $\mathbb{K}\subset H$ \cite{DT}.

\subsection{Frobenius $2$-algebras and Heisenberg doubles}
Let $H=(H,m,\D,1,\varepsilon,S)$ be a finite-dimensional Hopf algebra with invertible antipode.

We consider the Heisenberg double $\cH(H)$ of $H$. Let $V$ and $U$ be representations of $\cH(H)$ and of $\cH(H)^{op}$, respectively. 

In this situation, the four canonical elements defined by \eqref{can}--\eqref{can'} give the operators
\begin{alignat*}{2}
    \al&=\tau \tilde{T}: U\otimes U\rightarrow U\otimes U,&\quad \bt&=\tau T:V\otimes V\rightarrow V\otimes V, \\
    \gm&=\tau T^{\prime}: U\otimes V\rightarrow V\otimes U, &\quad \brg&=\tau T^{\prime\prime}: V\otimes U\rightarrow U\otimes V,
\end{alignat*}
where $\tau$ denotes the flip map of tensor products. Then we have a Frobenius $2$-algebra $(*,U,V,\al,\bt,\gm,\brg)$ in the semistrict monoidal $2$-category $\cC=\Sigma \mathbf{Vect}_{\mathbb{K}}$ (where $\mathbf{Vect}_{\mathbb{K}}$ denotes the category of finite-dimensional vector spaces over $\mathbb{K}$).

In fact, the eight axioms \eqref{diag:1}--\eqref{diag:8} in this case are equivalent to the eight mixed-pentagon relations \eqref{ptg}--\eqref{ptg'} (see Lemma~\ref{lem}). Moreover, the $R$-matrix given by \eqref{eq:R} is related to the embedded $R$-matrix $\hat{\mathcal{R}}\in \left(\cH(H)\otimes \cH(H)^{op}\right)^2$ as follows.

\begin{prop}\label{ex:kash}
  If $R$ is the $R$-matrix of the Frobenius $2$-algebra $(*,U,V,\al,\bt,\gm,\brg)$ in $\Sigma \mathbf{Vect}_{\mathbb{K}}$, then we have
  \begin{equation}
    R=\tau_{X,X}\hat{\mathcal{R}}
  \end{equation}
  as operators on $X\otimes X$, where $X=V\otimes U$.
\end{prop}
\begin{proof}
From \eqref{kash}, it follows that
$$
R=\tau_{X,X}T^{\prime\prime}_{14}T_{13}\tilde{T}_{24}T^{\prime}_{23}=\tau_{X,X}\hat{\mathcal{R}},
$$
where the first equality follows from the following calculation.
\begin{center}
\iflatexml
\includegraphics{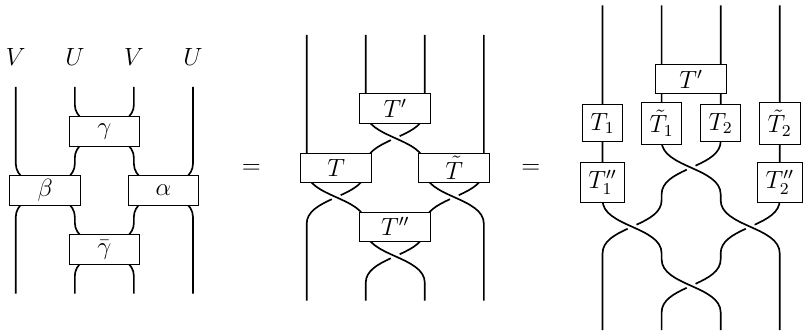}
\else
\includegraphics[trim={120bp 519bp 104bp 112bp},
  clip]{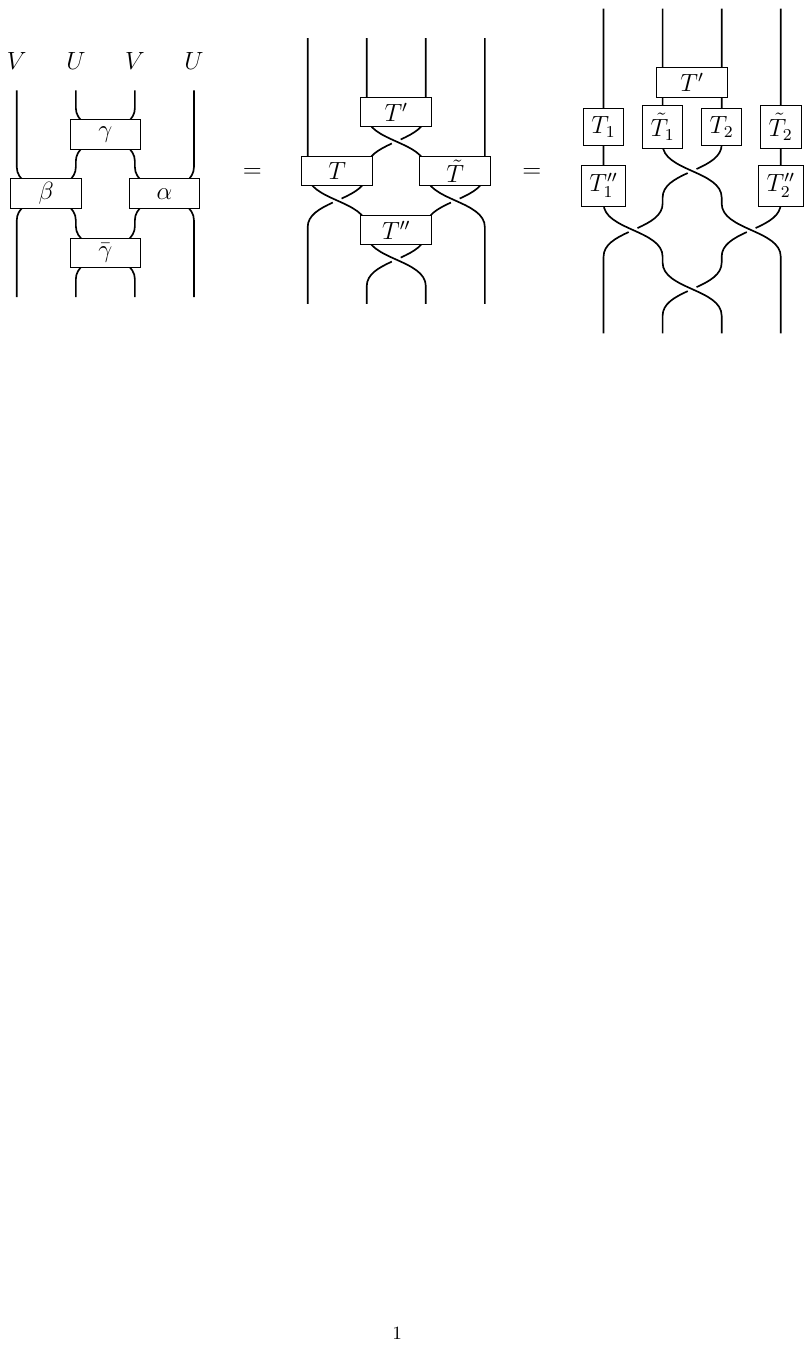}
\fi
\end{center}
\end{proof}

\begin{lem}\label{lem}
Let $V$, $W$ and $U$ be finite-dimensional vector spaces. Assume given linear mappings $F:V\otimes W\rightarrow V\otimes W$, $G:V\otimes U\rightarrow V\otimes U$ and $K:W\otimes U\rightarrow W\otimes U$.
\begin{enumerate}
  \item   The relation $$K_{23}G_{13}F_{12}=F_{12}K_{23}$$ holds if and only if
  \begin{center}
    \iflatexml
\includegraphics{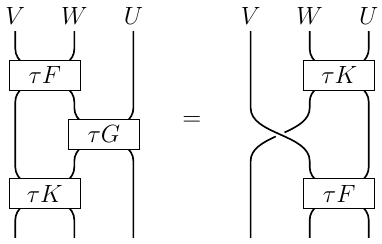}
\else
\includegraphics[
  trim={218bp 566bp 211bp 109bp},
  clip
]{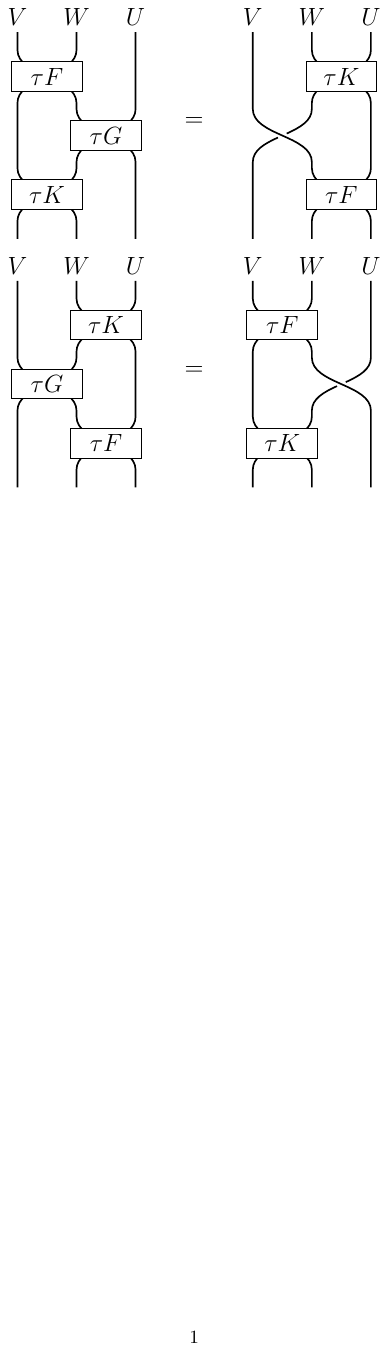}
\fi
  \end{center}
\item The relation $$F_{12}G_{13}K_{23}=K_{23}F_{12}$$ holds if and only if
\begin{center}
\iflatexml
\includegraphics{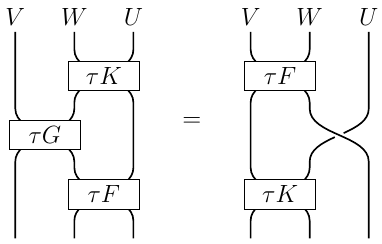}
\else
\includegraphics[
  trim={218bp 447bp 211bp 228bp},
  clip
]{lemma.pdf}
\fi
\end{center}
\end{enumerate}
\end{lem}

\begin{proof}
In the diagrams, we write $F=F_1\otimes F_2$, and similarly
for $G$ and $K$, with summation understood.
  \begin{enumerate}
    \item The following diagrammatic calculation proves the first claim.
    % 上段の等式
\iflatexml
\begin{center}
  \includegraphics{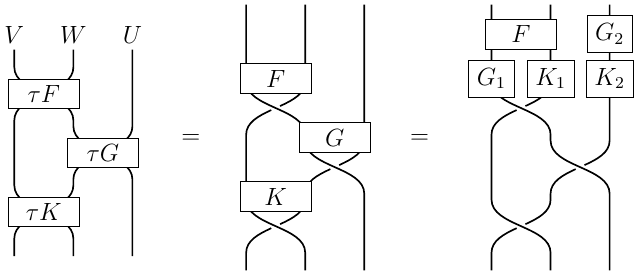}
\end{center}
\else
\[
  \includegraphics[
    page=1,
    trim=157bp 553bp 148bp 107bp,
    clip
  ]{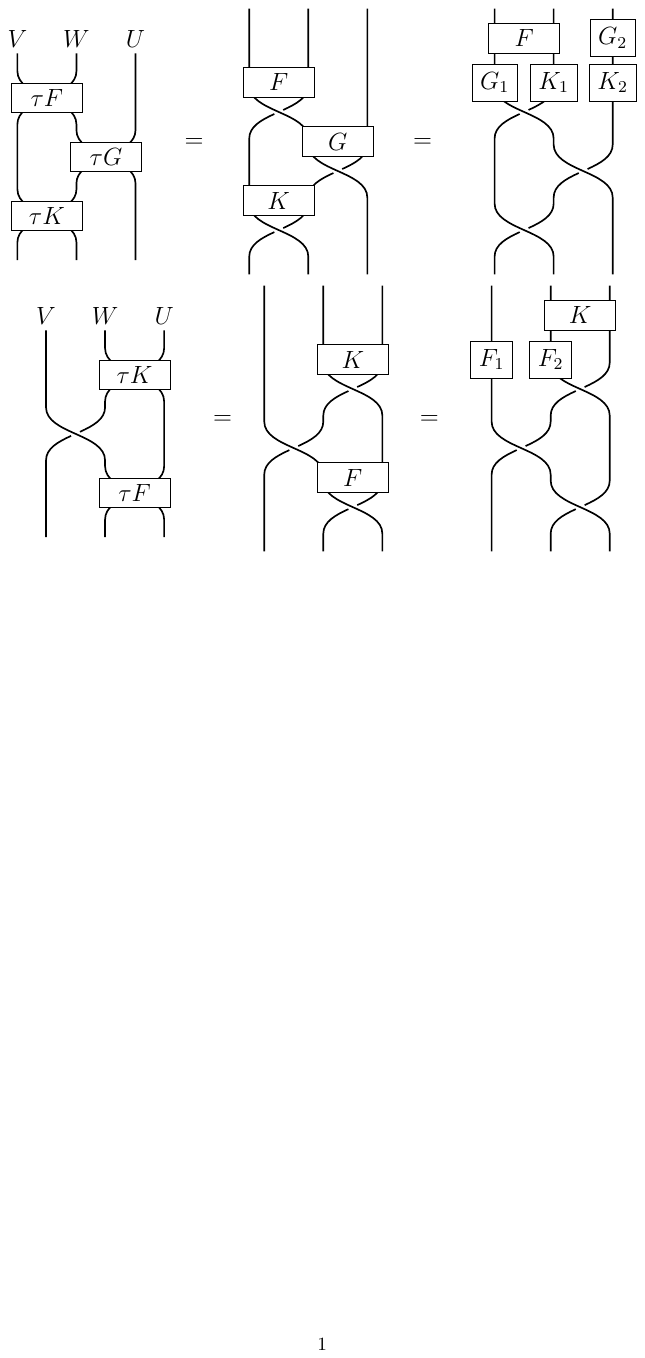}
\]
\fi

% 下段の等式
\iflatexml
\begin{center}
  \includegraphics{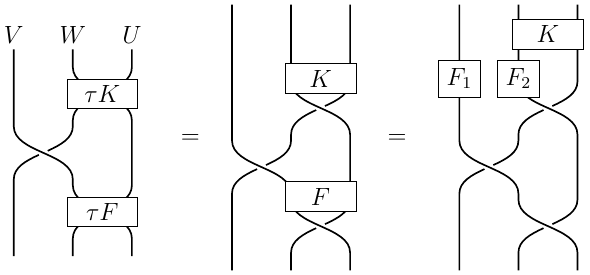}
\end{center}
\else
\[
  \includegraphics[
    page=1,
    trim=171bp 420bp 158bp 240bp,
    clip
  ]{picture.pdf}
\]
\fi
\item Similarly, we have the second claim.
  \end{enumerate}
\end{proof}

\end{document}